\documentclass[12pt,a4paper]{article}

\usepackage{graphicx}

\usepackage{amsmath}
\usepackage{amsfonts}
\usepackage{amssymb}             
\usepackage{amsxtra}             
\usepackage{amsbsy}              
\usepackage{natbib}
\usepackage[font=small,labelfont=bf,margin=2em]{caption}
\usepackage[makeroom]{cancel}

\allowdisplaybreaks

\newcommand{\asin}{\ensuremath{\mathrm{asin}}}

\newcommand{\acos}{\ensuremath{\mathrm{acos}}}
\newcommand{\atan}{\ensuremath{\mathrm{atan}}}

\newcommand{\const}{\ensuremath{\mathrm{const}}}

\DeclareMathAlphabet{\mathbfl}{OML}{cmm}{b}{it} 
\DeclareFontFamily{OT1}{pzc}{}
\DeclareFontShape{OT1}{pzc}{b}{it}{<-> s * [1.200] pzcmi7t}{}
\DeclareMathAlphabet{\mathopr}{OT1}{pzc}{b}{it}

\newcommand{\vect}[1]{\ensuremath{\mathbfl{ #1 }}}    
\newcommand{\vectg}[1]{\ensuremath{\boldsymbol{ #1}}} 
\newcommand{\tp}{\ensuremath{^\mathsf{T}}}

\newcommand{\DS}{\ensuremath{\displaystyle}}
\newcommand{\TS}{\ensuremath{\textstyle}}

\newcommand{\half}{\ensuremath{{\textstyle \!\frac{1}{2}}}}
\newcommand{\third}{\ensuremath{{\textstyle \!\frac{1}{3}}}}
\newcommand{\quart}{\ensuremath{{\textstyle \!\frac{1}{4}}}}
\newcommand{\fifth}{\ensuremath{{\textstyle \!\frac{1}{5}}}}
\newcommand{\sixth}{\ensuremath{{\textstyle \!\frac{1}{6}}}}

\newcommand{\grd}[1]{\ensuremath{\boldsymbol \nabla} #1}

\newcommand{\bra}{\ensuremath{\langle}}
\newcommand{\ket}{\ensuremath{\rangle}}

\newcommand{\Dpar}[2]{\ensuremath{\frac{\partial #1}{\partial #2}}}

\newcommand{\dpar}[2]{\ensuremath{\partial_{#2} #1\,}}

\newcommand{\lvecc}[2]{\ensuremath{( #1 , #2 )}}
\newcommand{\rvecc}[2]{\ensuremath{
              \left( \begin{array}{c} \!\!#1\!\! \\
                                      \!\!#2\!\! \end{array} \right) }}

\newcommand{\rveccc}[3]{\ensuremath{
               \left( \begin{array}{c} \!\!#1\!\! \\
                                       \!\!#2\!\! \\
                                       \!\!#3\!\! \end{array} \right) }}

\newcommand{\erg}{\ensuremath{w}}            
\newcommand{\pow}{\ensuremath{\Phi}}         
\newcommand{\rin}{\ensuremath{I}}            
\newcommand{\irr}{\ensuremath{Q}}            
\newcommand{\rad}{\ensuremath{L}}            
\newcommand{\zdis}{\ensuremath{\zeta}}
\newcommand{\dist}{\ensuremath{\ell}}

\renewcommand{\bra}{\ensuremath{<\!}}
\renewcommand{\ket}{\ensuremath{\!>}}
\renewcommand{\Re}{\ensuremath{\mathcal{R}}}
\renewcommand{\Im}{\ensuremath{\mathcal{I}}}
\newcommand{\1}{\ensuremath{\boldsymbol{1}}}

\newcommand{\uect}[1]{\ensuremath{\vect{\hat{#1}}}}
\newcommand{\uectg}[1]{\ensuremath{\vectg{\hat{#1}}}}

\newcommand{\inc}{\ensuremath{_\mathrm{in}}}
\newcommand{\sca}{\ensuremath{_\mathrm{sc}}}
\newcommand{\obs}{\ensuremath{_\mathrm{obs}}}
\newcommand{\ret}{\ensuremath{_\mathrm{ret}}}
\newcommand{\pol}{\ensuremath{_\mathrm{p}}}

\newcommand{\Fill}{\hspace*{\fill}}

\renewcommand{\theequation}{\arabic{section}.\arabic{equation}}

\DeclareMathAlphabet{\compf}{OT1}{cmss}{m}{n} 
\DeclareMathAlphabet{\vectf}{OT1}{cmss}{bx}{n} 

\begin{document}
\bibliographystyle{plainnat}

\title{Thomson Scattering in the Solar Corona}

\author{Bernd Inhester, Max-Planck-Institut f\"ur Sonnensystemforschung}
\date{\today}
\maketitle

The basis for the application of Thomson scattering to the analysis of
coronagraph images has been laid decades ago
\citep{Schuster:1879,Minnaert:1930,vandeHulst:1950}. Even though the basic
formulation is undebated, a discussion has grown in recent years about the
spatial distribution of Thomson scatter sensitivity in the corona and
the inner heliosphere. These notes are
an attempt to clarify the understanding of this topic.
\\
We reformulate the classical scattering calculations in a more transparent way
using modern SI-compatible quantities extended to field correlation matrices.
The resulting concise formulation is easily extended to the case of relativistic
electrons.
\\
For relativistic electrons we calculate the Stokes parameters of the scattered
radiation and determine changes in degree and orientation of its polarisation,
blue-shift and radiant intensities depending on the electron velocity
magnitude and direction.
We discuss the probability to see these relativistic effects in white-light
coronagraph observations of the solar corona.
\\
Many mathematical and some basic physical ingredients
are made explicit in several chapters of the appendix.

\section{Introduction -- a brief view on history}

The observation of the polarisation of the solar coronal brightness are among
the earliest manifestations of Thomson scattering. In fact, the first
observations and part of their correct interpretation were made decades before
Thomson scattering and even the electron were known.

The first successful observation seem to have been made by
Fran\c{c}ois Arago in southern France on the occasion of the 1842 eclipse
\citep{Harvey:2014}. His brief observation
was followed by a number of other reports from researchers observing
at subsequent eclipses
\footnote{A compilation of these early observations can be found in
  \citep{Ranyard:1879}. In fact, F. Arago gave the first report of
  the polarisation of coronal light. The Italian and polish
  astronomers Pietro Secchi and
  Adam Pra\`zmowski were among the first to determine the
  correct orientation of the polarisation from their 1860 eclipse
  observations. But all reports were qualitative so far.
  G.K. Winter (1871) seems to have been the first to measure the
  degree of polarisation quantitatively.  It is his observation which
  Schuster (1979) refers to in his theoretical explanation.}.
These observations were interpreted by \cite{Schuster:1879} in terms of Sun
light scattered at small particles in the solar corona. Schuster's work was
very much inspired by prior calculations of \cite{Rayleigh:1871} on the
scattering of Sun light by particles in the Earth's atmosphere to explain the
polarisation of the sky brightness.

At least for the corona, there was no idea at the time as to which particles 
were responsible for the scattering.
Therefore Schuster simply adopted the differential scattering
cross section derived by Rayleigh for scattering
sources much smaller in size than the wavelength of the scattered light. Today
we know that this scattering cross section applies much better to the corona
than to the Earth's atmosphere for which it was first derived. Since the
electron was unknown at the time, the magnitude of the cross section as well as
the number density of coronal scatterers were unknown. However, Schuster derived
the ratio of the polarisation in directions tangential and radial to the Sun's
centre for which the absolute cross section is not required. The ratio he
calculated for various distances $r$ from the Sun centre agreed with the poor
observations known at the time.
In fact, the integrals $C(r)$ and $C(r)-A(r)$ as they are called
today, directly go back to Schuster's paper.

It took another 23 years until Thomson proposed the existence of the electron
from cathode ray experiments in 1896 and further eleven years to formulate
what we know today as Thomson scattering \citep{Thomson:1907}. As coronal
polarisation observations became more precise, it became evident that
Schuster's calculations had to be refined. In 1930, \cite{Minnaert:1930}
extended them by taking the solar limb darkening into account. This involved
two more integrals, termed $D(r)$ and $D(r)-B(r)$ by Minnaert. This notation
has been popularised by \citep{Billings:1966} and is still used today in
coronal physics.

For a long time, coronal brightness observations were one of the few
confirmations of Thomson's scattering theory. It was not before 1958 that
ionospheric scattering experiments with radio waves by \cite{Bowles:1958}
provided another verification. However his measurements revealed unexpected
spectral details which could be explained only a few years later.
At a wavelength of the scattered wave larger than the plasma Debye length the
scattered signal is spectrally modified by the collective plasma response to
its own thermal fluctuations \citep[e.g.,][]{Hutchinson:2002}.
In the corona, the Debye length is typically a few cm, 
much larger than an optical wavelength, and similar effects do not
occur in coronagraphy.
Active laboratory experiments of Thomson scattering had to wait
for the invention of the laser. They were first reported by
\cite{FioccoThompson:1963}.

A relatively new aspect of Thomson scattering in the corona is the
contribution from relativistic electrons. Even though the topic was first
raised already decades ago by \cite{Molodensky:1973}, it received little
attention so far. To compensate for this deficiency, this review devotes a
relatively large part to this topic. For the solar corona its effect may
be marginal except for seldom events when the corona is locally extremely
heated and energised during strong flares. In these cases, however, 
observed anomalies of the scattering signal may give hints on
the local electron velocity distribution.

Thomson and Compton scattering is also relevant in laboratory plasmas where it
is used as an important diagnostic tool \citep{Hutchinson:2002}. It also
occurs in astrophysical objects of all sizes. Since they are often
unresolved, the spectral and polarimetric characteristics of the observed
light yields important additional information about these objects. Examples
are protoplanetary disks of young stars
\citep{WoodEtal:1993,BrownWood:1994,VinkEtal:2002,Oudmaijer2007}, to accretion
disks around active galactic nuclei
\citep{SunyaevTitarchuk:1985,AntonnucciMiller:1985,WolfHenning:1999}, and
supernova clouds \citep{WangWheeler:2008,Hoffman:2014}.

Almost 80 years after Minnaert's refinements of the
Thomson scattering formulae for the solar corona,
the instruments employed
in coronagraphy have again undergone considerable further improvements
and time may have come to look for more details in the data which
are not included in the classical theory.
Also, the sensitivity of conventional Thomson scattering for viewing
geometries which deviate largely from conventional Earth-bound and
small field-of-view conditions have become an issue recently 
\citep[e.g.,][]{VourlidasHoward:2006,HowardDeforest:2012,DeforestHoward:2013}.
The space craft of the STEREO mission and also of the future SOLAR ORBITER and
SOLAR PROBE missions are all equipped with ordinary and partly with wide-angle
coronagraphs \citep{HowardStereo:2008} and provide or will provide views
onto the solar corona with quite different fields-of-view, from
different perspectives and closer distances from the centre of the
Sun compared to conventional, Earth-bound observations.

The scattering calculations are sometimes not easy to visualise due to their
geometric complexity. Therefore even modern reviews of the topic follow the
original approach of \citep{Schuster:1879,Minnaert:1930} when rederiving the
Thomson scattering response from the corona. In this paper we attempt a more
modern and hopefully more transparent approach which may more easily be
extended to more complex situations when the surface radiance from the Sun (or
a star) is more involved. For example, Sun spots may matter when Thomson
scattering is observed closely above the solar limb and also for star coronae
above huge star spots or with embedded polarised light sources.

There is sometimes confusion about the relevant physical terms
needed to describe photon fluxes and quantities derived from them.
We will use the official SI radiometric terms to give our calculations a sound
physical basis.
The SI quantities differ slightly from the quantities commonly
used in astrophysics, but they are favourable here because they have
systematic relativistic transformations.
For readers which are not familiar with the SI quantities, we
explain the relevant terms in an initial chapter. The next chapter provides a
rederivation of the classical scattering expressions. In chapter 3 we evaluate
them in line-of-sight integrals over some simplified but instructive coronal
density distributions. The fourth chapter extends the classical
calculations to relativistic electrons and presents the major deviations
in polarisation degree and orientation, intensity and frequency compared
to the classical non-relativistic case.
All mathematical derivations are detailed in the appendix starting from
textbook level. This hopefully enables the interested reader to follow
all calculations. Readers who find the appendix helpful, might also want
to consult \cite{SaitoEtal:1970} for extended calculations on
classical coronal Thomson scattering and the introduction by
\cite{Prunty:2014} on scattering at relativistic electrons in the lab.

\section{Radiation basics}
\setcounter{equation}{0}

Given the electric field of a directed monochromatic wave
\[
\vect{E}_\vect{k}(\vect{r},t)
=\Re[\vect{Z}_\vect{k}\,e^{i(\vect{k}\tp\vect{r}-ckt)}]
\]
the mean wave energy density (including the wave magnetic field) and the
mean Poynting energy flux in the direction $\vect{\hat{k}}$ of the
wave propagation are
\begin{equation}
  \begin{aligned}
  W_\vect{k}
  =& \epsilon_0 \bra\vect{E}_\vect{k}\tp(\vect{r},t)
                    \vect{E}_\vect{k}(\vect{r},t)\ket
  = \frac{\epsilon_0}{2} \vect{Z}_\vect{k}\tp
                         \vect{Z}_\vect{k}^*
   &[\mathrm{J/m^3}]
\\
  \vect{S}_\vect{k}
  =& c\,W(\vect{k}) \,\vect{\hat{k}}
   &[\mathrm{W/m^2}]
  \end{aligned}
\label{ergdensity_discr}\end{equation}
where $\epsilon_0$ is the vacuum dielectricity and $c$ the speed of light.
The average $\bra\dots\ket$ is over the wave phase and introduces an
additional factor 1/2 if the squared wave electric field is replaced by the
squared wave amplitude.

Real wave fields have a finite directional and spectral width. The spectral
distribution of the wave power does not matter much for our purposes here
because Thomson scattering is wavelength independent over a wide range of
wavelengths and in addition coronagraphs integrate over a wide 
wavelength range. We will therefore often ignore the magnitude of 
the wave vector $\vect{k}$.

However we have to be concerned about the directional distribution of the
Poynting flux. A real wave field can be thought of as being made up of many
wave packets of different wave vectors $\vect{k}$. In this case we have to replace
$\bra\vect{E}_\vect{k}\tp\vect{E}_\vect{k}\ket$ above by the power spectral
density in the sense of Wiener-Khinchine \citep[see also
 appendix~\ref{App:Wiener-Kintchine}]{Papoulis:1981}. We
define
\begin{gather*}
  \vect{\tilde{E}}_{V(\vect{r})}(\vect{k},t)
 =e^{ickt}\int_{V(\vect{r})} \vect{E}(\vect{r}',t)
                e^{i\vect{k}\tp\vect{r}'}\;d^3\vect{r}'   
\end{gather*}
as a tapered Fourier transform with the taper window centred at $\vect{r}$ and
with edge lengths larger than the electric field correlation length.
Then the expectation value of the power spectral density
is given by the Fourier transform of the spatial correlation
$R(\Delta\vect{r},t)$
\begin{gather}
   w(\vect{k})=\epsilon_0
   \lim_{V\rightarrow\infty}\frac{1}{V}\bra\vect{\tilde{E}}\tp_V(\vect{k},t)
       \vect{\tilde{E}}^*_V(\vect{k},t)\ket\;
  = \epsilon_0 \int \!R(\Delta\vect{r},t)\;e^{-i\vect{k}\tp
               \Delta\vect{r}}\;d^3 \Delta\vect{r}
\qquad[\mathrm{J}]
\label{PowSpDensity}\\
\text{where}\quad R(\Delta\vect{r},dt)
=\;\bra\vect{E}\tp\!(\vect{r},t)\vect{E}(\vect{r}+\Delta\vect{r},t)\ket
\nonumber\end{gather}
Due to random correlations, the expectation value
$\bra\vect{\tilde{E}}_{V}\tp\vect{\tilde{E}}^{*}_{V}\ket$ does for a large
window size not increase with $V^2$ but only proportional to $V$ so that
the limit is well defined. The spatial power spectral
density of the electric wave field can be used for a more general definition
of the spectral densities of energy and Poynting flux compared to
(\ref{ergdensity_discr})
\begin{gather}
  \begin{aligned}
  \erg(\vect{k})\,d^3\vect{k}
  =& \epsilon_0 \lim_{V\rightarrow\infty} \frac{1}{V}\bra\vect{\tilde{E}}_V\tp(\vect{k})
                   \vect{\tilde{E}}^*_V(\vect{k})\ket d^3\vect{k}
   &[\mathrm{J/m^3}]
  \\
  \vect{s}(\vect{k})\,d^3\vect{k}
  =& c\,\erg(\vect{k})\,\vect{\hat{k}}\,d^3\vect{k}
   &[\mathrm{W/m^2}]
  \end{aligned}
\label{ergdensity_cont}\end{gather}
Note the expectation value $\bra\dots\ket$ implies a time
averaging over many wave periods $2\pi/ck$.

The magnitude of the Poynting flux at a given wave vector $\vect{k}$
is the spectral radiance
\[
\rad_\mathrm{spec}(\vect{k})=c\,\erg(\vect{k})
  \qquad[\mathrm{W/m^2/nm^{-3}}]
\]
White-light coronagraphs integrate over a wide spectral range so that only the
ordinary radiance matters which is, however, still selective to the direction
$\vect{\hat{k}}$. Using $d^3\vect{k}=k^2\,dk\,d\Omega(\vect{\hat{k}})$
and choosing appropriate $k$-integration bounds (depending on wavelength
passband $\Delta k$ of the instrument) we have for the relevant radiance
\begin{equation}
 \rad(\vect{\hat{k}})
=c\int_{\Delta k} \erg(k\vect{\hat{k}})\, k^2\,dk
= \int_{\Delta k} |\vect{s}(k\vect{\hat{k}})|\, k^2\,dk
  \qquad[\mathrm{W/m^2/sr}]
\label{radiance}\end{equation}
which collects all photons within the passband in direction $\vect{\hat{k}}$.
More intuitively, the radiance and derived
quantities are often expressed by the respective photon flux density.
It is obtained after dividing $\rad$
by the photon energy $ck\hbar$, e.g.,
\begin{equation*}
\rad_\mathrm{phot}(\vect{\hat{k}})
=\frac{1}{\hbar}\int \erg(k\vect{\hat{k}})\, k\,dk
  \qquad[\mathrm{photons/s/m^2/sr}]
\end{equation*}
Integrating (\ref{radiance}) over the 
directions of the relevant solid angle $Omega$ yields the local 
irradiance
\begin{gather}
\irr = \int_{\Omega}\rad(\vect{\hat{k}}) d\Omega(\vect{\hat{k}})
    =c \int_{\Omega}\int_{\Delta k}
          \erg(k\vect{\hat{k}})\, k^2\,dk\;d\Omega(\vect{\hat{k}})
     =c \int \erg(\vect{k})\, d^3\vect{k}
\qquad[\mathrm{W/m^2}]
\label{irradiance}\\
  =c\epsilon_0 R(0) =c\epsilon_0 \bra\vect{E}\tp\!(\vect{r},t)\vect{E}(\vect{r},t)\ket
\label{Irr_Esq}\end{gather}
where $d\Omega(\vect{\hat{k}})$ is the solid angle
element around the flux direction $\vect{\hat{k}}$.
In the last line we assumed that there is no relevant light emission 
outside of the wavelength passband $\Delta k$ and the solid angle 
$\Omega$ so that we could extend the integration boundaries to 
$\pm\infty$ and $4\pi$, respectively.
The last line then follows from (\ref{PowSpDensity}) and 
constitutes a version of Parseval's theorem.
It is important to keep in mind that the irradiance characterises
the local field fluctuations irrespective of the propagation direction of
the light which causes the fluctuations.
Obviously, $\irr$ is the trace of a more general correlation matrix
\[
 \vect{Q}=c\epsilon_0\bra\vect{E}(\vect{r},t)\vect{E}\tp(\vect{r},t)\ket
\]
which will become relevant in our scattering calculations below because
it retains to some extend the polarisation and propagation information of the
wave field.


The above quantities only depend on the local photon field fluctuations.
To assess transport of energy by photons, in particular in association with a
measurement process, we need in addition quantities which reference an area
element or its normal direction.
The power of the photon wave field emitted from (or received on) a surface
element $d\vect{A}$, e.g., the area of an emitting surface element or the
aperture area of a detecting instrument is obtained by a similar angular
integration of the radiance (\ref{radiance}) as in (\ref{irradiance}),
however, weighted with the projection $\vect{\hat{k}}\tp d\vect{A}$ of
the area element into direction $\vect{\hat{k}}$.
This way we obtain the radiant flux through the area element
\begin{equation}
 d\pow = \int_\Omega
      \rad(\vect{\hat{k}}) \;\vect{\hat{k}}\tp d\vect{A}
      \;d\Omega(\vect{\hat{k}})
    = \int_\Omega
      \rad(\vect{\hat{k}}) \,\cos\zdis \,d\Omega(\vect{\hat{k}})\;dA 
   \qquad[\mathrm{W}]
\label{power}\end{equation}
Here $\zdis$ is the angle between $\vect{\hat{k}}$ and the
normal of $d\vect{A}$ . If $d\vect{A}$ is the area of a detector pixel, 
$d\pow$ is the power received by the pixel.
For an emitting surface, $d\pow$ is the power radiated from the
area element.
An emitting surface is Lambertian when $\rad(\vect{\hat{k}})$ is a constant for
all emission directions $\vect{\hat{k}}$. More generally, $\rad$ often depends
only on the angle $\zdis$ with respect to the surface normal. In these cases
$\rad$ can be expanded in powers of $\cos\zdis$. We will therefore often
replace the argument $\vect{\hat{k}}$ of $\rad$ by $\cos\zdis$.
For a receiving instrument the effective radiance on the detector surface might
include a $\cos\zeta$ dependence due to vignetting.
The quantity $d\pow/dA$ is the net photon flux density along the normal
direction of the area element. It has the same units [W/m$^2$] as the
irradiance but is physically different: while $Q$ cannot become negative,
$d\pow/dA$ can if the area normal is reversed.

While $d\pow$ integrates the radiation of all directions, the integrand
in (\ref{power})
\begin{equation}
d\rin(\vect{\hat{k}})
=\frac{d\pow}{d\Omega(\vect{\hat{k}})}
=\rad(\vect{\hat{k}})\;\vect{\hat{k}}\tp d\vect{A}
\qquad[\mathrm{W/sr}]
\label{rin}\end{equation}
represents the radiant intensity into a specific direction $\vect{\hat{k}}$.
It vanishes for directions normal to the emitting or receiving
surface $d\vect{A}$.

\begin{figure}[b]
  \Fill
  \includegraphics[viewport=50 30 680 470,clip,width=7.5cm]
  {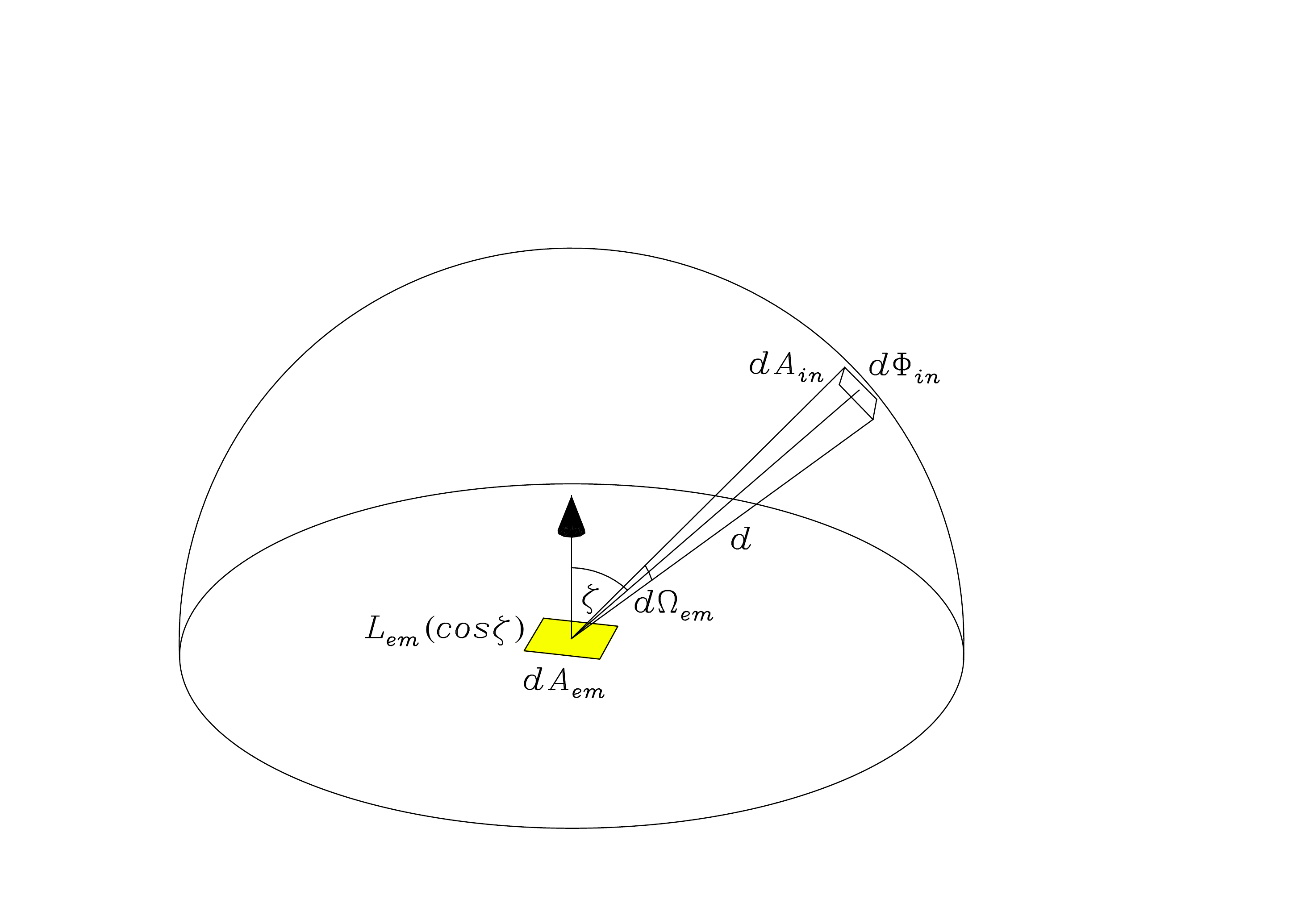}
  \includegraphics[viewport=50 30 680 470,clip,width=7.5cm]
  {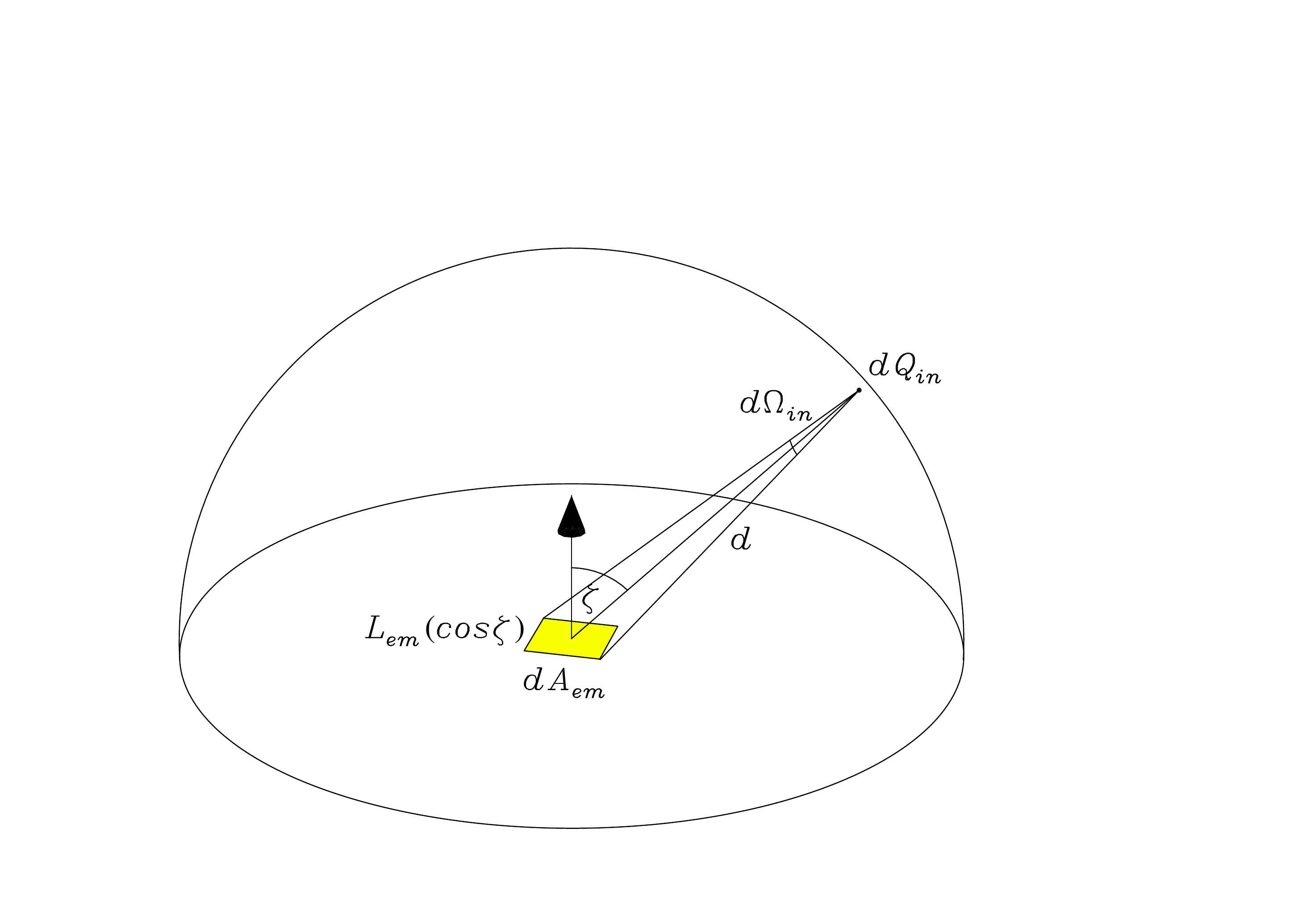}
  \Fill
  \caption{Sketch to illustrate the difference between emitted radiance $\rad_\mathrm{em}$,
    incident irradiance $d\irr\inc$ and incident power $d\pow\inc$. 
    \label{Fig:Radiometry}}
\end{figure}

\begin{table}
\Fill
\begin{tabular}{llll}
name & variable in astrophysics & SI equivalent & SI unit\\  
intensity, brightness
& $I(\uect{k})$
& $L(\uect{k})$
& W m$^{-2}$sr$^{-1}$ \\
energy density
& $U=\frac{1}{c}\int_{4\pi} I(\uect{k}) d\Omega$
& $\frac{1}{c}Q$
& Ws m$^{-3}$\\
flux in direction $\uect{a}$
& $F(\uect{a}) =  \int_{4\pi} I(\uect{k})\; \uect{k}\tp\uect{a} \;d\Omega$
& $\frac{d\Phi}{dA}$ & W m$^{-2}$ \\
\end{tabular}
\Fill
\caption{Table of commonly used radiometric quantities in astrophysics
  and their SI-compatible counterpart. All astrophysics quantities are
  often used in their spectral variant, i.e., per frequency or wave length
  interval. The direction $\uect{a}$ in the flux definition is the normal
  of the area element $d\vect{A}$ in the SI definition of $\pow$.}
\end{table}

So far all quantities and their relations were local. Of interest is the
situation were we have an emitting and an incident side.
We illustrate the relations between either side by two examples. The
quantities are
marked by subscripts ``em'' and ``in'' depending on which site they relate to.
Assume as in Fig.~\ref{Fig:Radiometry} that a surface $dA_\mathrm{em}$ is
emitting into the entire half space $\Omega_\mathrm{em}=2\pi$. A small part of
this flux is incident on an instrument at distance $d$ in direction
$\vect{\hat{k}}$ with aperture area $d\vect{A}_\mathrm{in}$ pointing exactly
in direction $-\vect{\hat{k}}$. Then the power collected by the instrument is
confined to the infinitesimal solid emission angle
$d\Omega(\vect{\hat{k}})=d\Omega_\mathrm{em}= dA_\mathrm{in}/d^2$. We have
according to (\ref{power})
\begin{align}
 d\pow_\mathrm{em} 
    &= \rad_{\mathrm{em}}(\vect{\hat{k}}) \,\cos\zdis\,dA_\mathrm{em}
   \;d\Omega_\mathrm{em}(\vect{\hat{k}})
 &\text{flux emitted into}\; d\Omega_\mathrm{em}
\label{dPow_em}\\
=d\pow_\mathrm{in} 
    &= \rad_{\mathrm{em}}(\vect{\hat{k}}) \,\cos\zdis
       \frac{dA_\mathrm{em}}{d^2}\,dA_\mathrm{in} 
     = \rad_\mathrm{em}(\vect{\hat{k}}) \,d\Omega_\mathrm{in} dA_\mathrm{in}
 &\text{flux collected from}\;d\Omega_\mathrm{in}
\label{dPow_in}\end{align}
where $d\Omega_\mathrm{in}=\cos\zdis\,dA_\mathrm{em}/d^2$ is the solid angle
subtended by the emitting surface at the observing instrument.
\begin{figure}[t]
  \Fill
  \parbox{7.5cm}{\includegraphics[viewport=50 30 680 470,clip,width=7.5cm]
  {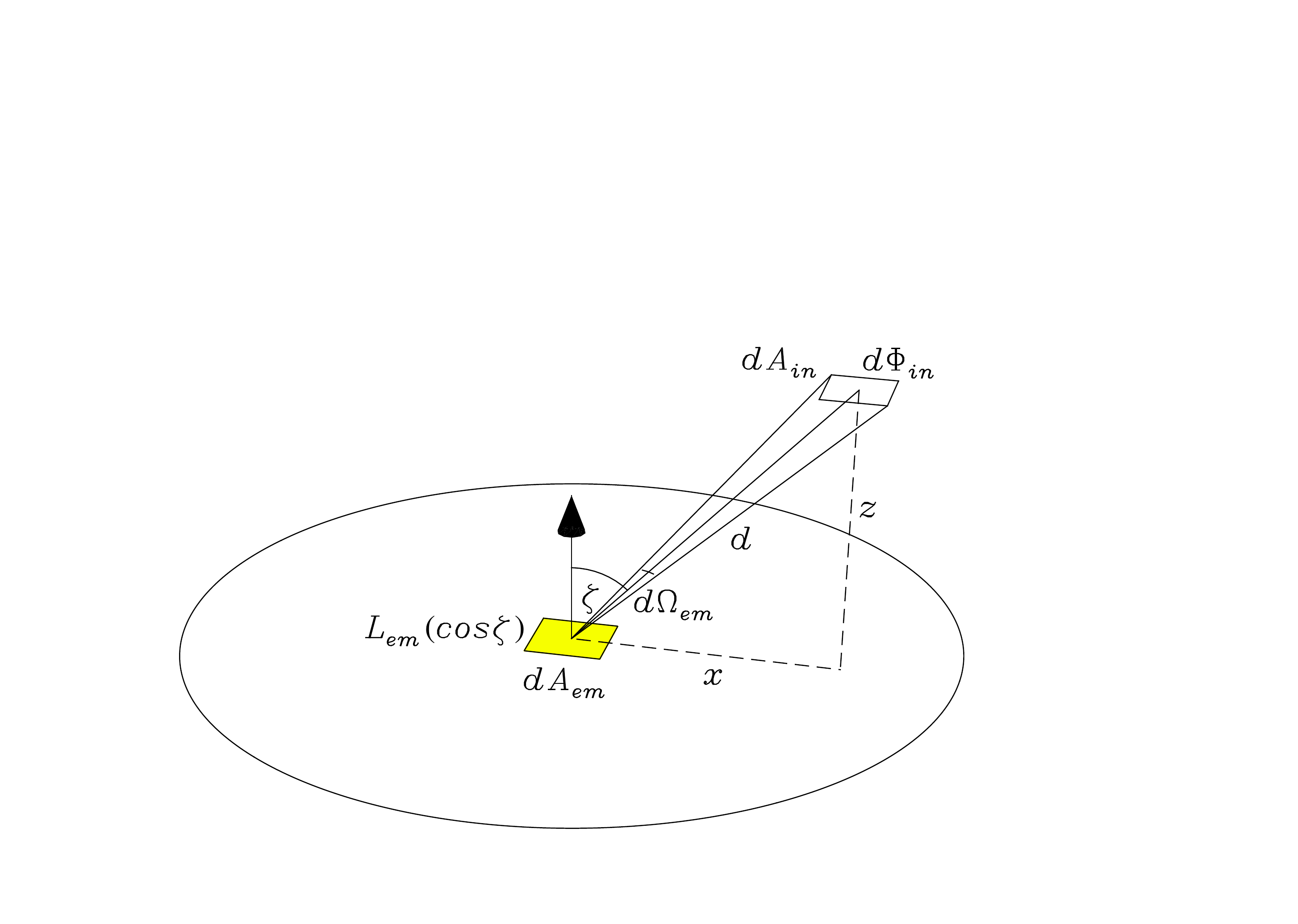}}
  \parbox{7.5cm}{\caption{Sketch to illustrate $\cos^4\zdis$ dependence
    of $d\pow_\mathrm{in}$ when $d\vect{A}_\mathrm{em}$ and
    $d\vect{A}_\mathrm{in}$ are parallel and their distance $z$ constant.
    \label{Fig:cos4law}}}
  \Fill
\end{figure}
If instead we point the detector area $d\vect{A}_\mathrm{in}$ vertically
parallel to $-d\vect{A}_\mathrm{em}$ (see Fig.~\ref{Fig:cos4law}), we have a
reduced effective emission angle $d\Omega_\mathrm{em}=\vect{\hat{k}}\tp
d\vect{A}_\mathrm{in}/d^2 =\cos\zdis\,dA_\mathrm{in}/d^2$. If in addition we
keep the vertical distance between emitter and receiver constant at a height
$z$, their mutual distance becomes $d=z/\cos\zdis$. Altogether, the received
power will in this setup be $d\pow_\mathrm{in}= \rad_{\mathrm{em}}
dA_\mathrm{em}dA_\mathrm{in}\cos^2\zdis /d^2 =\rad_{\mathrm{em}}
dA_\mathrm{em}dA_\mathrm{in}\cos^4\zdis /z^2$ instead of (\ref{dPow_in}).
The $\cos^4\zdis$ dependence represents the ideal vignetting for a
pinhole camera.
For large surfaces $dA_\mathrm{em}$ and imaging
instruments, $d\Omega_\mathrm{in}$ is often constrained by the solid angle of
the instrument resolution rather than by the size of $dA_\mathrm{em}$. In this case,
$d\Omega_\mathrm{in}=A_\mathrm{pixel}/f^2$ depends on the focal length $f$
and the pixel area $A_\mathrm{pixel}$ of
the instrument rather than on distance $d$ and the received power
$d\pow_\mathrm{in}$ per pixel does not change with distance form the
emitting surface. From (\ref{dPow_in}), the ratio
\begin{equation}
  d\irr_\mathrm{in}=\frac{d\pow_\mathrm{in}}{dA_\mathrm{in}}
 = \rad_{\mathrm{em}}(\vect{\hat{k}}) \,\cos\zdis \;\frac{dA_\mathrm{em}}{d^2}
 = \rad_{\mathrm{em}}(\vect{\hat{k}}) d\Omega_\mathrm{in}
   \qquad[\mathrm{W/m^2}]
\label{dIrr_radiance}
\end{equation}
is the irradiance produced by photons from the small cone
$d\Omega_\mathrm{em}=dA_\mathrm{em}/d^2$ around $\vect{\hat{k}}$
at the site of the instrument.
It is the relevant incident irradiance for a scattering particle
at the location of the instrument in our example.

As a second example, consider the emitting surface replaced by a point source.
Now, the emitting radiance $\rad_\mathrm{em}$ is not suitable any more to
describe the source. However, we expect that we measure a radiant flux
in direction $\uect{k}$
proportional to the solid angle $d\Omega_\mathrm{em}
=dA_\mathrm{in}/d^2$ from photons which propagate inside this solid angle and
hit the detector surface $dA_\mathrm{in}$ at distance $d$.
Again we assume that the receiving aperture area $d\vect{A}_\mathrm{in}$
points exactly to $-\vect{\hat{k}}$ in direction to the source.
We can therefore define a radiant intensity (\ref{rin}) of
\begin{equation}
 \rin_\mathrm{in}(\vect{\hat{k}})
=\frac{d\pow_\mathrm{in}}{d\Omega_\mathrm{em}}
=\frac{d\pow_\mathrm{in}}{dA_\mathrm{in}}\;d^2
= \irr_\mathrm{in}\;d^2
  \qquad[\mathrm{W/sr}]
\label{dRin_re}\end{equation}
which  characterises the directional emission pattern of the point
source. The radiant intensity is the relevant quantity to
describe the far field of a scattering particle.

At first sight, there seems to be an inconsistency of units in (\ref{dRin_re})
because irradiance $Q$ has units of [W/m$^2$]. The problem can be traced
back to the fact that $dA_\mathrm{em}/d^2$ was considered to be a solid
angle. Since $dA_\mathrm{em}$ is an area, $d^2$ formally must have the units
[m$^2$/sr]. This unusual interpretation must be kept in mind when an area
and the solid angle it subtends in some distant centre are compared.

The entire emitted power from the point source is
\[
 \pow_\mathrm{em}
=\pow_\mathrm{in}
=\int \rin_\mathrm{in}(\vect{\hat{k}}) d\Omega_\mathrm{em}(\vect{\hat{k}})
\]
An isotropic source with $\rin_\mathrm{in}=\mathrm{const}$ emits a power
$\pow_\mathrm{em}=4\pi \rin_\mathrm{in}$ and produces a local irradiance 
according to (\ref{dRin_re}) of $\irr_\mathrm{in}=\pow_\mathrm{em}/4\pi d^2$.


\section{Electrons at rest -- classical Thomson scattering in the corona}
\setcounter{equation}{0}

In the following we will apply the above expressions to the scattering of Sun
light at coronal electrons. The emitting area element will be extended to the
visible solar surface, the receiving area element is replaced by the scattering
electron. We will switch to a spherical coordinate system with its origin
in the scattering site $\vect{r}$ at distant $r$ from the solar centre and 
its zenith axis aligned with the radial direction from the Sun.
Variable $d$ will continue to be the distance from the emitting surface
element to the scattering site but because of the spherical
symmetry, $\pow_\mathrm{in}$ and $\irr_\mathrm{in}$ will from now on depend on
the distance $r$.

\subsection{Irradiance of Sun light}

\begin{figure}[b]
\Fill
\includegraphics[width=6cm]{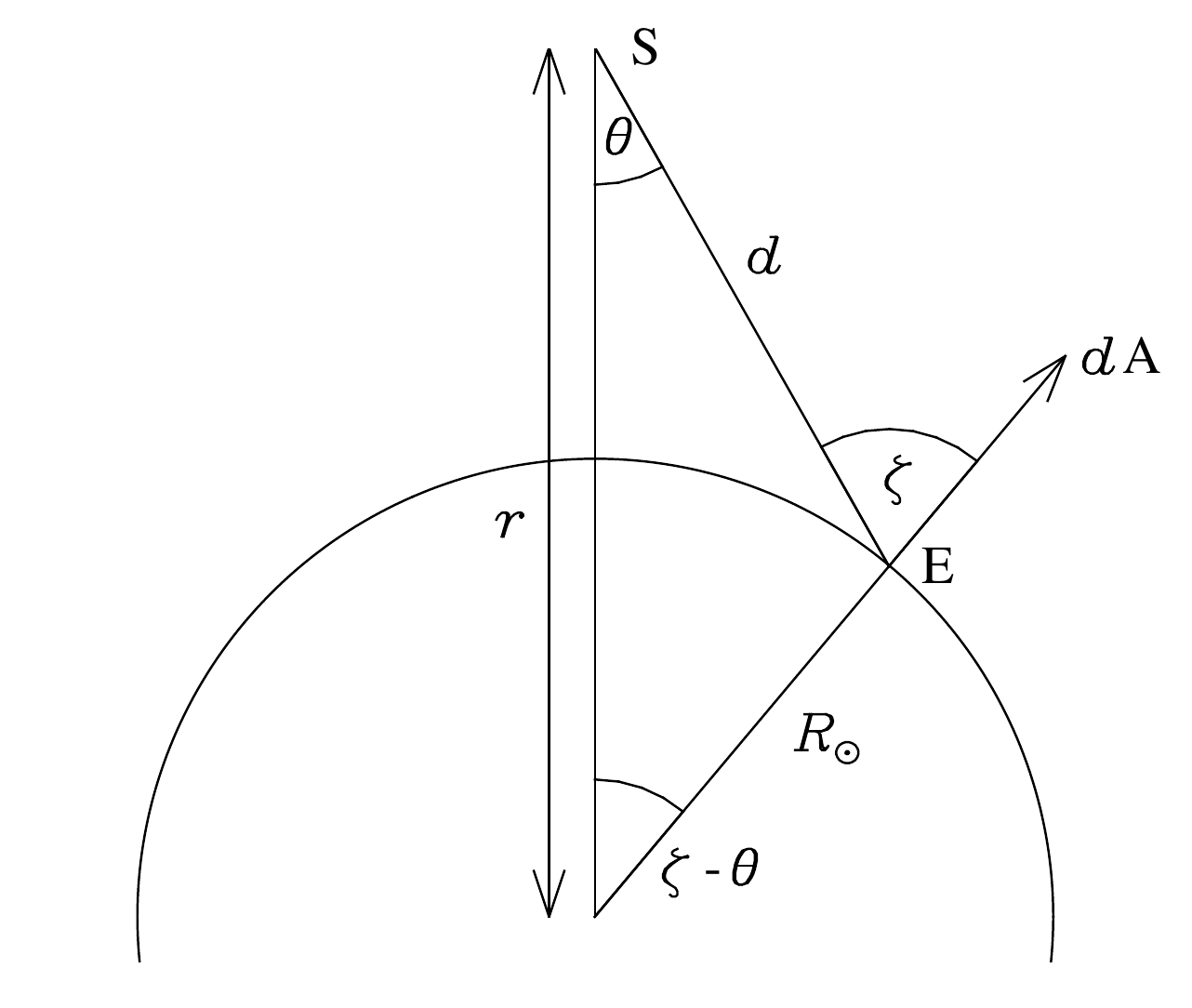}
\includegraphics[width=6cm]{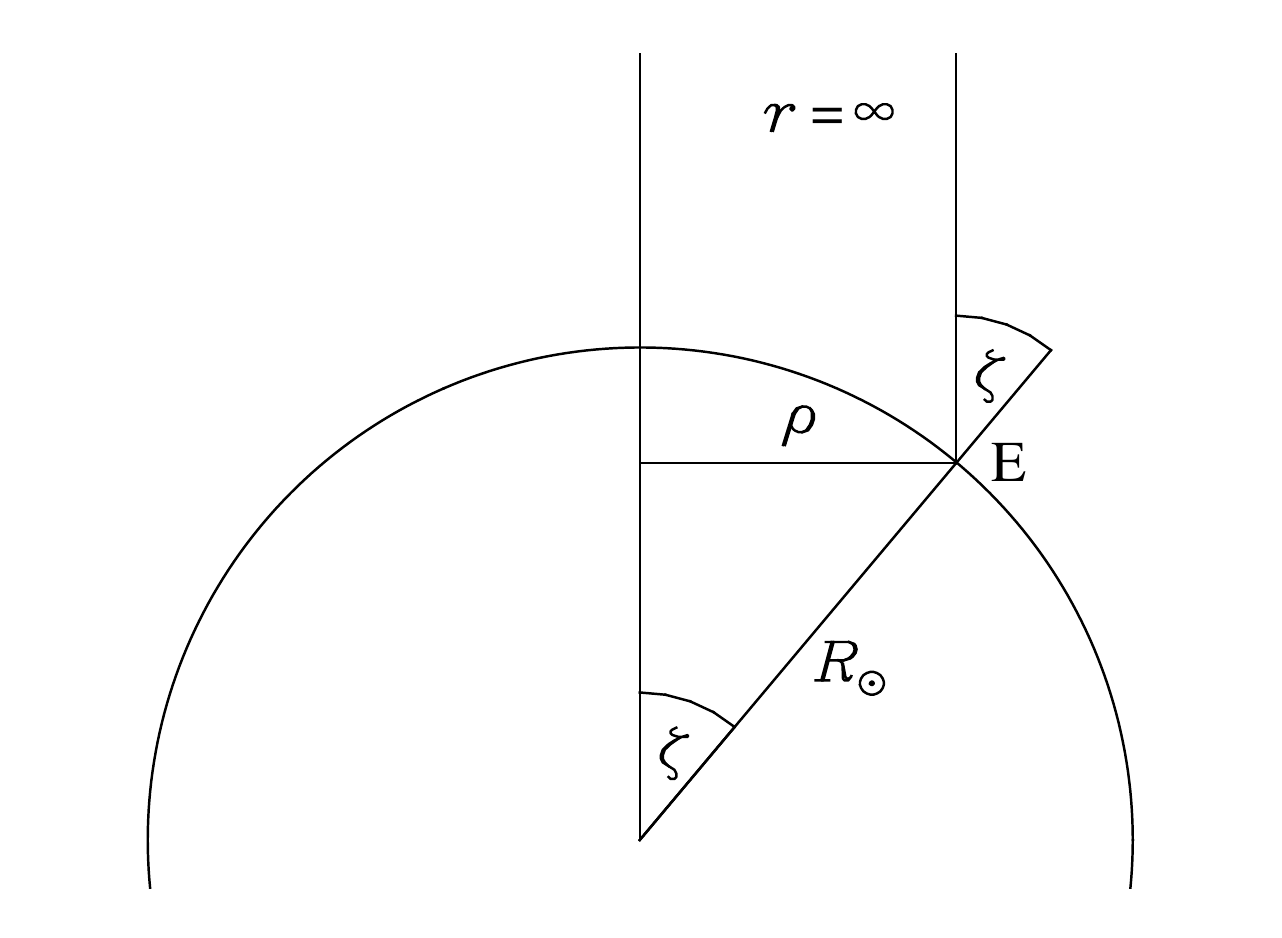}
\Fill
\caption{Sketch illustrating the geometry of the illumination of a point $S$ at
  finite distance $r$ (left) from the Sun's centre and at infinite distance
  (right). \label{Fig:ScaGeom2D}}
\end{figure}

The irradiance incident from the solar surface is obtained from
(\ref{dIrr_radiance}) by extending $\Omega_\mathrm{in}=\Omega(r)$
over the part of the surface visible from distance $r$. 
In spherical coordinates $d\Omega=\sin\theta d\theta d\phi$ 
(for the geometry see Fig.~\ref{Fig:ScaGeom2D}), where $\theta$ and 
$\phi$ are the spherical zenith and azimuth angles.
\begin{gather}
 \irr_\mathrm{in}(r)
 = \int_0^{2\pi} \int_0^{\theta_\mathrm{max}}
    \rad(\cos\zdis) \;\sin\theta d\theta d\phi
 = 2\pi\int_0^{\theta_\mathrm{max}} \rad(\cos\zdis) \;\sin\theta d\theta
 \nonumber\\
  =2\pi\int^1_{\cos\theta_\mathrm{max}} \mspace{-10mu}
             \rad(\cos\zdis) \; d\cos\theta
 \qquad[\mathrm{W/m^2}]
 \label{Irr_integral}\end{gather}
The ignorable azimuth angle $\phi$ could readily be integrated over because
of the cylindrical symmetry about the zenith axis of the spherical
coordinate system.

In (\ref{Irr_integral}), $\zdis$ is the zenith angle of the radiance beam
direction with the surface normal. We have to find its relation with the
integration variable $\theta$.
By the law of sines we have at a distance $r$ and for a solar radius
$R_\odot$ (see Fig.~\ref{Fig:ScaGeom2D},
note $\sin\zdis=\sin(\pi-\zdis$))
\begin{gather}
  \frac{\sin\zdis}{r}=\frac{\sin\theta}{R_\odot}
  ,\qquad
  \frac{1}{r}=\frac{\sin\theta_\mathrm{max}}{R_\odot}
\label{sinpsi}\\
  \text{so that}\quad
  \cos\zdis=\sqrt{1-(\frac{r}{R_\odot})^2\sin^2\theta}
           =\sqrt{1-(\frac{r}{R_\odot})^2+(\frac{r}{R_\odot})^2\cos^2\theta}
\label{cospsi}\end{gather}
where the second equation in (\ref{sinpsi}) refers to the view from the
observer onto the solar limb with maximum $\theta$ such that $\zdis=\pi/2$.
Equation
(\ref{cospsi}) is used in the following to express the argument $\cos\zdis$ in
terms of the integration variable $\cos\theta$.

For the Sun, the surface radiance is approximately given by
\citep[e.g.,][]{NeckelLabs:1994,Neckel:1996}
\begin{equation}
   \rad(\cos\zdis)=\rad_\odot(1-u + u\cos\zdis)
\label{Rad_Sun}\end{equation}
with $\rad_\odot$ the radiance in vertical direction. The limb darkening
parameter $u$ has been empirically determined to about 0.6 in the optical
wavelength
range. If we insert (\ref{Rad_Sun}) into (\ref{Irr_integral}) we find
\begin{equation}
  \irr_\mathrm{in}(r) = \rad_\odot[(1-u)I_0(r)+uI_1(r)]
\label{Irr_0}\end{equation}
which contains the two integrals
\begin{gather}
  I_0(r)=2\pi\int_{\cos\theta_\mathrm{max}(r)}^1
         \mspace{-15mu}d\cos\theta,
  \quad
  I_1(r)=2\pi\int_{\cos\theta_\mathrm{max}(r)}^1
         \mspace{-15mu}\cos\zdis\,d\cos\theta
\label{IntegralsI_01}\end{gather}
The integrals are calculated in the appendix~\ref{App:MinnaertCoeff}
after expressing $\cos\zdis$ in terms of $\cos\theta$ by means of
(\ref{cospsi}). Inserting the analytical expressions (\ref{I_0})
and (\ref{I_1}) for the integrals in (\ref{Irr_0}) gives
\begin{gather}
  \irr_\mathrm{in}(r)
    = 2\pi \rad_\odot
       [(1-u)(1-\cos\theta_\mathrm{max})
         +\frac{u}{2} \;\left[1
     -\frac{\cos^2\theta_\mathrm{max}}{\sin\theta_\mathrm{max}}
        \ln(\frac{1+\sin\theta_\mathrm{max}}{\cos\theta_\mathrm{max}})\right]]
 \nonumber\\
     = 2\pi \rad_\odot
       [(1-\frac{u}{2})(1-\cos\theta_\mathrm{max})
         +\frac{u}{2}\cos\theta_\mathrm{max} \;\left[1
     -\frac{\cos\theta_\mathrm{max}}{\sin\theta_\mathrm{max}}
        \ln(\frac{1+\sin\theta_\mathrm{max}}{\cos\theta_\mathrm{max}})\right]]
\label{Irr_r}\end{gather}
For $r\rightarrow\infty$ we have $\theta_\mathrm{max}\rightarrow R_\odot/r$
and (see appendix~\ref{App:MinnaertCoeff})
\begin{gather*}
  \irr_\mathrm{in}(r)
  \xrightarrow{r\rightarrow\infty}
       \rad_\odot[(1-u)I^\infty_0+uI^\infty_1]\\
  \simeq\pi \rad_\odot[(1-u)\theta^2_\mathrm{max}
                              + u {\TS\frac{2}{3}}\theta^2_\mathrm{max}]
       =\pi \rad_\odot(1-\frac{u}{3})\theta^2_\mathrm{max}
       =\frac{\pi R_\odot^2}{r^2} \bar{\rad}_\odot
\end{gather*}
where $\bar{\rad}_\odot=\rad_\odot(1-u/3)$ is the average radiance
of the solar disk.
This is consistent with the integration of the original expression in
(\ref{Irr_integral}) in the far-distance limit $r\rightarrow \infty$
such that $\sin\theta\simeq\theta\simeq \rho/r$ and
$\cos\zdis=\sqrt{1-\sin^2\zdis}\simeq\sqrt{1-(\rho/R_\odot)^2}$
(see Fig.~\ref{Fig:ScaGeom2D})
\begin{gather}
\irr_\mathrm{in}(r)
  \xrightarrow{r\rightarrow\infty}\frac{2\pi}{r^2}
\int_0^{R_\odot} \rad(\sqrt{1-(\frac{\rho}{R_\odot})^2})\; \rho d\rho
\nonumber\\
=\pi (\frac{R_\odot}{r})^2 \int_0^1 \rad(\sqrt{1-(\frac{\rho}{R_\odot})^2}) \;
      d(\frac{\rho}{R_\odot})^2
\qquad\text{substitute} \; 1-(\frac{\rho}{R_\odot})^2=\cos^2\zdis=x
\nonumber\\
=\pi (\frac{R_\odot}{r})^2 \rad_\odot \int_0^1 [1-u + ux^{1/2}] dx
=\pi (\frac{R_\odot}{r})^2 \rad_\odot
  [\left.(1-u)x\right|_0^1+\frac{2}{3}u\left.x^{3/2}\right|_0^1]
 \nonumber\\
 =\pi (\frac{R_\odot}{r})^2 \rad_\odot [(1-u)+\frac{2}{3}u]
 =\pi (\frac{R_\odot}{r})^2 \bar{\rad}_\odot
\label{Irr_inf}\end{gather}
The power radiated from the entire solar surface (luminosity) is
\[
 \pow_\odot
=4\pi r^2\;\irr_\mathrm{in}(r)
=4\pi^2 R_\odot^2 \bar{\rad}_\odot
=\pi \bar{\rad}_\odot\times\text{solar surface}
\qquad[\mathrm{W}]
\]

\subsection{Anisotropy of solar irradiance}
\label{sec:AnisotricIrradiation}

In the previous chapter we have only calculated the scalar irradiance of Sun 
light at
a given distance from Sun. For treating the scattering adequately we also need
to know how the electric field fluctuations of the light from the solar
surface are oriented. In order to assess this property, we have to extend the
scalar radiance and irradiance to space tensors. The concept is
well known in optics to characterise the correlation and polarisation of
electromagnetic wave fields \citep[see e.g.,][]{Wolf:2007}. Most often
this concept is applied to beams of light propagating in a well
defined direction for which
the field correlation can be described by a 2$\times$2
coherency matrix spanning the plane normal to the beam
propagation direction.
Since we consider here a spatially extended source with light from
different directions, a 3$\times$3 matrix is required instead. We first extend
the definition of the power spectral density (\ref{PowSpDensity}) in an
obvious way
\begin{gather}
   \lim_{V\rightarrow\infty}\frac{1}{V}
   \bra\vect{\tilde{E}}_V(\vect{k})\vect{\tilde{E}}_V^{*\mathsf{T}}(\vect{k})\ket\;
  =\int \!\vect{R}(\Delta\vect{r})\;e^{-i\vect{k}\tp \Delta\vect{r}}\;d^3 \Delta\vect{r}
\label{PowSpDensity3x3}\\
\vect{R}(\Delta\vect{r})=\;\bra\vect{E}(\vect{r},t)\vect{E}\tp(\vect{r}+\Delta\vect{r},t)\ket
\nonumber\end{gather}
where the correlation matrix $\vect{R}$ is symmetric and has a trace of
$R$ as defined in (\ref{PowSpDensity}). The according radiance matrix is
in analogy to (\ref{radiance})
\begin{equation*}
 \vect{\rad}(\vect{\hat{k}})
=c\epsilon_0
   \lim_{V\rightarrow\infty}\frac{1}{V}\int\! 
   \bra\vect{\tilde{E}}_V(k\vect{\hat{k}})
       \vect{\tilde{E}}_V^{*\mathsf{T}}(k\vect{\hat{k}})\ket k^2\,dk
\end{equation*}
Since the radiance selects only those wave field components from the spectrum
which propagate along $\vect{\hat{k}}$, we have
$\vect{\rad}(\vect{\hat{k}})\vect{\hat{k}}=0$.
Hence $\vect{\rad}(\vect{\hat{k}})$ spans only the subspace perpendicular to
$\vect{\hat{k}}$ and could be reduced to a 2$\times$2 submatrix which has the
same properties as a conventional coherency matrix (see
appendix~\ref{app:Polarisation}). Finally, we need the same generalisation
for the irradiance from (\ref{irradiance}) and (\ref{Irr_Esq})
\begin{gather}
\vect{\irr}
= \int_{4\pi} \vect{\rad}(\vect{\hat{k}}) d\Omega(\vect{\hat{k}})
= c\epsilon_0 \lim_{V\rightarrow\infty}\frac{1}{V}\int\!  
   \bra\vect{\tilde{E}}_V(\vect{k})
       \vect{\tilde{E}}_V^{*\mathsf{T}}(\vect{k})\ket \,d^3\vect{k}
\label{irradiance3x3}\\
  =c\epsilon_0 \vect{R}(0)
  =c\epsilon_0 \bra\vect{E}(\vect{r},t)\vect{E}\tp(\vect{r},t)\ket
\nonumber\end{gather}
All these tensorialised quantities reduce to their scalar analogues
used in the previous chapter by forming the trace.

In principle, we now have to repeat the integration (\ref{Irr_integral}) as in
the previous chapter, however for each matrix element separately. For our
problem the radiance $\vect{\rad}(\vect{\hat{k}})$ in (\ref{irradiance3x3}) is
non-zero only for a limited cone $\Omega(r)$ of directions $\vect{\hat{k}}$
from the visible solar surface to the scattering site $\vect{r}$. Therefore
the integration is effectively over $\Omega(r)$ rather than over $4\pi$ as in
the general case in (\ref{irradiance3x3}). The integration is also greatly
simplified by the assumption that the radiance from the solar surface is
unpolarised \citep{KempEtal:1987}. The radiance matrix is then
(see appendix~\ref{app:Polarisation})
\begin{equation}
 \vect{\rad}(\vect{\hat{k}})
 =\frac{1}{2}L(\cos\zdis)\,
  (\vect{\hat{e}}_1\vect{\hat{e}}_1\tp+\vect{\hat{e}}_2\vect{\hat{e}}\tp_2)
 =\frac{1}{2}L(\cos\zdis)\,(\boldsymbol{1}-\vect{\hat{k}}\vect{\hat{k}}\tp)
\label{radiance3x3}\end{equation}
where $L(\cos\zdis)$ is the scalar radiance (\ref{Rad_Sun}),
$\vect{\hat{e}}_1$ and $\vect{\hat{e}}_2$ are two mutually orthogonal
polarisation vectors spanning the plane perpendicular to $\vect{\hat{k}}$
and $\boldsymbol{1}$ is the identity matrix.
In case that $\vect{\rad}$ is polarised all matrix elements
$\vect{\hat{e}}_i\vect{\hat{e}}\tp_j$ are needed and weighted with
different coefficients related to the Stokes parameters.
We then cannot use the second form in (\ref{radiance3x3}). 
But for the unpolarised incident radiation
assumed here, the second form is favourable because $\vect{\hat{e}}_1$ and
$\vect{\hat{e}}_2$ do not need to be specified.
The final expression for the irradiance matrix at the scattering site is
then from (\ref{irradiance3x3}) and (\ref{radiance3x3})
\begin{equation}
\vect{\irr}_{\mathrm{in}}(\vect{r})
 =\frac{1}{2}\int_{\Omega}
 L(\cos\zdis)\,(\boldsymbol{1}-\vect{\hat{k}}\vect{\hat{k}}\tp)
d\Omega(\vect{\hat{k}})
\label{Irr_1}\end{equation}
where instead of $4\pi$ we only integrate the directions $\uect{k}$ over
the solar surface $\Omega(r)$ visible from distance $r$.

\begin{figure}[b]
\hspace*{\fill}
\parbox{7cm}{\includegraphics[viewport= 120 80 610 569, clip, width=7cm]
            {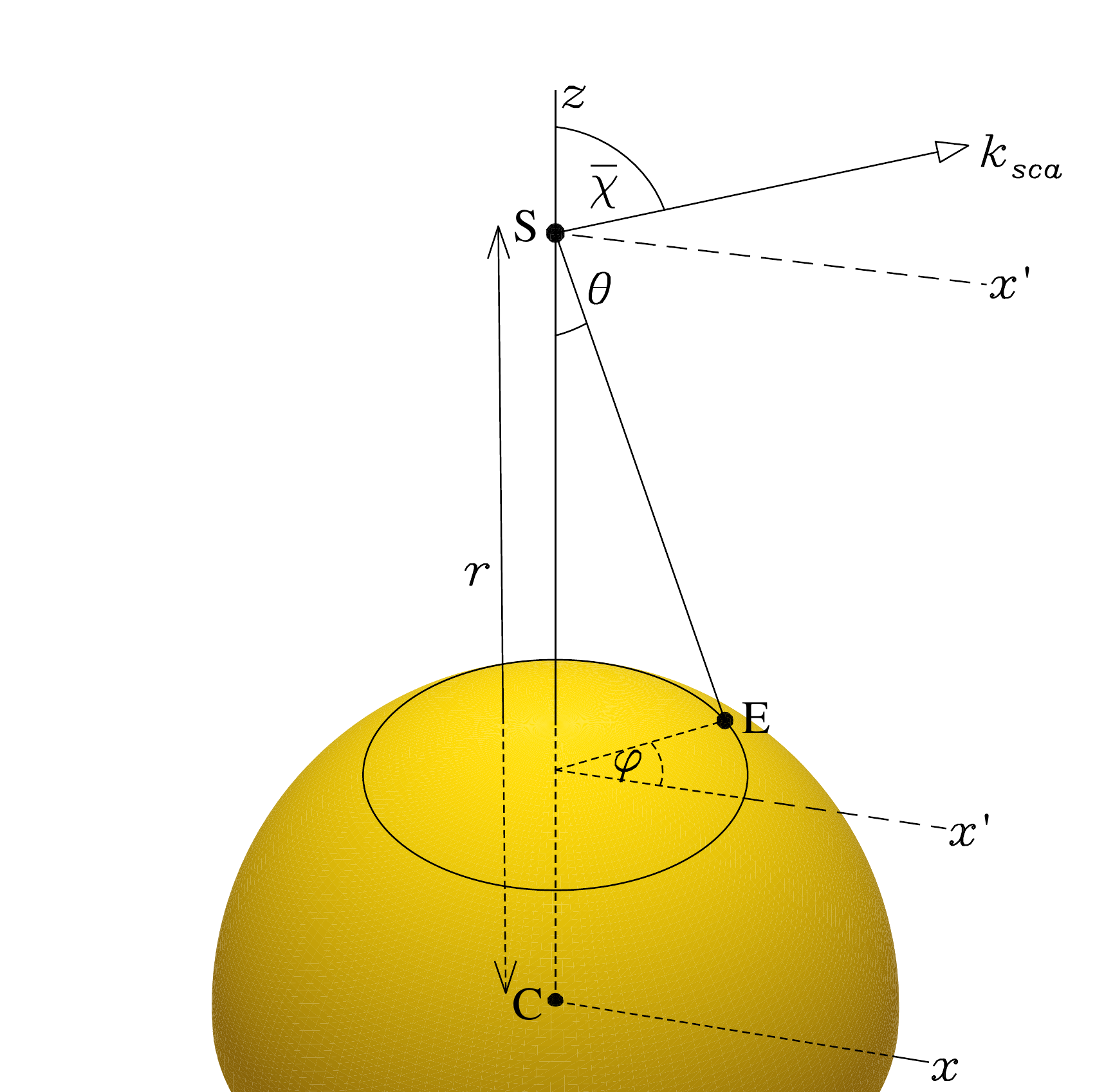}}\hspace{1ex}
\parbox{8cm}{\caption{Sketch illustrating the geometry of scattering at point
    S. The centre of the coordinate system is the solar centre C, the
    radiation originates from a point E on the solar surface at distance $r$
    along the $z$ axis. The scattered ray $\vect{k}_\mathrm{sc}$ lies in the
    $\uect{x},\uect{z}$ plane which defines the $x$-axis. The dashed axes $x'$ are drawn to
    help the eye and are all parallel to the $x$ axis. The $y$-axis is normal
    to $\vect{\hat{x}}$ and $\vect{\hat{z}}$.
    \label{Fig:ScaGeom3D}}}
\hspace*{\fill}
\end{figure}

In the coordinate system of Fig.~\ref{Fig:ScaGeom3D} we can write explicitly
the propagation direction $\vect{\hat{k}}$ of a given beam from a point on the
solar surface to the scattering site (points E and S in
Fig.~\ref{Fig:ScaGeom3D}) as
\begin{equation}
 \vect{\hat{k}}=
   \rveccc{\cos\phi \sin\theta}{\sin\phi \sin\theta}{\cos\theta}
\label{ScaGeom3DCoords}\end{equation}
We can now express the radiance matrix elements in the
integrand of (\ref{Irr_1}) for different locations on the solar surface. 
From (\ref{radiance3x3}) and (\ref{ScaGeom3DCoords}) we find
\begin{align*}
 \vect{\hat{x}}\vect{\rad}(\vect{\hat{k}})\vect{\hat{x}}
 &=\frac{1}{2}L(\cos\zdis)\,(1-\cos^2\phi\sin^2\theta)
 \\
 \vect{\hat{y}}\vect{\rad}(\vect{\hat{k}})\vect{\hat{y}}
 &=\frac{1}{2}L(\cos\zdis)\,(1-\sin^2\phi\sin^2\theta)
 \\
 \vect{\hat{z}}\vect{\rad}(\vect{\hat{k}})\vect{\hat{z}}
 &=\frac{1}{2}L(\cos\zdis)\,(1-\cos^2\theta)
 \\
 \vect{\hat{x}}\vect{\rad}(\vect{\hat{k}})\vect{\hat{y}}
=\vect{\hat{y}}\vect{\rad}(\vect{\hat{k}})\vect{\hat{x}}
 &=\frac{1}{2}L(\cos\zdis)\,\cos\phi\sin\phi\sin^2\theta)
 \\
 \vect{\hat{x}}\vect{\rad}(\vect{\hat{k}})\vect{\hat{z}}
=\vect{\hat{z}}\vect{\rad}(\vect{\hat{k}})\vect{\hat{x}}
 &=-\frac{1}{2}L(\cos\zdis)\,\cos\phi\sin\theta\cos\theta
 \\
 \vect{\hat{y}}\vect{\rad}(\vect{\hat{k}})\vect{\hat{z}}
=\vect{\hat{z}}\vect{\rad}(\vect{\hat{k}})\vect{\hat{y}}
 &=-\frac{1}{2}L(\cos\zdis)\,\sin\phi\sin\theta\cos\theta
\end{align*}
Note that the off-diagonal elements are all odd in $\phi$ and will therefore
vanish in the subsequent integration over the visible solar surface.
Note also that
$\vect{\hat{x}}\vect{\hat{x}}\tp
+\vect{\hat{y}}\vect{\hat{y}}\tp
+\vect{\hat{z}}\vect{\hat{z}}\tp$ is the unit matrix and
$\mathrm{trace}({\vect{\rad}})$ reproduces the scalar $L(\cos\zdis)$ which
was integrated in the previous chapter.

The contributions of the emission from all points like E in
Fig.~\ref{Fig:ScaGeom3D} to the irradiance at point S at distance $r$ can now
be summed up. Only elements which are even in $\phi$ survive the integration
in azimuth angle so that we readily obtain after the $\phi$ integration
using (\ref{Irr_1}) and (\ref{ScaGeom3DCoords})
\begin{gather}
  \vect{\irr}_{\mathrm{in}}(\vect{r})
 =\frac{1}{2}\int_{\Omega}
  L(\cos\zdis)\,(\boldsymbol{1}-\vect{\hat{k}}\vect{\hat{k}}\tp)
  \,\sin\theta d\theta d\phi
\label{Irr_gen}\\  
 =\frac{2\pi}{2}\int_0^{\theta\mathrm{max}} \rad(\cos\zdis)
  \begin{pmatrix}
    \half+\half\cos^2\theta & 0 & 0 \\
    0 & \half+\half\cos^2\theta & 0 \\
    0 & 0 & 1-\cos^2\theta
  \end{pmatrix}
  d\cos\theta
\label{Irr3x3}\end{gather}
where $\sin^2\theta$ was replaced by $1-\cos^2\theta$ throughout.
If we insert the limb-darkened solar radiance (\ref{Rad_Sun}) into
(\ref{Irr3x3}) we get
\begin{gather*}
  \vect{\irr}_{\mathrm{in}}(\vect{r})
=\frac{\rad_\odot}{2}
[(1-u)\begin{pmatrix}\frac{I_0+J_0}{2} & 0 & 0 \\
                     0 & \frac{I_0+J_0}{2} & 0 \\
                     0 & 0 & I_0-J_0 
       \end{pmatrix}
  +u\begin{pmatrix}\frac{I_1+J_1}{2} & 0 & 0 \\
                   0 & \frac{I_1+J_1}{2} & 0 \\
                   0 & 0 & I_1-J_1 
       \end{pmatrix}]
\end{gather*}
where we introduced two new integrals $I_1$ and $J_1$.
The integrals of the $\theta$-independent terms in the matrix elements
in eq.~\ref{Irr3x3}) agree with the known integrals $I_0$, $I_1$.
The contributions of the $\cos^2\theta$ terms in the matrix elements
lead to two new integrals 
\begin{gather}
  J_0(r)=2\pi\int_{\cos\theta_\mathrm{max}(r)}^1\mspace{-15mu}
                   \cos^2\theta\,d\cos\theta,
  \quad
  J_1(r)=2\pi\int_{\cos\theta_\mathrm{max}(r)}^1\mspace{-15mu}
                    \cos\zdis\cos^2\theta \,d\cos\theta
\label{IntegralsJ_01}\end{gather}
The integrals are calculated in the appendix~\ref{App:MinnaertCoeff}
again after expressing $\cos\zdis$ in terms of $\cos\theta$
as in (\ref{cospsi}). If instead of (\ref{Rad_Sun}) we wanted to
use limb-darkening laws with higher powers in $\cos\zdis$ we
obtain corresponding integrals $I_n$ and $J_n$ for which we
also derive analytic expressions in appendix~\ref{App:MinnaertCoeff}.
Finally we can write the non-zero elements of the local irradiance
matrix as
\begin{align*}
   \irr_{\mathrm{in},xx}(r)
 = \irr_{\mathrm{in},yy}(r)
 =&\; \quart \rad_\odot ((1-u)(I_0+J_0) + u(I_1+J_1))
\\
  \irr_{\mathrm{in},zz}(r)
 =&\;\half \rad_\odot ((1-u)(I_0-J_0) + u(I_1-J_1))
\end{align*}
A combination of the integrals introduced by \citep{Minnaert:1930}
and often preferred in the literature is
\begin{gather}
  C=\frac{I_0+J_0}{2\pi}, \qquad D=\frac{I_1+J_1}{2\pi} \nonumber\\
  A=\frac{3J_0-I_0}{2\pi},\qquad B=\frac{3J_1-I_1}{2\pi}\nonumber\\
\irr_{\mathrm{in},xx}=\irr_{\mathrm{in},yy}
 =\frac{\pi \rad_\odot}{2}((1-u)C + uD)
  \label{Irr_xy_Minnaert}\\
   \irr_{\mathrm{in},xx}-\irr_{\mathrm{in},zz}
 =\quart \rad_\odot
   ((1-u)[(I_0+J_0)-2(I_0-J_0)]
\nonumber\\\hspace*{8em}
 + u[(I_1+J_1)-2(I_1-J_1)])
\nonumber\\
    =\frac{\rad_\odot}{4}((1-u)(3J_0-I_0) + u(3J_1-I_1))
\nonumber\\
    =\frac{\pi \rad_\odot}{2}((1-u)A + uB)
\label{Irr_xz_Minnaert}\\
\text{or}\quad
 \vect{\irr}_{\mathrm{in}}(\vect{r})
=\frac{\pi \rad_\odot}{2}
 \big[((1-u)C + uD)\vectg{1}
    - ((1-u)A + uB)\vect{\hat{r}}\vect{\hat{r}}\tp\big]
\label{Irr_in_Minnaert}\end{gather}
where in the last row we replaced $\vect{\hat{z}}$ of the local coordinate system (see
Fig.~\ref{Fig:ScaGeom3D}) by the more general unit vector $\vect{\hat{r}}$
from the solar centre to the point for which the irradiance is calculated. The
coefficients $A$, $B$, $C$ and $D$ depend only on the distance $r$ from the
solar centre. Some useful properties of these coefficients are derived in
appendix~\ref{App:MinnaertCoeff}.

To this end we are able to characterise the field fluctuations at a distance
$r$ (point S in Fig.~\ref{Fig:ScaGeom3D}) from the solar centre. It is not
surprising that due to symmetry their polarisations in $x$ and $y$ are
identical. As also expected for $r\rightarrow\infty$,
\begin{gather}
\irr_{\mathrm{in},xx}=\irr_{\mathrm{in},yy}
\rightarrow \frac{\pi \rad_\odot}{2}
     ((1-u)-\frac{2}{3}u)(\frac{R_\odot}{r})^2
=\frac{\pi}{2} (\frac{R_\odot}{r})^2 \bar{\rad}_\odot
\label{Irr_xy_inf}\\
\irr_{\mathrm{in},zz}=
\frac{\pi \rad_\odot}{2}((1-u)(C-A)-u(D-B))\rightarrow r^{-4}
\label{Irr_zz_inf}\end{gather}
i.e., the radial element decreases much faster then the tangential elements.
Each of the latter approach half the total asymptotic irradiance
(\ref{Irr_inf}).

\subsection{Scattering of solar irradiance}
\label{sec:ScattSunLight}

We start again with scalar variables to discuss the scattering process in
general. The principle will then be extended to the matrix quantities derived
in the previous chapter. We assume as in (\ref{dIrr_radiance}) that the
irradiance $d\irr_\mathrm{in}=\rad(\uect{k}\inc)d\Omega\inc$ 
describes the field fluctuations of a narrow beam in direction $\uect{k}\inc$ 
incident at the scattering site $\vect{r}$. 
If the total scattering cross section from a single scatterer
is $\sigma$, the total number of scattered photons is equivalent to the totally
scattered power (the hat on $\hat{\pow}$ indicates normalisation to a single
scattering centre)
\begin{equation}
  d\hat{\pow}
  =\sigma d\irr_\mathrm{in}
  =\sigma \rad(\vect{\hat{k}}_\mathrm{in}) d\Omega_\mathrm{in}
  \qquad[\mathrm{W}]
\label{dIrr_sca1}\end{equation}
The single scattering centre is the source of a scattered radiant intensity
which describes the angular distribution of the scattered photons. If we only
count the number of scattered photons in a special direction
$\vect{\hat{k}}_\mathrm{sc}$, we have to replace the total scattering cross
section $\sigma=\int (d\sigma/d\Omega\sca)\;d\Omega\sca$ by the 
differential cross
section $d\sigma/d\Omega\sca$. This yields a scattered radiant intensity in the
scattering direction $\vect{\hat{k}}_\mathrm{sc}$ of
\begin{equation}
  d\rin(\vect{r},\vect{\hat{k}}_\mathrm{sc})
 =\frac{d\Phi\sca}{d\Omega\sca}
 =\frac{d\sigma}{d\Omega\sca}(\vect{\hat{k}}_\mathrm{in},
                              \vect{\hat{k}}_\mathrm{sc})
                      d\irr_\mathrm{in}(\vect{r})
 =\frac{d\sigma}{d\Omega\sca}(\vect{\hat{k}}_\mathrm{in},
                          \vect{\hat{k}}_\mathrm{sc})
                      \rad(\vect{\hat{k}}_\mathrm{in})
                          d\Omega_\mathrm{in}
  \qquad[\mathrm{W/sr}]
\label{dRin_sca1}\end{equation}
Integrating over all scattering directions yields again the total scattered
power (\ref{dIrr_sca1}). 
We collect the scattered photons at 
$\vect{r}_\mathrm{obs}=\vect{r}+\ell\vect{\hat{k}}_\mathrm{sc}$
in an aperture area $\vect{A}_\mathrm{aperture}$ with normal in
$-\uect{k}\sca$. The aperture then subtends a solid angle
$d\Omega_\mathrm{sc}= A_\mathrm{aperture}/\ell^2$ at the scattering site and
the instrument integrates over all scattering directions
$\vect{\hat{k}}_\mathrm{sc}$ inside this solid angle. Hence from the single
scattering centre we obtain a radiant flux in the instrument at distance
$\ell$ equivalent to the power
\begin{gather}
  d\hat{\pow}(\vect{r}_\mathrm{obs})
 =d\Omega_\mathrm{sc}\,
  d\rin(\vect{r},\vect{\hat{k}}_\mathrm{sc})
 =\frac{A_\mathrm{aperture}}{\ell^2}\,
  d\rin(\vect{r},\vect{\hat{k}}_\mathrm{sc})
  \qquad[\mathrm{W}]
\label{dPow_sngsca}\end{gather}
which corresponds to an irradiance at the instrument of
\begin{gather}
 d\irr_\mathrm{sc}(\vect{r}_\mathrm{obs})
 =\frac{d\hat{\pow}(\vect{r}_\mathrm{obs})}{A_\mathrm{aperture}}
 =\frac{d\rin(\vect{r},\vect{\hat{k}}_\mathrm{sc})}{\ell^2}\,
 =\frac{d\sigma}{\ell^2d\Omega\sca}(\vect{\hat{k}}_\mathrm{in},
                                \vect{\hat{k}}_\mathrm{sc})\;
                           \rad(\vect{\hat{k}}_\mathrm{in})
                           d\Omega_\mathrm{in}
\nonumber\\                           
 =\frac{d\sigma}{\ell^2d\Omega\sca}(\vect{\hat{k}}_\mathrm{in},
                           \vect{\hat{k}}_\mathrm{sc})\;
                           d\irr_\mathrm{in}
  \qquad[\mathrm{W/m^2}]
\label{dIrr_sngsca}\end{gather}
Note the difference between $d\sigma/d\Omega\sca$ in (\ref{dRin_sca1}) and
$d\sigma/\ell^2d\Omega\sca$ in (\ref{dIrr_sngsca}). 
While the former has the units
[m$^2$/sr] the latter is the dimensionless ratio of two areas, namely the
influx area $d\sigma$ from which incident photons are redirected into the
solid angle $d\Omega\sca$ around the scattering direction
$\vect{\hat{k}}_\mathrm{sc}$ which is subtended by the detection area $\ell^2
d\Omega\sca$ at distance $\ell$. Since the number of photons is preserved, this
area ratio is just the ratio of scattered and incident irradiances. 
The problem here is the same as with $d^2$ in (\ref{dRin_re}). 
Formally we require that $\ell^2$ in (\ref{dIrr_sngsca}) has the units 
[m$^2$/sr] so that $\ell^2 d\Omega\sca$ represents an area in [m$^2$].

A distribution of many scattering centres results in an extended radiating
volume. We call $N_e$ the local number density of the scattering centres. Then
the number of scatterers which is visible in the instrument is $N_e(\vect{r})
\ell^2 d\Omega\obs d\ell$ where $d\ell$ is the depth of the scattering 
volume at
$\vect{r}$ along the line-of-sight direction $\vect{\hat{k}}_\mathrm{sc}$. The
solid angle $d\Omega\obs$ is the smaller of either the angle which the 
scattering
cloud subtends at the instrument or the angle of the instrument's pixel field
of view. Hence, $d\Omega\obs$ plays the same role with respect to the projected
area of the scattering volume as $d\Omega_\mathrm{in}$ with respect to the
solar surface element $dA_\mathrm{em}$. For an imaging coronagraph which well
resolves the cloud of scattering centres, the solid angle of the volume
visible to a pixel is often limited by the pixel resolution of the instrument
rather than the size of the cloud. In this case $d\Omega\obs=
A_\mathrm{pixel}/f^2$ where $A_\mathrm{pixel}$ is the physical pixel area and
$f$ the instrument focal length. Instead of (\ref{dPow_sngsca}), the photon
flux per pixel for an observer at
$\vect{r}_\mathrm{obs}=\vect{r}+\ell\vect{\hat{k}}_\mathrm{sc}$ in the case of
volume scattering is
\begin{gather}
 d\pow(\vect{r}_\mathrm{obs})=
    N_e(\vect{r}) \ell^2 d\Omega\obs\,d\ell\; 
 d\hat{\pow}_\mathrm{sc}(\vect{r}_\mathrm{obs})=
    N_e(\vect{r}) \ell^2 d\Omega\obs\,d\ell\; 
               d\Omega_\mathrm{sc}\,dI(\vect{r},\vect{\hat{k}}_\mathrm{sc})
\nonumber\\
           =\frac{A_\mathrm{pixel}A_\mathrm{aperture}}{f^2} N_e(\vect{r})\;d\ell\;
                         dI(\vect{r},\vect{\hat{k}}_\mathrm{sc})
  \qquad[\mathrm{W}]
\label{dPow_volsca}\end{gather}
Different from the $d\hat{\pow}$ in the single scattering case 
(\ref{dPow_sngsca}), $d\pow_\mathrm{sc}$ in (\ref{dPow_volsca}) has
no more explicit dependence on distance $\ell$ (there is still an implicit
dependence through $N_e(\vect{r})$, $d\sigma/d\Omega\sca$ and
$dI(\vect{r},\vect{\hat{k}}_\mathrm{sc})$, though). 
The units of $\ell^2$ [m$^2$/sr]
in (\ref{dIrr_sngsca}) are now adopted by $f^2$ since $f^2
d\Omega\obs$ is the pixel area in [m$^2$].

We have not specified the nature of the scattering yet. Thomson scattering is
one of the simplest possible scattering mechanisms. This is why Rayleigh
(1871) could derive it with few basic assumptions (the particle is at rest,
unbound and much smaller than $2\pi/k$) and without making any further guess
about the nature of the particle. As we have seen, the solar irradiance at the
scattering site at $\vect{r}$ is anisotropic, i.e., it has different field
strengths in tangential and radial direction. Thomson scattering of these
fluctuations modifies this incident polarisation further. The scattering is
due to the dipole excitation of free electrons such that the radiating dipole
axis is directed along the incident electric field $\vect{E}_\mathrm{in}$.
Provided the driving field strength $E_\mathrm{in}$ is well below $m_ec^2k/e$
so that the electron is not oscillating with a relativistic velocity, the
orientation of the scattered field is just the projection of the incident
field $\vect{E}_\mathrm{in}$ into the transverse polarisation modes of the
scattered wave which propagates in direction $\vect{\hat{k}}_\mathrm{sc}$.
The scattered electric field at some distance $\ell$ from the scattering
electron at $\vect{r}$ is therefore \citep{Jackson:1998}
\begin{equation}
  \vect{E}_\mathrm{sc}
  (\vect{r}+\ell\vect{\hat{k}}_\mathrm{sc},t)
  =-\frac{r_e}{\ell}\mathcal{P}_{\vect{\hat{k}}_\mathrm{sc}}
  \vect{E}_\mathrm{in}(\vect{r},t-\frac{\ell}{c})
\label{Thomson_E}\end{equation}
where $r_e$ is the classical electron radius
\[
  r_e=\frac{e^2}{4\pi\epsilon_0 m_e c^2}=2.8179\:10^{-15}\mathrm{m}
\]
at which the Coulomb potential energy equals the electron rest mass energy $m_e
c^2$. The minus sign in (\ref{Thomson_E}) accounts for the phase shift of
$\pi$ between the electron oscillation and the driving field, the time
retardation $\ell/c$ for the phase delay of the scattered field after
travelling the distance $\ell$. Both will be irrelevant for the stationary
irradiances calculated here. Finally,
$\mathcal{P}_{\vect{\hat{k}}_\mathrm{sc}}$ is a projection normal to the
propagation direction $\vect{\hat{k}}_\mathrm{sc}$ \citep{Hutchinson:2002},
defined as
\begin{gather}
\mathcal{P}_{\vect{\hat{k}}_\mathrm{sc}}
=(\boldsymbol{1}-\vect{\hat{k}}_\mathrm{sc}\vect{\hat{k}}\tp_\mathrm{sc})
\quad\text{which satisfies}
\nonumber\\
 \mathcal{P}\tp_{\vect{\hat{k}}_\mathrm{sc}}=
 \mathcal{P}_{\vect{\hat{k}}_\mathrm{sc}},
 \qquad
 \mathcal{P}^2_{\vect{\hat{k}}_\mathrm{sc}}=
 \mathcal{P}_{\vect{\hat{k}}_\mathrm{sc}},
 \qquad
 \mathcal{P}_{\vect{\hat{k}}_\mathrm{sc}}\vect{\hat{k}}_\mathrm{sc}
 =0,
 \quad\text{and}\quad
 \mathcal{P}_{\vect{\hat{k}}_\mathrm{sc}}\vect{\hat{p}}
 =\vect{\hat{p}}
\label{PolProj}\end{gather}
for every polarisation direction $\vect{\hat{p}}$ of the
scattered radiation since it is normal to $\vect{\hat{k}}_\mathrm{sc}$.

The basic Thomson scattering law (\ref{Thomson_E}) can easily be extended
to describe the scattering effect on the irradiance matrix,
\begin{gather}
 \vect{\irr}_\mathrm{sc}(\vect{r}+\ell\vect{\hat{k}}_\mathrm{sc})
 =c\epsilon_0
 \bra\vect{E}_\mathrm{sc}
        (\vect{r}+\ell\vect{\hat{k}}_\mathrm{sc},t)
     \vect{E}^{*\mathsf{T}}_\mathrm{sc}
        (\vect{r}+\ell\vect{\hat{k}}_\mathrm{sc},t)\ket
\nonumber\\     
 =c\epsilon_0 \frac{r_e^2}{\ell^2}
  \bra\mathcal{P}_{\vect{\hat{k}}_\mathrm{sc}}
         \vect{E}_\mathrm{in}
           (\vect{r},t-\frac{\ell}{c})
     (\mathcal{P}_{\vect{\hat{k}}_\mathrm{sc}}
         \vect{E}_\mathrm{in})\tp
           (\vect{r},t-\frac{\ell}{c})\ket
\nonumber\\
 =c\epsilon_0 \frac{r_e^2}{\ell^2}
      \mathcal{P}_{\vect{\hat{k}}_\mathrm{sc}}\!\!
     \bra\vect{E}_\mathrm{in}
         \vect{E}\tp_\mathrm{in}\ket\!(\vect{r})\,
      \mathcal{P}_{\vect{\hat{k}}_\mathrm{sc}}
 =\frac{r_e^2}{\ell^2}
      \mathcal{P}_{\vect{\hat{k}}_\mathrm{sc}}
       \vect{\irr}_\mathrm{in}(\vect{r})\,
      \mathcal{P}_{\vect{\hat{k}}_\mathrm{sc}}
\label{Irr_Qsc}\end{gather}
Obviously, for sufficiently large distance $\ell$ from the
scattering source the scattered waves which
contribute to $\vect{\irr}_\mathrm{sc}$ all propagate in direction
$\vect{\hat{k}}_\mathrm{sc}$. Therefore
$\vect{\irr}_\mathrm{sc}\vect{\hat{k}}_\mathrm{sc}=0$ and the 2$\times$2
submatrix spanning the space perpendicular to $\vect{\hat{k}}_\mathrm{sc}$
characterises the polarisation state of the scattered beam.
Note that the reverse reasoning is not generally true. Contrary to
$\vect{\rad}$, the irradiance $\vect{\irr}$ may be superposed of wave fields
from different directions and $\vect{\irr}\vect{\hat{k}}=0$ does not
necessarily imply that these waves all propagate along $\vect{\hat{k}}$. To
estimate polarisation properties from a general 3$\times$3 correlation matrix
has been an issue for a long time \citep[see, e.g.][]{EllisEtal:2005}.
In the scattering case and at a position $\vect{r}_\mathrm{obs}=\vect{r}+
\ell\vect{\hat{k}}_\mathrm{sc}$ in the far field of the scatterer,
the irradiance component polarised along
$\vect{\hat{p}}\perp\vect{\hat{k}}_\mathrm{sc}$ is therefore simply
obtained from the matrix element
\footnote{A circular polarisation cannot be derived from the irradiance matrix
  as defined here. It requires a complex hermitian $\vect{\irr}$ which is
  easily obtained if restrict our definitions to a monochromatic wave field
  Fourier-transformed with respect to time. Since circular polarisation is not
  an issue here, we rather tried to keep all field matrices real and
  symmetric.}
\begin{equation}
 \vect{\hat{p}}\tp\vect{\irr}_\mathrm{sc}(\vect{r}_\mathrm{obs})\vect{\hat{p}}
=\frac{r_e^2}{\ell^2}
      \uect{p}\tp\mathcal{P}_{\vect{\hat{k}}_\mathrm{sc}}
       \vect{\irr}_\mathrm{in}(\vect{r})\,
      \mathcal{P}_{\vect{\hat{k}}_\mathrm{sc}}\uect{p}
=\frac{r_e^2}{\ell^2}
      \uect{p}\tp\vect{\irr}_\mathrm{in}(\vect{r})\,\uect{p}
\label{Irr_pQscap}\end{equation}
where we got rid of the scattering projection 
$\mathcal{P}_{\vect{\hat{k}}_\mathrm{sc}}$ by means of
(\ref{PolProj}). Furthermore, we can finally use (eq \ref{dRin_re}, with
$\ell^2$ instead of $d^2$) to obtain the
scattered radiant intensity for polarisation in direction
$\uect{p}$
\begin{equation}
 \rin_{\vect{\hat{p}}}(\vect{r},\vect{\hat{k}}_\mathrm{sc})
 =\ell^2\vect{\hat{p}}{\tp}
    \vect{\irr}_\mathrm{sc}(\vect{r}+\ell\vect{\hat{k}}_\mathrm{sc})
    \vect{\hat{p}}
 = r_e^2 \vect{\hat{p}}\tp \vect{\irr}_\mathrm{in}(\vect{r})\vect{\hat{p}}
\label{Rinp}\end{equation}
This last relation is all we need to relate the scattered radiant intensity
to the radiance incident at the electron. 

Assume a single incident unpolarised beam 
which propagates along $\vect{\hat{k}}_\mathrm{in}$
to the scattering site at $\vect{r}$. It produces an irradiance
matrix at $\vect{r}$ of
\begin{equation}
  d\vect{\irr}_\mathrm{in}
 =\frac{1}{2}d\irr_\mathrm{in}(\vect{\hat{e}}_1\vect{\hat{e}}\tp_1+
                            \vect{\hat{e}}_2\vect{\hat{e}}\tp_2)
 =\frac{1}{2}d\irr_\mathrm{in}(\boldsymbol{1}-
         \vect{\hat{k}}_\mathrm{in}\vect{\hat{k}}\tp_\mathrm{in})
\label{Irr_2}\end{equation}
where $\vect{\hat{e}}_1$ and $\vect{\hat{e}}_2$ form a polarisation base
perpendicular to the propagation direction $\vect{\hat{k}}_\mathrm{in}$.
Because $\vect{\hat{p}}\perp\vect{\hat{k}}_\mathrm{sc}$,
the linear polarisation component of the scattered irradiation along
$\vect{\hat{p}}$ is using (\ref{Irr_pQscap}) and (\ref{Irr_2})
\begin{gather}
  \vect{\hat{p}}\tp d\vect{\irr}_\mathrm{sc}\vect{\hat{p}}
 =\frac{r_e^2}{\ell^2}\;
      \vect{\hat{p}}\tp d\vect{\irr}_\mathrm{in}\vect{\hat{p}}
\nonumber\\
 =\frac{r_e^2}{2\ell^2}
      ((\vect{\hat{p}}\tp\vect{\hat{e}}_1)^2
       +\vect{\hat{p}}\tp\vect{\hat{e}}_2)^2)\;
      d\irr_\mathrm{in}
 =\frac{r_e^2}{2\ell^2}
      (1-(\vect{\hat{k}}_\mathrm{in}\tp\vect{\hat{p}})^2)\;
      d\irr_\mathrm{in}
\label{Irr_pQinp}\end{gather}
By comparison with (\ref{dIrr_sngsca}) we find for the differential Thomson
cross section of an unpolarised incident beam scattered at a single electron
into light polarised in direction $\vect{\hat{p}}$
\begin{equation}
\frac{d\sigma_{T,\vect{\hat{p}}}}{d\Omega\sca}
(\vect{\hat{k}}_\mathrm{in},\vect{\hat{k}}_\mathrm{sc})
 = \frac{r_e^2}{2}
  ((\vect{\hat{e}}_1\tp\vect{\hat{p}})^2
  +(\vect{\hat{e}}_2\tp\vect{\hat{p}})^2)
 = \frac{r_e^2}{2}
  (1-(\vect{\hat{k}}_\mathrm{in}\tp\vect{\hat{p}})^2)
  \qquad[\mathrm{m^2/sr}]
\label{dThomson_pol}\end{equation}
Recall that $\vect{\hat{e}}_1$ and $\vect{\hat{e}}_2$ are two orthogonal
polarisation directions of the incident radiation normal to
$\vect{\hat{k}}_\mathrm{in}$ and $\vect{\hat{p}}$ is an arbitrary polarisation
direction of the scattered radiation normal to $\vect{\hat{k}}_\mathrm{sc}$.
If we take $\vect{\hat{p}}$ normal to the scattering plane, then
$\vect{\hat{k}}_\mathrm{in}\tp\vect{\hat{p}}=0$ and the differential cross
section for this polarisation is $r^2_e/2$. If we rotate $\vect{\hat{p}}$ by
$\pi/2$ into the scattering plane the angle between
$\vect{\hat{k}}_\mathrm{in}$ and $\vect{\hat{p}}$ is just $\chi\pm\pi/2$ where
$\chi$ is the scattering angle between $\vect{\hat{k}}_\mathrm{in}$ and
$\vect{\hat{k}}_\mathrm{sc}$ so that
$\vect{\hat{k}}_\mathrm{in}\tp\vect{\hat{p}}=\pm\sin\bar{\chi}$. The
differential cross section for this polarisation is therefore
$(r^2_e\cos^2\chi)/2$.
The sum of the two polarised irradiances yields the total scattered irradiance,
therefore the differential Thomson cross section for unpolarised
photons is
\[
\frac{d\sigma_T}{d\Omega\sca}(\vect{\hat{k}}_\mathrm{in},
                        \vect{\hat{k}}_\mathrm{sc})
 = \frac{r_e^2}{2}(1+\cos^2\chi)
\qquad[\mathrm{m^2/sr}]
\]
Integration over all scattering directions produces the total
Thomson cross section
\[
\sigma_T
=\frac{r_e^2}{2}\int_0^\pi\int_0^{2\pi}(1+\cos^2\chi)\sin\chi d\phi d\chi
=\pi r_e^2\int_{-1}^{1}(1+\cos^2\chi)d\cos\chi
=\frac{8\pi}{3} r_e^2
\qquad[\mathrm{m^2}]
\]

After these preliminaries, we can proceed with our specific scattering
problem. Inserting the irradiance matrix (\ref{Irr_gen}) from the previous 
chapter into (\ref{Rinp}) we can write the general expression
for the radiant intensity scattered from a
single electron at $\vect{r}$ into direction $\vect{\hat{k}}_\mathrm{sc}$
and polarised along $\uect{p}$
\begin{gather}
 \rin_{\vect{\hat{p}}}(\vect{r},\vect{\hat{k}}_\mathrm{sc})
 = r_e^2 \vect{\hat{p}}\tp \vect{\irr}_\mathrm{in}(\vect{r})\vect{\hat{p}}
 \nonumber\\
 = \frac{r_e^2}{2} \int_{\Omega(r)} \rad(\cos\zdis)\;
 [1-(\vect{\hat{p}}\tp\uect{k}\inc)^2] \;d\Omega(\uect{k}\inc)
\label{Rin_sca2}\end{gather}
To evaluate the integral
we maintain the geometry in Fig.~\ref{Fig:ScaGeom3D}. But now the situation is
slightly complicated by the additional direction $\vect{\hat{k}}_\mathrm{sc}$
of the scattered beam from the scattering site S towards the observer in the
$\uect{x},\uect{z}$ plane. The scattered beam makes the angle $\bar{\chi}$
with the radial vector $\vect{\hat{z}}$ from solar centre to the scattering
site. We will call this angle the mean scattering angle and the
$\uect{x},\uect{z}$ plane is the mean scattering plane. The actual scattering
angle for a photon from some point E on the solar surface scattered at S to
the observer could largely deviate from $\bar{\chi}$, but we will not need the
actual angle explicitly in the calculation, at least for simple Thomson
scatter.

For the observer looking in direction $-\vect{\hat{k}}_\mathrm{sc}$ a natural
base for his polariser orientation is $\uect{p}_\mathrm{tan}$,
$\uect{p}_\mathrm{rad}$ defined such that they form a right-handed orthogonal
system with $-\vect{\hat{k}}_\mathrm{sc}$ as third direction. We define
generally
\begin{equation}
  \uect{p}_\mathrm{tan}=\frac{\uect{k}\sca\times\uect{r}}{\sin\bar{\chi}},
  \qquad
  \uect{p}_\mathrm{rad}=\uect{p}_\mathrm{tan}\times\uect{k}\sca
  =\frac{(\1-\uect{k}\sca\uect{k}\tp\sca)\uect{r}}{\sin\bar{\chi}}
\label{PolBase0}\end{equation}
In the Cartesian coordinates of Fig.~\ref{Fig:ScaGeom3D} we have
\[
  -\vect{\hat{k}}_\mathrm{sc}
 =\rveccc{-\sin\bar{\chi}}{0}{-\cos\bar{\chi}},
 \quad
  \vect{\hat{p}}_\mathrm{tan}
 =\rveccc{0}{-1}{0},
 \quad
  \vect{\hat{p}}_\mathrm{rad}
 =\rveccc{-\cos\bar{\chi}}{0}{\sin\bar{\chi}}
\]
From the observer's point of view $\vect{\hat{p}}_\mathrm{tan}$ is always
tangential to the solar surface and $\vect{\hat{p}}_\mathrm{rad}$ always
points away from the projected centre of the Sun. Note the mean scattering
angle $\bar{\chi}$ varies from 0 (forward scatter) to $\pi$ (backscatter). The
observed polarised irradiance components can now easily be determined from
(\ref{Rin_sca2}) and (\ref{Irr_in_Minnaert})
\begin{align}
 \rin_\mathrm{tan}(\vect{r},\vect{\hat{k}}_\mathrm{sc})
 &= r_e^2 \;\vect{\hat{p}}_\mathrm{tan}\tp\vect{\irr}_\mathrm{in}(\vect{r})
         \vect{\hat{p}}_\mathrm{tan}
  = r_e^2 \;Q_{\mathrm{in},yy}(r)
  = \frac{\pi \rad_\odot r_e^2}{2}\,((1-u)C(r) + uD(r))
\label{Irr_tan_Minnaert}\\         
 \rin_\mathrm{rad}(\vect{r},\vect{\hat{k}}_\mathrm{sc})
 &= r_e^2 \;\vect{\hat{p}}_\mathrm{rad}\tp\vect{\irr}_\mathrm{in}(\vect{r})
         \vect{\hat{p}}_\mathrm{rad}
  = r_e^2 \;(Q_{\mathrm{in},xx}(r)\cos^2\bar{\chi}
         +Q_{\mathrm{in},zz}(r)\sin^2\bar{\chi})
\nonumber\\         
 &=\frac{\pi \rad_\odot r_e^2}{2}\,
          \big[(1-u)C(r) + uD(r))
             - (1-u)A(r) + uB(r))\sin^2\bar{\chi}\big]
\label{Irr_rad_Minnaert}\end{align}
Keeping the projection properties of Thomson scattering in mind,
(\ref{Irr_tan_Minnaert}) and (\ref{Irr_rad_Minnaert}) are intuitively clear:
$\rin_\mathrm{tan}$ measures the polarisation normal to the scattering plane
and is independent from the mean scattering angle and proportional to the solar
irradiance at the scattering site polarised in this direction.
$\rin_\mathrm{rad}$ measures the polarisation in the scattering plane and is
proportional to the local irradiance projected normal to the line-of-sight in
the scattering plane. Expressions similar to (\ref{Irr_tan_Minnaert}) and
(\ref{Irr_rad_Minnaert}) were already mentioned by \citet[][eqs 14 and
  15]{vandeHulst:1950}. In the literature the
polarised and total components of the radiant intensity (recall that
$\irr_{\mathrm{in},xx}=\irr_{\mathrm{in},yy}$, eq.~\ref{Irr_xy_Minnaert})
\begin{gather*}
\rin_\mathrm{pol}=\rin_\mathrm{tan}-\rin_\mathrm{rad}
  = r_e^2 (\irr_{\mathrm{in},xx}-\irr_{\mathrm{in},zz})\sin^2\bar{\chi}
\\  
\rin_\mathrm{tot}=\rin_\mathrm{tan}+\rin_\mathrm{rad}
  = 2 r_e^2 \irr_{\mathrm{in},xx} - \rin_\mathrm{pol}
\end{gather*}
are often preferred instead of (\ref{Irr_tan_Minnaert}) and 
(\ref{Irr_rad_Minnaert}). 
Recall that Thomson scattering contributes only very little to the
complication of these expressions. The Minnaert's coefficients already arise
in (\ref{Irr_in_Minnaert}) when the irradiance of the solar light is
calculated at the scattering site.

It is instructive to consider the case of a single beam from the centre of the
solar disk. This corresponds to the limit
$r\rightarrow\infty$ of the expressions (\ref{Irr_tan_Minnaert}) and
(\ref{Irr_rad_Minnaert}) such that the finite solid angle $\Omega(r)$
subtended by the solar disk shrinks to zero.
In this case $\vect{\hat{k}}_\mathrm{in}=\vect{\hat{z}}$ and the
beam produces radiant intensities (\ref{Irr_pQinp})
with polarisations
\begin{gather}
   d\rin_\mathrm{tan}
   = r_e^2 \vect{\hat{p}}_\mathrm{tan}\tp d\vect{\irr}_\mathrm{in}
           \vect{\hat{p}}_\mathrm{tan} 
   = \frac{r_e^2}{2} d\irr_\mathrm{in}
           (1-(\vect{\hat{k}}_\mathrm{in}\tp\vect{\hat{p}}_\mathrm{tan})^2)
   = \frac{r_e^2}{2} d\irr_\mathrm{in}
\nonumber\\   
   d\rin_\mathrm{rad}
   = r_e^2 \vect{\hat{p}}_\mathrm{rad}\tp d\vect{\irr}_\mathrm{in}
           \vect{\hat{p}}_\mathrm{rad} 
   = \frac{r_e^2}{2} d\irr_\mathrm{in}
           (1-(\vect{\hat{k}}_\mathrm{in}\tp\vect{\hat{p}}_\mathrm{rad})^2)
\nonumber\\   = r_e^2 d\irr_\mathrm{in}
           (1-(\vect{\hat{z}}\tp\vect{\hat{p}}_\mathrm{rad})^2)
   = \frac{r_e^2}{2} d\irr_\mathrm{in} \cos^2\chi
\nonumber\intertext{which gives a polarisation degree of}
P =\frac{d\rin_\mathrm{pol}}{dI_\mathrm{tot}}
  =\frac{d\rin_\mathrm{tan}-dI_\mathrm{rad}}{dI_\mathrm{tan}+dI_\mathrm{rad}}
  =\frac{\sin^2\chi}{1+\cos^2\chi}
\label{PolDegree}\end{gather}
Note that $\uect{k}\inc\tp\vect{\hat{p}}_\mathrm{tan}=0$ because $\uect{k}\inc$
is normal to the scattering plane.
As we pointed out above the same polarisation degree is approached by
(\ref{Irr_tan_Minnaert}) and (\ref{Irr_rad_Minnaert}) for large distances $r$
from the Sun except that $\chi$ is replaced by $\bar{\chi}$. The reason is, as
we have seen in (\ref{Irr_zz_inf}), that $C-A$ and $D-B$ decrease rapidly 
with $r$.
The asymptotically scattered radiant intensity is obtained if we
insert the asymptotic value (\ref{Irr_inf}) for $d\irr_\mathrm{in}$.
From (\ref{Irr_xy_inf}) and (\ref{Irr_zz_inf}) it is clear,
that the power then resides only in the elements
$\irr_{\mathrm{in},xx}=\irr_{\mathrm{in},yy}$ of the
irradiance matrix and
\begin{equation}
   d\rin_\mathrm{tan} = \frac{\pi r_e^2}{2}(\frac{R_\odot}{r})^2\bar{L}
   ,\qquad
   d\rin_\mathrm{rad} = \frac{\pi r_e^2}{2}(\frac{R_\odot}{r})^2\bar{L}
    \cos^2\chi 
\label{Rad_inf}\end{equation}

The final radiant flux into the pixel of an ideal instrument is obtained as in
(\ref{dPow_volsca}) by multiplying the radiant intensity
(\ref{Irr_tan_Minnaert}) or (\ref{Irr_rad_Minnaert}) scattered from each
electron with the number of electrons $N_e(\vect{r})$, integrating over the
line-of-sight and multiplying the appropriate instrument
parameters
\begin{equation}
  \pow_\mathrm{p}(\vect{r}_\mathrm{obs})
  =\frac{A_\mathrm{pixel}A_\mathrm{aperture}}{f^2}
     \int_\mathrm{LOS} \rin_\mathrm{p}(\vect{r},\vect{\hat{k}}_\mathrm{sc})
     \;N_e(\vect{r})\,d\ell
  \qquad[\mathrm{W}]
\label{pow_LOS}\end{equation}
where $\vect{r}=\vect{r}_\mathrm{obs}-\ell\vect{\hat{k}}_\mathrm{sc}$
in the integrand has to be treated as a function of distance $\ell$
for fixed $\vect{r}\obs$. 
The subscript p in (\ref{pow_LOS}) stands for the polarisations
``tan'' or ``pol'' or any linear combination of the two, like
$\pow_\mathrm{rad}
= \pow_\mathrm{tan}-\pow_\mathrm{pol}$ or 
$\pow_\mathrm{tot}
=\pow_\mathrm{tan} + \pow_\mathrm{rad}
=2\pow_\mathrm{tan} - \pow_\mathrm{pol}$. 
An assumption made in (\ref{pow_LOS}) is that the scattering is incoherent
\footnote{The term ``incoherent'' scatter is used slightly differently in
  different communities: in the strict sense, the scattering is incoherent if
  the power of the scattered wave scales linearly with the number $N_e$ of
  scattering electrons because their positions are sufficiently random. In
  laboratory plasma physics Thomson scattering is termed coherent when the wave
  length of the scattered wave exceeds the Debye length. Then two-particle
  correlations in (\ref{Irr_Qsc}) start to matter due to the plasma response
  to thermal field fluctuations. However in thermal equilibrium the total
  cross section is still $\propto N_e$.},
i.e., two-electron correlations in
$\bra\vect{E}_\mathrm{sc}\vect{E}^{*\mathsf{T}}_\mathrm{sc}\ket$ of
(\ref{Irr_Qsc}) average off because of random phase relations between the
electric field scattered at different randomly positioned particles. In this
case we can treat the contribution of each electron to the scattered power
independently. In the corona this assumption is well justified for white-light
wavelengths.


\begin{figure}
\hspace*{\fill}
\includegraphics[viewport=20 20 500 510, clip, width=7cm]{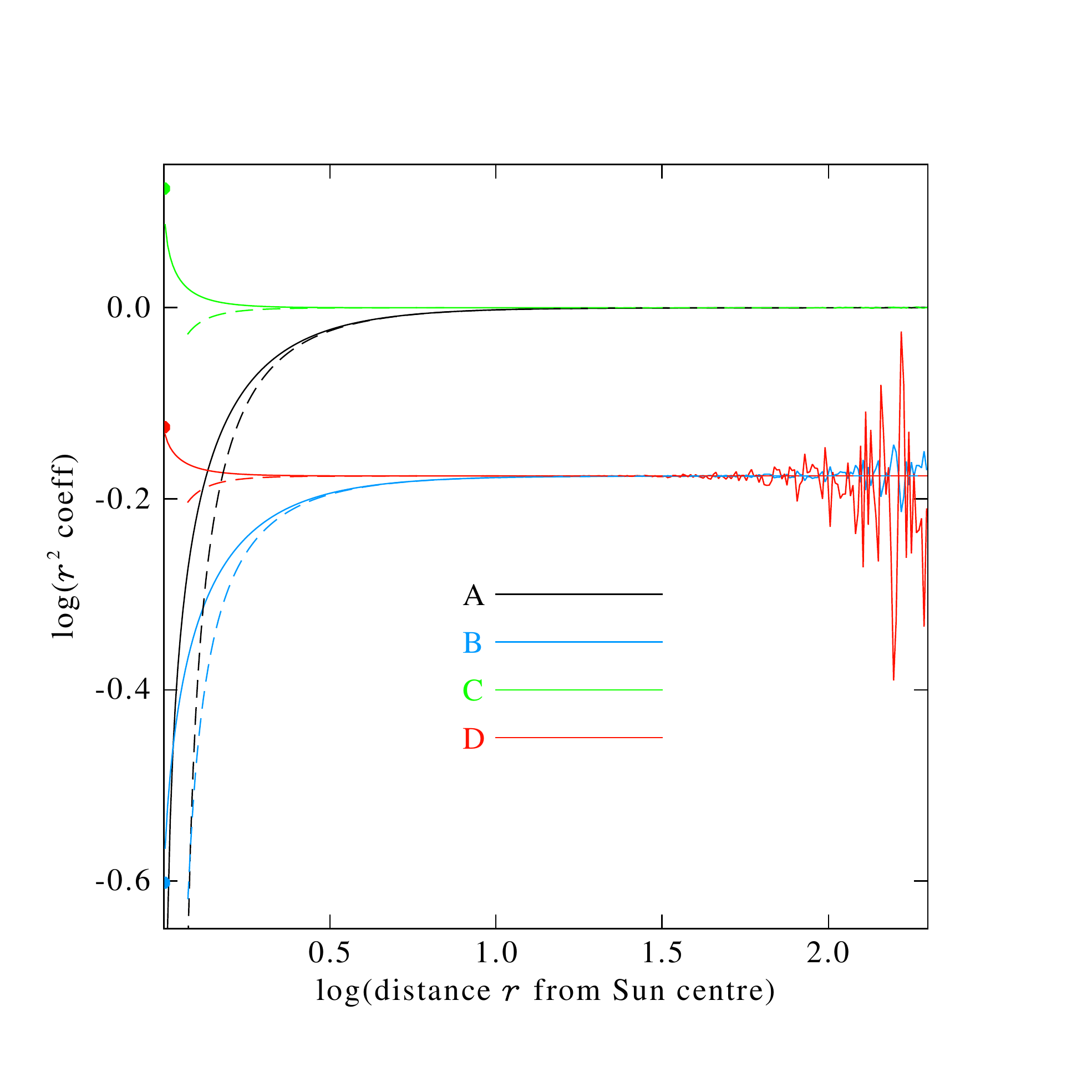}
\hspace{2em}
\includegraphics[viewport=20 20 500 510, clip, width=7cm]{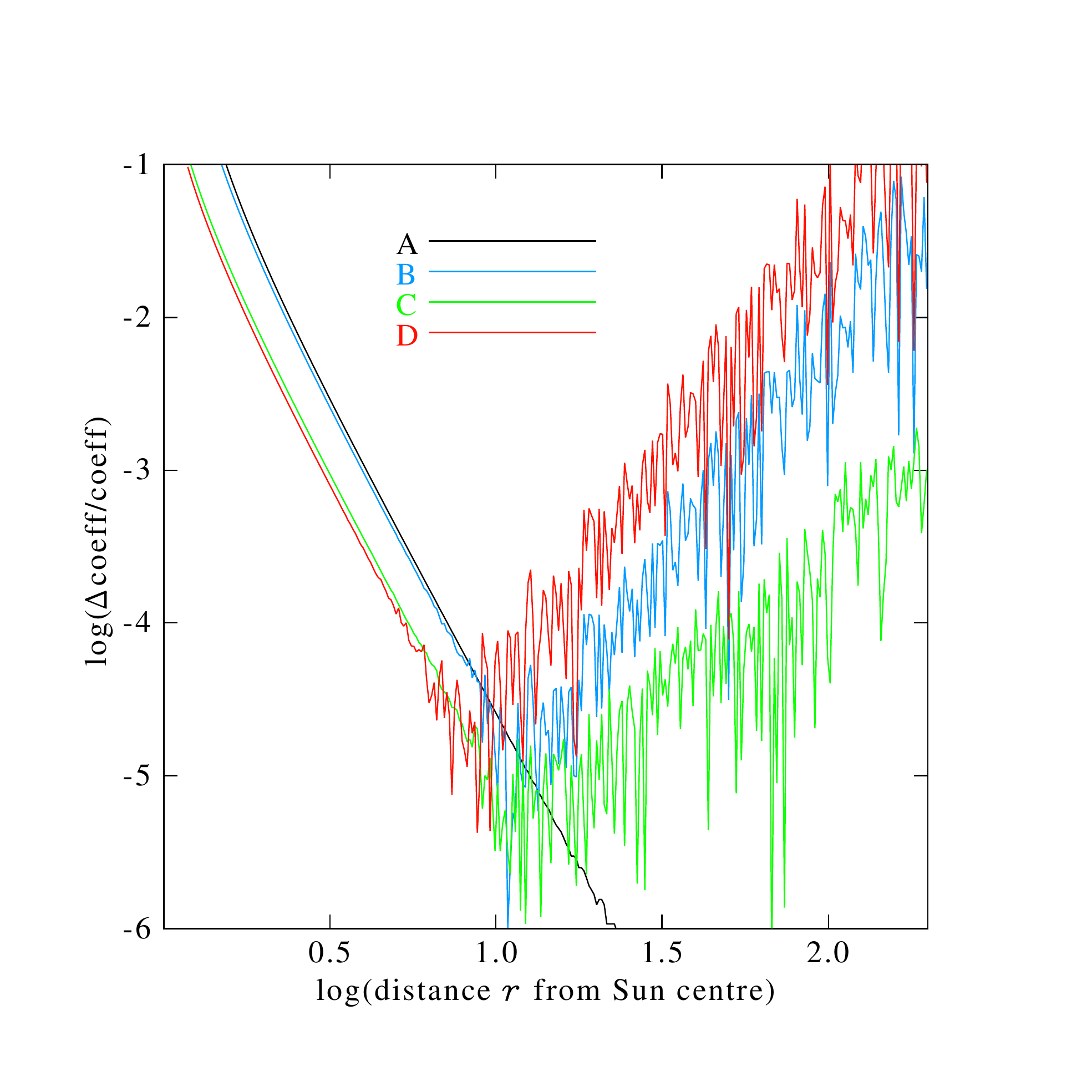}
\hspace*{\fill}
\caption{Radial variation of Minnaert's coefficients $A$, $B$, $C$ and $D$
  with distance from the solar centre. Asymptotically, the coefficients
  decrease as $r^{-2}$, the left diagram therefore shows the coefficients
  multiplied with $r^2$. The dashed curves represent the respective asymptotic
  expansions. The coloured dots on the ordinate represents the respective
  value for $r=R_\odot$ except for $A(r=R_\odot)=0$. In the right diagram we
  display the logarithm of the absolute relative difference between the
  coefficients calculated in single precision and the asymptotic approximation.
\label{Fig:MinnaertCoeff}}\end{figure}

In the left diagram of Fig.~\ref{Fig:MinnaertCoeff} we show the variation of
Minnaert's coefficients with distance $r$. For comparison, their respective
asymptotic series expansion as derived in 
chapter~\ref{app:LimitInfty} of the appendix is also displayed (dashed). 
For large $r$ all coefficients have to decrease as $r^{-2}$ to reproduce the
analogous decrease of the solar radiation at distances for which the Sun can be
treated as a point source. At these large distances, the coefficients,
especially 
$B$ and $D$, are prone to numerical roundoff errors. These numerical errors are
even more dominant when the radial polarisation $\rin_\mathrm{rad}$ is
calculated for which the combinations $C-A$ and $D-B$ of the coefficients are
required which decrease as $r^{-4}$. In the right diagram we show the
relative error between the coefficients and their asymptotic expansions. The
error below $r\simeq 10 R_\odot$ is caused by the insufficient expansion which
takes only the two lowest order terms into account. Above $r\simeq 10 R_\odot$
the error is due to the numerical instability of the full expressions for the
coefficients $B$, $D$ and $C$ (here calculated with single precision). 
In view of the fact that the asymptotic terms are also much simpler to 
calculate, we suggest to switch to the latter when the argument $r$ exceeds 
about $10\;R_\odot$.

\begin{figure}
\hspace*{\fill}  
\includegraphics[viewport=20 10 681 369, clip, width=10cm]{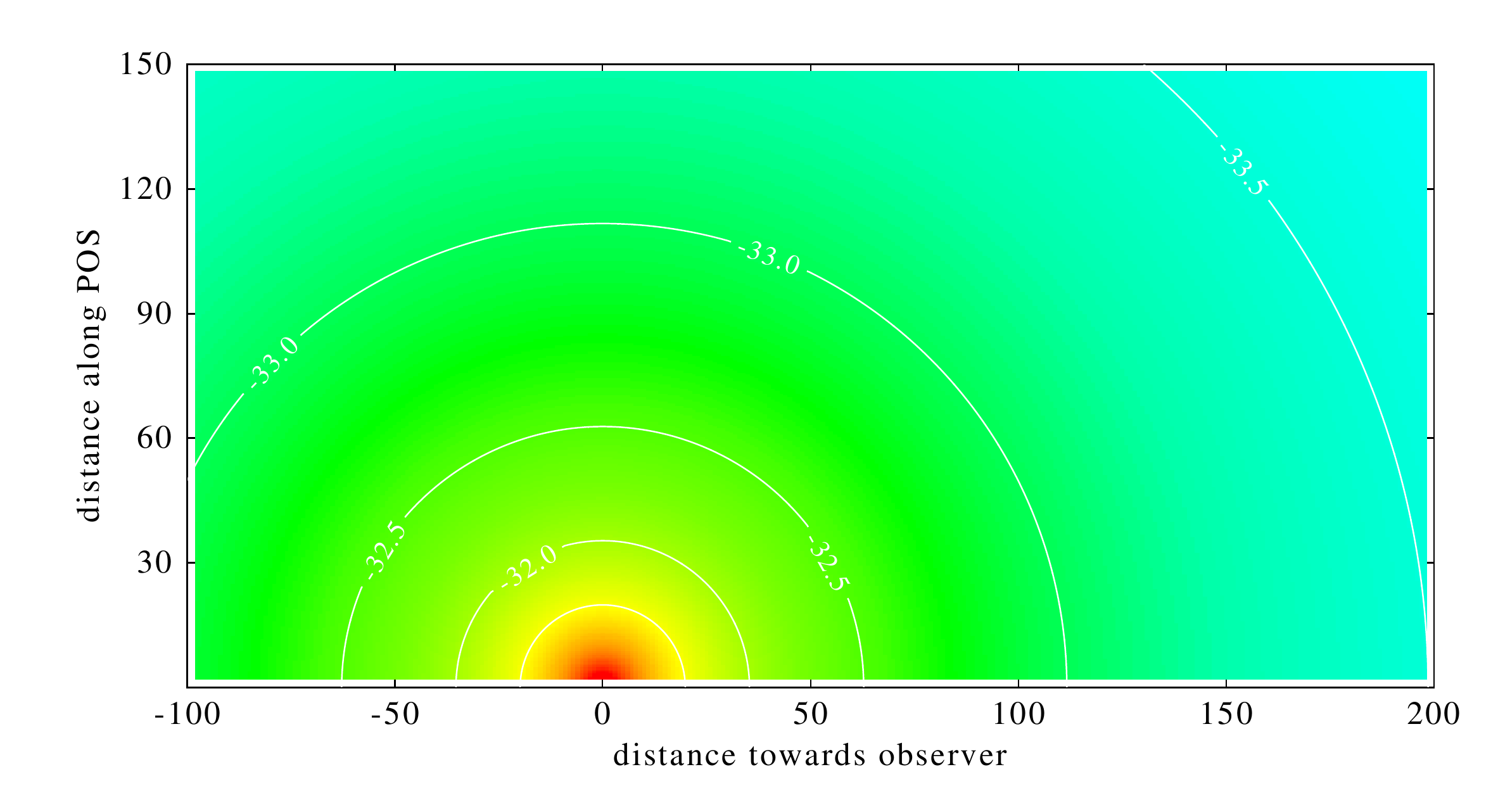}
\hspace*{\fill}\\\hspace*{\fill}  
\includegraphics[viewport=20 10 681 369, clip, width=10cm]{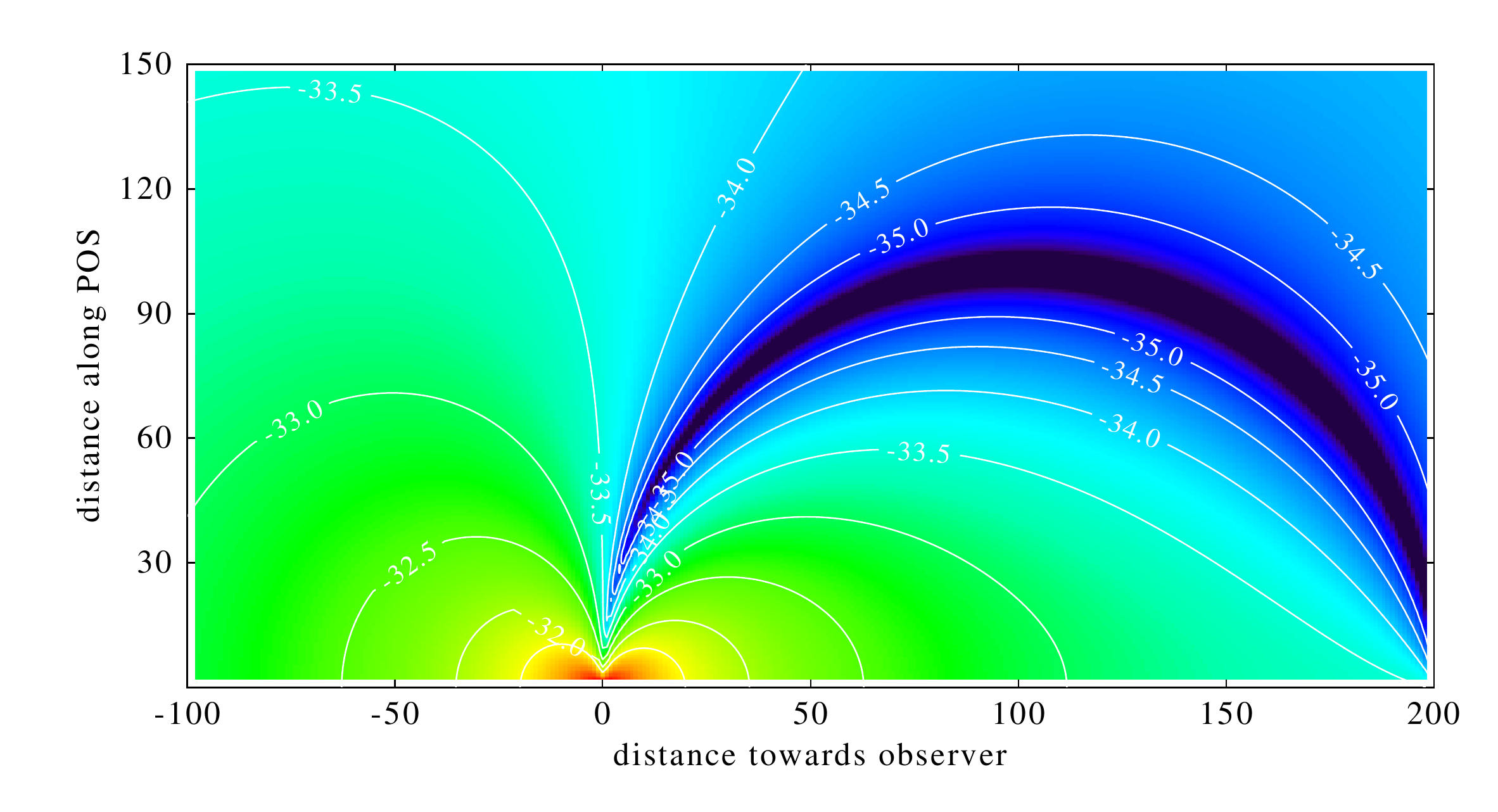}
\hspace*{\fill}
\caption{
  Radiant intensities $\rin_\mathrm{tan}$ (top) and $\rin_\mathrm{rad}$
  (bottom) Thomson scattered from a single electron and polarised in 
  tangential and radial direction. The distribution of the radiant intensities
  is displayed in the mean scattering plane according to
  (\ref{Irr_tan_Minnaert}) and (\ref{Irr_rad_Minnaert}), respectively.
  The Sun is at the origin, the
  spatial units are $R_\odot$, the observer is assumed at (200,0) $R_\odot$.
  The contour labels refer to the decadic logarithm of the radiant intensity
  [W/sr] for a unit mean solar radiance, i.e., setting
  $\rad_\odot=(1-u/3)^{-1}$ W/m$^2$/sr.}
\label{Fig:Rin_tanrad}\end{figure}

\begin{figure}
\hspace*{\fill}  
\includegraphics[viewport=20 10 681 369, clip, width=10cm]{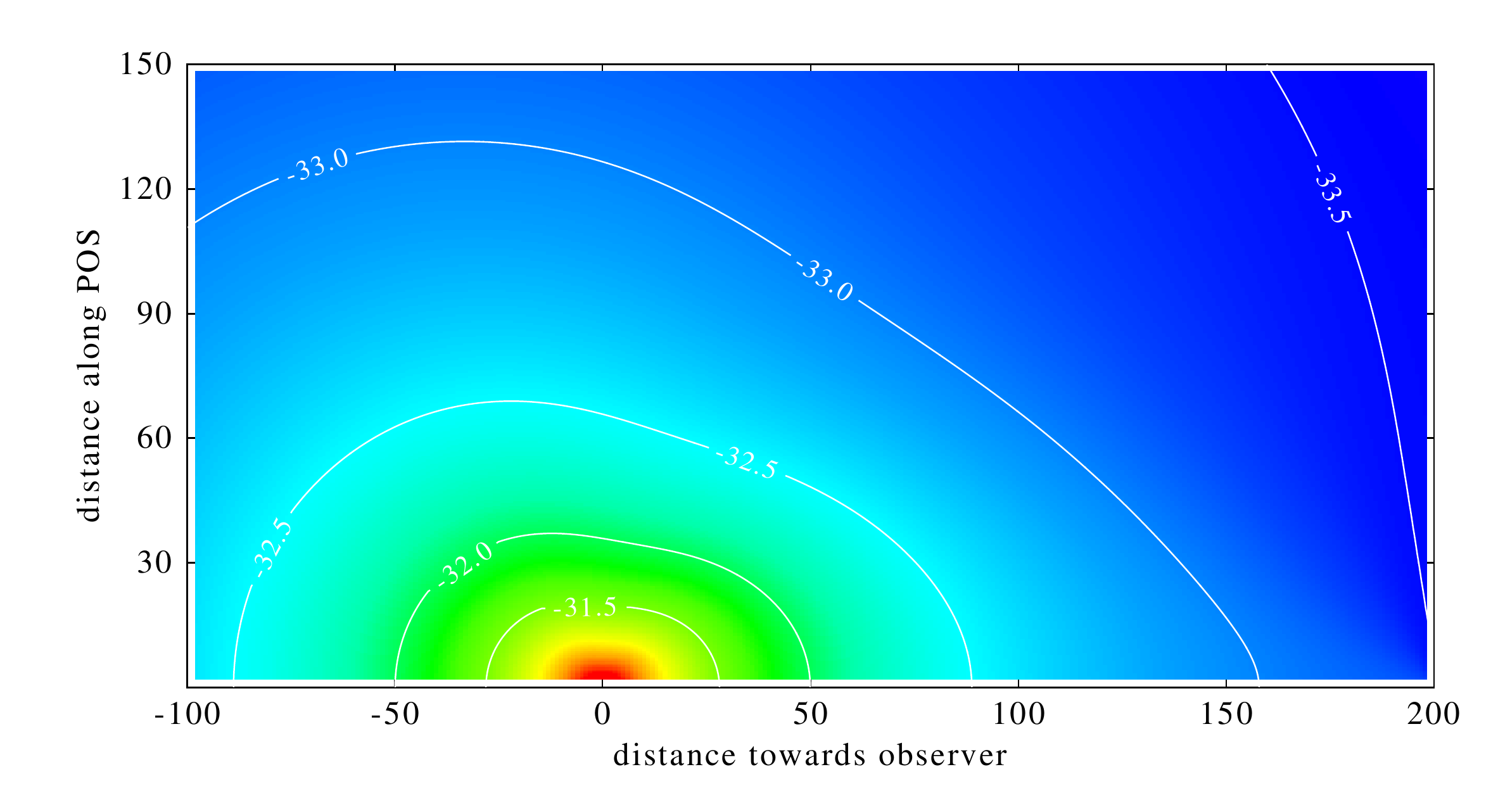}
\hspace*{\fill}\\\hspace*{\fill}  
\includegraphics[viewport=20 10 681 369, clip, width=10cm]{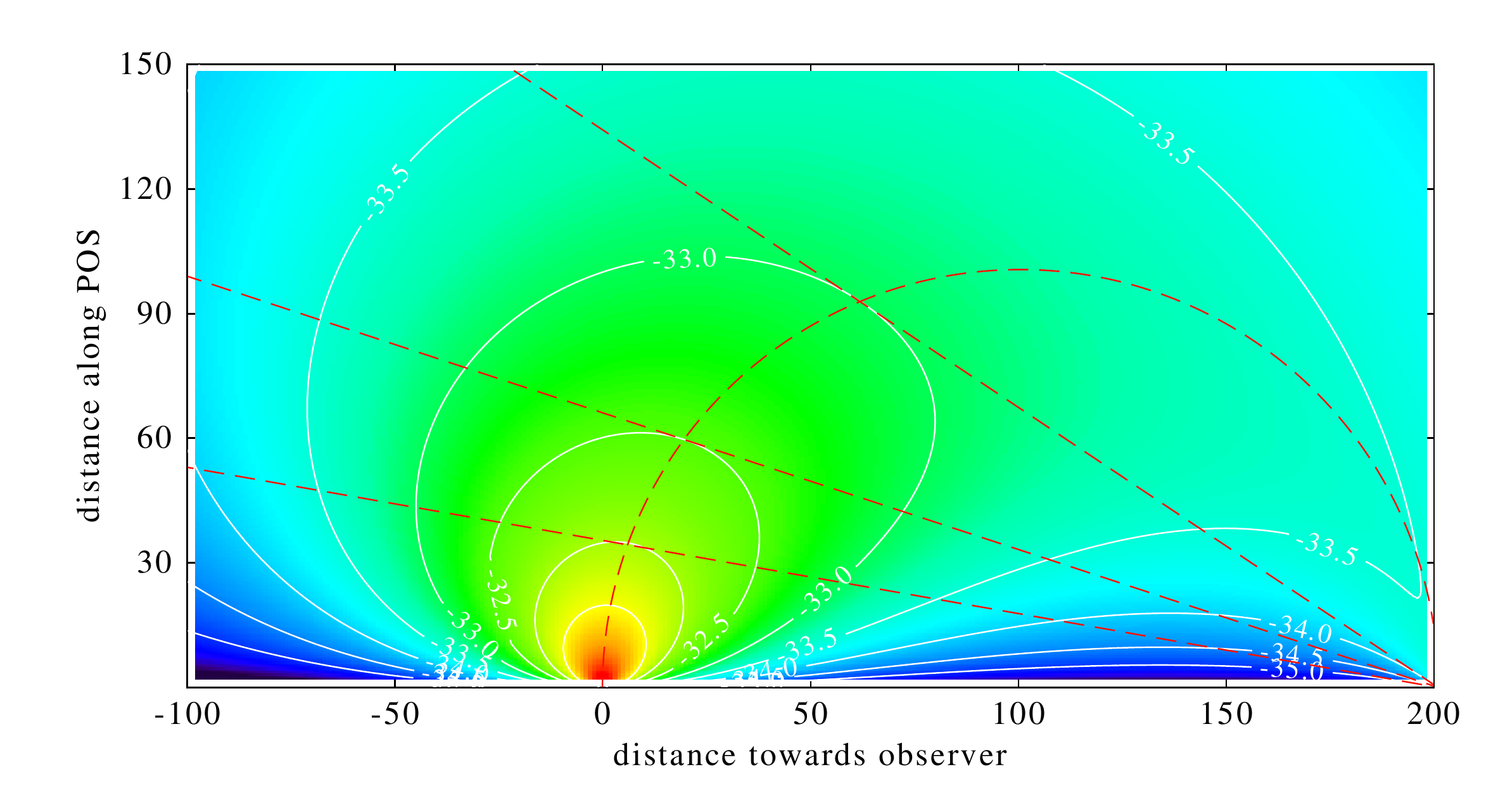}
\hspace*{\fill}
\caption{Total and polarised radiant intensities
  $\rin_\mathrm{tan}+\rin_\mathrm{rad}$ and
  $\rin_\mathrm{tan}-\rin_\mathrm{rad}$, respectively, Thomson scattered from
  a single electron. Parameters are as in Fig.~\ref{Fig:Rin_tanrad}, in
  particular the Sun is at the origin, the spatial units are $R_\odot$ and the
  observer is assumed at (200,0) $R_\odot$.}
\label{Fig:Rin_totpol}\end{figure}

In Fig.~\ref{Fig:Rin_tanrad} we show the spatial distribution of the radiant
intensity scattered from a single electron in the mean scattering plane around
the Sun. The contours of the tangential component
$\rin_\mathrm{tan}(\vect{r})$ (upper panel) are concentric around the Sun 
because it does not depend on the mean scattering angle $\bar{\chi}$ but 
only on the radial distance $r$.
This is different for $\rin_\mathrm{rad}(\vect{r})$ (lower panel) which for
$\bar{\chi}=\pi/2$ only represents the small radial $\irr_{\mathrm{sc},zz}$ 
component (see
eqs.~ \ref{Irr_xy_Minnaert} and \ref{Irr_xz_Minnaert}) of the Sun's incident
irradiance. The latter is due to the Sun's finite apparent size and
$\irr_{\mathrm{sc},zz}$ rapidly decreases with distance $r$. The condition 
$\bar{\chi}=\pi/2$ defines the ``Thomson scattering sphere''
\citep{VourlidasHoward:2006}. In the spatial distribution $\rin_\mathrm{rad}$ 
this surface is  marked by a deep depletion.

This minimum of $\rin_\mathrm{rad}$ is filled by forming combinations of 
$\rin_\mathrm{tan}$ and $\rin_\mathrm{rad}$ which are shown in
Fig.~\ref{Fig:Rin_totpol}. Here we display the combinations
$\rin_\mathrm{rad}\pm\rin_\mathrm{rad}$, the total ($+$) and the polarised 
($-$) components of the radiant intensity. For
$\rin_\mathrm{tot}$, the depletion of $\rin_\mathrm{rad}$
at $\bar{\chi}=\pi/2$ leads to a flattening of the contours at this 
mean scattering angle, while $\rin_\mathrm{pol}$ is 
naturally suppressed in the forward and backward scattering direction, i.e. at 
$\bar{\chi}=0$ and $\pi/2$. At these scattering angles the tangential and
radial polarisations become indistinguishable and $\rin_\mathrm{tan}$ and
$\rin_\mathrm{rad}$ become equal. 
A halo-CME might be considerably dimmed in polarised brightness 
relative to the near-Sun coronal background and it should be much better 
visible in radial polarisation.
The exact limits $\bar{\chi}=0$ or $\pi$ are, of course, hidden behind a 
coronagraph by the occulter. 

When the observer receives a signal from a certain line-of-sight he in general
does not know how $N_e(\vect{r}(\ell))$ is distributed. With the
distribution of the radiant intensity as the only a-priori information at
hand, we might be tempted to argue that the most probable scattering site is
where $\rin(\vect{r}(\ell))$ maximises for this line-of-sight as function of
$\ell$. It is easily confirmed that the radiant intensity for all
polarisations except $\rin_\mathrm{rad}$ maximises at a point near
$\bar{\chi}=\pi/2$. The locus of these points agrees therefore
with the Thomson sphere mentioned above. In
the bottom diagram of Fig.~\ref{Fig:Rin_totpol} we have drawn as an example
three such lines-of-sights as dashed lines from the observer at (200,0) and the
Thomson sphere as a dashed circle. The line-of-sight beams
tangentially touch the contour of their highest radiant intensity
at the Thomson sphere. In this way,
$\bar{\chi}=\pi/2$ is defined graphically as the locus of the largest
scattering probability on each individual line-of-sight. From the
distance between the line-of-sight and its maximum intensity contour
at $\bar{\chi}=\pi/2$ we can infer the slow decrease of
the radiant intensity from its maximum along a line-of-sight. It is
obviously much slower for $\rin_\mathrm{tot}$ than for $\rin_\mathrm{pol}$
as was noted by \cite{HowardDeforest:2012}. They
also pointed out that the reason for the maximum of the radiant intensity on
the Thomson sphere is its comparatively rapid radial decrease which can be
traced back
to the $r$-dependence of Minnaert's coefficients. It therefore does not 
reflect any peculiarity of Thomson scattering.

It is difficult to say how much significance the Thomson sphere has for
practical observations. It has to be kept in mind, that the respective radiant
intensity per electron must still be multiplied with the density $N_e$ before
the line-of-sight signal can be integrated. The fact that the density can vary
by orders of magnitude considerably reduces the relevance of the 
line-of-sight variation of the radiant intensity. Moreover, the radiant
intensity also decreases in radial direction. A plasma cloud propagating away
from the Sun with $\bar{\chi}$ well away from $\pi/2$ produces a somewhat
attenuated scattering signal and will therefore be visible out to a slightly
shorter radial distance until its brightness contrast is drowned in noise.
A propagation well off the Thomson sphere will therefore probably not 
completely prevent the detection of the cloud.

\section{Integration of simple coronal density models}
\setcounter{equation}{0}

\begin{figure}
\hspace*{\fill}
\parbox{10cm}{\includegraphics[width=10cm]{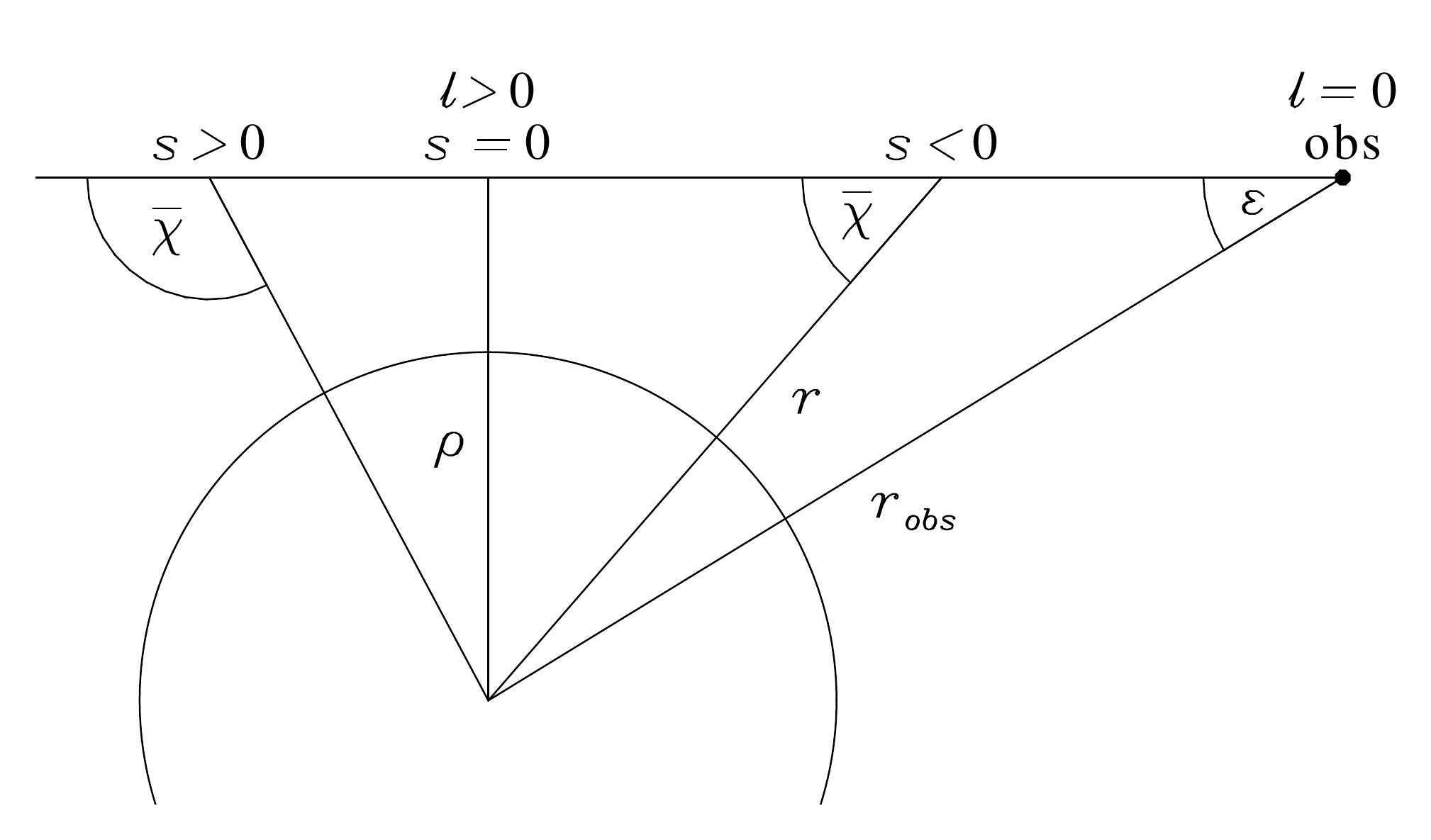}}
\parbox{6cm}{\caption{Geometry of the line-of-sight.
  We use two alternative length parameters along the line-of-sight: $\ell$ is
  the distance from the observer, $s$ is the distance from the point of
  closest approach of the line-of-sight to the Sun.
  The angle $\bar{\chi}$ is the mean scattering angle at the scattering
  site $\vect{r}$. \label{Fig:ScaGeom_LOS}}}
\hspace*{\fill}
\end{figure}

In some simple cases, the line-of-sight integrals (\ref{pow_LOS}) can be
calculated analytically but in most practical cases numerical methods are
needed. Either way, the line-of-sight integration should be arranged
suitably. For that purpose we replace the line-of-sight parameter
$\ell$ by a new parameter $s$ which measures the (signed) geometrical distance
along the line-of-sight from the point of closest approach to the solar
centre at $\bar{\chi}=\pi/2$. The new line-of-sight parameter
$s=\ell+s_\mathrm{obs}$ ranges from $s=s_\mathrm{obs}<0$ to
$s\rightarrow +\infty$.
The distance of the line-of-sight at $s=0$ from the solar centre is $\rho$. 
Then we have the following relations between $s$ and $\bar{\chi}$ along a
line-of-sight specified by either $\rho$ or $\varepsilon$
(see Fig.~\ref{Fig:ScaGeom_LOS}) 
\begin{equation}
  \begin{gathered}
  r=\sqrt{\rho^2+s^2},\qquad
  s=-r\cos\bar{\chi},\qquad
  \rho=r\sin\bar{\chi}\\
  \text{especially},\;
  s_\mathrm{obs}=-r_\mathrm{obs}\cos\varepsilon,\qquad
  r_\mathrm{obs}=\rho/\sin\varepsilon
  \end{gathered}
\label{rsrhochi}\end{equation}
Here $r_\mathrm{obs}$ is the distance of the observer from the Sun centre and
$\varepsilon$ is the elongation of the line-of-sight path as seen from the 
observer. In (\ref{pow_LOS}) we have to read $r$ and
$\bar{\chi}$ as functions of $s$ and the line-of-sight constant $\rho$
\[
  r(s)=\sqrt{\rho^2+s^2},\qquad
  \bar{\chi}(s)=\acos(\frac{s}{r(s)})
\]
For practical calculations we want avoid the infinite upper integration
boundary of $s$. We therefore substitute $s\longrightarrow \bar{\chi}$
as integration variable.
The Jacobian of this variable transformation is
\begin{equation}
\frac{ds}{d\bar{\chi}}
= - (\frac{dr}{d\bar{\chi}})\cos\bar{\chi} + r\sin\bar{\chi}
= - \rho(\frac{d}{d\bar{\chi}}\frac{1}{\sin\bar{\chi}})\cos\bar{\chi} + \rho
= \rho(\frac{\cos^2\bar{\chi}}{\sin^2\bar{\chi}} + 1)
= \frac{\rho}{\sin^2\bar{\chi}}
\label{dsdchi}\end{equation}
We insert the Minnaert expressions (\ref{Irr_tan_Minnaert}) and
(\ref{Irr_rad_Minnaert}) for the radiant intensities in the integrand of
(\ref{pow_LOS}) with $r(\bar{\chi})=\rho/\sin\bar{\chi}$ and use the above
variable transformation to express $N_e(\vect{r}(\ell))=N_e(r,\bar{\chi})$. 
Then the observed radiant fluxes (\ref{pow_LOS}) into a pixel 
in various polarisations read
\begin{gather}
  \pow_\mathrm{p}
= \frac{A_\mathrm{aperture}A_\mathrm{pixel}}{f^2} \;
  \frac{\pi r_e^2 \rad_\odot}{2}
  K_\mathrm{p}(\rho)
\label{Phi_K}\quad\text{for p=``tan'', ``rad'', ``pol'', ``tot''}\\    
K_\mathrm{tan}(\rho) = \rho  \int_{\varepsilon}^{\pi} 
    N_e(r,\bar{\chi})\;((1-u)C(r)+uD(r))\;\frac{d\bar{\chi}}{\sin^2\bar{\chi}}
\qquad[\mathrm{1/m^2}]
\nonumber\\
K_\mathrm{pol}(\rho) = \rho  \int_{\varepsilon}^{\pi} 
    N_e(r,\bar{\chi})\;((1-u)A(r)+uB(r))\;d\bar{\chi}
\nonumber\\
K_\mathrm{tot}  = 2K_\mathrm{tan}-K_\mathrm{pol},\qquad
K_\mathrm{rad}  =  K_\mathrm{tan}-K_\mathrm{pol}
\nonumber\\
\text{where}\quad r(\bar{\chi})=\frac{\rho}{\sin\bar{\chi}}\quad
\text{has to be used in the integrands}
\nonumber\end{gather}
These equations will serve as the basis for the numerical and
analytical line-of-sight integrations below.

\subsection{Axially symmetric coronal density}

To simplify the above integrations in (\ref{Phi_K}) we will make two
assumptions in this section: \\
\hspace*{2em} 1) \parbox[t]{14cm}{$N_e(r,\bar{\chi})$ only depends on the
  distance from the solar centre $r$}\\
\hspace*{2em} 2) \parbox[t]{14cm}{we approximate the lower integration
  boundary in (\ref{Phi_K}) by $\varepsilon=0$ assuming an infinite
  distance $r_\mathrm{obs}$ of the observer.}\\[0.5ex]
Condition 1 may be relaxed in that the dependence $N_e(r)$
may differ in each azimuthal scattering plane. However, even this assumption
is unrealistic for the coronal density distribution at a given time. Yet it
sometimes may be meaningful when a long time mean or background density is
considered \citep{HayesEtal:2001,QuemeraisLamy:2002}.
Since (\ref{pow_LOS}) is linear in density $N_e$, any averaging of
coronagraph images is likewise implicitly applied to the density $N_e(r)$.
This way image data can be generated for which the averaged $N_e$ may come
close to condition 1.

Condition 2 may be satisfied by an appropriate combination of measured data if
we add to the radiant flux per pixel $\pow(\varepsilon)$ observed
at finite distance $r_\mathrm{obs}$ the data from
$\pow(\pi-\varepsilon)$ observed in the exact opposite direction.
Since the dependence of the integrand on the mean
scattering angle $\bar{\chi}$ is only through
$\sin\bar{\chi}=\sin(\pi-\bar{\chi})$, the sum of these observations
represents the integral over the entire line-of-sight from $s=-\infty$ to
$+\infty$ or from a virtual space craft at infinite distance from Sun made at
$\varepsilon\simeq 0$. In most cases, e.g., for observations from a distance 
of 1 AU, these ``anti-sunward'' observations $\pow(\pi-\varepsilon)$
will not be considered to make a significant contribution and condition 2 is
therefore not very restrictive.

Accepting the above assumptions, the integration (\ref{Phi_K}) becomes
equivalent to an Abel transformation we denote by $f(\rho)=\mathcal{A}(g(r))$
(see chapter~\ref{App:AbelTrans} in the appendix).
Multiplying $1=(\rho/r\sin\bar{\chi})^2$ to the integrand of $K_\mathrm{pol}$
of (\ref{Phi_K}), we can write the line-of-sight integrals as
\begin{gather}
  K_\mathrm{tan}(\rho)
  =\mathcal{A}[N_e(r)\;((1-u)C(r)+uD(r))]
\label{Ktan_rho}\\    
  K_\mathrm{pol}(\rho)
  =\;\rho^2\; \mathcal{A}[N_e(r)\;\frac{(1-u)A(r)+uB(r)}{r^2}]
\label{Kpol_rho}\end{gather}
The inversion of the two above Abel transforms independently 
yields the density $N_e(r)$ from the two polarisation
observations $K_\mathrm{tan}$ and $K_\mathrm{pol}$ by
\begin{gather}
  N_e(r)
=\frac{1}{(1-u)C(r)+uD(r)}\;\mathcal{A}^{-1}[K_\mathrm{tan}(\rho)]
\nonumber\\ \hspace*{8ex}
=\frac{r^2}{(1-u)A(r)+uB(r)}
 \;\mathcal{A}^{-1}[\frac{K_\mathrm{pol}(\rho)}{\rho^2}]
\label{UniqueN}\end{gather}
In principle, this relation could serve as a
check of the assumption that $N_e$ has spherical symmetry.
For real measurements, however, (\ref{UniqueN}) is never satisfied because
both, $K_\mathrm{tan}(\rho)$ and $K_\mathrm{pol}(\rho)$, are contaminated by
Rayleigh scattering and defraction at dust particles and stray light produced
inside the instrument \citep{KoutchmyLamy:1985,LlebariaEtal:2010}.
The latter two contributions are often assumed unpolarised and therefore
contribute only to $K_\mathrm{tan}(\rho)$. Rayleigh scattering at irregularly
formed dust particles has an increasing polarisation with
scattering angles deviating from forward scattering. As a result, it yields
a non-negligible contribution also to $K_\mathrm{pol}(\rho)$ at
distances $\rho>4$ to 5 $R_\odot$ \citep[e.g.,][]{LevasseurEtal:2001}.
In many studies, (\ref{UniqueN}) is therefore rather used to separate
the Thomson scattered signal (K-corona) from the Rayleigh scattered
contribution (F-corona) and from instrument stray light. Another way to 
distinguish the Thomson scattered part from the rest is by spectroscopy.
The large Doppler shift in scattering at fast electrons smears out the
Fraunhofer lines which remain detectable in the other contributions.

\begin{figure}
 \Fill
 \parbox{7cm}{\includegraphics[height=8cm]{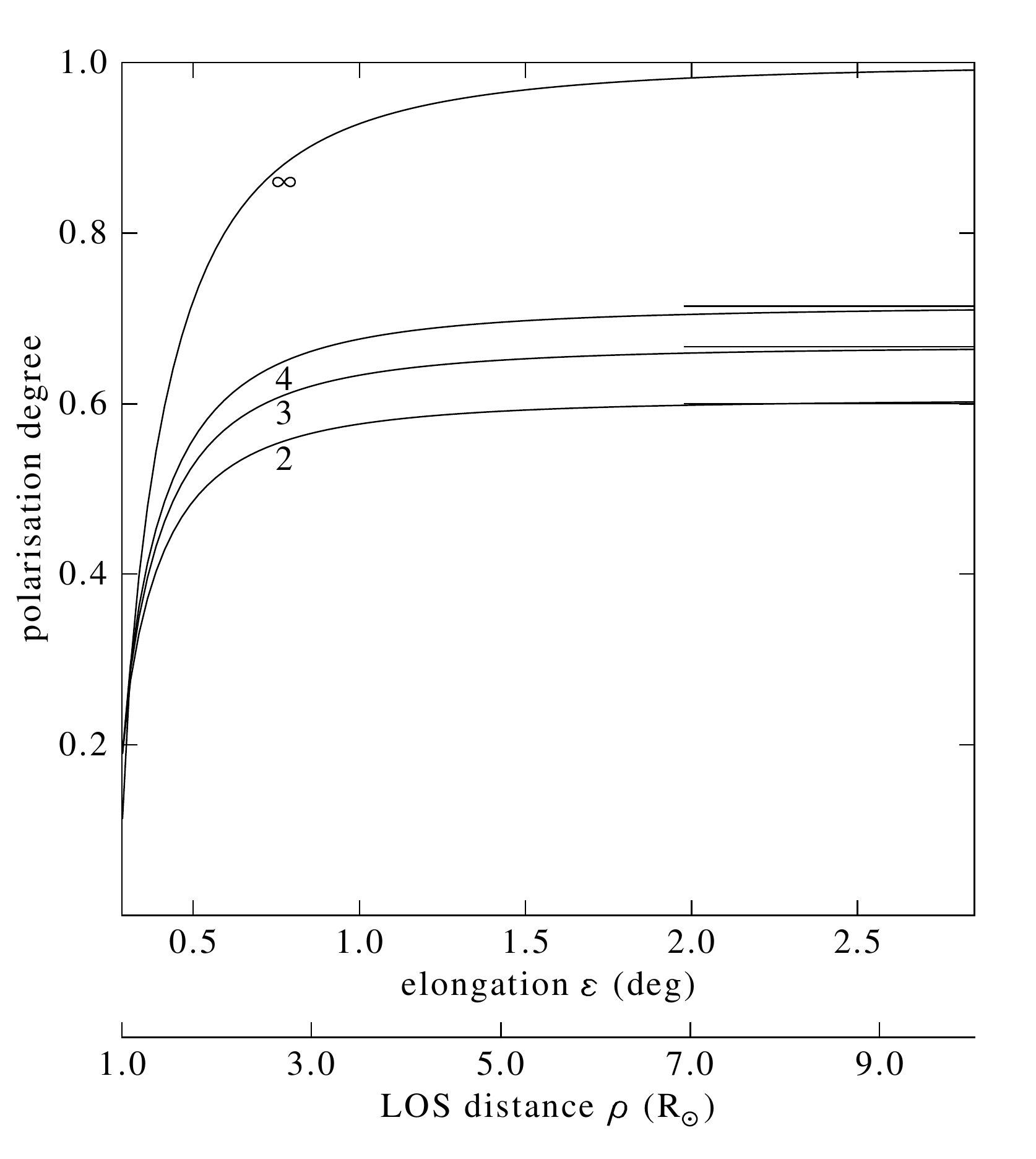}}
 \parbox{8cm}{\caption{The polarisation degree from a spherically symmetric electron
   corona with a density $N_e(r)=N (R_\odot/r)^{-\gamma}$ for different
   values of $\gamma$. The formal limit $\gamma\rightarrow\infty$ can of
   of course never be observed because the rapid drop of the density
   for large $\gamma$ eventually reduces the absolute signal strength
   below measurability.}\label{Fig:PolDegree}}
 \Fill
\end{figure}

Some additional considerations how to calculate the inverse
Abel transformation numerically and analytically can be found in
appendix~\ref{App:AbelTrans}.
As a simple application, let us determine the polarisation degree for a
power-law coronal electron density. This was already treated by
\citep{vandeHulst:1950}.
Even \cite{Schuster:1879} made the first calculations of this ratio because
it is independent from the then unknown scattering cross section
and the radial dependence of the polarisation degree was a first
test of the scattering theory in those days.
Because of the complex radial dependence of the Minnaert's coefficients
in (\ref{Ktan_rho}) and (\ref{Kpol_rho}), the integrals still have to be 
evaluated numerically, but at a large
enough distance $r$ we can use the asymptotic dependence for the Minnaert
coefficients (see
chapter~\ref{app:LimitUnity} in the appendix)
  \begin{gather*}
    (1-u)C(r)+uD(r) \xrightarrow{\rho\rightarrow\infty}
    (1-u/3)L_\odot(\frac{R_\odot}{r})^2
  \\
    (1-u)A(r)+uB(r) \xrightarrow{\rho\rightarrow\infty}
    (1-u/3)L_\odot(\frac{R_\odot}{r})^2
  \end{gather*}
If we further assume a power law dependence for the density of
$N_e(r)=N (R_\odot/r)^{-\gamma}$ we find
  \begin{gather*}
   K_\mathrm{tan}(\rho) \xrightarrow{\rho\rightarrow\infty}
   (1-u/3)L_\odot N R_\odot^{2+\gamma}  \mathcal{A}[\frac{1}{r^{2+\gamma}}]
   \\
   K_\mathrm{pol}(\rho) \xrightarrow{\rho\rightarrow\infty}
   (1-u/3)L_\odot N R_\odot^{2+\gamma}\rho^2
   \mathcal{A}[\frac{1}{r^{4+\gamma}}]
  \end{gather*}
For the Abel transform of a power law we have (see chapter~\ref{app:ExpInvPow}
in the appendix)
  \begin{gather*}
   \mathcal{A}(r^{-\alpha})=\frac{2\pi}{\alpha-1}
    \frac{1}{\DS B(\frac{\alpha}{2},\frac{1}{2})}
    \rho^{1-\alpha},
   \quad
   \mathcal{A}^{-1}(\rho^{-\beta})=\frac{\beta}{2\pi}
    B(\frac{\beta+1}{2},\frac{1}{2})
    r^{-1-\beta}
  \end{gather*}
where $B(x,y)$ is the beta function (\ref{BetaFctn}).
This yields
  \begin{gather*}
   K_\mathrm{tan}(\rho) \xrightarrow{\rho\rightarrow\infty}
   (1-u/3)L_\odot N R_\odot^{2+\gamma}
    \frac{2\pi}{\gamma+1}
    \frac{1}{B(\frac{\gamma+2}{2},\frac{1}{2})}
    \rho^{-\gamma-1}
   \\
   K_\mathrm{pol}(\rho) \xrightarrow{\rho\rightarrow\infty}
   (1-u/3)L_\odot N R_\odot^{2+\gamma}
    \frac{2\pi}{\gamma+3}
    \frac{1}{B(\frac{\gamma+4}{2},\frac{1}{2})}
    \rho^{-\gamma-1}
    \\
   \frac{K_\mathrm{tan}}{K_\mathrm{pol}} \xrightarrow{\rho\rightarrow\infty}
    \frac{\gamma+3}{\gamma+1}
    \frac{B(\frac{\gamma+4}{2},\frac{1}{2})}{B(\frac{\gamma+2}{2},\frac{1}{2})}
   =\frac{\gamma+3}{\gamma+1}
    \frac{\Gamma(\frac{\gamma+4}{2})}{\Gamma(\frac{\gamma+5}{2})}
    \frac{\Gamma(\frac{\gamma+3}{2})}{\Gamma(\frac{\gamma+2}{2})}
   =\frac{\gamma+3}{\gamma+1}
    \frac{\frac{\gamma+2}{2}}{\frac{\gamma+3}{2}}
   =\frac{\gamma+2}{\gamma+1}
  \end{gather*}
For the asymptotic polarisation degree we find
  \begin{equation}
  P=\frac{K_\mathrm{pol}}{K_\mathrm{tot}}
   =\frac{K_\mathrm{pol}}{2K_\mathrm{tan}-K_\mathrm{pol}}
   =\frac{1}{2\frac{K_\mathrm{tan}}{K_\mathrm{pol}}-1}
   =\frac{\gamma+1}{2(\gamma+2)-\gamma-1}
   =\frac{\gamma+1}{\gamma+3}
  \label{AsymPol}\end{equation}
In Fig.~\ref{Fig:PolDegree} we show the degree $P$ of a mere electron corona
as function of the line-of-sight distance $\rho$ for different power laws of
the coronal density. The polarisation degree is calculated from
(\ref{Ktan_rho}) and (\ref{Kpol_rho}), their asymptotic value (\ref{AsymPol})
is marked by a horizontal line.
Again, the measured polarisation degree differs from Fig.~\ref{Fig:PolDegree}
\citep[e.g.,][]{SaitoEtal:1970,KoutchmyLamy:1985, BadalianEtal:1993} because
of additional contributions from scattering at dust and from instrument stray
light. Since the electron density drops rapidly with distance from the Sun,
the relative influence of these additional sources increase with distance.
They contribute mainly to the unpolarised signal and as a result the
measured polarisation degree drops with distance $\rho$ beyond 1.5 to 2
$R_\odot$.

\subsection{CME-like density perturbation}
\label{Sec:ObserveCME}

In this section discuss the observation of a CME-like
density perturbation and how a varying width of the CME may
modify our estimate of its propagation direction, its column mass 
density and of the entire CME mass.
From a single image alone, the width of a CME in heliographic longitude cannot
be perceived. Attempts have been made to use two (or more) images from the
STEREO space craft from different perspectives or to use two different
polarisations from a single perspective to make a guess of the propagation
direction and the width of a CME \citep[e.g.,][]{MierlaEtal:2011}.
As the width estimate often rather crude, so will be our CME model.
\begin{figure}
\hspace*{\fill}
  \includegraphics[width=8cm]{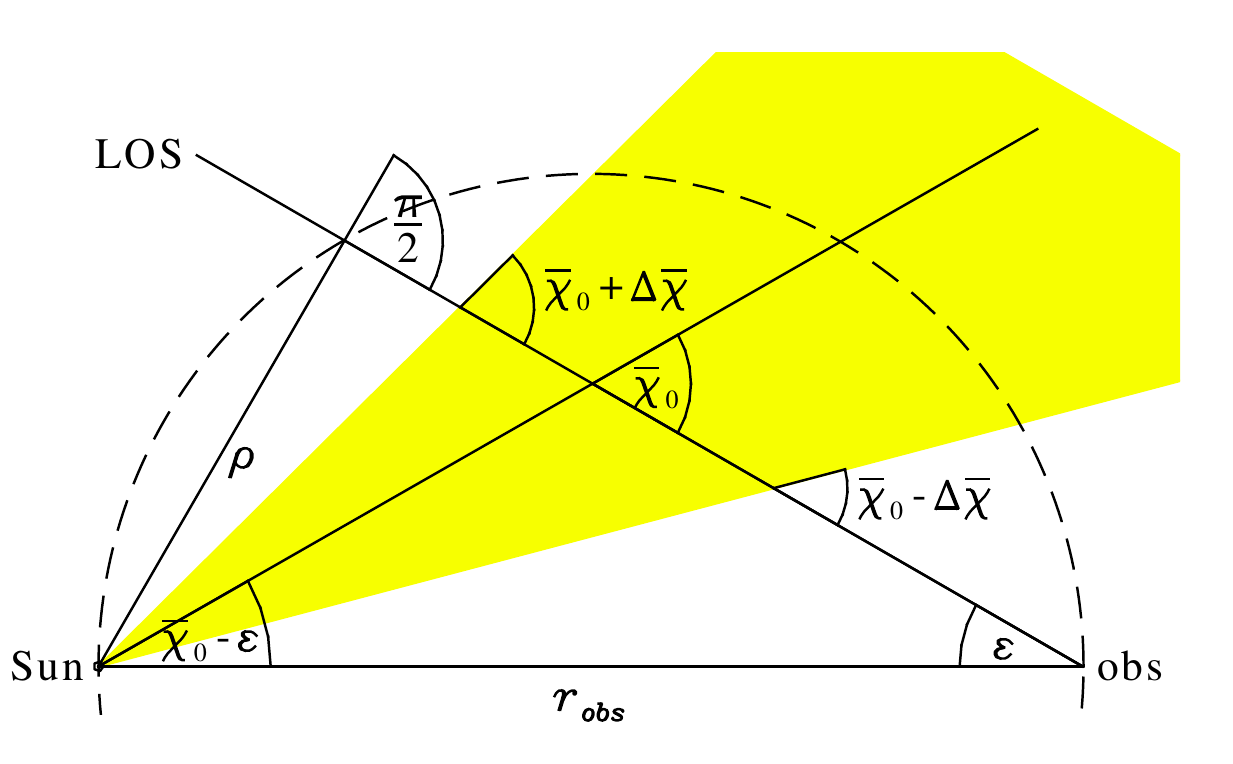}
\hspace*{\fill}
\caption{Geometry of the model CME discussed in the text. The CME centre
  propagates from the Sun at the lower left at an angle $\bar{\chi}_0-\varepsilon$
  with respect to the Sun-observer line. The CME cone of width $\pm\Delta\bar{\chi}$
  around the CME centre axis is shaded in yellow. The line-of-sight starting
  at the observer (at the lower right) makes an angle of the elongation
  $\varepsilon$ with the Sun-observer line.
    \label{Fig:CMEintegr}}
\end{figure}
We assume for a given line-of-sight at elongation $\varepsilon$
(see Fig.~\ref{Fig:CMEintegr}) a CME the centre of which crosses the
line-of-sight at a central mean scattering angle $\bar{\chi}_0$. Along the 
line-of-sight, the electron density may be distributed like a Gaussian
in the variable $\bar{\chi}-\bar{\chi}_0$ with width $\Delta\bar{\chi}$.
In the case that the background has been successfully eliminated
from observed data by forming difference images, the density perturbation
responsible for the residual signal power is then
\begin{gather}
  N_e(r,\bar{\chi})
= N_0 \frac{R^2_\odot}{r^2} g(\bar{\chi};\bar{\chi}_0,\Delta\bar{\chi}),
\quad
 g(\bar{\chi};\bar{\chi}_0,\Delta\bar{\chi})= 
  \frac{1}{\sqrt{\pi}\Delta\bar{\chi}}
   \exp(-(\frac{\bar{\chi}-\bar{\chi}_0}{\Delta\bar{\chi}})^2)
 \label{CMEModel}\end{gather}
where we used once again $r=\rho/\sin\bar{\chi}$.
This model CME has a column density integral of
\begin{gather}
   n_\mathrm{col}=\int_{-\infty}^\infty N_e(r,\bar{\chi}) \;ds
 = N_0 \int_{-\infty}^\infty \frac{R^2_\odot\sin^2\bar{\chi}}{\rho^2}  g(\bar{\chi})
   \frac{\rho d\bar{\chi}}{\sin^2\bar{\chi}}
\nonumber\\   
 = N_0 \frac{R^2_\odot}{\rho}
   \int_{-\infty}^\infty g(\bar{\chi})\;d\bar{\chi}
 = N_0 \frac{R^2_\odot}{\rho}
\qquad[\mathrm{1/m^2}]
\label{Ncol}\end{gather}
Its geometry is sketched in Fig.~\ref{Fig:CMEintegr}.
For $\bar{\chi}_0=\pi/2$ the intersection of the CME centre with 
the line-of-sight is located on the Thomson sphere, for 
$\bar{\chi}_0<\pi/2$ it is inside and for
$\bar{\chi}_0>\pi/2$ it is outside of the Thomson sphere. 
The CME propagation angle with respect to the Sun-observer line is
$\bar{\chi}_0-\varepsilon=\bar{\chi}_0-\asin(\rho/r_\mathrm{obs})$.

The line-of-sight integrations to be performed in (\ref{Phi_K}) are then
\begin{gather}
  K_\mathrm{tan}(\rho)
= N_0 R^2_\odot\rho  \int_{\varepsilon}^{\pi} 
    ((1-u)C(r)+uD(r)) \;g(\bar{\chi})\;\frac{d\bar{\chi}}{r^2\sin^2\bar{\chi}}
\nonumber\\
= n_\mathrm{col}  \int_{\varepsilon}^{\pi} 
  ((1-u)C(r)+uD(r)) \;g(\bar{\chi})\;d\bar{\chi}
\label{K_tan}\\
  K_\mathrm{pol}(\rho)
= N_0 R^2_\odot\rho \int_{\varepsilon}^{\pi} 
  ((1-u)A(r)+uB(r)) \;g(\bar{\chi})\;\frac{d\bar{\chi}}{r^2}
\nonumber\\    
= n_\mathrm{col}\int_{\varepsilon}^{\pi} 
    ((1-u)A(r)+uB(r)) \;g(\bar{\chi})\sin^2\bar{\chi}\,d\bar{\chi}
\label{K_pol}\\    
\nonumber\end{gather} where we used again $r\sin\bar{\chi}=\rho$. For
$\Delta\bar{\chi}\rightarrow 0$ these expressions become independent on the
special shape function $g(\bar{\chi})$
\begin{gather}
\lim_{\Delta\bar{\chi}\rightarrow 0} K_\mathrm{tan} = n_\mathrm{col}\;
((1-u)C(\frac{\rho}{\sin\bar{\chi}_0})+uD(\frac{\rho}{\sin\bar{\chi}_0}))
\nonumber\\ \lim_{\Delta\bar{\chi}\rightarrow 0} K_\mathrm{pol} =
n_\mathrm{col}\;
((1-u)A(\frac{\rho}{\sin\bar{\chi}_0})+uB(\frac{\rho}{\sin\bar{\chi}_0}))\sin^2\bar{\chi}_0
\nonumber\end{gather} and given that $\bar{\chi}_0$ is known the observations
can be inverted to yield an estimate of the column density $n_\mathrm{col}$.
These relations have often been used for CME mass estimates which are obtained 
by summing the column densities pixel 
by pixel and multiplying with a mean coronal ion mass per electron charge.
\begin{figure}[t]
\hspace*{\fill}
  \includegraphics[width=7cm]{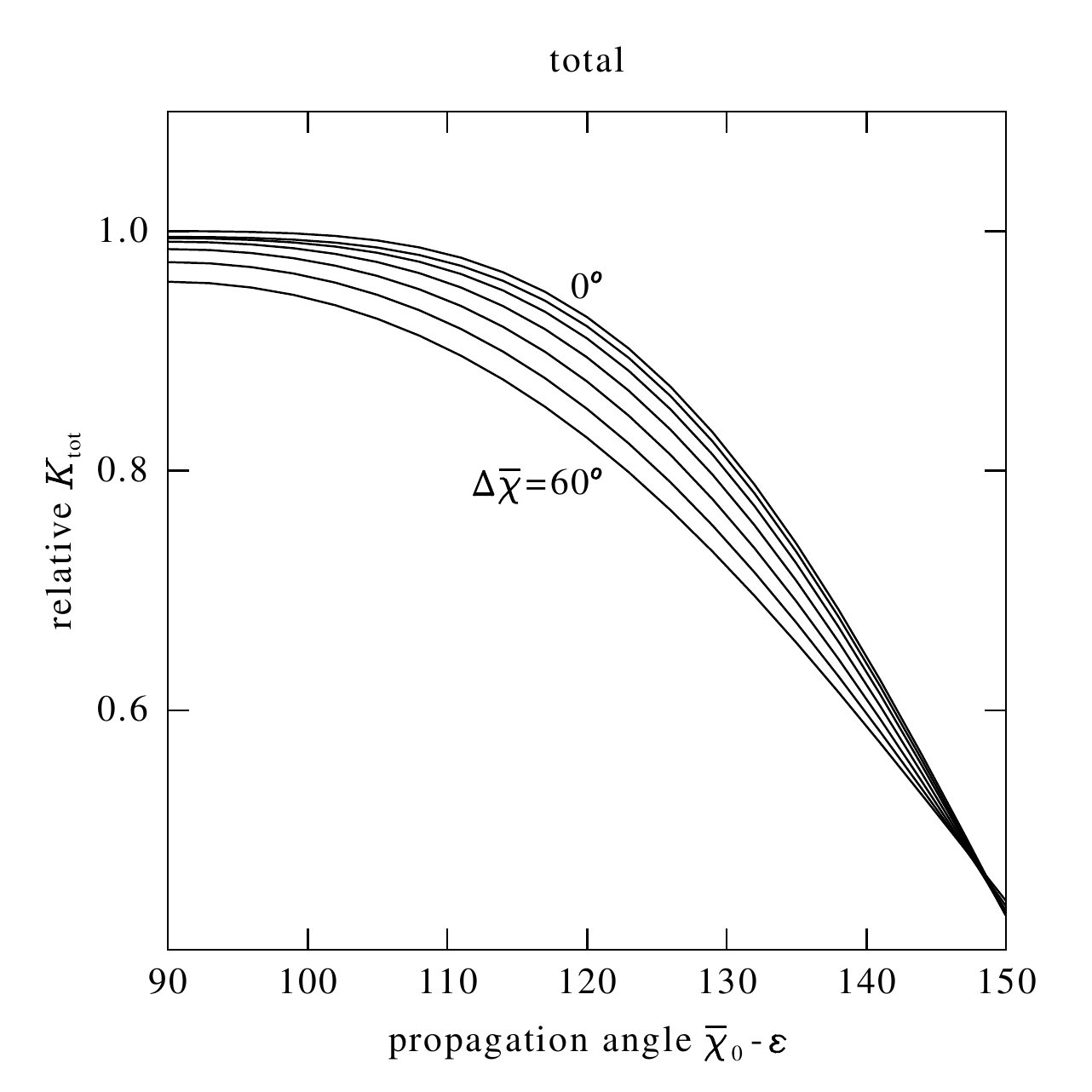}
\hspace*{\fill}
  \includegraphics[width=7cm]{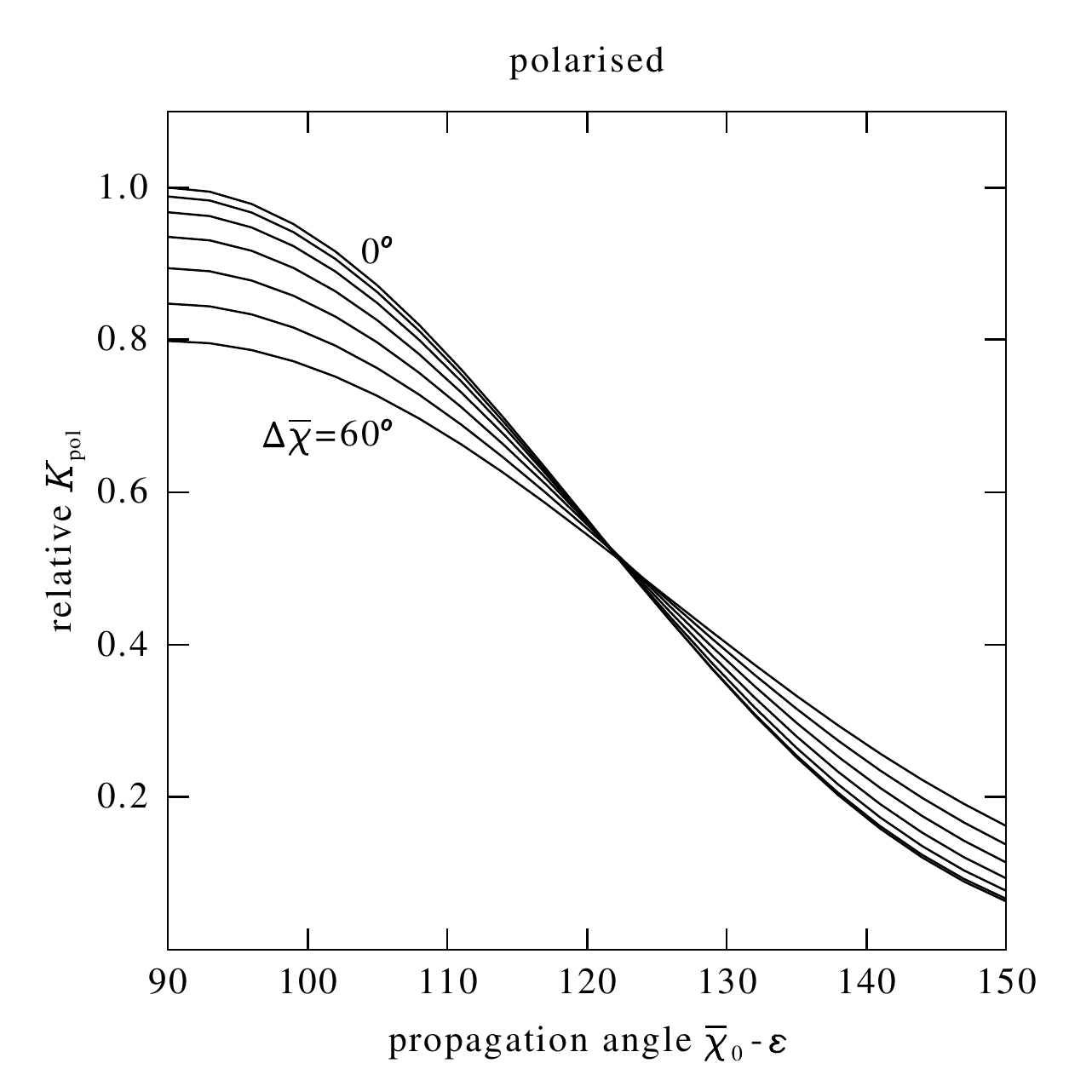}
\hspace*{\fill}
\caption{Plot of $K(\bar{\chi}_0,\Delta\bar{\chi},\rho)$ for a CME cone with Gaussian
  cross section vs the propagation angle $\bar{\chi}_0-\varepsilon$ of the CME
  and for different widths $\Delta\bar{\chi}$. The widths are varied from
  $\Delta\bar{\chi}=$ 0$^\circ$ to 60$^\circ$ in steps of 10$^\circ$ and the
  line-of-sight distance to the solar centre was chosen to $\rho=$ 5
  $R_\odot$. The $K$-values are normalised by $K(\rho,\pi/2,0)$ of a CME
  density concentrated entirely on the Thomson sphere. The left diagram
  displays the total signal, the right diagram the polarised
  signal.
    \label{Fig:MassErr}}
\end{figure}

\begin{figure}
\hspace*{\fill}
\parbox{7cm}{\includegraphics[width=7cm]{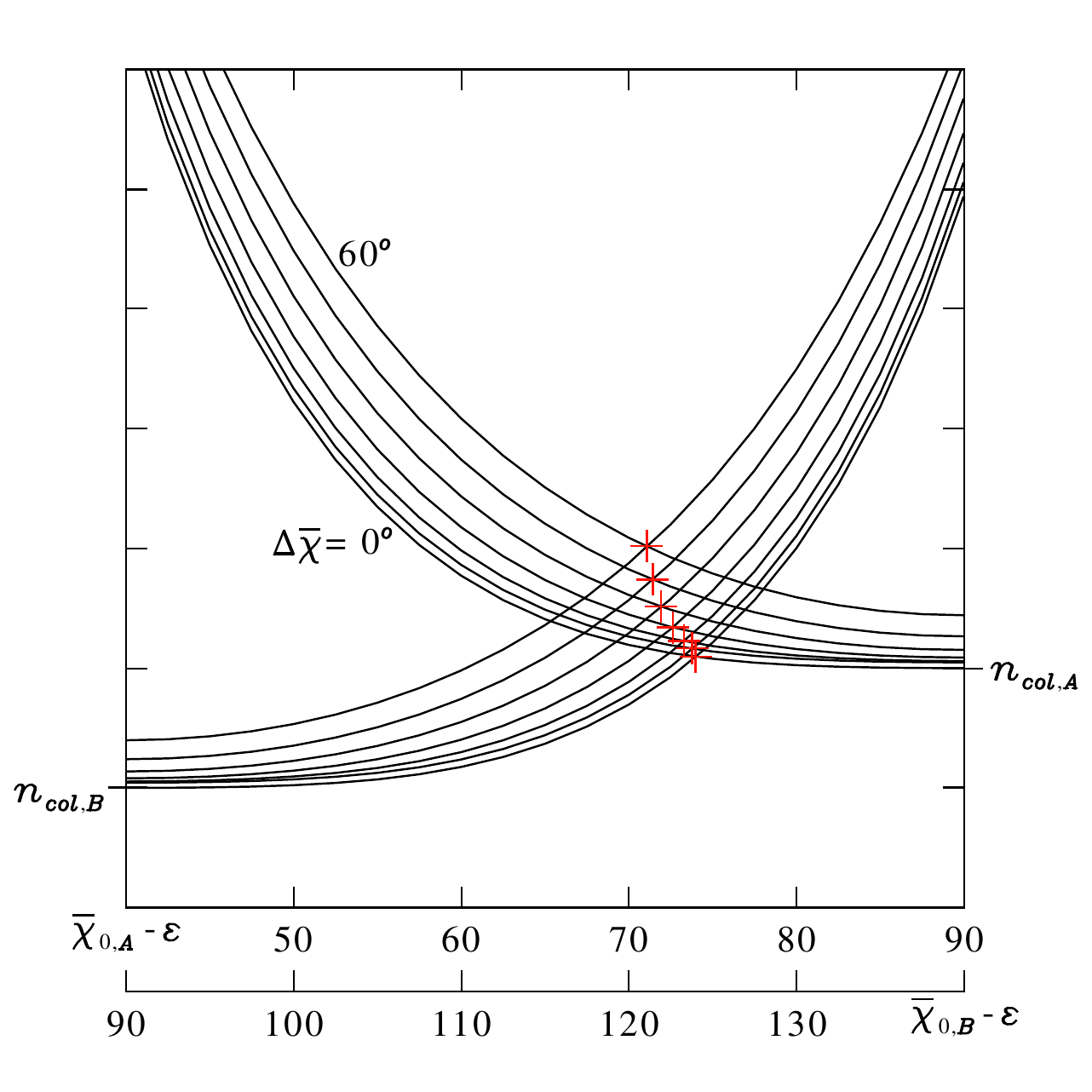}}
\hspace*{\fill}
\parbox{9cm}{\caption{Example of the determination of the propagation angle
    and a consistent column mass density from the observation of
    $K_\mathrm{tot}$ form two vantage points $A$ and $B$ at $\rho=5R_\odot$
    for both of them. The two observers are $50^\circ$ apart. The curves
    represent the column mass density obtained from
    $K_\mathrm{tot}(\rho;\bar{\chi}_0,\Delta\bar{\chi})$ as function of the
    propagation angle with respect to spacecraft A (decreasing curves) and B
    (increasing curves) and for different assumed widths $\Delta\bar{\chi}$.
    $n_\mathrm{col, A}$ and $n_\mathrm{col, B}$ are the respective minimal
    column mass density corresponding to $K_\mathrm{tot}(\rho; \pi/2,0)$. The
    consistent propagation angle and column mass density is obtained from the
    intersection of the curve from A and B with the same assumed width
    $\Delta\bar{\chi}$. The tick marks on the ordinate are in units of 0.1
    $n_\mathrm{col, A}$. For further an explanation see the text.
    \label{Fig:MassErr_example}}}
\hspace*{\fill}
\end{figure}

In Fig.~\ref{Fig:MassErr} we show the total
$K_\mathrm{tot}(\rho;\bar{\chi}_0,\Delta\bar{\chi})$ and the polarised signal
$K_\mathrm{pol}(\rho;\bar{\chi}_0,\Delta\bar{\chi})$ to be observed for such
an idealised CME cone on a line-of-sight with $\rho=5R_\odot$ (i.e., for
constant elongation $\varepsilon$) and for different $\bar{\chi}_0$ and
$\Delta\bar{\chi}$. In both cases we normalise
$K(\rho;\bar{\chi}_0,\Delta\bar{\chi})$ by the respective limit
$K(\rho;\pi/2,0)$ obtained for a CME density entirely concentrated on the
Thomson sphere. This choice $\bar{\chi}_0=\pi/2$ and $\Delta\bar{\chi}=0$ was
the general assumption for CME mass estimates before the STEREO era. As noted
by \cite{VourlidasHoward:2006} this assumption could result in a considerable
underestimate of $n_\mathrm{col}$ when the true propagation angle
$\bar{\chi}_0$ differs from $\pi/2$. Fig~.\ref{Fig:MassErr} shows that a
neglect of a finite CME width $\Delta\bar{\chi}$ even enhances this
underestimate by up to 20\% when $\bar{\chi}_0$ is less than 50$^\circ$ off
the Thomson sphere for $K_\mathrm{tot}$ and less than 30$^\circ$ off for
$K_\mathrm{pol}$.

With the stereo information from two space craft, either the propagation
direction could be determined by triangulation or by adjusting the propagation
angles relative to the two space craft until the mass estimate from both space
craft is consistent \citep{ColaninnoVourlidas:2009}. This latter method can be
applied graphically by plotting the estimated mass vs propagation
angle dependence for both view points into one diagram with the propagation
angles appropriately shifted by the heliospheric longitude difference of the
tow observing spacecraft (assuming both space craft have the same distance
from Sun). The intersection of these curves yields the consistent
mass and the associated propagation angles with respect to each
spacecraft.
An example for such a diagram is shown in Fig.~\ref{Fig:MassErr} for different
assumed CME widths.
Two observers A and B with a 50$^\circ$ heliographic longitude difference measure
$K_\mathrm{tot}(\rho)$ at the same $\rho=5 R_\odot$ and
corresponding to $n_\mathrm{col, A}$ and $n_\mathrm{col, B}$, respectively,
under the assumption $\bar{\chi}_0=\pi/2$ and $\Delta\bar{\chi}=0$.
The intersection for curves with the same assumed width are marked with a
red cross. Each width yields a slightly different propagation
direction and a different column mass estimate.
If the finite width of the CME is ignored and $\Delta\bar{\chi}=0$
is assumed (lowest curves in Fig.~\ref{Fig:MassErr}),
the column density could still be appreciably underestimated.
The consistent propagation angle varies within $\pm 5^\circ$ in the idealised
case treated here.

To apply this method to just a column integral is only justified here
because we use an idealised cone as CME model. However,
the method could in principle be extended to the sum over all pixels
illuminated by a CME and the respective CME mass estimates from both
view point could be used instead of the
column density integrals. A slight complication arises because there is a
difference between $\bar{\chi}_0$ and the propagation angle 
$\bar{\chi}_0-\varepsilon$.
For $\rho=5 R_\odot$ and $r_\mathrm{obs}=200 R_\odot$ as in our example
$\varepsilon$ is less than a degree and $\bar{\chi}_0$ practically 
agrees with the propagation angle. 
For larger $\varepsilon$, as they occur in heliospheric
imager observations this difference cannot be neglected anymore. It varies
with $\rho$ and has to be taken into account when column density
integrals are summed to estimate the total CME mass.

\begin{figure} \hspace*{\fill}
\parbox{7cm}{\includegraphics[width=7cm]{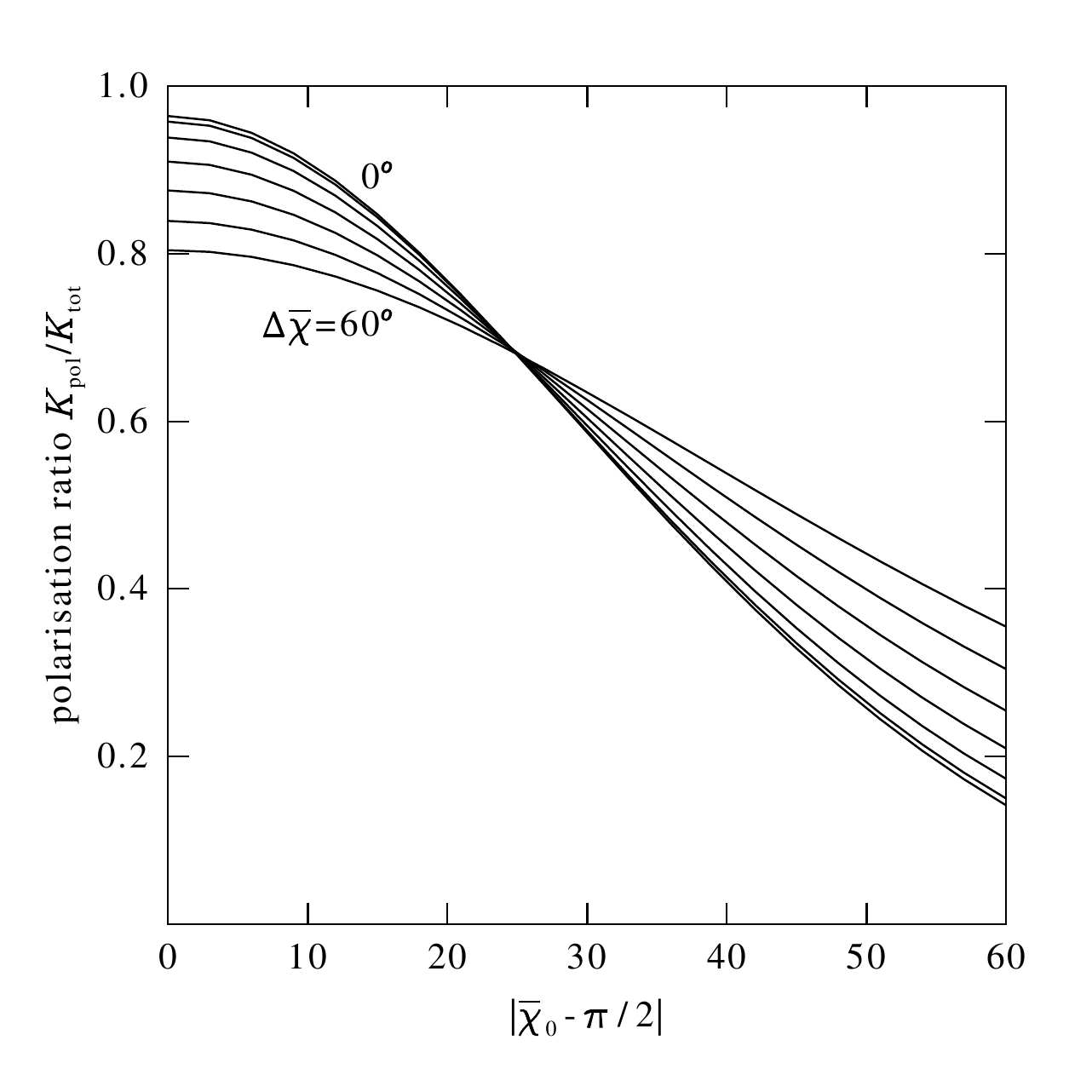}}
\hspace*{\fill} \parbox{9cm}{\caption{Plot of the polarisation ratio
  $K_\mathrm{pol}(\bar{\chi}_0,\Delta\bar{\chi},\rho)$ to
  $K_\mathrm{tot}(\bar{\chi}_0,\Delta\bar{\chi},\rho)$ for a CME cone of
  Gaussian cross section vs the angle $\bar{\chi}_0-\pi/2$ off the plane of the
  sky for different widths from $\Delta\bar{\chi}=$ 0 to 60$^\circ$ in steps of
  10$^\circ$ and for $\rho=$ 5 $R_\odot$. \label{Fig:PolErr}}} \hspace*{\fill}
\end{figure}

A similar error could affect the polarisation ratio method
\citep{MoranDavila:2004} which uses the ratio $K_\mathrm{pol}/K_\mathrm{tot}$
to estimate the scattering position $|\bar{\chi}_0-\pi/2|$ off the Thomson
sphere. For small elongations $\varepsilon$, this corresponds to the CME
propagation angle off the plane of the sky. Again a vanishing CME width
$\Delta\bar{\chi}$ is traditionally assumed. In Fig.~\ref{Fig:PolErr}, we
display this ratio for varying cone widths. A neglect of the width could again
lead to a wrong estimate of $|\bar{\chi}_0-\pi/2|$. E.g., a wide CME with
$\Delta\bar{\chi}=60^\circ$ propagating at $\bar{\chi}_0=0$ yields a ratio of
0.8 which could be interpreted as a CME propagating at 20$^\circ$ if the width
is ignored.
From Fig.~\ref{Fig:PolErr} we see that this way the propagation angle could
be overestimated by up to 20$^\circ$ for $K_\mathrm{pol}/K_\mathrm{tot}>0.7$
and may be underestimated even more for $K_\mathrm{pol}/K_\mathrm{tot}<0.6$,
all depending on the true width $\Delta\bar{\chi}$.

A major simplification of (\ref{K_tan}) and (\ref{K_pol}) is obtained if
the column density integrals are evaluated on line-of-sights with
large $\rho=r_\mathrm{obs}\sin\varepsilon$, i.e. for large spacecraft
distances and sufficiently large elongations $\varepsilon$.
Since $r\ge\rho$ we have from appendix~\ref{app:LimitInfty}    
\[
 \lim_{r\rightarrow\infty} A(r)
=\lim_{r\rightarrow\infty} C(r)
=\theta^2_\mathrm{max}\simeq\frac{R^2_\odot}{r^2},
\quad
 \lim_{r\rightarrow\infty} B(r)
=\lim_{r\rightarrow\infty} D(r)
=\frac{2}{3}\theta^2_\mathrm{max}\simeq\frac{2R^2_\odot}{3r^2}
\]
and setting $r=\rho/\sin\bar{\chi}$, we obtain
\begin{gather}
  K_\mathrm{tan}(\rho)
\xrightarrow{\rho\rightarrow\infty}
 n_\mathrm{col}\;(1-u/3) \frac{R^2_\odot}{\rho^2}
\int_{\varepsilon}^{\pi} g(\bar{\chi})\sin^2\bar{\chi} d\bar{\chi}
\label{Ktan}\\
  K_\mathrm{pol}(\rho)
\xrightarrow{\rho\rightarrow\infty}
 n_\mathrm{col}\;(1-u/3) \frac{R^2_\odot}{\rho^2}
\int_{\varepsilon}^{\pi} g(\bar{\chi})\sin^4\bar{\chi} d\bar{\chi}
\label{Kpol}\\
  K_\mathrm{tot}(\rho)
=2K_\mathrm{tan}(\rho)-K_\mathrm{pol}(\rho)
\hspace*{13em}\nonumber\\\hspace*{6em}
\xrightarrow{\rho\rightarrow\infty}
 n_\mathrm{col}\;(1-u/3) \frac{R^2_\odot}{\rho^2}
\int_{\varepsilon}^{\pi} g(\bar{\chi})(1-\cos^4\bar{\chi})d\bar{\chi}
\label{Ktot}\end{gather}
The interpretation of these formulas is straight forward. The factor $1-u/3$
arises because for the Sun as a point source the radiation form the entire Sun
matters rather than its central radiance $\rad_\odot$ (see also
eq.~\ref{Irr_inf}).
The observed $K(\rho)$ effectively decreases with $\rho^{-3}$ because
$n_\mathrm{col}\propto\rho^{-1}$ (see eq~\ref{Ncol}).
Both the solar irradiance and the density decrease with
$r^{-2}=\sin^2\bar{\chi}/\rho^2$ while the length of the line-of-sight section
which intersects the CME cone grows as $\rho/\sin^2\bar{\chi}$ (see
eq.~\ref{dsdchi}). Together this yields the $\sin^2\bar{\chi}/\rho^3$
dependence of the tangential signal $K_\mathrm{tan}$.
The $\bar{\chi}$ dependence of $K_\mathrm{pol}$ and $K_\mathrm{tot}$ is
further modified by the additional dependence of the radiant intensity for
these polarisations on the mean scattering angle through the Thomson
scattering cross section \cite{HowardDeforest:2012}.

\section{Electrons in motion -- relativistic effects}
\setcounter{equation}{0}
\label{Sec:ElecMotion}

So far, we have neglected relativistic effects. Most coronagraphs operate at
optical wavelengths and integrate over a wide wavelength range. The Compton
wavelength shift at these wavelengths is tiny and is only of the order of
$\lambda/\lambda_\mathrm{Compton}=\mathcal{O}(10^{-5})$. However, even if the
energy of the observed photons is small, relativistic effects may come into
play due to the finite energy of the electrons. For a temperature of 10$^{6}$
K, the ratio $\beta$ of the electron velocity to the speed of light for a
thermal electron is $\mathcal{O}(10^{-2})$ which is not too far away from speeds
where relativistic effects matter.
Electron beams with higher energy are likely to exist in the solar corona
at least sporadically close to X-ray flares and in the source region of
type radio bursts.

\subsection{Expected effects -- a qualitative discussion}

It can immediately be seen by an argument from \cite{Molodensky:1973} that a
finite electron energy has an influence. Due to aberration, a relativistic
electron moving to or away from the centre of the Sun will see the Sun's size
in its rest frame at a reduced or an enhanced viewing angle
$\theta_\mathrm{max}$, respectively, compared to the value of (\ref{sinpsi})
we found for an electron at rest. The scattering at an electron moving in one
of these directions is therefore to some extent equivalent to the scattering
at a stationary electron but at a different apparent distance from the Sun.
Since the scattering takes place in the electron rest frame, the polarisation
properties of the scattered radiation will correspond to the respective
apparent distance, except that the observer will see the scattered light
coming from the electron's true (i.e., retarded) distance from the Sun in his
own rest frame.

\begin{figure}
\hspace*{\fill}
\parbox{7cm}{\includegraphics[width=7.5cm]{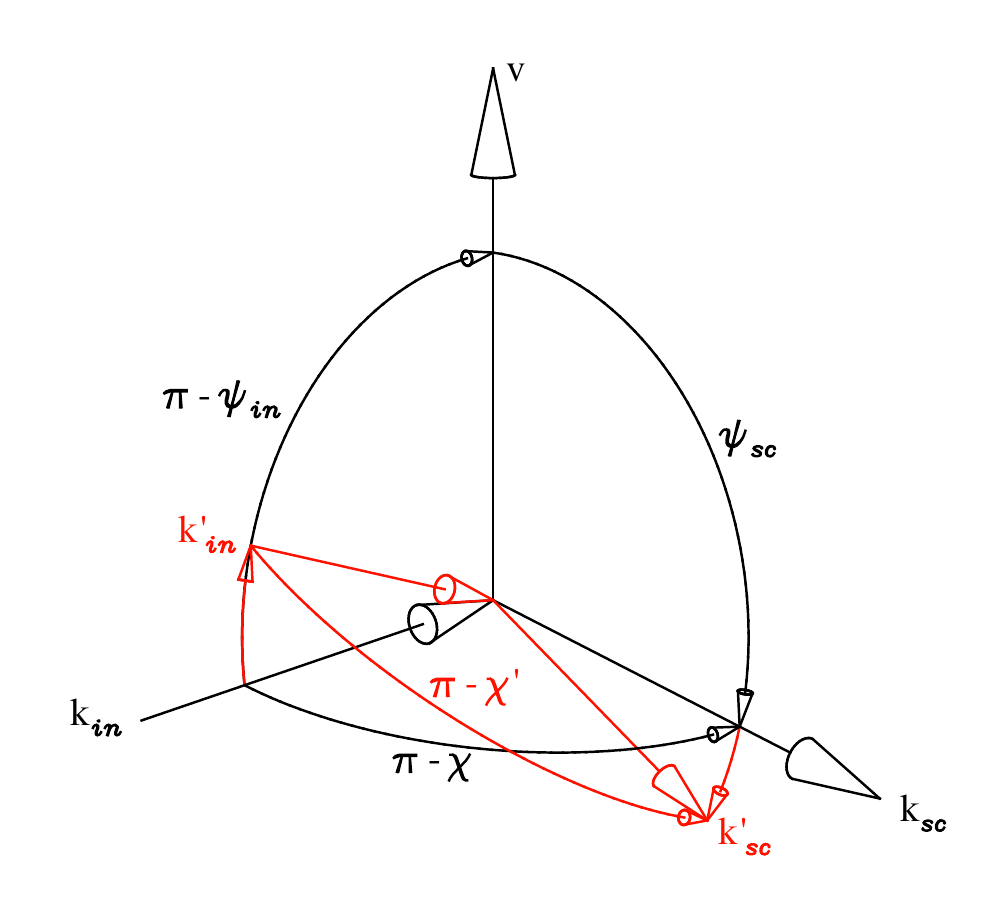}}
\hspace*{\fill}
\parbox{9cm}{\caption{Illustration of the scattering of a photon
    at an electron moving with a relativistic velocity.
    The incident and scattered wave vectors are $\vect{k}_\mathrm{in}$
    and $\vect{k}_\mathrm{sc}$, respectively. If the electron is at rest,
    they span the scattering plane with the scattering angle $\chi$
    between them. If the electron is in motion its velocity $\vect{v}$
    spans two aberration planes with each of the wave vectors. The aberrated
    wave vectors $\vect{k}'_\mathrm{in}$ and $\vect{k}'_\mathrm{sc}$ in
    the electron rest frame both lie in the same respective plane. These
    wave vectors then span the scattering plane in the rest frame of the
    electron (drawn in red).
    \label{Fig:ComptScScene}}}
\hspace*{\fill}
\end{figure}

Therefore even if the photon energy is moderate in the electron rest frame and
Thomson scattering applies, the transformations into and out of the electron
rest frame can make a substantial difference in the energy and polarisation of
the scattered photon compared to Thomson scatter at an electron at rest.
Before we present the calculations in detail, we will first give a qualitative
description of how to approach the problem. Consider an electron with a
velocity $\vect{v}=c\vectg{\beta}$ in the Sun's reference frame $S$. Let
$\nu_\mathrm{in}$ be the frequency of the incident photon and
$\psi_\mathrm{in}$ the angle between its propagation direction
$\vect{\hat{k}}_\mathrm{in}$ and $\vect{v}$ as in
Fig.~\ref{Fig:ComptScScene}. 
In general, we will denote quantities in the electron rest frame $S'$ by a
dash attached to the equivalent variable name used in
the Sun's rest frame $S$. Then $\nu'$ and $\psi'_\mathrm{in}$
are the respective parameters seen in the electron rest frame $S'$. They
transform as (see eqs.~\ref{FrqTran} and \ref{Aberr_c} in
appendix~\ref{App:SpecRel})
\begin{gather}
   \nu'
  =\frac{\nu_\mathrm{in}}{D(\vect{\hat{k}}_\mathrm{in},\vectg{\beta})}
  =\nu_\mathrm{in}\gamma(1 - \beta \cos\psi_\mathrm{in})
 \label{IncidentFshift}\\
   \cos\psi'_\mathrm{in}=\frac{\cos\psi_\mathrm{in}
            - \beta}{1 - \beta \cos\psi_\mathrm{in}}
 \label{IncidentAberr}\end{gather}
where $\gamma=1/\sqrt{1-\beta^2}$ is the Lorentz factor and
$D(\vect{\hat{k}}_\mathrm{in},\vectg{\beta})$ is commonly called the
frequency shift factor (note it is larger than one for a redshift).
The maximum frequency upshift is achieved for $\psi_\mathrm{in}=\pi$ when
$\vect{k}_\mathrm{in}$ and $\vect{v}$ are antiparallel. Then the upshift
amounts to a factor $\nu'/\nu_\mathrm{in}=\sqrt{(1+\beta)/(1-\beta)}$.
Unless $\beta$ is close to unity, incident white light photons with a wave
length around $5\;10^{-7}$ m will remain far from the Compton regime in the
rest frame of the electron. For the photon momentum to matter the photon 
must be transformed to a wave length near
$\lambda_C = 2\pi\hbar/m_e c = 2.4\;10^{-12}$ m
equivalent to a frequency blueshift of 
$\nu'/\nu_\mathrm{in}\simeq 2\gamma\simeq 10^5$. We can therefore well 
approximate the photon scattering in the electron rest frame $S'$ 
by elastic Thomson scattering.

After being scattered, the photon has to be transformed back into the
Sun's reference frame $S$. We call $\psi_\mathrm{sc}$ the angle between the
direction of $\vect{v}$ and the direction $\vect{\hat{k}}_\mathrm{sc}$ from
the scattering site to the observer in the Sun's rest frame.
Given $\vect{\hat{k}}_\mathrm{sc}$ and its angle $\psi_\mathrm{sc}$ with the
electron velocity, the scattered photon transforms according to
(see eqs.~\ref{FrqTran_inv} and \ref{Aberr_c} in
appendix~\ref{App:SpecRel})
\begin{gather}
   \nu_\mathrm{sc}
  =\frac{\nu'}{D(\vect{\hat{k}}_\mathrm{sc},-\vectg{\beta})}
  =\nu'\;D(\vect{\hat{k}}_\mathrm{sc},\vectg{\beta})
  =\frac{\nu'}{\gamma(1 - \beta \cos\psi_\mathrm{sc})}
\label{ScatteredFshift}\\
   \cos\psi'_\mathrm{sc}=\frac{\cos\psi_\mathrm{sc}
            - \beta}{1 - \beta \cos\psi_\mathrm{sc}}
\label{ScatteredAberr}\end{gather}
The second equation determines the direction $\vect{\hat{k}'}_\mathrm{sc}$
into which the photon has to be scattered in the electron rest frame to
reach the observer.

The geometry of the scattering process is illustrated in
Fig.~\ref{Fig:ComptScScene}.
The incident direction $\vect{\hat{k}}_\mathrm{in}$ and
$\vect{v}$ span the incident aberration plane which also contains
$\vect{\hat{k}'}_\mathrm{in}$, however, tilted with respect
to $\vect{\hat{k}}_\mathrm{in}$
according to (\ref{IncidentAberr}).
Similarly, we have a scattering aberration plane formed by
$\vect{\hat{k}}_\mathrm{sc}$ and $\vect{v}$ also containing
$\vect{\hat{k}'}_\mathrm{sc}$.
These planes differ in general from the scattering plane formed
by $\vect{\hat{k}}_\mathrm{in}$ and $\vect{\hat{k}}_\mathrm{sc}$ unless
$\vect{v}$ lies in the scattering plane. Except for this latter case,
the scattering plane in the electron's rest frame formed by 
$\vect{\hat{k}}'_\mathrm{in}$ and $\vect{\hat{k}}'_\mathrm{sc}$ is
inclined with respect to the scattering plane in the Sun's rest 
frame. Since the former plane (drawn in red in Fig.~\ref{Fig:ComptScScene})
is relevant for the scattering process, 
the observed polarisation in the Sun's reference frame is in
general inclined with respect to an observation with the electron at rest.

The total observed frequency shift from (\ref{IncidentFshift}) and
(\ref{ScatteredFshift}) is
\begin{equation}
   \nu_\mathrm{sc}=\nu_\mathrm{in}
   \frac{1 - \beta \cos\psi_\mathrm{in}}
        {1 - \beta \cos\psi_\mathrm{sc}}
  =\nu_\mathrm{in}\frac{D(\vect{\hat{k}}_\mathrm{sc},\vectg{\beta})}
                       {D(\vect{\hat{k}}_\mathrm{in},\vectg{\beta})}
\label{FrqBlueShft}\end{equation}
Whether the frequency is red- or blueshifted depends on the angles
$\psi_\mathrm{in}$ and $\psi_\mathrm{sc}$. The locus of electron
velocities $\vectg{\beta}$ which produce a given Doppler shift  
$\Delta\nu=\nu_\mathrm{sc}-\nu_\mathrm{in}$
is given by the plane in velocity space
\begin{equation}
  \vectg{\beta}\tp(
  \nu_\mathrm{sc}\vect{\hat{k}}_\mathrm{sc}
 -\nu_\mathrm{in}\vect{\hat{k}}_\mathrm{in})
=\Delta\nu
\label{FrqContPlanes}\end{equation}
The planes for different $\Delta\nu$ are not parallel but
all planes intersect along the line given by
$\vectg{\beta}\tp\vect{\hat{k}}_\mathrm{sc}
=\vectg{\beta}\tp\vect{\hat{k}}_\mathrm{in}=1$. Hence, any
real particle with $\beta<1$ produces a unique Doppler shift.
It turns out that the headlight effect makes the scattering 
of photons in an upshift direction much more probable than in a direction
which downshifts the frequency
(for the headlight or searchlight effect, see appendix~\ref{App:SpecRel}).
If we assume an isotropic distribution of incident photons, the scattering
electron will see the photons coming preferentially from ahead.
But Thomson scattering is only mildly anisotropic so that the photons
are scattered again more or less isotropically in the electron rest frame.
In the lab frame, however, these photons appear beamed preferentially in
forward direction. For $\beta$ close to unity this effect is very pronounced
and we can estimate the mean Doppler shift factor by integrating
(\ref{FrqBlueShft}) over the entire sphere of unit directions
$\vect{\hat{k}}_\mathrm{in}$ while we confine $\psi_\mathrm{sc}$ to about
zero. We then obtain as an estimate for the mean frequency shift
\[
 \frac{\nu_\mathrm{sc}}{\nu_\mathrm{in}}
 \simeq\frac{2\pi}{4\pi}\int_{-\pi}^{\pi}\frac{1 - \beta \cos\psi_\mathrm{in}}
            {1 - \beta}\sin\psi_\mathrm{in} d\psi_\mathrm{in}
 =\frac{(1+\beta)}{2(1-\beta^2)}\int_{-1}^{1}(1 - \beta \cos\psi_\mathrm{in})
            d\cos\psi_\mathrm{in}
 \simeq 2\gamma^2
\]
For $\beta\rightarrow 1$, the Lorentz factor may greatly exceed unity and the
photons are on average strongly blueshifted (inverse Compton effect). By
repeated scattering the photon energy may rise eventually until Thomson
scattering in the electron frame becomes invalid. Thomson scattering then has
to be replaced by the more general Compton scattering process which takes
proper account of the momentum and energy exchange between electron and
photon. This way the cold photons and hot electrons may eventually come to
an equilibrium. The inverse Compton process occurs in hot coronae of,
e.g., active galactic nuclei but neither multiple scattering nor a huge
$\gamma$ are likely in the solar corona.

\subsection{The details -- a quantitative treatment}

A quantitative evaluation of the radiant intensity in the case that the
scattering electron has a relativistic velocity proceeds similarly as for the
electron at rest. We recall that for the case $\beta=0$ in
chapter~\ref{sec:AnisotricIrradiation} the irradiance at a distance $r$ from
the solar centre was (\ref{Irr3x3})
\begin{equation}
 \vect{\irr}_{\mathrm{in}}(r)
 =\int_{\Omega(r)}
   \vect{\rad}(\vect{\hat{k}}_\mathrm{in})
     \;d\Omega(\vect{\hat{k}}_\mathrm{in})
\label{IrrMat_beta0}\end{equation}
where the integration over $\Omega(r)$ covers all photon propagation
directions $\vect{\hat{k}}_\mathrm{in}$ from the solar surface which can
directly reach the scattering site $\vect{r}$. The radiance matrix of each
beam of the unpolarised radiation from the solar surface was expressed in
terms of two mutually orthogonal polarisation vectors $\vect{\hat{e}}_1$ and
$\vect{\hat{e}}_2$ normal to $\vect{\hat{k}}_\mathrm{in}$ as
\begin{equation}
 \vect{\rad}(\vect{\hat{k}}_\mathrm{in})
 =\frac{\rad(\vect{\hat{k}}_\mathrm{in})}{2}
 (\vect{\hat{e}}_1{\vect{\hat{e}}_1}\tp
 +\vect{\hat{e}}_2{\vect{\hat{e}}_2}\tp)
 =\frac{\rad(\vect{\hat{k}}_\mathrm{in})}{2}
 (\boldsymbol{1}-\vect{\hat{k}}_\mathrm{in}{\vect{\hat{k}}}\tp_\mathrm{in})
\label{RadMat_beta0}\end{equation}
Here $\rad(\vect{\hat{k}}_\mathrm{in})$ is the scalar radiance
(\ref{Rad_Sun}) of the visible solar disk in the Sun's rest frame in the
direction $\vect{\hat{k}}_\mathrm{in}$ to the scattering site $\vect{r}$
which makes an
angle $\zdis$ with the local solar surface normal. This was our approach in
(\ref{Irr_integral}) and (\ref{Irr3x3}). The integration over the visible
solar surface was performed in a spherical coordinate system centred at
$\vect{r}$ and with its zenith axis through the centre of the Sun so that the
integration became analytically tractable.

In the case of a moving electron, we need the irradiance in the rest frame of
the electron. This is the same as above, however, all variables have to be
transformed into the electron rest frame, i.e., 
\begin{equation}
 \vect{\irr'}_{\mathrm{in}}
 =\int_{\Omega'} \vect{\rad}'(\vect{\hat{k}'}_\mathrm{in})
 \;d\Omega'(\vect{\hat{k}'}_\mathrm{in})
\label{Irr_Integ1}\end{equation}
The direction vectors $\vect{\hat{k}'}_\mathrm{in}$ in the electron rest frame
and $\vect{\hat{k}}_\mathrm{in}$ in the Sun's rest frame are related by
aberration (\ref{KvecTran2}).
Likewise, $\Omega'$ denotes the aberrated solid angle of feasible directions
$\vect{\hat{k}'}_\mathrm{in}$ and
$\vect{\rad}'(\vect{\hat{k}'}_\mathrm{in})$ is the radiance of an unpolarised
beam incident in direction $\vect{\hat{k}'}_\mathrm{in}$ from the solar
surface to the electron, however, transformed to the electron rest frame. The
incident beam is unpolarised as in the Sun's rest frame but the
polarisation plane is tilted according to the aberrated propagation
direction such that the polarisation base vectors $\vect{\hat{e}'}_1$ and
$\vect{\hat{e}'}_2$ are orthogonal to $\vect{\hat{k}'}_\mathrm{in}$
(see \cite{CockeHolm:1972} or chapter~\ref{app:EFieldTransform} in
the appendix). Then like (\ref{RadMat_beta0}) above we have
\begin{equation}
 \vect{\rad}'(\vect{\hat{k}'}_\mathrm{in})
 =\frac{\rad'(\vect{\hat{k}'}_\mathrm{in})}{2}
 (\vect{\hat{e}'}_1{\vect{\hat{e}'}_1}{}\tp
 +\vect{\hat{e}'}_2{\vect{\hat{e}'}_2}{}\tp)
 =\frac{\rad'(\vect{\hat{k}'}_\mathrm{in})}{2}
 (\boldsymbol{1}-\vect{\hat{k}'}_\mathrm{in}
                 \vect{\hat{k}'}{\tp_\mathrm{in}})
\label{IrrMat_beta}\end{equation}
In the appendix (chapter~\ref{app:SpecRel_radiance}) we illustrate how
the shape $\Omega'(\vect{\hat{k}}'_\mathrm{in})$ and the apparent radiance
distribution $L'(\vect{\hat{k}}'_\mathrm{in})$ from the solar surface
change with increasing $\beta$.
The scalar radiance in the electron frame is related to the respective
distribution in the Sun's rest frame by the transformation
(see e.g., \cite{McKinley:1980,EriksenGron:1992,WeiskopfEtal:1999}
or chapter~\ref{app:SpecRel_radiance} in the appendix)
\begin{equation}
 \rad'(\vect{\hat{k}'}_\mathrm{in})
 =\frac{\rad(\vect{\hat{k}}_\mathrm{in})}
        {D^4(\vect{\hat{k}}_\mathrm{in},\vectg{\beta})}
\label{IrrScal_beta}\end{equation}
where $D$ is the frequency shift factor (\ref{IncidentFshift}) and
$\rad(\vect{\hat{k}}_\mathrm{in})$ is the scalar radiance distribution
(\ref{RadMat_beta0}) in the Sun's rest frame.

Upon substituting the integration variable $\vect{\hat{k}'}_\mathrm{in}$
for integration over $\Omega'$ in the electron rest frame by the
unaberrated direction $\vect{\hat{k}}_\mathrm{in}$ over the Sun's disk
$\Omega$ in the solar rest frame we obtain from (\ref{Irr_Integ1}),
(\ref{IrrMat_beta}) and (\ref{IrrScal_beta})
\begin{equation}
 \vect{\irr'}_{\mathrm{in}}
 =\half\int_{\Omega(r)}
   \frac{\rad(\vect{\hat{k}}_\mathrm{in})}
         {D^2(\vect{\hat{k}}_\mathrm{in},\vectg{\beta})}
 (\boldsymbol{1}-\vect{\hat{k}'}_\mathrm{in}
                 \vect{\hat{k}'}{\tp_\mathrm{in}})\;
                d\Omega(\vect{\hat{k}}_\mathrm{in})
\label{Irr_Integ2}\end{equation}
Note that the Jacobian of this transformation yields 
$d\Omega'(\vect{\hat{k}}'_\mathrm{in})
=D^2(\vect{\hat{k}}_\mathrm{in},\vectg{\beta})
d\Omega(\vect{\hat{k}}_\mathrm{in})$
(see eq.~\ref{dOmegaTran} in appendix~\ref{App:SpecRel}).
We can therefore integrate in the Sun's frame of reference as in
chapter~\ref{sec:AnisotricIrradiation} except that we have to bend the photon
direction in the radiance matrix of the incident field into the aberrated
direction and divide its field energy density by
$D^2(\vect{\hat{k}}_\mathrm{in},\vectg{\beta})$.

The details of the transformation of the propagation directions and the
associated polarisation into the electron rest frame and back is more
involved. We will introduce a local orthonormal base attached to each of the
aberration planes (see Fig.~\ref{Fig:ComptScScene}) and mark the respective
base vectors by subscripts ``in'' and ``sc'' for the incident and the
scattering aberration plane, respectively.
The incident aberration plane is spanned by the electron velocity direction
$\uectg{\beta}$ and the propagation direction
$\vect{\hat{k}}_\mathrm{in}$ of the incident photon. The attached orthogonal
base vectors $\uectg{\beta}$, $\vectg{\hat{\mu}}_\mathrm{in}$ and
$\vectg{\hat{\nu}}_\mathrm{in}$ form a right-handed system such that
$\vectg{\hat{\nu}}_\mathrm{in}$ is normal to the incident aberration plane.
With the angle $\psi_\mathrm{in}$ between $\vect{\hat{k}}_\mathrm{in}$ and
$\uectg{\beta}$, we define
\begin{equation}
 \vectg{\hat{\nu}}_\mathrm{in}
=\frac{\uectg{\beta}\times\vect{\hat{k}}_\mathrm{in}}
      {\sin\psi_\mathrm{in}}
\qquad
 \vectg{\hat{\mu}}_\mathrm{in}
=\vectg{\hat{\nu}}_\mathrm{in}\times\uectg{\beta}
\label{Aberr_nin}\end{equation}
We restrict
$\psi_\mathrm{in}=\acos(\vect{\hat{k}}_\mathrm{in}\tp\uectg{\beta})$ to
values $0\dots\pi$ so that $\sin\psi_\mathrm{in}$ is never negative.
The aberration plane is the same in both frames $S$ and $S'$. Its normal
$\vectg{\hat{\nu}}_\mathrm{in}$ is therefore unaffected
by the transformation into the electron rest frame. This clearly
holds also for $\uectg{\beta}$ and therefore 
$\vectg{\hat{\mu}}_\mathrm{in}$ is also invariant.

The vector $\vect{\hat{k}}_\mathrm{in}$ lies in the aberration plane
and is transformed according to 
\begin{equation}
   \vect{\hat{k}}_\mathrm{in}
  =\cos\psi_\mathrm{in}\uectg{\beta}
  +\sin\psi_\mathrm{in}\vectg{\hat{\mu}}_\mathrm{in},
\quad\xrightarrow{S\rightarrow S'}\quad
   \vect{\hat{k}'}_\mathrm{in}
  =\cos\psi'_\mathrm{in}\uectg{\beta}
  +\sin\psi'_\mathrm{in}\vectg{\hat{\mu}}_\mathrm{in},
\label{Aberr_kin}\end{equation}
where $\cos\psi'_\mathrm{in}$ follows from (\ref{IncidentAberr})
and $\sin\psi'_\mathrm{in}$ accordingly (see eq.~\ref{Aberr_s} in
appendix~\ref{App:SpecRel}).

We introduce a similar right-handed orthogonal base $\uectg{\beta}$,
$\vectg{\hat{\mu}}_\mathrm{sc}$ and $\vectg{\hat{\nu}}_\mathrm{sc}$ for the
scattering aberration plane spanned by $\uectg{\beta}$ and
$\vect{\hat{k}}_\mathrm{sc}$ and with normal $\vectg{\hat{\nu}}_\mathrm{sc}$.
Definitions (\ref{Aberr_nin}) and (\ref{Aberr_kin}) can straight forwardly
be adapted with subscript ``in'' replaced by ``sc''.

Finally, we have to define two orthogonal polarisation directions
$\vect{\hat{p}}_1$ and $\vect{\hat{p}}_2$ of the scattered beam which span the
plane-of-sky through the scattering site $\vect{r}$ so that they are
orthogonal to $\vect{\hat{k}}_\mathrm{sc}$.
We chose $\vect{\hat{p}}_1$ along the normal of the aberration plane
and $\vect{\hat{p}}_2=\uectg{\nu}\sca\times\uect{k}\sca$.
The signs are chosen so that $\vect{\hat{p}}_1$, $\vect{\hat{p}}_2$
and the view direction $-\vect{\hat{k}}_\mathrm{sc}$ form a right-handed
orthogonal system.
This way, the observer's plane-of-sky can
easily be transformed from the Sun's rest frame to that of the electron. With
these requirements we have
\begin{gather}
  \hspace*{4em}                   
  \vect{\hat{p}}_1=\vectg{\hat{\nu}}_\mathrm{sc},\qquad
  \vect{\hat{p}}_2=\cos\psi_\mathrm{sc}\vectg{\hat{\mu}}_\mathrm{sc}
                  -\sin\psi_\mathrm{sc}\uectg{\beta}
\label{PolBase}\\   
\xrightarrow{S\rightarrow S'}\quad
  \vect{\hat{p}'}_1=\vectg{\hat{\nu}}_\mathrm{sc},\qquad
  \vect{\hat{p}'}_2=\cos\psi'_\mathrm{sc}\vectg{\hat{\mu}}_\mathrm{sc}
                   -\sin\psi'_\mathrm{sc}\uectg{\beta}
\label{PolBase_}\end{gather}
where $\cos\psi'_\mathrm{sc}$ is related to $\cos\psi_\mathrm{sc}$ in frame
$S$ by (\ref{ScatteredAberr}).

With this polarisation base the irradiance matrix (\ref{Irr_Integ2}) can be
reduced to the coherency matrix $\vect{\rin}'$ for the far-field beam in the
scattering direction $\vect{\hat{k}}'_\mathrm{sc}$ (Recall that $\vect{\irr}$
is 3$\times$3 while $\vect{\rin}$ is 2$\times$2). In complete analogy to
(\ref{Rin_sca2}) but in the electron rest frame $S'$, the elements
$\vect{\rin}'$ of can be written down explicitly
\begin{gather}
\rin'_{ij}(\vect{r},\vect{\hat{k}'}_\mathrm{sc})
= {\ell'}^2\vect{\hat{p}'}_i{}\tp\vect{\irr'}_{\mathrm{sc}}
          (\vect{r}+\ell'\vect{\hat{k}'}_\mathrm{sc})\vect{\hat{p}'}_j
    = r_e^2 \vect{\hat{p}'}_i{}\tp
            \vect{\irr'}_{\mathrm{in}}(\vect{r})
            \vect{\hat{p}'}_j
\nonumber\\
=\frac{r_e^2}{2}\int_\Omega \frac{\rad(\vect{\hat{k}}_\mathrm{in})}
                         {D^2(\vect{\hat{k}}_\mathrm{in},\vectg{\beta})}
 (\delta_{ij}-(\vect{\hat{p}'}_i{}\tp\vect{\hat{k}'}_\mathrm{in})
              (\vect{\hat{p}'}_j{}\tp\vect{\hat{k}'}_\mathrm{in}))
   \;d\Omega(\vect{\hat{k}}_\mathrm{in})
\label{Irr_Integ3}\\
\text{where}
\qquad
  \vect{\hat{p}'}_1{}\tp\vect{\hat{k}'}_\mathrm{in}
  =\sin\psi'_\mathrm{in}\;(\vectg{\hat{\nu}}_\mathrm{sc}\tp
                           \vectg{\hat{\mu}}_\mathrm{in})
  \hspace*{15em}
\label{pk1_in}\\
  \vect{\hat{p}'}_2{}\tp\vect{\hat{k}'}_\mathrm{in}
  =\sin\psi'_\mathrm{in}\cos\psi'_\mathrm{sc}\;
        (\vectg{\hat{\mu}}_\mathrm{sc}\tp
         \vectg{\hat{\mu}}_\mathrm{in})
  -\cos\psi'_\mathrm{in}\sin\psi'_\mathrm{sc}
\label{pk2_in}\end{gather}
Contrary to (\ref{Rin_sca2}), we now need all four
elements of $\rin'_{ij}$ (in fact only three since $\vect{\rin}'$ is
symmetric) because we cannot expect that $\vect{\rin}'$ is diagonal
as it turned out for the electron at rest and the polarisation base
chosen in chapter~\ref{sec:ScattSunLight}.
For (\ref{pk1_in}) and (\ref{pk2_in}) we used definitions (\ref{PolBase_}) and
(\ref{Aberr_kin}) and the fact that $\uectg{\beta}\perp\vectg{\hat{\nu}}$
and $\uectg{\beta}\perp\vectg{\hat{\mu}}$ for both aberration planes
``in'' and ``sc''. The remaining non-zero scalar products of base vectors which
specify the incident and scattered aberration planes can be expressed 
entirely in terms of $\uectg{\beta}$, $\vect{\hat{k}}_\mathrm{in}$ and
$\vect{\hat{k}}_\mathrm{sc}$ using their definitions (\ref{Aberr_nin})
and the equivalent for the ``sc'' base. By insertion,
\begin{align}
 \vectg{\hat{\mu}}_\mathrm{sc}\tp\vectg{\hat{\mu}}_\mathrm{in}
&=(\vectg{\hat{\nu}}_\mathrm{sc}\times\uectg{\beta})\tp
 (\vectg{\hat{\nu}}_\mathrm{in}\times\uectg{\beta})
=-\uectg{\beta}\tp
 (\vectg{\hat{\nu}}_\mathrm{sc}\times
  \vectg{\hat{\nu}}_\mathrm{in}\times\uectg{\beta})
=\vectg{\hat{\nu}}\tp_\mathrm{sc}\vectg{\hat{\nu}}_\mathrm{in}
\label{Aberr_mumu}
\\
 \vectg{\hat{\nu}}_\mathrm{in}\tp\vectg{\hat{\nu}}_\mathrm{sc}
&=\frac{(\uectg{\beta}\times\vect{\hat{k}}_\mathrm{in})\tp
       (\uectg{\beta}\times\vect{\hat{k}}_\mathrm{sc})}
       {\sin\psi_\mathrm{in}\;\sin\psi_\mathrm{sc}}
=\frac{\vect{\hat{k}}_\mathrm{in}\tp\vect{\hat{k}}_\mathrm{sc}
      -(\uectg{\beta}\tp\vect{\hat{k}}_\mathrm{in})
       (\uectg{\beta}\tp\vect{\hat{k}}_\mathrm{sc})}
       {\sin\psi_\mathrm{in}\;\sin\psi_\mathrm{sc}}
=\frac{\vect{\hat{k}}_\mathrm{in}\tp
       (\boldsymbol{1}-\uectg{\beta}\uectg{\beta}\tp)
       \vect{\hat{k}}_\mathrm{sc}}
       {\sin\psi_\mathrm{in}\;\sin\psi_\mathrm{sc}}
\nonumber\\
&=\frac{\cos\chi - \cos\psi_\mathrm{in}\cos\psi_\mathrm{sc}}
      {\sin\psi_\mathrm{in}\;\sin\psi_\mathrm{sc}}
\label{Aberr_nunu}
\\
  \vectg{\hat{\nu}}_\mathrm{sc}\tp\vectg{\hat{\mu}}_\mathrm{in}
&=-\vectg{\hat{\nu}}_\mathrm{in}\tp\vectg{\hat{\mu}}_\mathrm{sc}
= \vectg{\hat{\nu}}\tp_\mathrm{sc}
 (\vectg{\hat{\nu}}_\mathrm{in}\times\uectg{\beta})
= \uectg{\beta}\tp(\vectg{\hat{\nu}}_\mathrm{sc}
                   \times\vectg{\hat{\nu}}_\mathrm{in})
\nonumber\\
&=\uectg{\beta}\tp
 \frac{(\vect{\hat{k}}_\mathrm{sc}\times\uectg{\beta})\times
       (\vect{\hat{k}}_\mathrm{in}\times\uectg{\beta})}
     {\sin\psi_\mathrm{sc}\;\sin\psi_\mathrm{in}}
=-\frac{(\vect{\hat{k}}_\mathrm{sc}\times\uectg{\beta})\tp
         \vect{\hat{k}}_\mathrm{in}}
       {\sin\psi_\mathrm{sc}\;\sin\psi_\mathrm{in}}
=\frac{\uectg{\beta}\tp
      (\vect{\hat{k}}_\mathrm{sc}\times\vect{\hat{k}}_\mathrm{in})}
      {\sin\psi_\mathrm{sc}\;\sin\psi_\mathrm{in}}
\label{Aberr_numu}\end{align}
The transformation of $I'_{i,j}$ into the observer frame now is straight
forward since the polarisation base $\vect{\hat{p}}_i$ and the electric field
transform alike \citep{CockeHolm:1972}, except that the field strength and the
distance $\ell'$ to the observer each have to be divided by
$D(\vect{\hat{k}'}_\mathrm{sc},-\vectg{\beta})
=1/D(\vect{\hat{k}}_\mathrm{sc},\vectg{\beta})$ for a transformation
$S'\rightarrow S$.
For the radiant intensity, this gives all together a factor
$D^4(\vect{\hat{k}}_\mathrm{sc},\vectg{\beta})$ (see
chapters~\ref{app:EFieldTransform} and eq.~\ref{RinTran} in
\ref{app:SpecRel_radiance} in the appendix).
\begin{gather}
I_{ij}(\vect{r},\vect{\hat{k}}_\mathrm{sc})
=\ell^2 \vect{\hat{p}}_i\vect{Q}_\mathrm{sc}\vect{\hat{p}}_j
=D^4(\vect{\hat{k}}_\mathrm{sc},\vectg{\beta})\;
\ell'^2 \vect{\hat{p}'}_i\vect{Q}'_\mathrm{sc}\vect{\hat{p}'}_j
\nonumber\\
=\frac{r_e^2}{2} D^4(\vect{\hat{k}}_\mathrm{sc},\vectg{\beta}) \int_\Omega
       \frac{\rad(\vect{\hat{k}}_\mathrm{in})}
                {D^2(\vect{\hat{k}}_\mathrm{in},\vectg{\beta})}
 (\delta_{ij}-(\vect{\hat{p}'}_i{}\tp\vect{\hat{k}'}_\mathrm{in})
              (\vect{\hat{p}'}_j{}\tp\vect{\hat{k}'}_\mathrm{in}))
   \;d\Omega(\vect{\hat{k}}_\mathrm{in})
\label{Radint_pp}\end{gather}
While (\ref{Radint_pp}) along with (\ref{pk1_in}) and (\ref{pk2_in}) are all
we need to calculate the scattered radiant intensity numerically, we can
further reduce (\ref{pk1_in}) and (\ref{pk2_in}) in terms of the equivalent
products $\uect{p}_1\tp\uect{k}\inc$ and $\uect{p}_2\tp\uect{k}\inc$ in the
observer frame. As result we find
(see chapter \ref{app:Derv_pkin_trans} of appendix~\ref{App:SpecRel})
\begin{gather}
 \uect{p}'{_i\tp}\uect{k}'\inc
=D(\uect{k}\inc,\vectg{\beta})\uect{p}_i\tp(\uect{k}\inc
-\frac{1-\cos\chi(\uect{k}\inc,\uect{k}\sca)}{1-\beta\cos\psi\sca}
  \uectg{\beta})
\label{pkin_trans}\end{gather}
Here, $\cos\chi(\uect{k}\inc,\uect{k}\sca)$ denotes the scattering
angle between $\uect{k}\inc$ and $\uect{k}\sca$ in the Sun's frame.
Moreover, we rederive our central result (\ref{Radint_pp}) and
(\ref{pkin_trans}) in a completely different and more tedious way avoiding
transformations between solar and electron frame. The derivation of this
alternative is also deferred to an appendix (see
appendix~\ref{App:Derv_altern}).

Our result (\ref{Radint_pp}) needs to be analysed further to produce
quantities which are actually observed like the Stokes parameters of
the scattered radiant intensity.
We now have to account for a more complicated polarisation state compared to
the situation when the electron is at rest because we also require the
non-diagonal elements of the radiant intensity coherency matrix.
In chapter~\ref{sec:ScattSunLight} the geometry
was simpler and $\rin_{ij}$ was diagonalised by the choice of the polarisation
base vectors $\vect{\hat{p}}_1\rightarrow\vect{\hat{p}}_\mathrm{tan}$ and
$\vect{\hat{p}}_2\rightarrow\vect{\hat{p}}_\mathrm{rad}$.
Now, the polarisation base vectors $\vect{\hat{p}}_1$ and $\vect{\hat{p}}_2$
had to be chosen in (\ref{PolBase}) according to the orientation of the
scattering aberration plane in order to enable the simple transformation of
the polarisation of the scattered beam.
In this special polarisation reference we express the radiant intensity
coherency matrix $\vect{\rin}$ in terms of Stokes parameters (see
chapter~\ref{app:Polarisation} in the appendix)
\begin{equation}
\vect{\rin}(\vect{\hat{k}}_\mathrm{sc})
= c\epsilon_0 r_e^2\vect{J}
= c\epsilon_0 \frac{r_e^2}{2}
 \begin{pmatrix}
  S_I+S_Q & S_U \\ S_U & S_I-S_Q
 \end{pmatrix}
\label{Radint_matrix}\end{equation}
By the choice of the constants, the elements of $\vect{J}$ have
the units of the electric field squared.
From the matrix $\vect{J}$ we can readily derive the
total intensity and the polarisation properties of the
radiant intensity scattered from the relativistic electron
\begin{equation}
  \rin_\mathrm{tot}
=c\epsilon_0 \,\frac{r_e^2}{2} \;\mathrm{trace}(\vect{J})
=c\epsilon_0 \,r_e^2\;S_I,
  \qquad
  \rin_\mathrm{pol}=P\rin_\mathrm{tot}
\label{Radint_tot}\end{equation}
where the polarisation degree $P$ and the orientation angle $\alpha$
of the major polarisation axis are 
\begin{equation}
P=\frac{\sqrt{S_Q^2+S_U^2}}{S_I},
\quad
\alpha
=\half\atan\frac{S_U}{S_Q}
\label{Radint_polang}\end{equation}
(see eqs.~\ref{PolRotation} and \ref{PolInvariants} in the
appendix~\ref{App:EMwaves}).

Note however that the polarisation angle $\alpha$ above is
measured in the Sun's rest frame in the plane-of-sky starting from
vector $\vect{\hat{p}}_1$ of our polarisation base.
In practical observations, the reference directions most often used are
$\vect{\hat{p}}_\mathrm{tan}$ and $\vect{\hat{p}}_\mathrm{rad}$
from (\ref{PolBase0}), i.e,
\[
\vect{\hat{p}}_\mathrm{tan}
=\frac{\vect{\hat{k}}_\mathrm{sc}\times\vect{\hat{r}}}{\sin\bar{\chi}}
,\qquad
\vect{\hat{p}}_\mathrm{rad}
=\vect{\hat{p}}_\mathrm{tan}\times\vect{\hat{k}}_\mathrm{sc}
\]
and $\bar{\chi}$ is the angle between $\vect{r}$ and
$\vect{\hat{k}}_\mathrm{sc}$ we termed the mean mean scattering angle.
The polarisation reference directions $\vect{\hat{p}}_\mathrm{tan}$ and
$\vect{\hat{p}}_1$ differ by an angle $\alpha_0$ which is given by
\begin{equation}
 \rvecc{\cos\alpha_0}{\sin\alpha_0}
=\rvecc{\vect{\hat{p}}_\mathrm{tan}\tp\vect{\hat{p}}_1}
       {\vect{\hat{p}}_\mathrm{tan}\tp\vect{\hat{p}}_2}
=\frac{1}{\sin\bar{\chi}}       
 \rvecc{\vect{\hat{r}}\tp(\vect{\hat{p}}_1\times\vect{\hat{k}}_\mathrm{sc})}
       {\vect{\hat{r}}\tp(\vect{\hat{p}}_2\times\vect{\hat{k}}_\mathrm{sc})}
=\frac{1}{\sin\bar{\chi}}       
 \rvecc{ \vect{\hat{r}}\tp\vect{\hat{p}}_2}
       {-\vect{\hat{r}}\tp\vect{\hat{p}}_1}
\label{cosin_alpha_0}\end{equation}
Here we made use of the above definition of $\vect{\hat{p}}_\mathrm{tan}$ and
the fact that $-\vect{\hat{k}}_\mathrm{sc}$, $\vect{\hat{p}}_1$ and
$\vect{\hat{p}}_2$ form a right-handed orthogonal base. A similar base
is formed by $-\vect{\hat{k}}_\mathrm{sc}$, $\vect{\hat{p}}_\mathrm{tan}$ and
$\vect{\hat{p}}_\mathrm{rad}$, except the latter is rotated by $\alpha_0$
with respect to the former.

 \subsection{Case $r\rightarrow\infty$ -- a single incident beam}

We now have all relations we need to determine the propertied of the scattered
radiant intensity. The relationship between the observable radiant intensity
matrix, the scattering geometry and $\vectg{\beta}$ can be made more
transparent if we first restrict to the scattering of a single beam from
direction $\vect{\hat{k}}_\mathrm{in}$ and an infinitesimal angular width
$d\Omega(\vect{\hat{k}}_\mathrm{in})$. Practically, this corresponds to the limit
$r\rightarrow\infty$ where the apparent Sun's disk shrinks to a point. 
For such an isolated incident beam we have, using 
(\ref{Radint_polang}), (\ref{Radint_matrix}) and (\ref{Radint_pp})
\begin{gather*}
 \begin{pmatrix}
  S_I+S_Q & S_U \\ S_U & S_I-S_Q
 \end{pmatrix}_{ij}
=\frac{2\rin_{ij}(\vect{\hat{k}}_\mathrm{sc})}
      {c\epsilon_0 r_e^2} 
=\frac{1}{c\epsilon_0} \frac{D^4(\vect{\hat{k}}_\mathrm{sc},\vectg{\beta})}
                            {D^2(\vect{\hat{k}}_\mathrm{in},\vectg{\beta})}
          \rad(\vect{\hat{k}}_\mathrm{in}) d\Omega\;
 [\delta_{i,j} - (\vect{\hat{p}'}_i{}\tp\vect{\hat{k}'}_\mathrm{in})
                 (\vect{\hat{p}'}_j{}\tp\vect{\hat{k}'}_\mathrm{in})]
\end{gather*}
where $\rad(\vect{\hat{k}}_\mathrm{in}) d\Omega$ is the beam irradiance.
Then
\begin{gather}
c\epsilon_0 S_I
 = \frac{1}{2}\frac{D^4(\vect{\hat{k}}_\mathrm{sc},\vectg{\beta})}
                   {D^2(\vect{\hat{k}}_\mathrm{in},\vectg{\beta})}
                   \rad(\vect{\hat{k}}_\mathrm{in})d\Omega\;
       (2-(\vect{\hat{k}'}{\tp_\mathrm{in}}\vect{\hat{p}'}_1)^2 
         -(\vect{\hat{k}'}{\tp_\mathrm{in}}\vect{\hat{p}'}_2)^2)
\nonumber\\
c\epsilon_0 S_Q
 = \frac{1}{2}\frac{D^4(\vect{\hat{k}}_\mathrm{sc},\vectg{\beta})}
                   {D^2(\vect{\hat{k}}_\mathrm{in},\vectg{\beta})}
                   \rad(\vect{\hat{k}}_\mathrm{in})d\Omega\;
       ((\vect{\hat{k}'}{\tp_\mathrm{in}}\vect{\hat{p}'}_2)^2 
       -(\vect{\hat{k}'}{\tp_\mathrm{in}}\vect{\hat{p}'}_1)^2)
\nonumber\\
c\epsilon_0 S_U
 = -\frac{D^4(\vect{\hat{k}}_\mathrm{sc},\vectg{\beta})}
         {D^2(\vect{\hat{k}}_\mathrm{in},\vectg{\beta})}
          \rad(\vect{\hat{k}}_\mathrm{in})d\Omega\;
  \vect{\hat{k}'}{\tp_\mathrm{in}}\vect{\hat{p}'}_1\;
  \vect{\hat{k}'}{\tp_\mathrm{in}}\vect{\hat{p}'}_2
\nonumber\\
P=\frac{\sqrt{S_Q^2+S_U^2}}{S_I}
 =\frac{\sqrt{
    ((\vect{\hat{k}'}{\tp_\mathrm{in}}\vect{\hat{p}'}_2)^2 
    -(\vect{\hat{k}'}{\tp_\mathrm{in}}\vect{\hat{p}'}_1)^2)^2
   +4(\vect{\hat{k}'}{\tp_\mathrm{in}}\vect{\hat{p}'}_1)^2
     (\vect{\hat{k}'}{\tp_\mathrm{in}}\vect{\hat{p}'}_2)^2}}
       {2-(\vect{\hat{k}'}{\tp_\mathrm{in}}\vect{\hat{p}'}_1)^2 
         -(\vect{\hat{k}'}{\tp_\mathrm{in}}\vect{\hat{p}'}_2)^2}
\nonumber\\
 =\frac{(\vect{\hat{k}'}{\tp_\mathrm{in}}\vect{\hat{p}'}_1)^2 
       +(\vect{\hat{k}'}{\tp_\mathrm{in}}\vect{\hat{p}'}_2)^2}
       {2-(\vect{\hat{k}'}{\tp_\mathrm{in}}\vect{\hat{p}'}_1)^2 
         -(\vect{\hat{k}'}{\tp_\mathrm{in}}\vect{\hat{p}'}_2)^2}
\label{PolDegree_Beam}\\
\frac{S_U}{S_Q}=\frac{-2\vect{\hat{k}'}{\tp_\mathrm{in}}\vect{\hat{p}'}_1\;
                       \vect{\hat{k}'}{\tp_\mathrm{in}}\vect{\hat{p}'}_2}
       {(\vect{\hat{k}'}{\tp_\mathrm{in}}\vect{\hat{p}'}_2)^2 
       -(\vect{\hat{k}'}{\tp_\mathrm{in}}\vect{\hat{p}'}_1)^2}
=\tan2\alpha
=\frac{2\tan\alpha}{1-\tan^2\alpha}
=\frac{2\sin\alpha\cos\alpha}{\cos^2\alpha-\sin^2\alpha}
\label{PolAngle_Beam}\end{gather}
The last line suggests to associate with yet unknown constant $a$
\begin{equation}
\vect{\hat{k}'}{\tp_\mathrm{in}}\vect{\hat{p}'}_1=a\sin\alpha,
\quad
\vect{\hat{k}'}{\tp_\mathrm{in}}\vect{\hat{p}'}_2=-a\cos\alpha,
\label{kp_a}\end{equation}
An alternatively possible choice would be
$\vect{\hat{k}'}{\tp_\mathrm{in}}\vect{\hat{p}'}_1=a\cos\alpha$ and
$\vect{\hat{k}'}{\tp_\mathrm{in}}\vect{\hat{p}'}_2=a\sin\alpha$
equivalent to shifting $\alpha$ by $\pi/2$.
We fix this ambiguity by requiring that $\alpha$ shall be zero
if $\vect{\hat{p}'}_1$ points normal to the scattering plane in
the electron rest frame and therefore also normal to  
$\vect{\hat{k}}'_\mathrm{in}$. This requires
$\vect{\hat{k}'}{\tp_\mathrm{in}}\vect{\hat{p}'}_1\propto\sin\alpha$.
The magnitude of the constant $a$ is determined from
\begin{gather*}
1=\cos^2\alpha+\sin^2\alpha
 =\frac{(\vect{\hat{k}'}{\tp_\mathrm{in}}\vect{\hat{p}'}_2)^2
       +(\vect{\hat{k}'}{\tp_\mathrm{in}}\vect{\hat{p}'}_1)^2}{a^2}
\\ 
 = \frac{\vect{\hat{k}'}{\tp_\mathrm{in}}
         (\boldsymbol{1}-\vect{\hat{k}'}_\mathrm{sc}
                         \vect{\hat{k}'}{\tp_\mathrm{sc}})
         \vect{\hat{k}'}_\mathrm{in}}{a^2}
 = \frac{1-(\vect{\hat{k}'}{\tp_\mathrm{in}}\vect{\hat{k}'}_\mathrm{sc})^2}
        {a^2}
 =\frac{\sin^2\chi'}{a^2}
\end{gather*}
We therefore find $a=\pm\sin\chi'$
where $\chi'(\vect{\hat{k}'}_\mathrm{in},\vect{\hat{k}'}_\mathrm{sc})$ is the
scattering angle of the beam in the electron rest frame. The alternating
signs do not matter here because they correspond to a shift of $\alpha$
by $\pi$ which does not matter for the polarisation angle.
Insertion into (\ref{PolDegree_Beam}) and (\ref{kp_a}) yields
\begin{equation}
P=\frac{\sin^2\chi'}{1+\cos^2\chi'},\qquad
\rvecc{\cos\alpha}{\sin\alpha}
=\frac{\pm 1}{\sin\chi'}
\rvecc{-\vect{\hat{k}'}{\tp_\mathrm{in}}\vect{\hat{p}'}_2}
      { \vect{\hat{k}'}{\tp_\mathrm{in}}\vect{\hat{p}'}_1}
\label{cosin_alpha}\end{equation}
For a single incident beam the polarisation degree $P$ therefore depends only
on the scattering angle $\chi'$ in the rest frame of the electron. A similar
result was obtained in (\ref{PolDegree}) for the electron at rest, except that
there the mean scattering angle $\bar{\chi}$ in the Sun's rest frame was
responsible.
The angle $\alpha$ of the major polarisation axis only depends on the 
orientation
of the incident beam projected in the plane-of-sky in the electron rest frame.
Using (\ref{cosin_alpha_0}) and (\ref{cosin_alpha}) we find for the deviation
$\alpha-\alpha_0$ of the major polarisation axis from the tangential direction
$\vect{\hat{p}}_\mathrm{tan}$
\begin{gather}
 \rvecc{\cos(\alpha-\alpha_0)}{\sin(\alpha-\alpha_0)}
=\rvecc{\cos\alpha\cos\alpha_0+\sin\alpha\sin\alpha_0}
       {\sin\alpha\cos\alpha_0-\cos\alpha\sin\alpha_0}
\nonumber\\
=\frac{\mp 1}{\sin\chi'\sin\bar{\chi}}
\rvecc{ \vect{\hat{k}'}{\tp_\mathrm{in}}\vect{\hat{p}'}_2\;
        \vect{\hat{r}}\tp\vect{\hat{p}}_2
       +\vect{\hat{k}'}{\tp_\mathrm{in}}\vect{\hat{p}'}_1\;
        \vect{\hat{r}}\tp\vect{\hat{p}}_1}
      {-\vect{\hat{k}'}{\tp_\mathrm{in}}\vect{\hat{p}'}_1\;
        \vect{\hat{r}}\tp\vect{\hat{p}}_2 
       +\vect{\hat{k}'}{\tp_\mathrm{in}}\vect{\hat{p}'}_2\;
        \vect{\hat{r}}\tp\vect{\hat{p}}_1}
\nonumber\\
=\frac{\mp 1}{\sin\chi'\sin\bar{\chi}}
\begin{pmatrix}
  \vect{\hat{k}'}{\tp_\mathrm{in}}\vect{\hat{p}'}_2 &
  \vect{\hat{k}'}{\tp_\mathrm{in}}\vect{\hat{p}'}_1 \\
 -\vect{\hat{k}'}{\tp_\mathrm{in}}\vect{\hat{p}'}_1 &
  \vect{\hat{k}'}{\tp_\mathrm{in}}\vect{\hat{p}'}_2
\end{pmatrix}
\rvecc{\vect{\hat{r}}\tp\vect{\hat{p}}_2}
      {\vect{\hat{r}}\tp\vect{\hat{p}}_1}
\label{cosin_dalpha}\end{gather}
Explicit expressions for the products
$\vect{\hat{k}'}{\tp_\mathrm{in}}\vect{\hat{p}'}_i$ were given in
(\ref{pk1_in}) and (\ref{pk2_in}). The other products required
can also be expressed
in terms of the directions $\uectg{\beta}$,
$\vectg{\hat{k}}_\mathrm{sc}$ and, since $r\rightarrow\infty$ here,
$\vectg{\hat{k}}_\mathrm{in}=\vectg{\hat{r}}$ if we
use (\ref{PolBase}) and (\ref{Aberr_nin}) 
\begin{gather*}
 \vect{\hat{r}}\tp\vect{\hat{p}}_1
=\vect{\hat{r}}\tp\vectg{\hat{\nu}}_\mathrm{sc}
=\frac{\vect{\hat{r}}\tp(\uectg{\beta}\times\vect{\hat{k}}_\mathrm{sc})}
 {\sin\psi_\mathrm{sc}}
\\
 \vect{\hat{r}}\tp\vect{\hat{p}}_2
=\vect{\hat{r}}\tp\vectg{\hat{\mu}}_\mathrm{sc}\cos\psi_\mathrm{sc}
       -\vect{\hat{r}}\tp\uectg{\beta}\sin\psi_\mathrm{sc}
\\
 \vect{\hat{r}}\tp\vectg{\hat{\mu}}_\mathrm{sc}
=\vect{\hat{r}}\tp(\vectg{\hat{\nu}}_\mathrm{sc}\times\uectg{\beta})
=-\frac{\vect{\hat{r}}\tp(\uectg{\beta}\times
       (\uectg{\beta}\times\vect{\hat{k}}_\mathrm{sc}))}
       {\sin\psi_\mathrm{sc}}
=\frac{\vect{\hat{r}}\tp
 (\boldsymbol{1}-\uectg{\beta}\uectg{\beta}\tp)
  \vect{\hat{k}}_\mathrm{sc}}
 {\sin\psi_\mathrm{sc}}
\end{gather*}
The limit $\beta=0$ can easily be verified. In the irradiance
matrix (\ref{Irr_Integ3}) we set
$D^2(\vect{\hat{k}}_\mathrm{in},\vectg{\beta})=1$ and all dashed
terms equal their
non-dashed counterparts, however $\uectg{\beta}$ may be arbitrary. This
is exactly what we had in (\ref{Rin_sca2}), except that there the polarisation
base was chosen to be aligned with $\vect{\hat{p}}_\mathrm{tan}$ and
$\vect{\hat{p}}_\mathrm{rad}$ while here, we align it with the scattering
aberration plane along $\vect{\hat{p}}_1$ and $\vect{\hat{p}}_2$, i.e., it is
rotated by an angle $\alpha_0$.
However, setting $\vect{\hat{p}'}_i=\vect{\hat{p}}_i$ in (\ref{cosin_dalpha})
the vector term becomes independent of the special orientation of the
polarisation base.
We assume that the beam comes from the disk centre so that
$\vect{\hat{k}'}_\mathrm{in}=\vect{\hat{k}}_\mathrm{in}=\vect{\hat{r}}$. Then
$\bar{\chi}=\chi'=\chi$ and from (\ref{cosin_dalpha})
\begin{gather*}
 \rvecc{\cos(\alpha-\alpha_0)}{\sin(\alpha-\alpha_0)}
=\frac{\mp 1}{\sin^2\chi}
\begin{pmatrix}
       \vect{\hat{k}}_\mathrm{in}\tp
        (\vect{\hat{p}}_2\vect{\hat{p}}_2\tp
        +\vect{\hat{p}}_1\vect{\hat{p}}_1\tp)
       \vect{\hat{r}}
       \\ 0
\end{pmatrix}
\\
=\frac{\mp 1}{\sin^2\chi}
\begin{pmatrix}
       \vect{\hat{k}}_\mathrm{in}\tp
        (\boldsymbol{1}-\vect{\hat{k}}_\mathrm{sc}\vect{\hat{k}}_\mathrm{sc}\tp)
       \vect{\hat{k}}_\mathrm{in}
       \\0
\end{pmatrix}
=\frac{\mp 1}{\sin^2\chi}
 \begin{pmatrix} \sin^2\chi \\ 0 \end{pmatrix}
=\begin{pmatrix}      \mp 1 \\ 0 \end{pmatrix}
\end{gather*}
or $\alpha-\alpha_0=0$ or $\pm\pi$, i.e., the major polarisation direction is
tangential. Since $\beta=0$ the result can not depend any more on the
orientation of $\uectg{\beta}$.

\subsection{Results}

The polarisation degree and tilt of the scattered beam, its frequency shift
and the total radiant intensity per electron depend on the mean scattering
angle $\bar{\chi}$ and on the electron velocity $\vectg{\beta}$. We can
present here only examples of the results for a few of these input
parameters. In this manuscript we concentrate on the dependence on the
direction $\uectg{\beta}$ and select three representative values of the
magnitude $\beta$. Concerning the scattering angle, we restrict to 
$\bar{\chi}=\pi/2$.

As in the coordinate system used in Fig.~\ref{Fig:ScaGeom3D}, the Sun centre
is in $-\uect{z}$ direction and $\vect{\hat{k}}_\mathrm{sc}$ along $\uect{x}$.
The directions of $\vectg{\beta}$ are given in spherical coordinates with the
zenith angle $\vartheta$ with respect to $\vect{\hat{z}}$ and a spherical
azimuth angle $\phi$, i.e., $\uectg{\beta}
=(\cos\phi\sin\vartheta,\sin\phi\sin\vartheta, \cos\vartheta)$.
Then $\uectg{\beta}$ is radially outward form the Sun centre, i.e.
parallel to $\vect{\hat{z}}$, for $\vartheta=0$ and it is parallel to
$\vect{\hat{k}}_\mathrm{sc}$ for $(\phi,\vartheta)=(0,\pi/2)$. The angle of
$\uectg{\beta}$ off the mean scattering plane spanned by the
$\uect{x}$ and $\uect{z}$ axes is $\asin(\sin\phi\sin\vartheta)$.

We have selected four quantities to demonstrate the effect of relativistic
electrons on scattering observations.
The quantities shown in the figures below are\\[1ex]
\begin{tabular}{ll}
 Polarisation degree &
  $P$ as in
  (\ref{Radint_polang}) \\[1ex]
 Polarisation tilt &
  $\alpha-\alpha_0$ as in
  (\ref{Radint_polang}) and (\ref{cosin_alpha_0}) \\[1ex]
 \parbox{4cm}{Total intensity\\ amplification} &
 $\DS \frac{I_\mathrm{tot}}{I_\mathrm{ref}}
 = \frac{I_{11}+I_{22}}{2I_\mathrm{ref}} $\\[0.7ex]
 &\hspace*{1.4em}
  $\DS = \frac{r_e^2 D^4(\vect{\hat{k}}_\mathrm{sc},\vectg{\beta})}{4I_\mathrm{ref}}
     \int_\Omega
      \frac{\rad(\vect{\hat{k}}_\mathrm{in})}
        {D^2(\vect{\hat{k}}_\mathrm{in},\vectg{\beta})}
      \big(2-\sum_{i=1}^2
             (\vect{\hat{p}'}_i{}\tp\vect{\hat{k}'}_\mathrm{in})^2\big)
   \;d\Omega(\vect{\hat{k}}_\mathrm{in}) $ \\
  Effective frequency shift &
$\DS \frac{\nu\sca}{\nu\inc}
=\frac{r_e^2 D^5(\vect{\hat{k}}_\mathrm{sc},\vectg{\beta})}{2 (I_{11}+I_{22})}
       \int_\Omega
       \frac{\rad(\vect{\hat{k}}_\mathrm{in})}
                {D^3(\vect{\hat{k}}_\mathrm{in},\vectg{\beta})}
       \big(2-\sum_{i=1}^2
              (\vect{\hat{p}'}_i{}\tp\vect{\hat{k}'}_\mathrm{in})^2\big)
  \;d\Omega(\vect{\hat{k}}_\mathrm{in}) $ \\
\end{tabular}\\[1ex]
They vary with $\uectg{\beta}$ and
deviate from the non-relativistic limit with increasing $\beta$.
For the total intensity amplification the reference intensity $I_\mathrm{ref}$
is the respective value for $\beta=0$. For the effective frequency shift
recall that
$D(\vect{\hat{k}}_\mathrm{sc},\vectg{\beta})/D(\vect{\hat{k}}_\mathrm{in},\vectg{\beta})$
is the frequency shift by scattering for the individual photon.

\begin{figure}
\parbox{1em}{\rotatebox{90}{$\vartheta$}}
\parbox{16.4cm}{
  \includegraphics[viewport=32 20 345 335,clip,width=8.2cm]{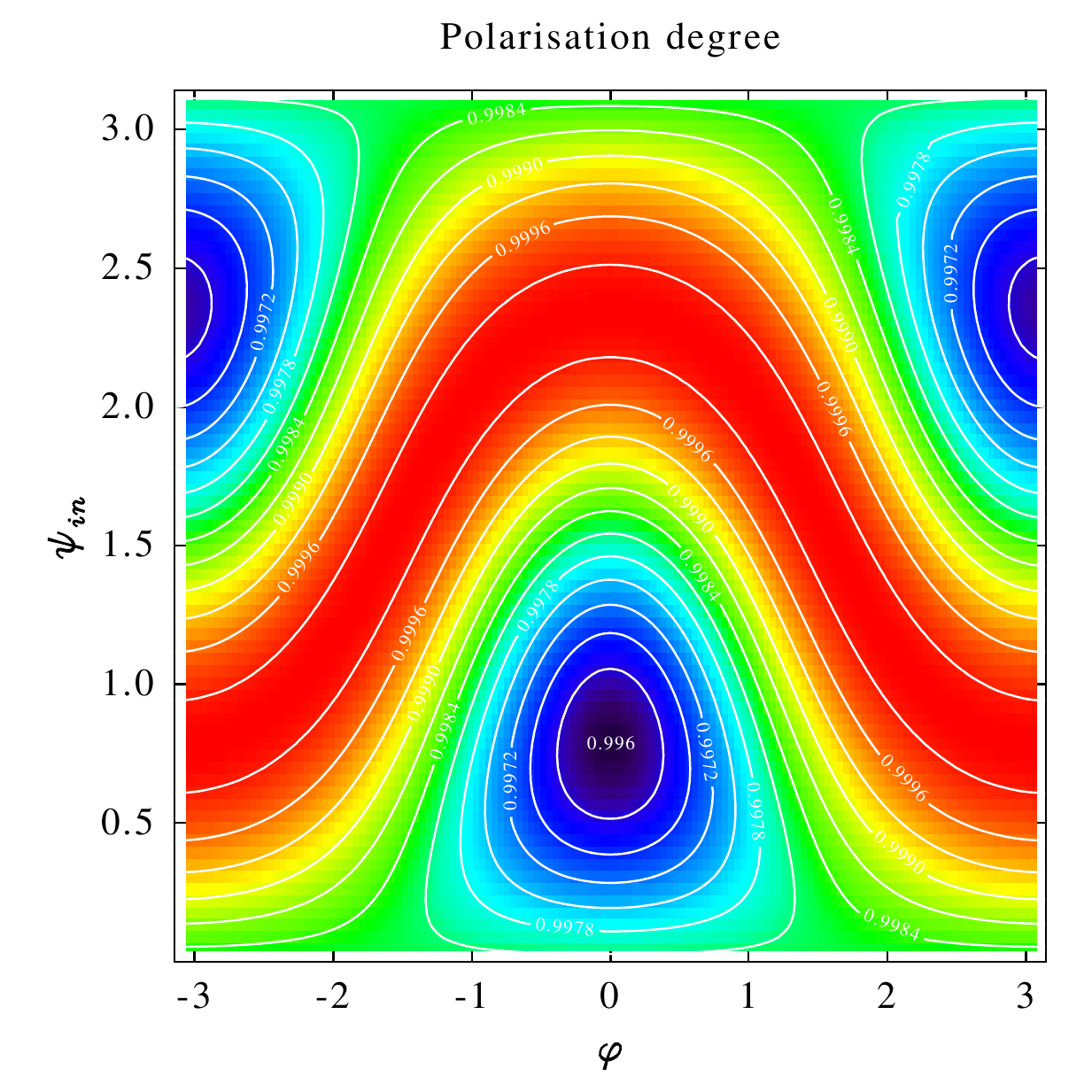}
  \includegraphics[viewport=32 20 345 335,clip,width=8.2cm]{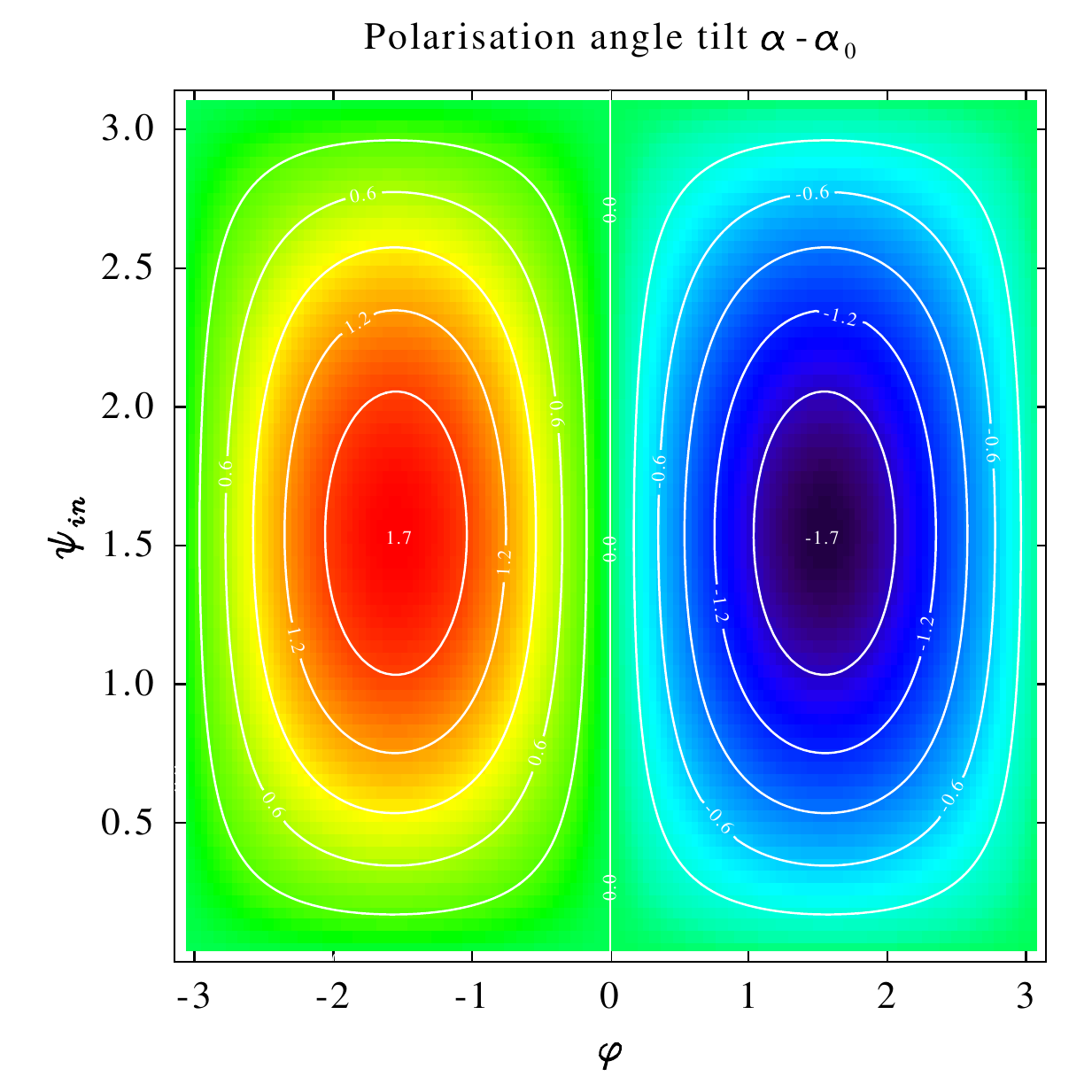}}\\
\parbox{1em}{\rotatebox{90}{$\vartheta$}}
\parbox{16.4cm}{
\includegraphics[viewport=32 20 345 335,clip,width=8.2cm]{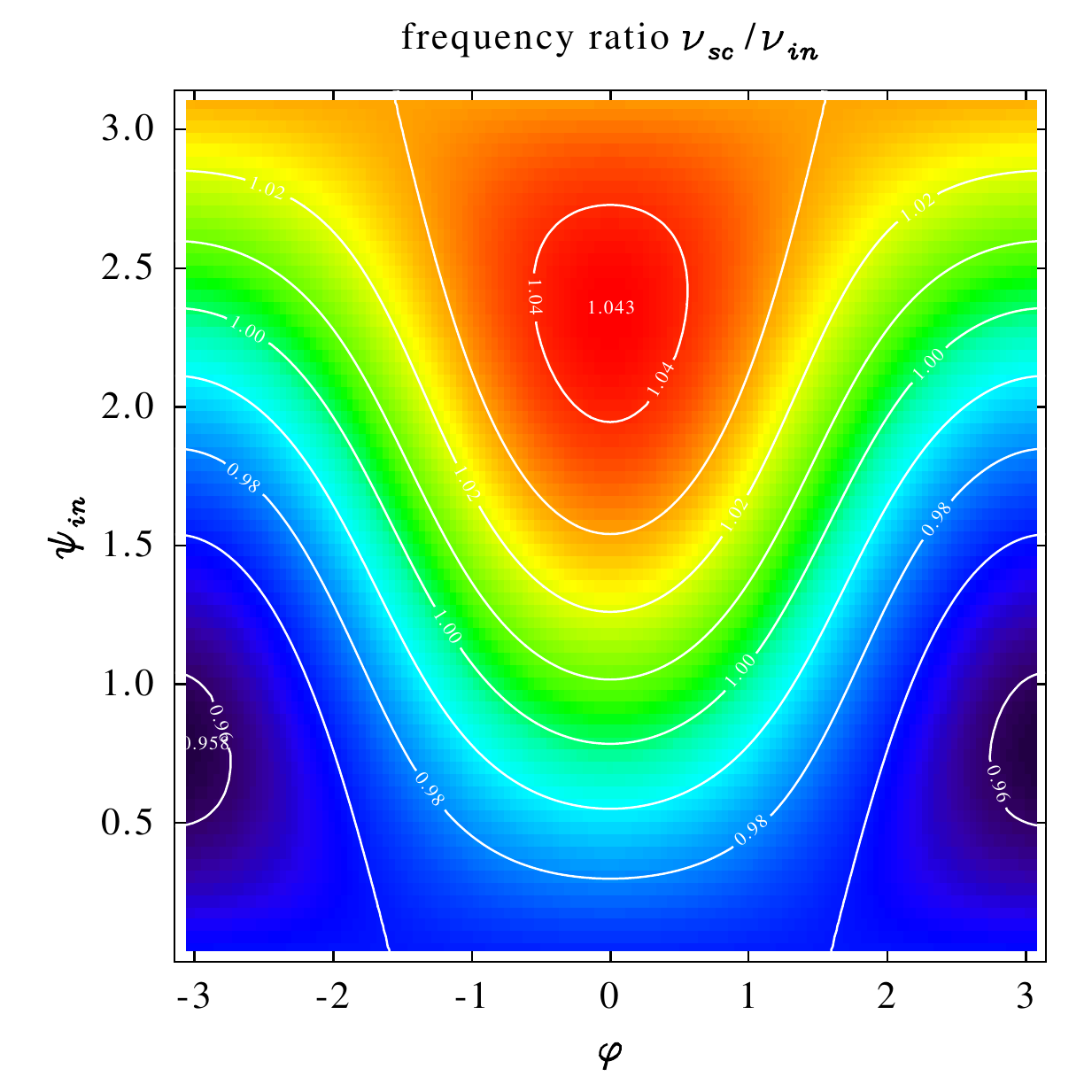}
\includegraphics[viewport=32 20 345 335,clip,width=8.2cm]{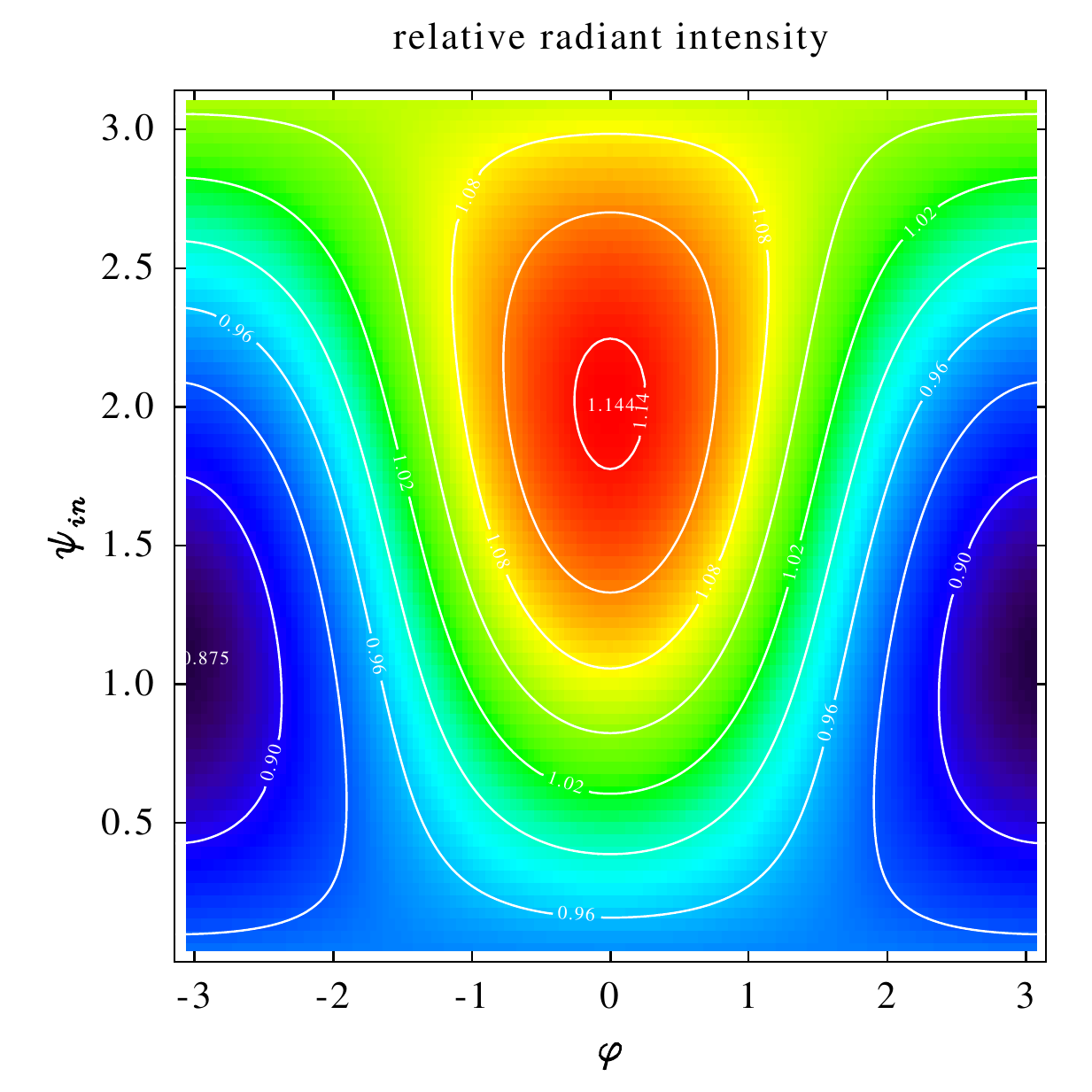}}\\
\hspace*{4.7cm}$\phi$\hspace*{8.1cm}$\phi$\\[-2ex]
\caption{For $\beta=0.03$ from top left to bottom right:
  Degree of polarisation $P$ (\ref{cosin_alpha}),
  deviation $\alpha-\alpha_0$ of the major
  polarisation direction from tangential in degrees (\ref{cosin_dalpha})
  frequency blueshift factor $\nu_\mathrm{sc}/\nu_\mathrm{in}$
  (\ref{FrqBlueShft}) and
  total intensity amplification per electron relative to the standard
  intensity $\pi r_e^2/2\,\bar{\rad}_\odot (R_\odot/r)^2$.
  The values refer to a single beam scattered at $\chi=\pi/2$.
  The angles $\phi$ and $\vartheta$ are the spherical angles of
  $\uectg{\beta}$ such that
  $\uectg{\beta}\parallel\vect{\hat{k}}_\mathrm{in}$ for
  $\vartheta=0$ and
  $\uectg{\beta}\parallel\vect{\hat{k}}_\mathrm{sc}$ for
  $\phi=0, \vartheta=\pi/2$.\label{Fig:Beam_03}}
\end{figure}

We first present the results for a single beam equivalent to a scattering site
at $r\rightarrow\infty$. There is just a single incident beam
$\vect{\hat{k}}_\mathrm{in}$ along the $\uect{z}$-direction and the angle
$\vartheta$ between $\vect{\hat{\beta}}$ and $\uect{k}\inc$ is exactly
$\psi_\mathrm{in}$.
For the scattering angle $\chi=\pi/2$ and for an electron at rest the
scattered beam has a polarisation degree of $P=1$ (see eq.~\ref{PolDegree}), a
tilt of the polarisation from the tangential direction of $\alpha-\alpha_0=0$,
a frequency shift of unity and a radiant intensity per electron of
$I_\mathrm{tot}=I_\mathrm{tan}=r_e^2/2\;\bar{L}_\odot(R_\odot/r)^2$ (see
eq~\ref{Rad_inf}). We will take the last value as reference $I_\mathrm{ref}$
for the intensity amplification presented for the relativistic case.

\begin{figure}
\hspace*{\fill}
\includegraphics[viewport=190 50 650 480,clip,width=5cm]{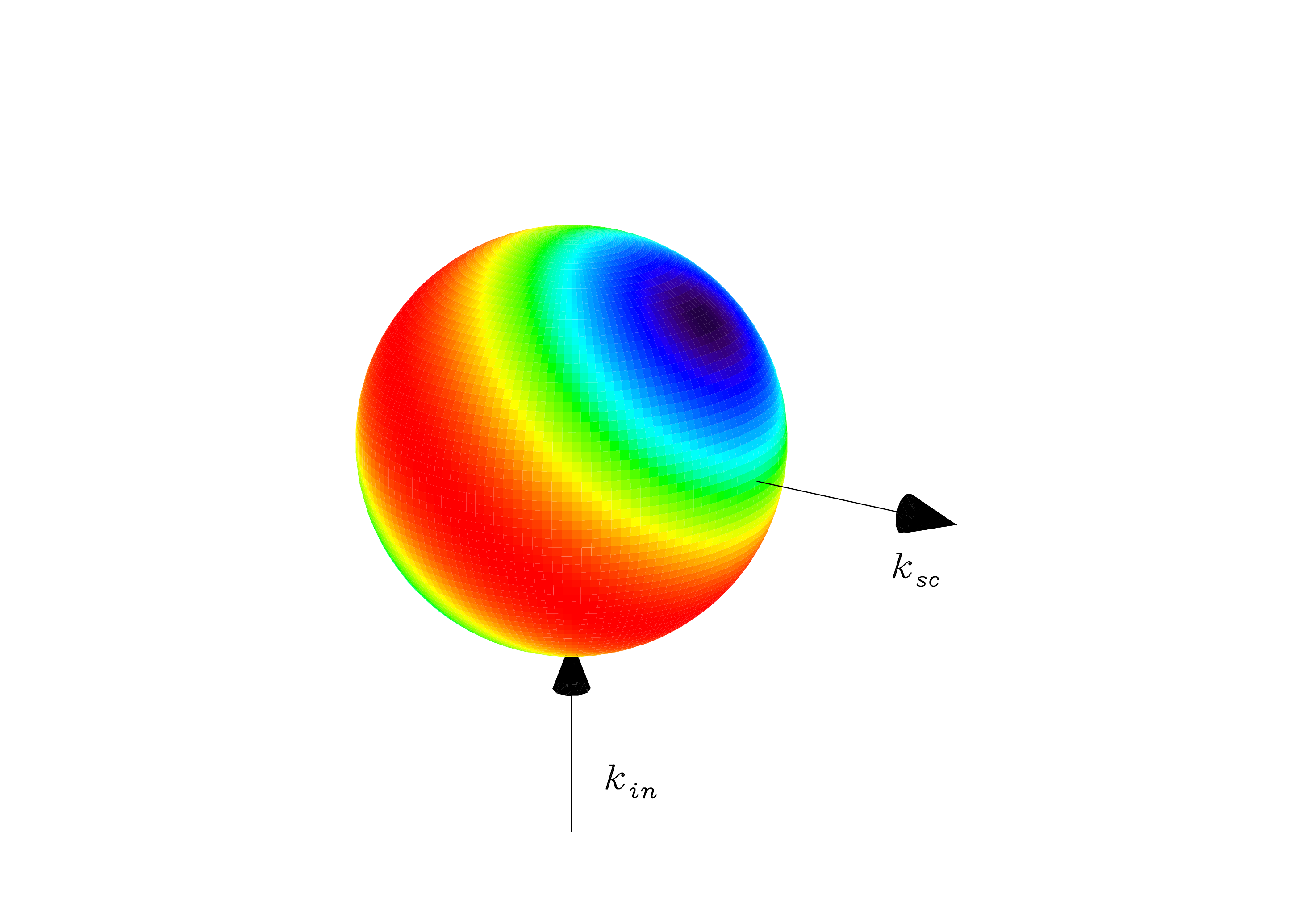}
\includegraphics[viewport=190 50 650 480,clip,width=5cm]{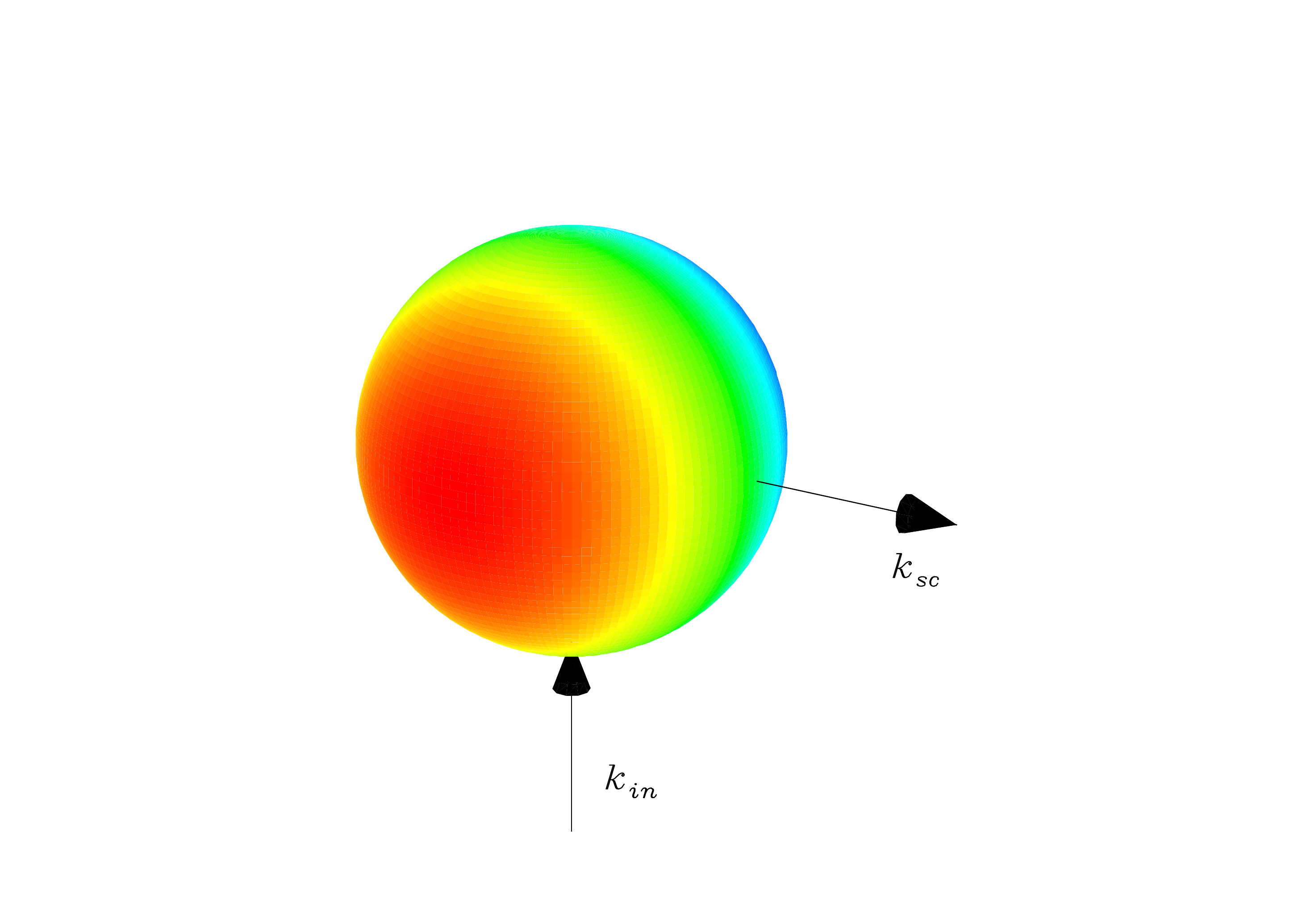}
\includegraphics[viewport=190 50 650 480,clip,width=5cm]{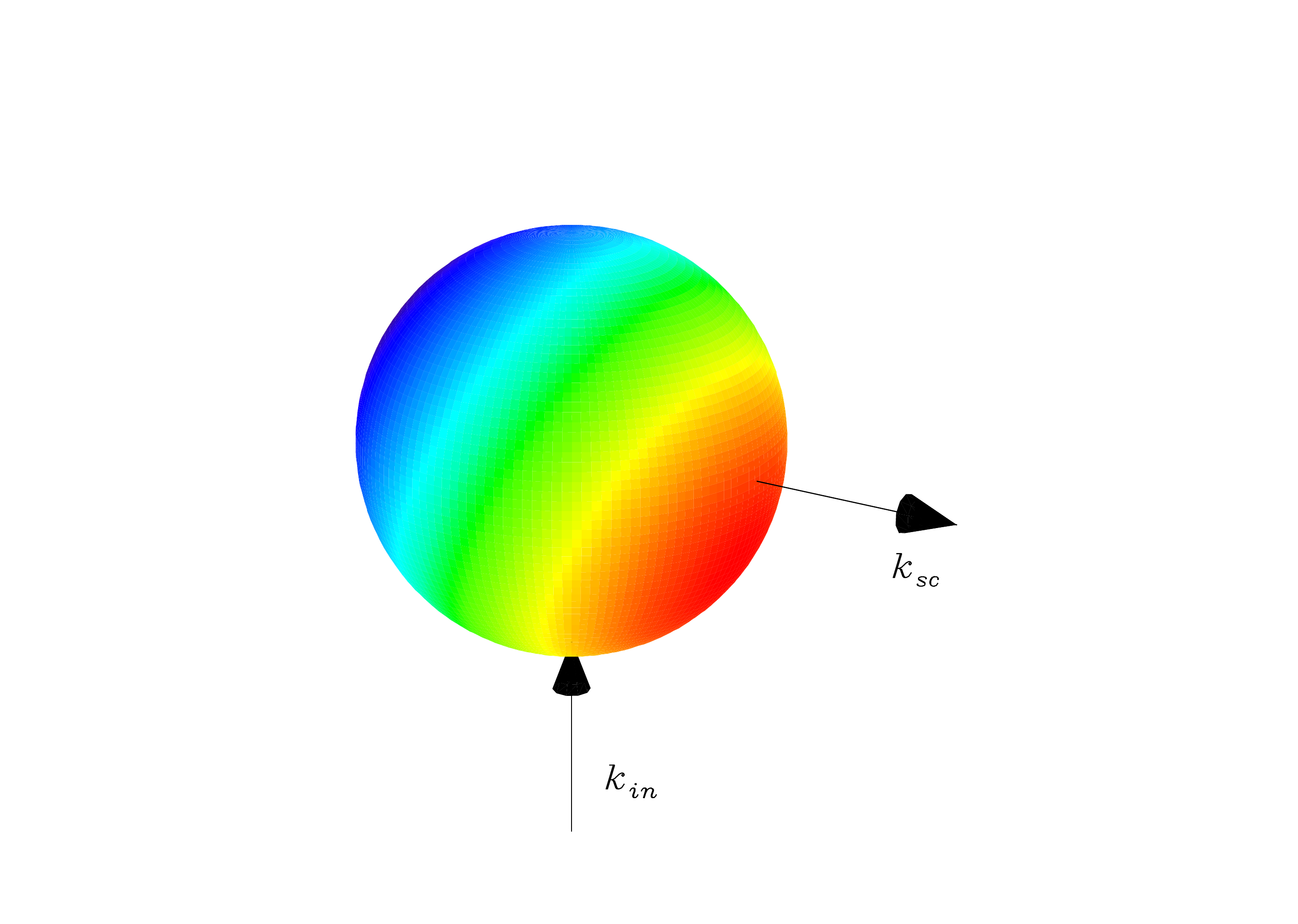}
\hspace*{\fill}
\caption{Results of Fig.~\ref{Fig:Beam_03} replotted on a 3D sphere to better
  illustrate the 3D variation with $\vect{\hat{\beta}}$ in space and their
  relation to the incident and scattered beam direction. From left to right:
  Degree of polarisation, deviation of the major polarisation direction from
  tangential and total intensity amplification per electron. The
  colour code is the same as in Fig.~\ref{Fig:Beam_03}.}
  \label{Fig:Beam_03_3D}
\end{figure}

On the top left panels of Figs.~\ref{Fig:Beam_03} we show the small deviations
from these non-relativistic values for $\beta$=0.03 which is about twice the
thermal speed of a coronal electron. The polarisation degree varies between
$P=0.996$ and 1 and the polarisation axis is titled away from the tangential
direction between $\pm 1.7^\circ$. The photon frequency is shifted by up to
$\pm$4\% and the total intensity is amplified by up to $\pm$11\%. All these
variations depend on the direction of $\uectg{\beta}$. To better
illustrate this dependency, we have replotted $P$, $\alpha-\alpha_0$ and
$I_\mathrm{tot}/I_\mathrm{ref}$ on a unit sphere in Fig.~\ref{Fig:Beam_03_3D}.

From this representation it can be seen that the minimum degree of the
polarisation is reached at directions of $\uectg{\beta}$ along
$\pm(\vect{\hat{k}}_\mathrm{sc}+\vect{\hat{k}}_\mathrm{in})$. For
$\uectg{\beta}$ in the plane normal to this axis we have $P=1$. As can
be easily checked for electron velocities in the plane
$\uectg{\beta}{}\tp(\vect{\hat{k}}_\mathrm{sc}+\vect{\hat{k}}_\mathrm{in})=0$
the scattering angle in the electron rest frame is $\chi'=\pi/2$.
The strongest polarisation tilt occurs for $\uectg{\beta}$ pointing
normal to the scattering plane. If $\uectg{\beta}$ lies in the
scattering plane, the aberration planes coincide with the scattering planes in
the Sun's frame and the scattering plane in the electron's frame is unchanged
from the Sun's frame.
The strongest frequency blueshift and intensity amplification occur when
$\uectg{\beta}$ points to $\vect{\hat{k}}_\mathrm{sc}-
\vect{\hat{k}}_\mathrm{in}$. For this electron velocity direction the
scattered photon frequency is upshifted twice because electron sees the
incident electron upshifted and the observer sees the electron scattered
emission upshifted.
The parameters $P$, $\nu_\mathrm{sc}/\nu_\mathrm{in}$ and
$I_\mathrm{tot}/I_\mathrm{ref}$ vary symmetrically with angle $\phi$, i.e.
there is symmetry for $\uectg{\beta}$ with respect to the scattering
plane in the Sun's rest frame. The polarisation tilt angle $\alpha-\alpha_0$
varies antisymmetrically instead. In general, we find that for this small
value of $\beta$ all four beam parameters change sign when
$\uectg{\beta}$ reverses sign which indicates that they are perturbed
only linearly by $\vectg{\beta}$.

\begin{figure}
\parbox{1em}{\rotatebox{90}{$\vartheta$}}
\parbox{16.4cm}{
  \includegraphics[viewport=32 20 345 335,clip,width=8.2cm]{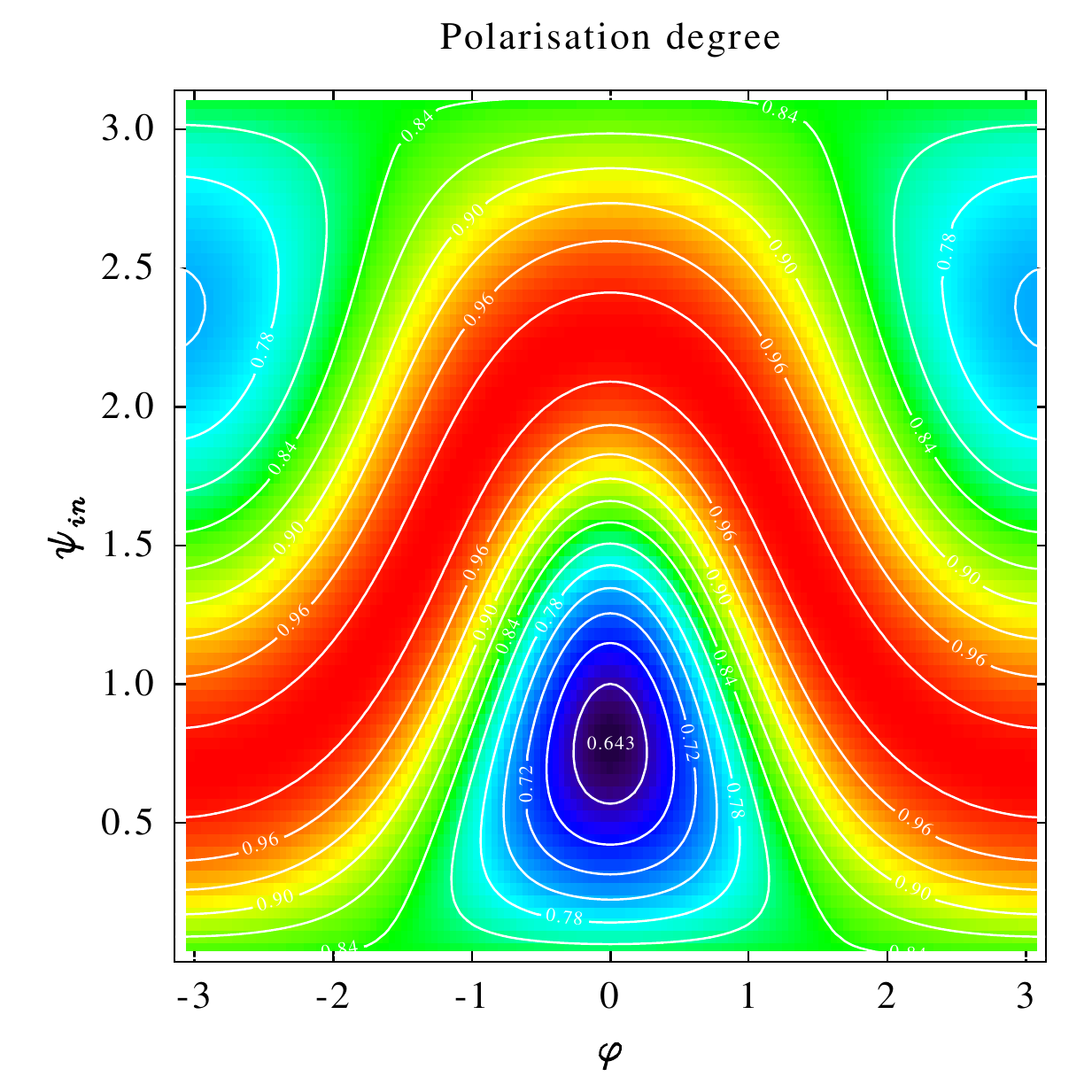}
  \includegraphics[viewport=32 20 345 335,clip,width=8.2cm]{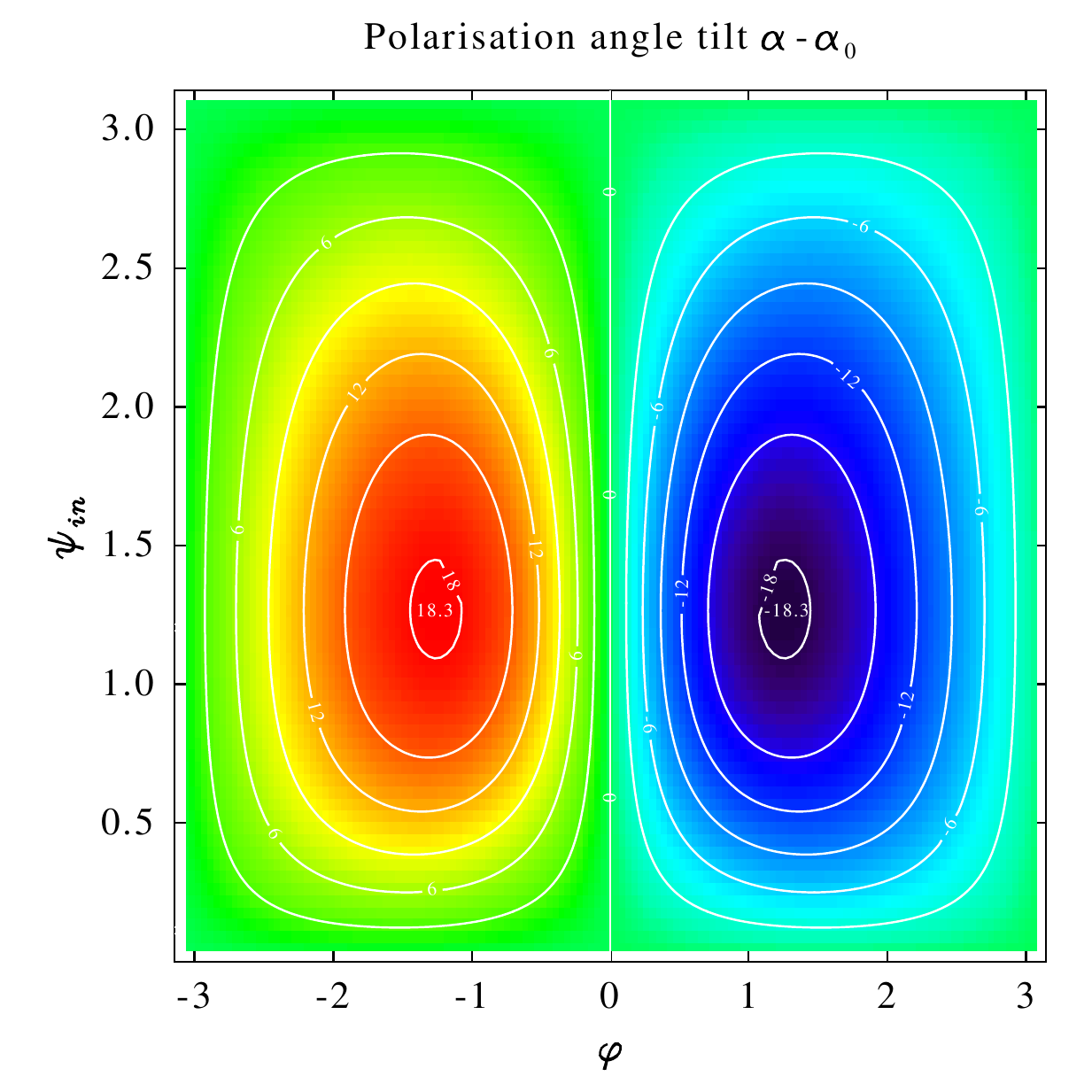}}\\
\parbox{1em}{\rotatebox{90}{$\vartheta$}}
\parbox{16.4cm}{
\includegraphics[viewport=32 20 345 335,clip,width=8.2cm]{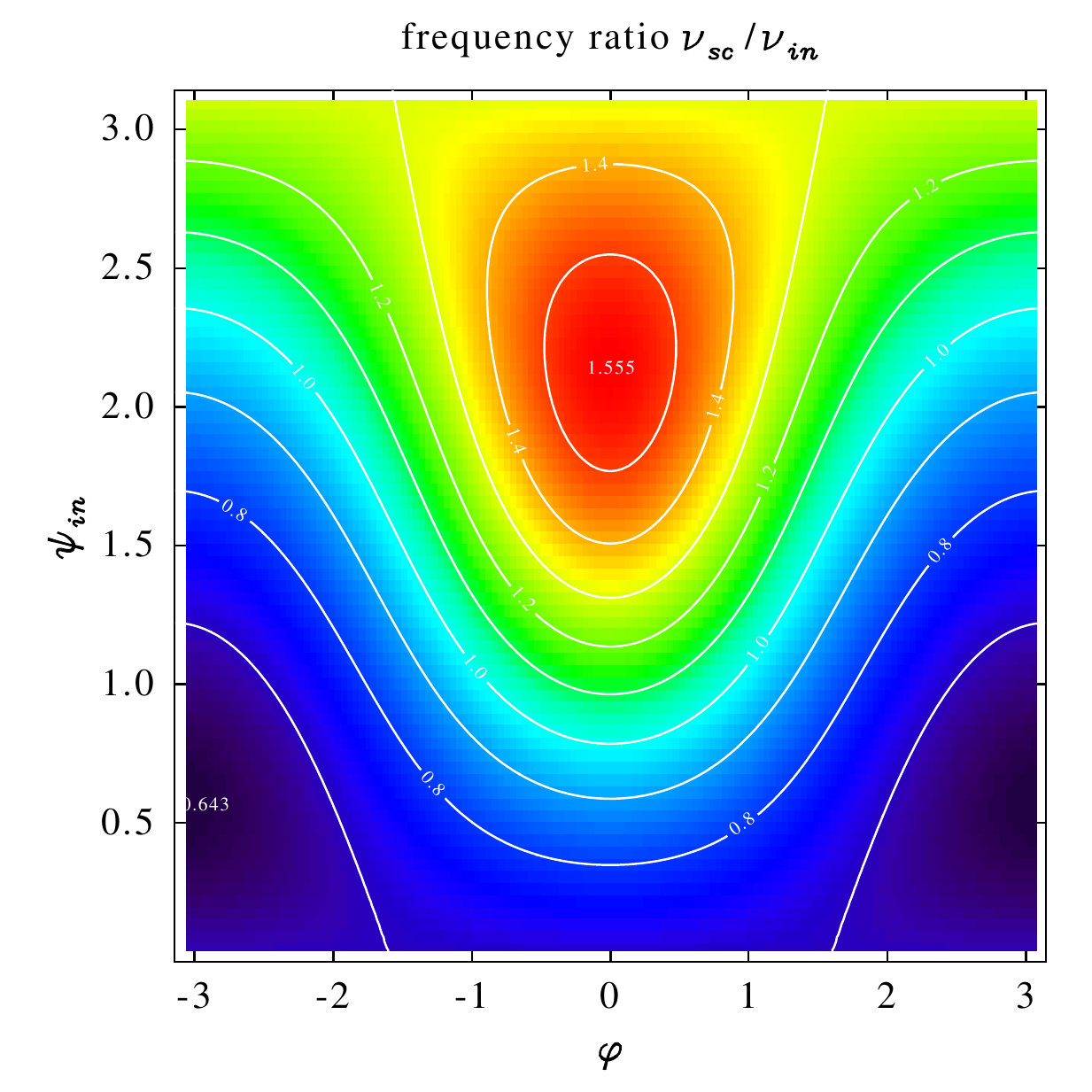}
\includegraphics[viewport=32 20 345 335,clip,width=8.2cm]{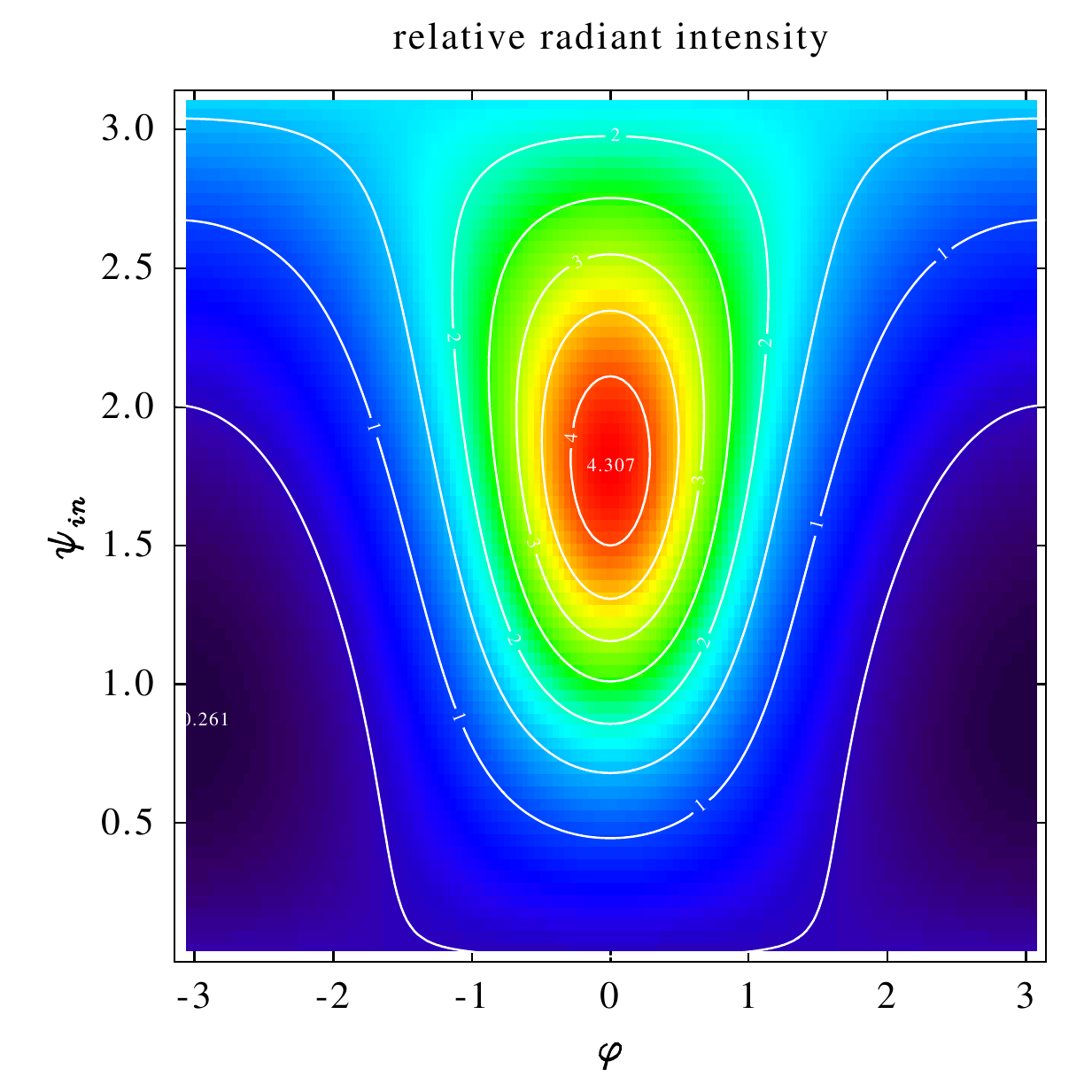}}\\
\hspace*{4.7cm}$\phi$\hspace*{8.1cm}$\phi$\\[-2ex]
\caption{Same as Fig.~\ref{Fig:Beam_03} but for $\beta=0.3$ From top left to
  bottom right: Degree of polarisation, deviation of the major polarisation
  direction from tangential in degrees, frequency blueshift factor and
  total intensity amplification per electron.}
  \label{Fig:Beam_30}.
\end{figure}

\begin{figure}
\parbox{1em}{\rotatebox{90}{$\vartheta$}}
\parbox{16.4cm}{
  \includegraphics[viewport=32 20 345 335,clip,width=8.2cm]{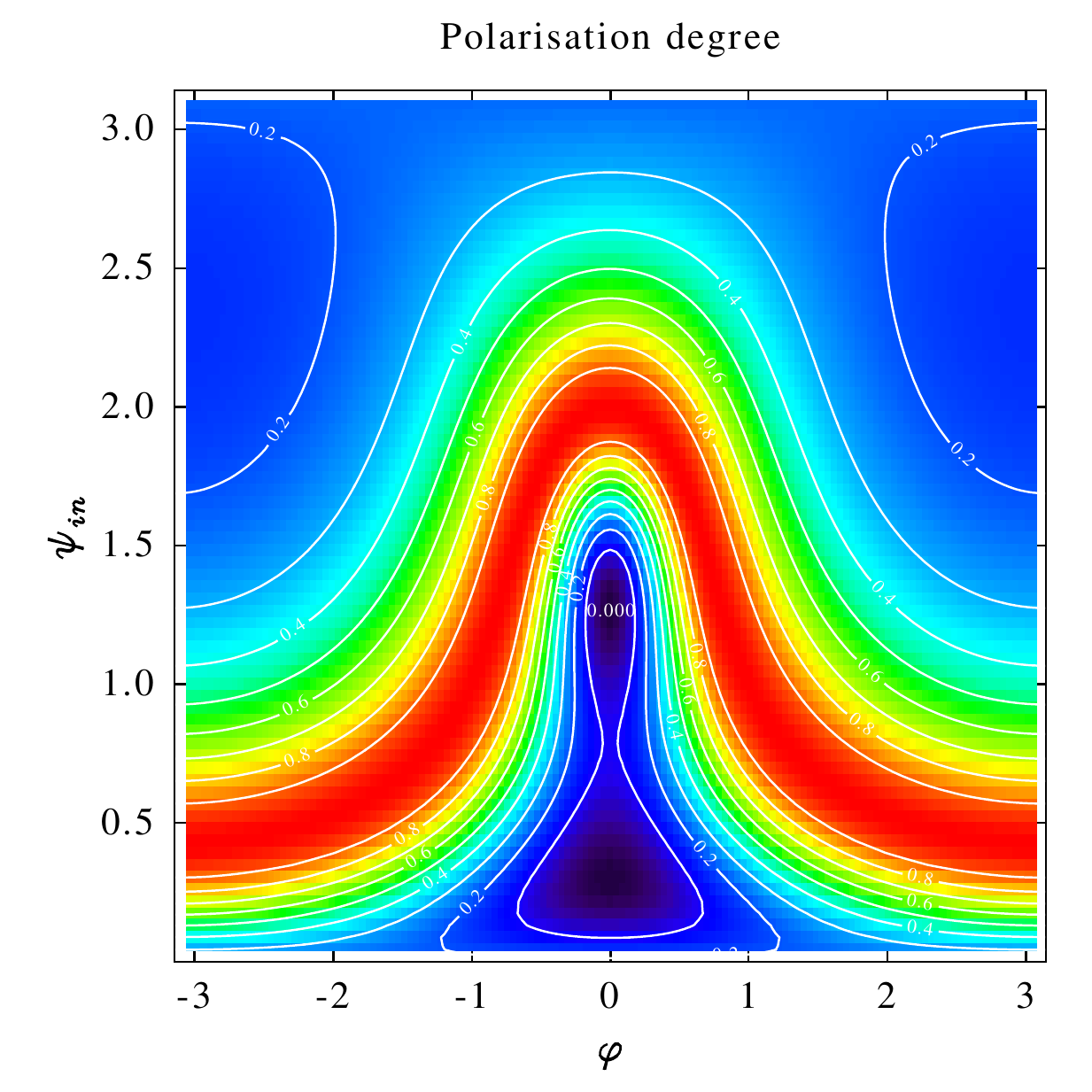}
  \includegraphics[viewport=32 20 345 335,clip,width=8.2cm]{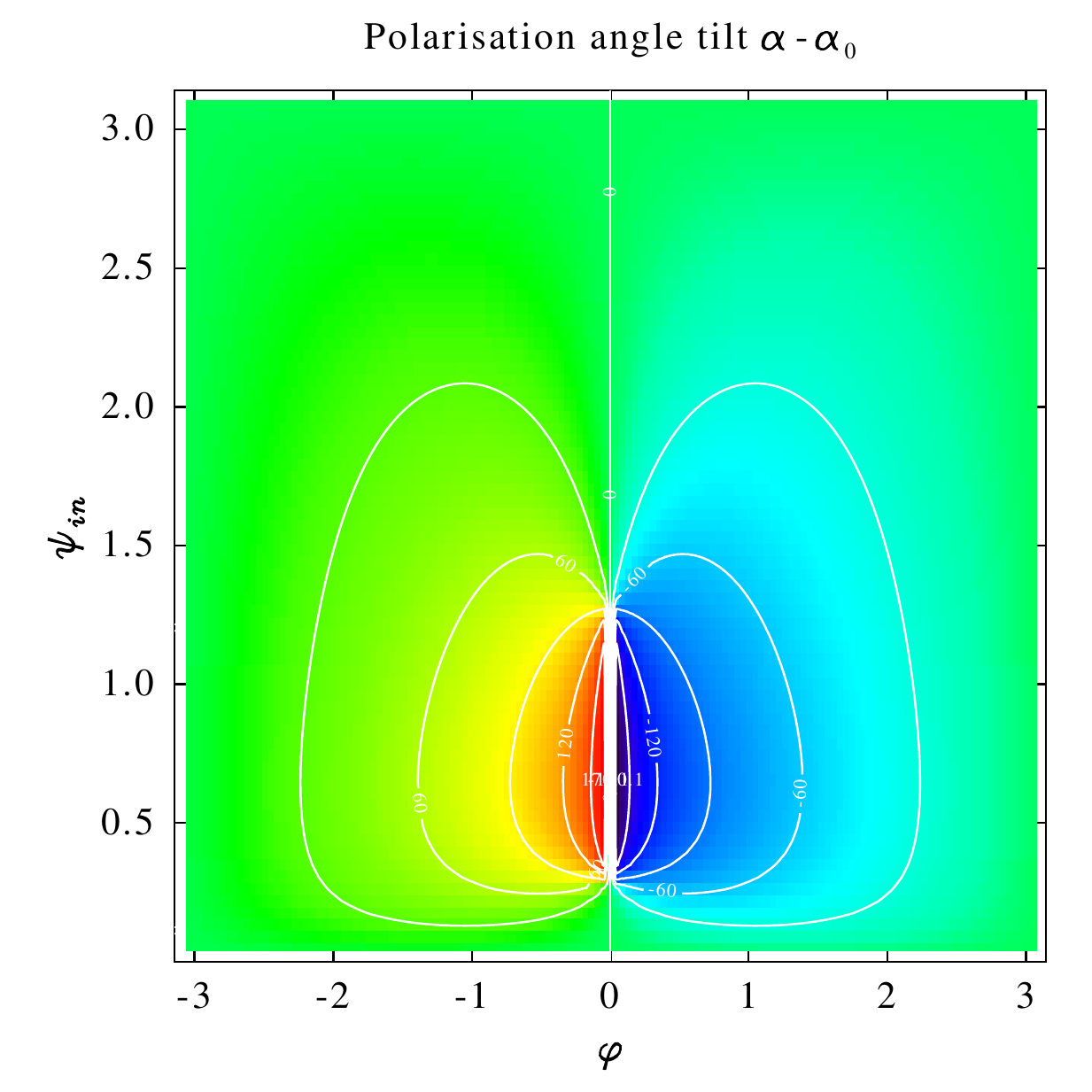}}\\
\parbox{1em}{\rotatebox{90}{$\vartheta$}}
\parbox{16.4cm}{
\includegraphics[viewport=32 20 345 335,clip,width=8.2cm]{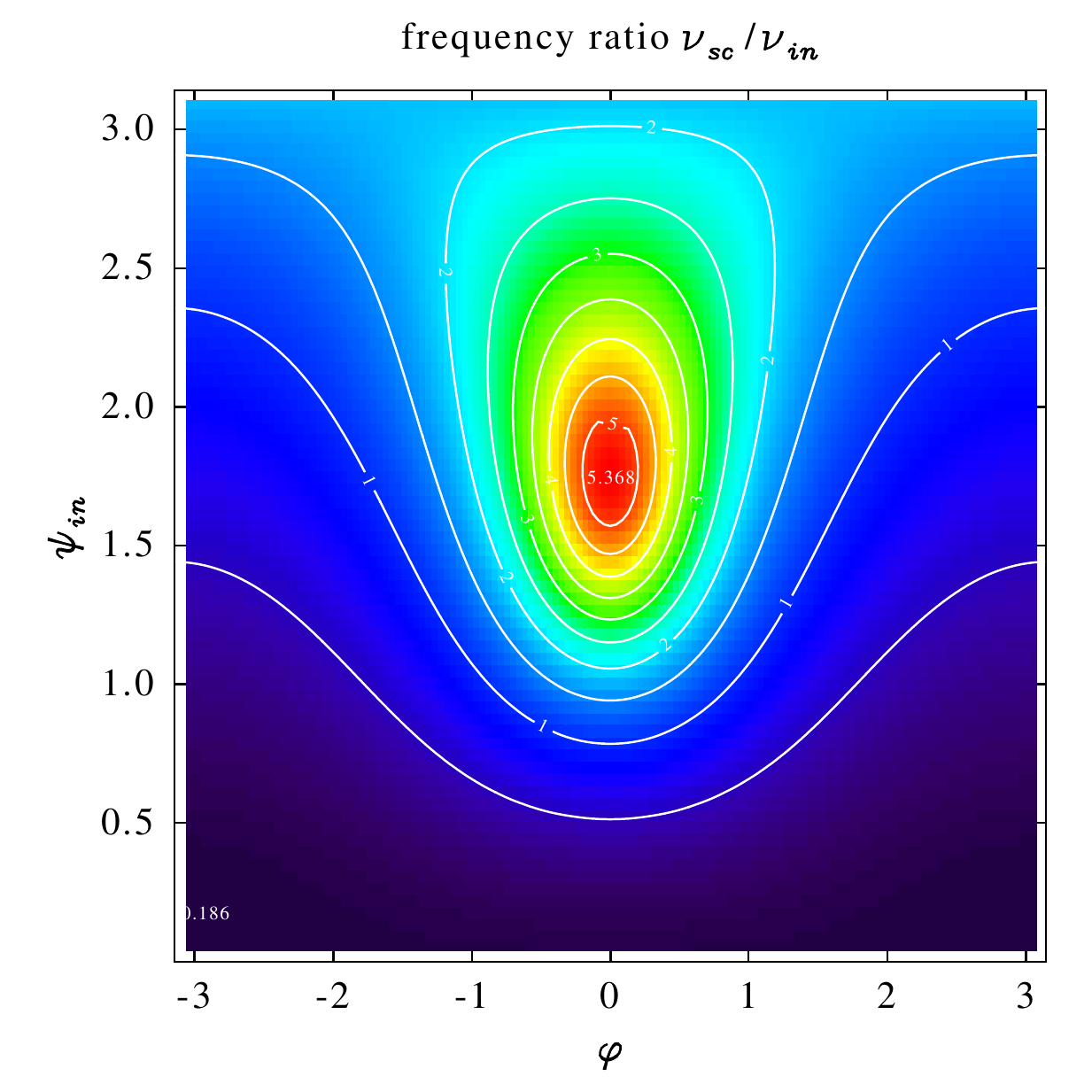}
\includegraphics[viewport=32 20 345 335,clip,width=8.2cm]{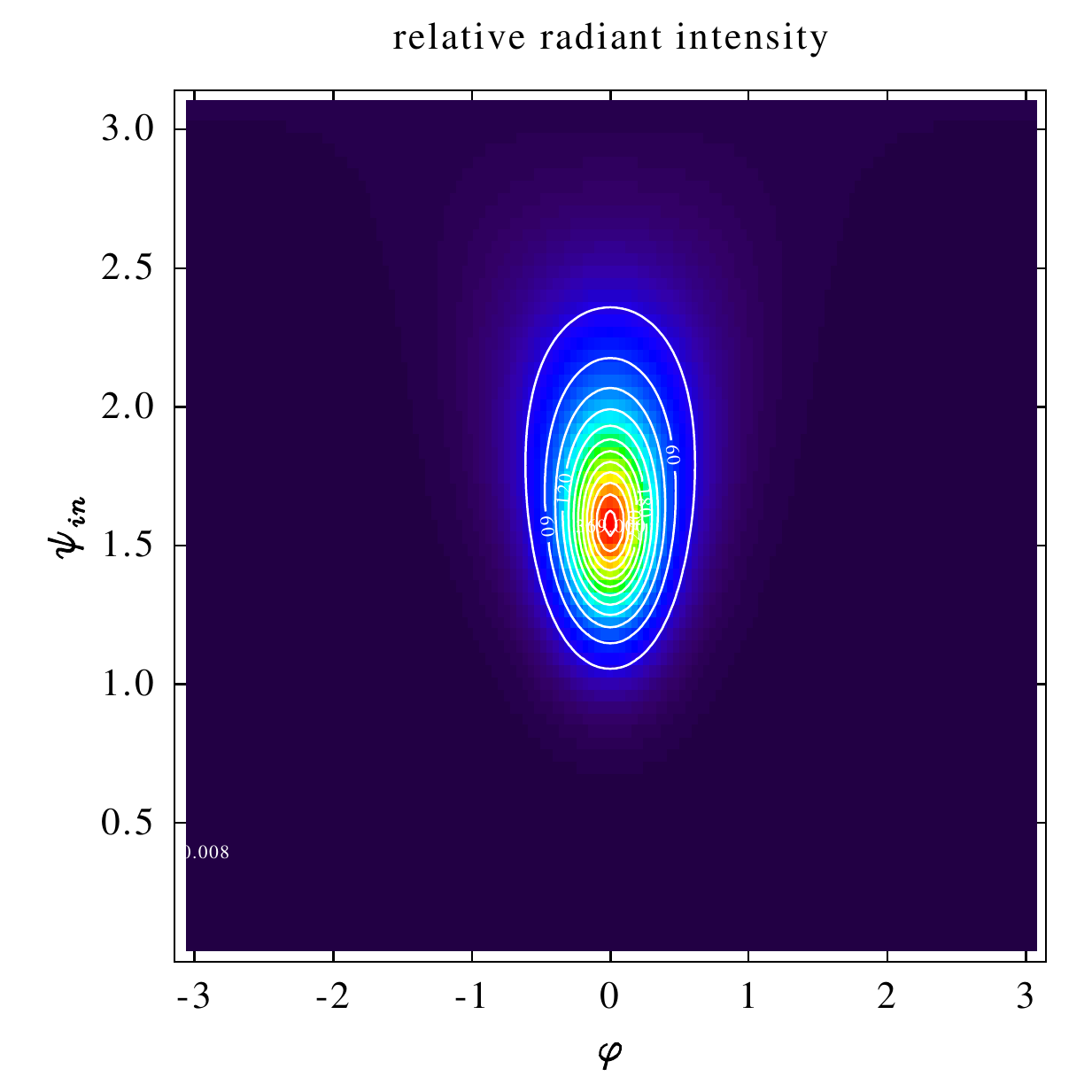}}\\
\hspace*{4.7cm}$\phi$\hspace*{8.1cm}$\phi$\\[-2ex]
\caption{Same as Fig.~\ref{Fig:Beam_03} but for $\beta=0.8$ From top left to
  bottom right: Degree of polarisation, deviation of the major polarisation
  direction from tangential in degrees, frequency blueshift factor and
  total radiant intensity amplification per electron.}
  \label{Fig:Beam_80}
\end{figure}


This antisymmetry with respect to $\pm\uectg{\beta}$ is lost when $\beta$ is
enhanced, only the (anti)symmetry with respect to the scattering plane
remains.
In Figs.~\ref{Fig:Beam_30} and \ref{Fig:Beam_80} we show the four beam
parameters for $\beta=0.3$ and 0.8, respectively.
The minimum degree of the polarisation reached is $P=0.643$ for $\beta=0.3$
and even $P=0$ for $\beta=0.8$. The direction of $\uectg{\beta}$ for this
minimal $P$ concentrates more and more on the scattering plane with directions
in the quadrant between $\vect{\hat{k}}_\mathrm{in}$ and
$\vect{\hat{k}}_\mathrm{sc}$. In the reverse direction the minimum
is much less pronounced with $P=0.75$ and 0.18 for $\beta=0.3$ and 0.8,
respectively. The surface $P=1$ has its normal inclined closer to
$\vect{\hat{k}}_\mathrm{in}$ as $\beta$ is enhanced and it is not dissolved
despite the relativistic speed of the electron.

The most extreme values of the polarisation tilt angle for $\beta=0.3$ are
$\alpha-\alpha_0=\pm 18.3^\circ$. They are assumed for $\uectg{\beta}$
not any more exactly normal to the scattering plane but inclined by 18$^\circ$
towards the quadrant between $\vect{\hat{k}}_\mathrm{in}$ and
$\vect{\hat{k}}_\mathrm{sc}$. For larger $\beta$ this tendency continues so
that for $\beta=0.8$ the extreme tilt angles reach $\alpha-\alpha_0=\pm
180^\circ$ in directions for $\uectg{\beta}$ which collapse on the
quadrant between $\vect{\hat{k}}_\mathrm{in}$ and
$\vect{\hat{k}}_\mathrm{sc}$.
Note that polarisation tilt angles of $\pm 180^\circ$ and $0^\circ$ are
equivalent so that the tilt angle rotates continuously through the tangential
direction as $\uectg{\beta}$ crosses the scattering plane at the
quadrant between $\vect{\hat{k}}_\mathrm{in}$ and $\vect{\hat{k}}_\mathrm{sc}$.

The frequency shift and the radiant intensity also concentrate their maximum
with increasing $\beta$ in the forward direction $\uectg{\beta}$ of the
electron. This complies with the headlight effect discussed qualitatively
above. For $\beta$=0.3, a maximum blueshift factor of 1.55 is obtained when
$\uectg{\beta}$ points only 122$^\circ$ away from
$\vect{\hat{k}}_\mathrm{in}$ and 32$^\circ$ from $\vect{\hat{k}}_\mathrm{sc}$.
In the reverse direction, the factor is 0.643 which corresponds to a red shift
of 1/0.643=1.55. For $\beta=0.8$ the maximum blueshift factor is 5.37 reached
when $\uectg{\beta}$ is almost parallel to $\vect{\hat{k}}_\mathrm{sc}$,
i.e., when the electron emission is seen with maximum blueshift. Accordingly,
the strongest redshift is by factor 0.186=1/5.37 in direction
$\uectg{\beta}\parallel\vect{\hat{k}}_\mathrm{in}$ when the scattering
electron sees the incoming photon with the strongest redshift.
The maximum radiant intensity is enhanced by a factor 4.30 above the
non-relativistic value for $\beta=0.3$ and even by 370 at $\beta=0.8$.
This strong amplification in the forward direction of the electron is
due to the headlight effect and is formally a consequence of the factor
$D^4(\vect{\hat{k}}_\mathrm{sc},\vectg{\beta})/D^2(\vect{\hat{k}}_\mathrm{in},\vectg{\beta})$
in (\ref{Radint_pp}). In forward direction, i.e. for $\psi\sca=0$ and $\psi\inc=0$,
this alone factor amounts to $\beta(1+\beta)/(1-\beta)^3=180$ for $\beta=0.8$.

\begin{figure}
\parbox{1em}{\rotatebox{90}{$\vartheta$}}
\parbox{16.4cm}{
  \includegraphics[viewport=32 20 345 335,clip,width=8.2cm]{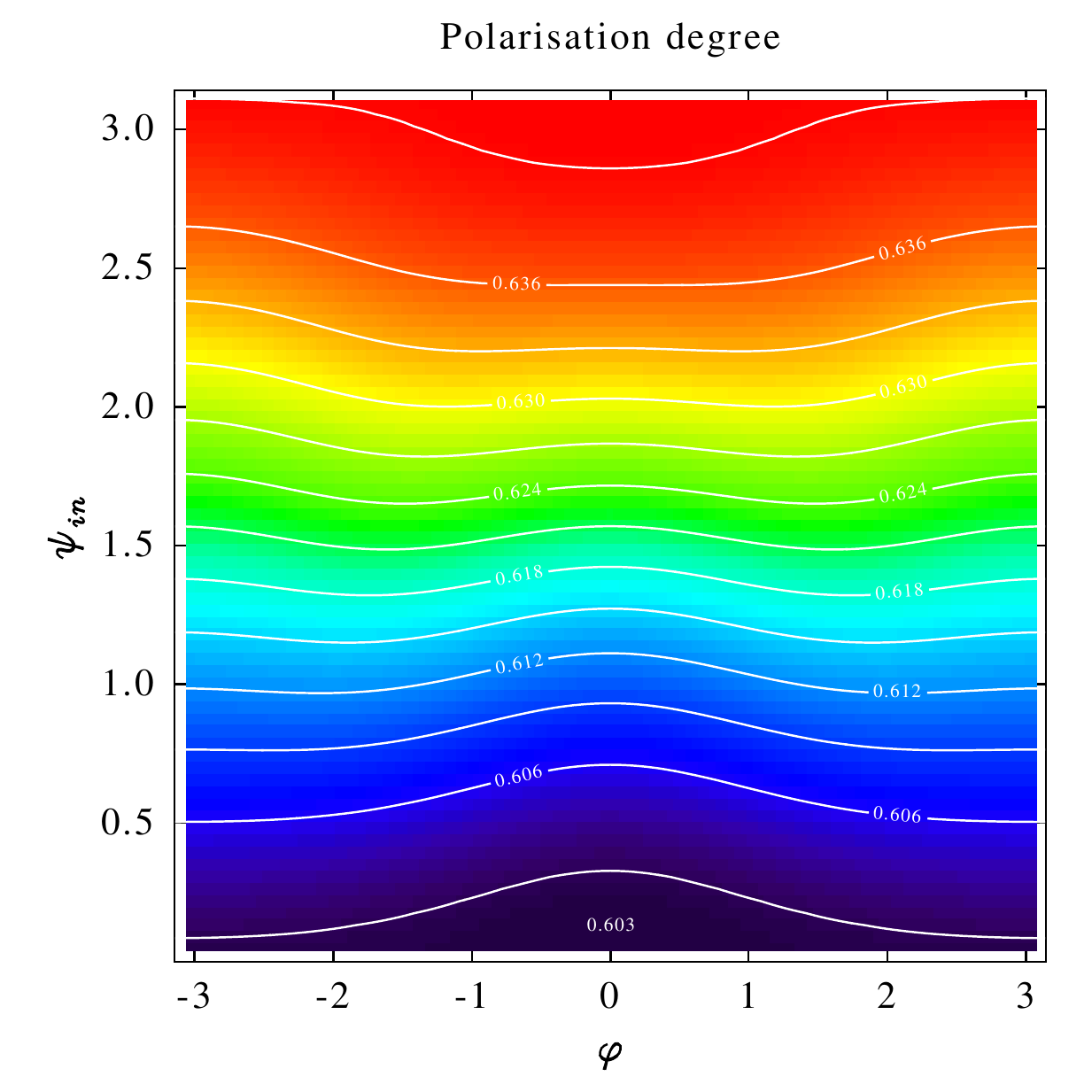}
  \includegraphics[viewport=32 20 345 335,clip,width=8.2cm]{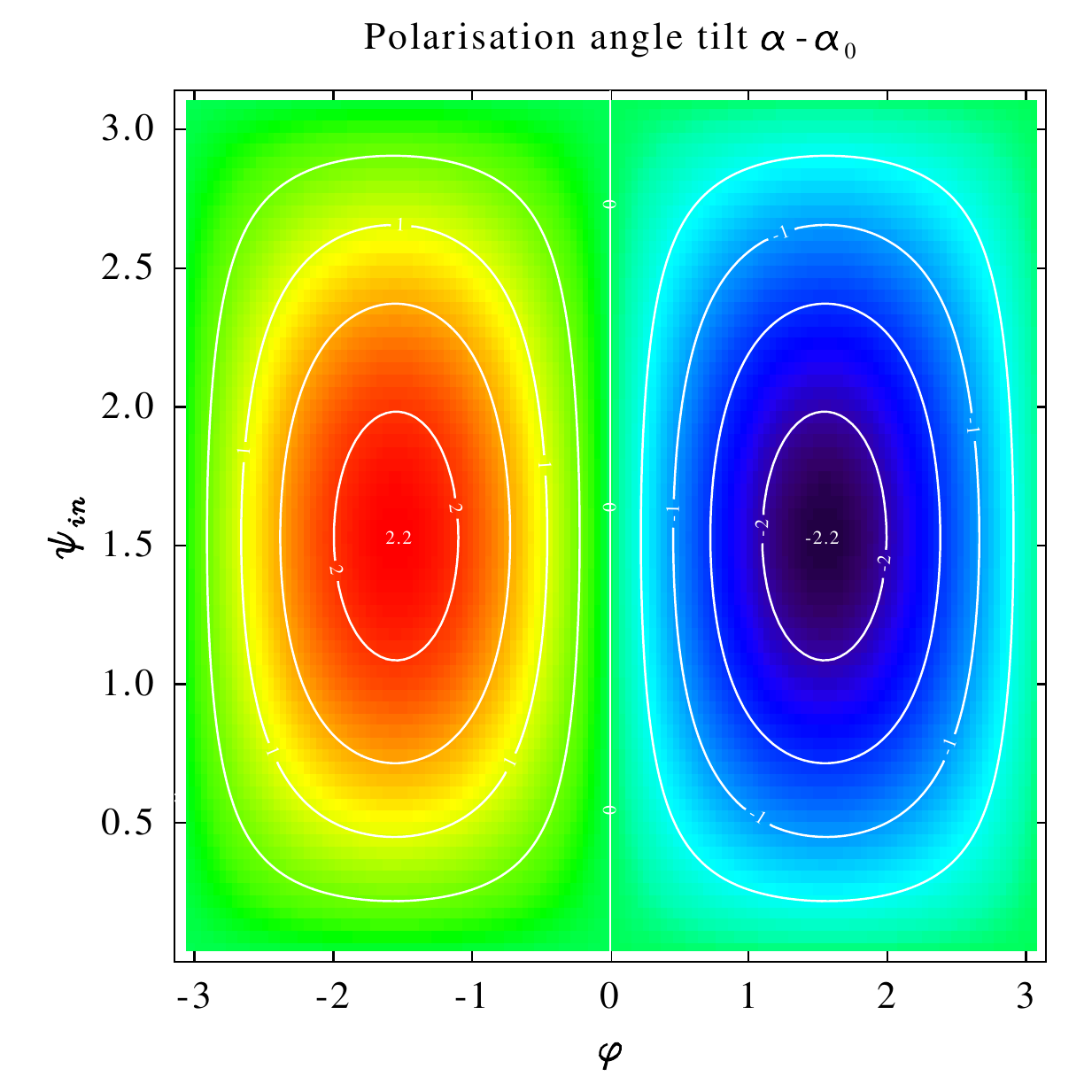}}\\
\parbox{1em}{\rotatebox{90}{$\vartheta$}}
\parbox{16.4cm}{
\includegraphics[viewport=32 20 345 335,clip,width=8.2cm]{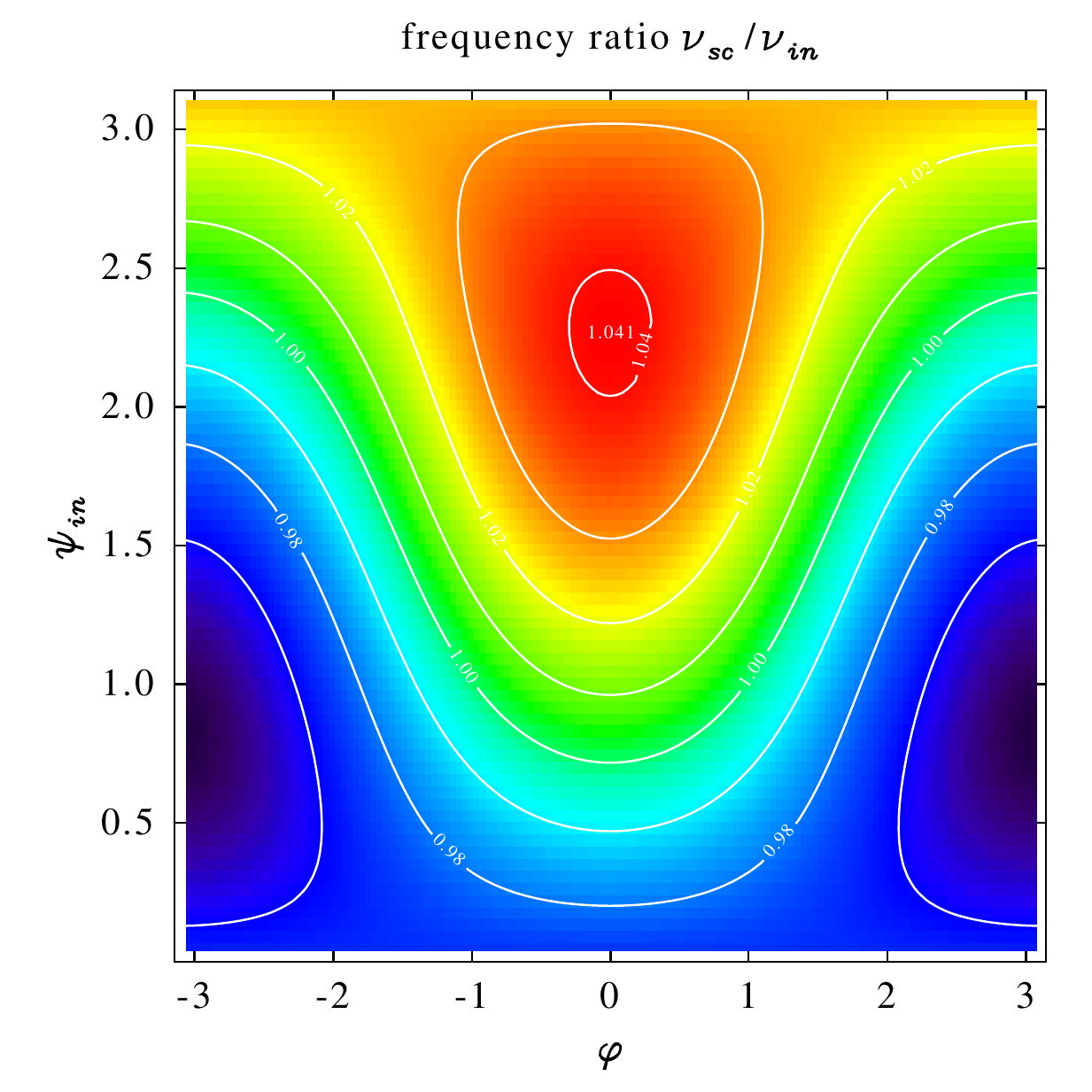}
\includegraphics[viewport=32 20 345 335,clip,width=8.2cm]{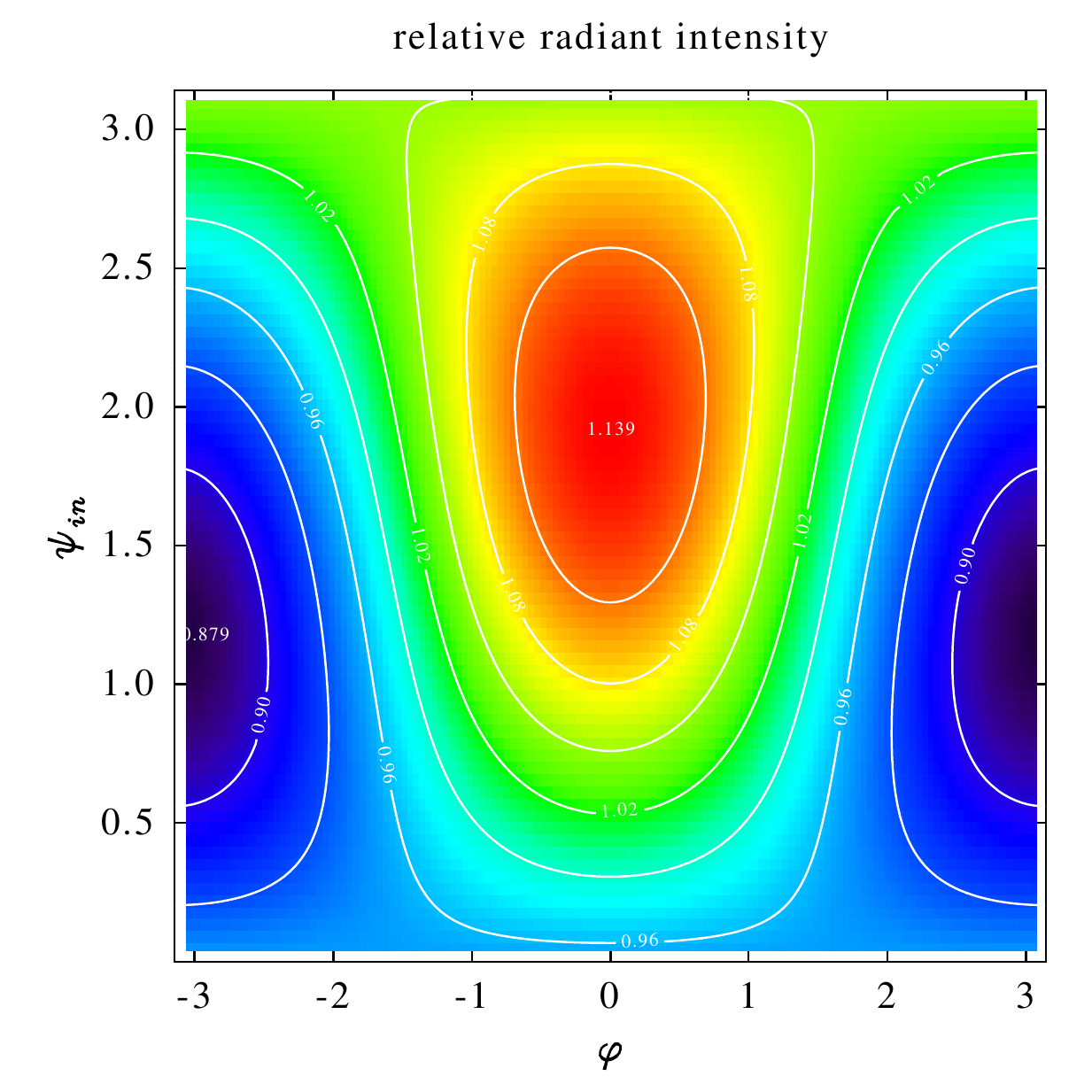}}\\
\hspace*{4.7cm}$\phi$\hspace*{8.1cm}$\phi$\\[-2ex]
\caption{For $\beta=0.03$ from top left to bottom right: Degree of
  polarisation $P$ and deviation $\alpha-\alpha_0$ of the major polarisation
  direction from tangential in degrees (\ref{Radint_polang}),
  intensity-weighted effective frequency blueshift factor 
  $\nu_\mathrm{sc}/\nu_\mathrm{in}$ and total radiant intensity amplification
  for an electron at
  rest. The values shown refer to a scattering electron at a distance $r=1.50
  R_\odot$ from Sun centre and at a scattering angle $\bar{\chi}=\pi/2$.
  The angles $\phi$ and $\vartheta$ are the spherical angles of
  $\uectg{\beta}$ as in Fig~\ref{Fig:Beam_03}.
  \label{Fig:FSun_03}}
\end{figure}

\begin{figure}
\parbox{1em}{\rotatebox{90}{$\vartheta$}}
\parbox{16.4cm}{
  \includegraphics[viewport=32 20 345 335,clip,width=8.2cm]{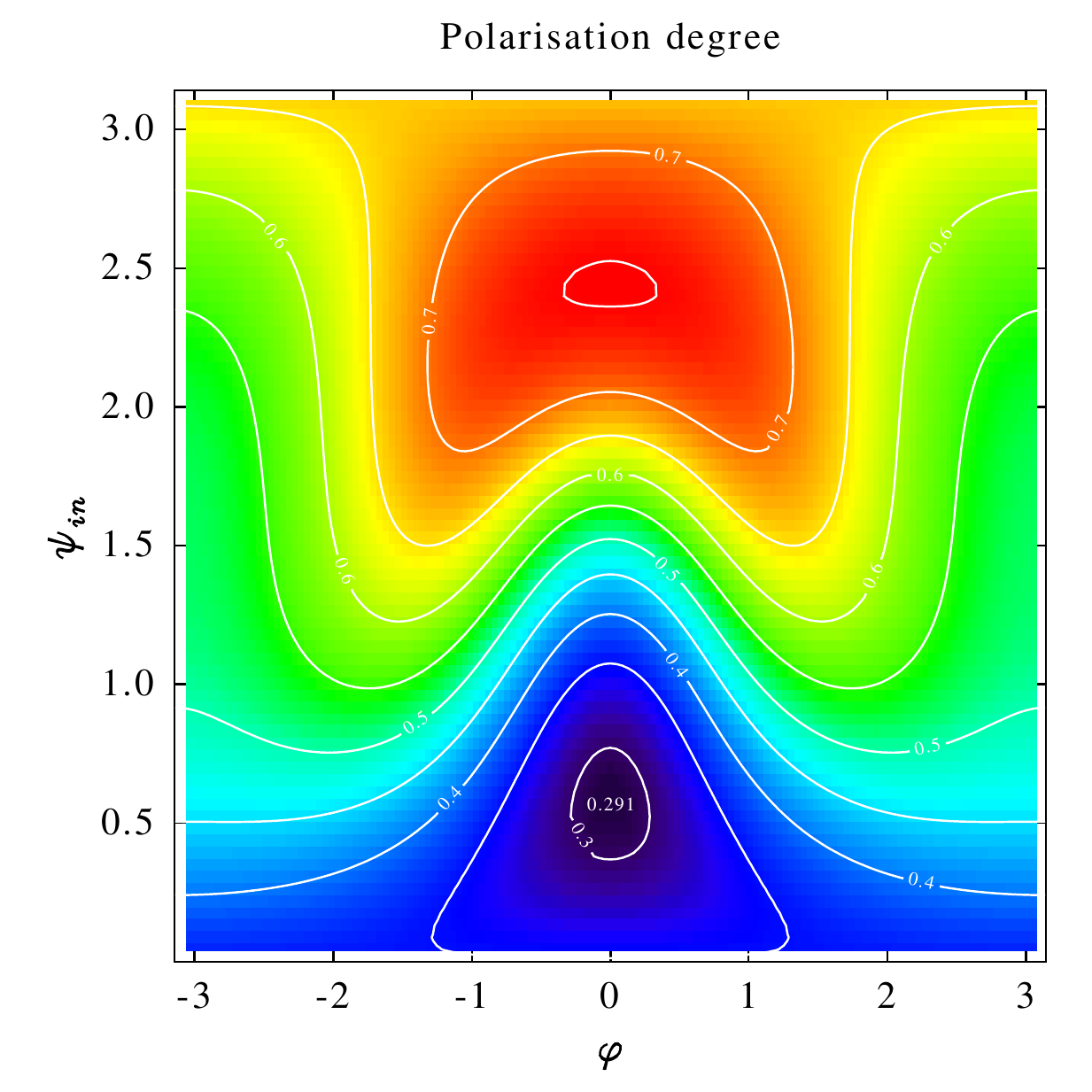}
  \includegraphics[viewport=32 20 345 335,clip,width=8.2cm]{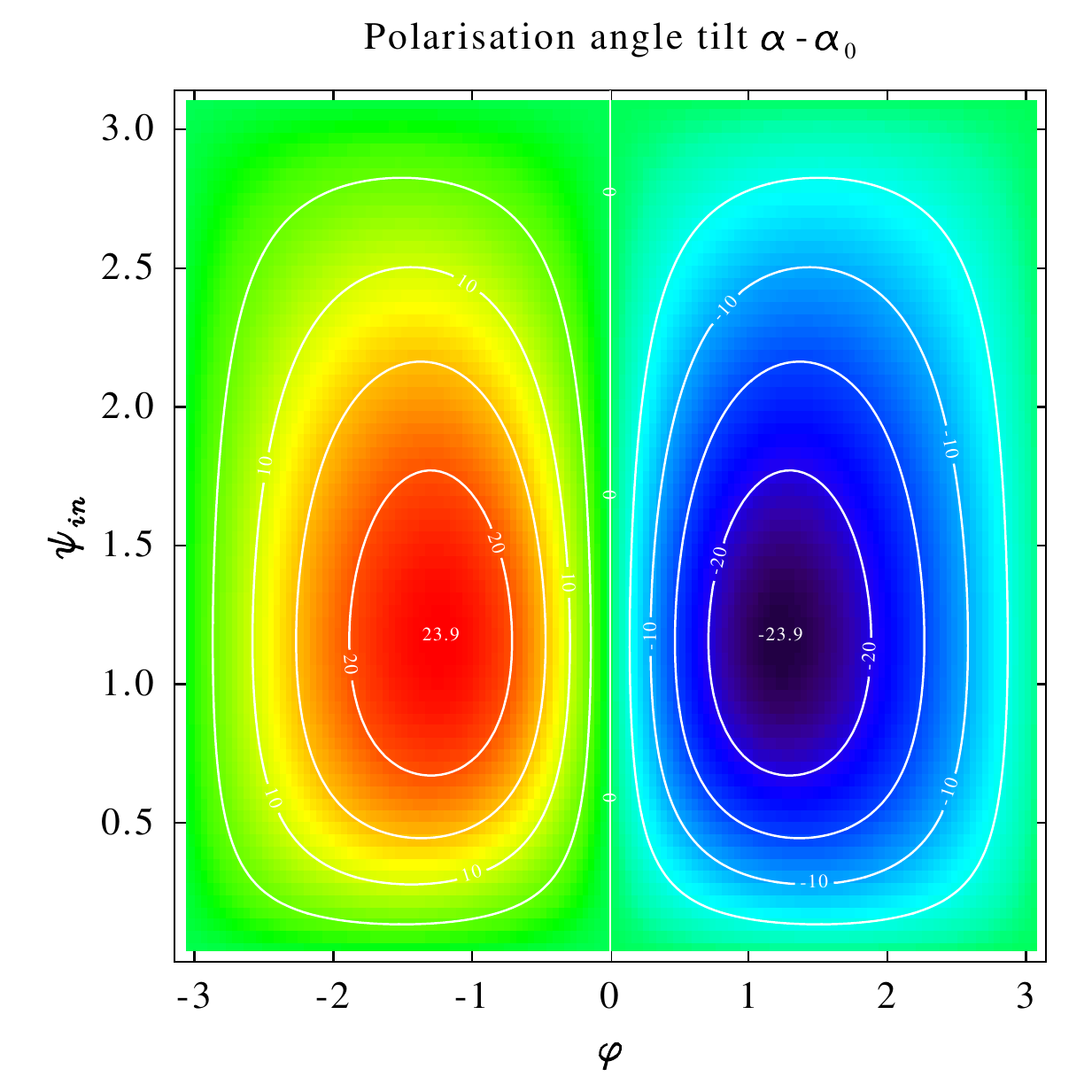}}\\
\parbox{1em}{\rotatebox{90}{$\vartheta$}}
\parbox{16.4cm}{
\includegraphics[viewport=32 20 345 335,clip,width=8.2cm]{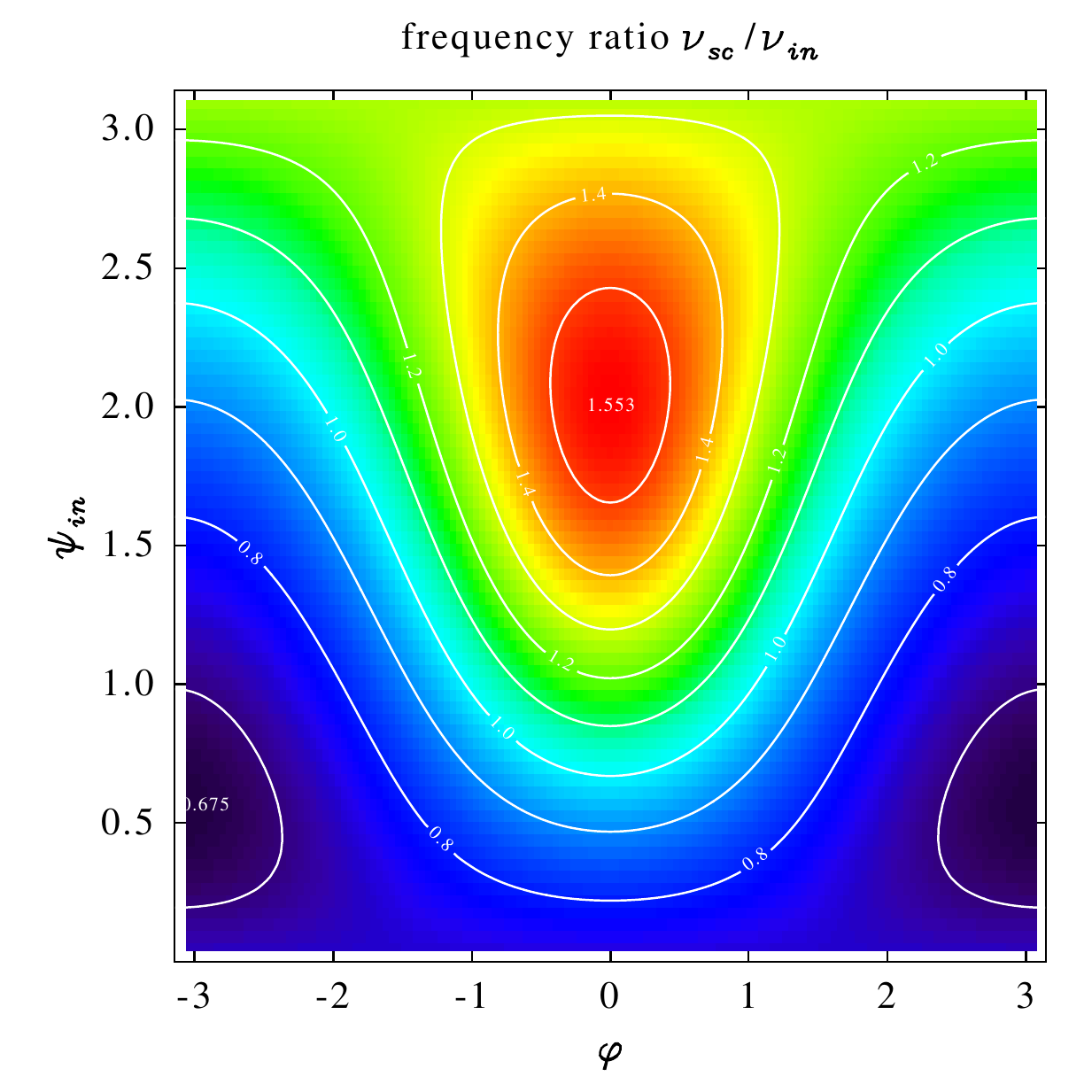}
\includegraphics[viewport=32 20 345 335,clip,width=8.2cm]{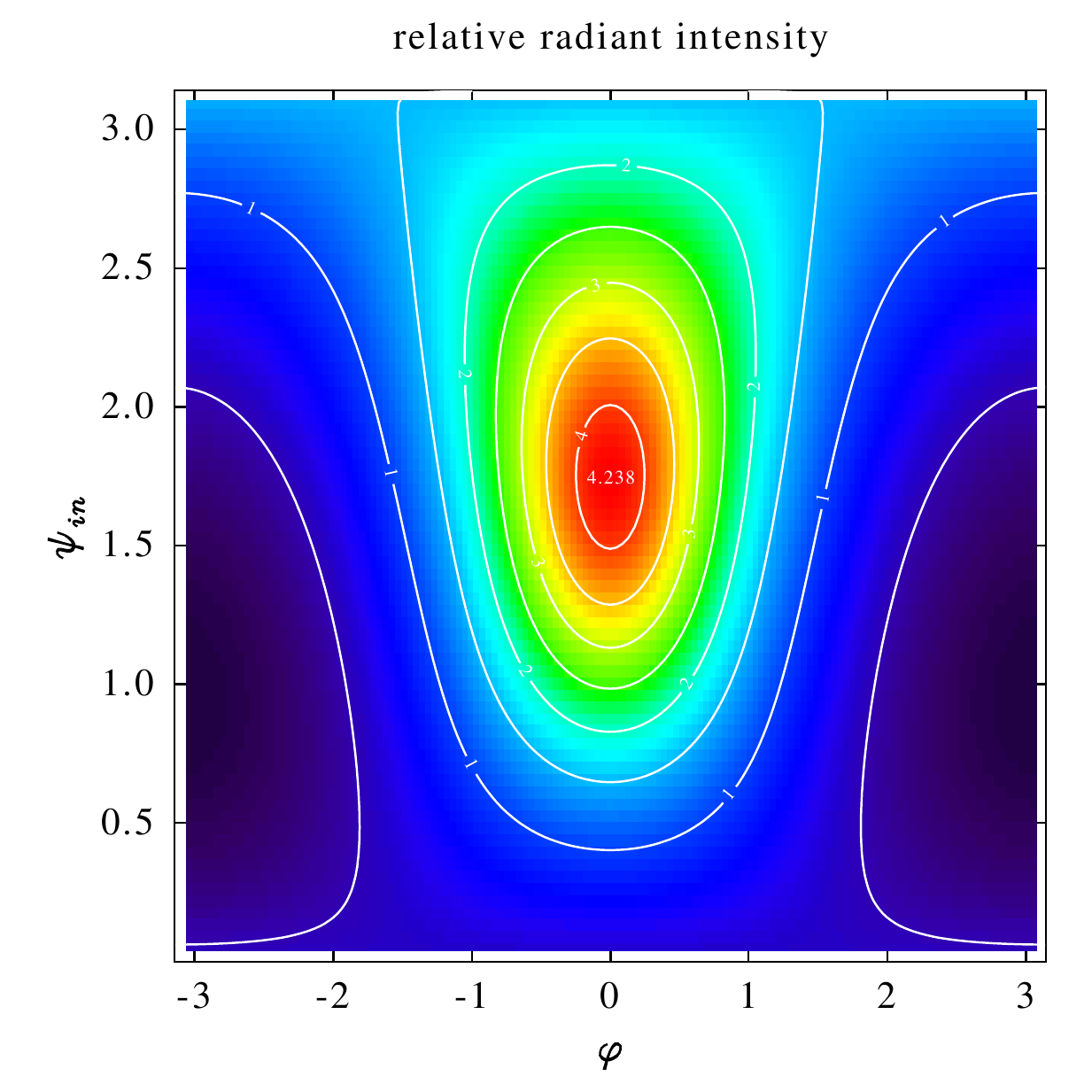}}\\
\hspace*{4.7cm}$\phi$\hspace*{8.1cm}$\phi$\\[-2ex]
\caption{Same as Fig.~\ref{Fig:FSun_03} but for $\beta=0.3$ From top left to
  bottom right: Degree of polarisation, deviation of the major polarisation
  direction from tangential in degrees, frequency blueshift factor and
  total radiant intensity amplification per electron.}
  \label{Fig:FSun_30}
\end{figure}

\begin{figure}
\parbox{1em}{\rotatebox{90}{$\vartheta$}}
\parbox{16.4cm}{
  \includegraphics[viewport=32 20 345 335,clip,width=8.2cm]{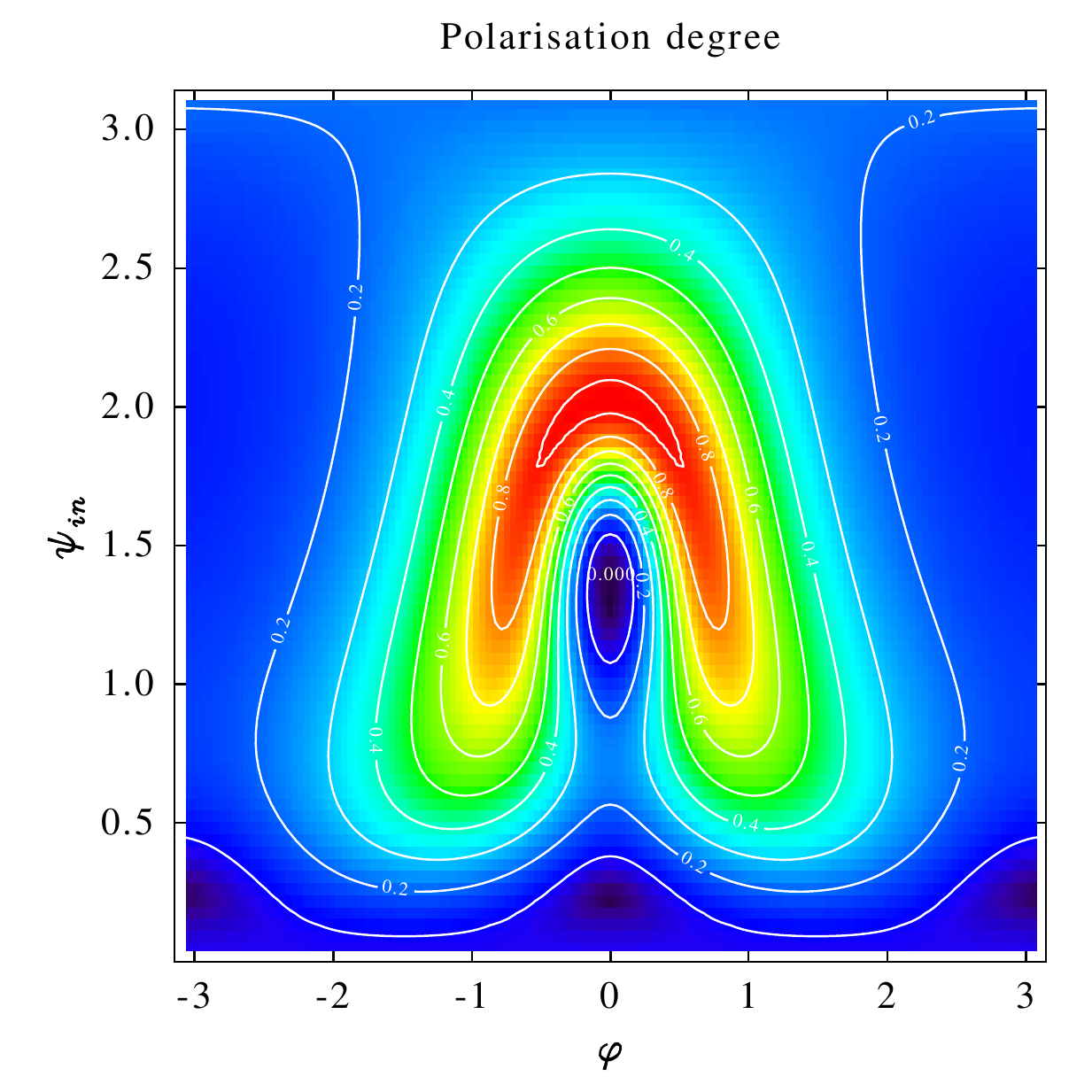}
  \includegraphics[viewport=32 20 345 335,clip,width=8.2cm]{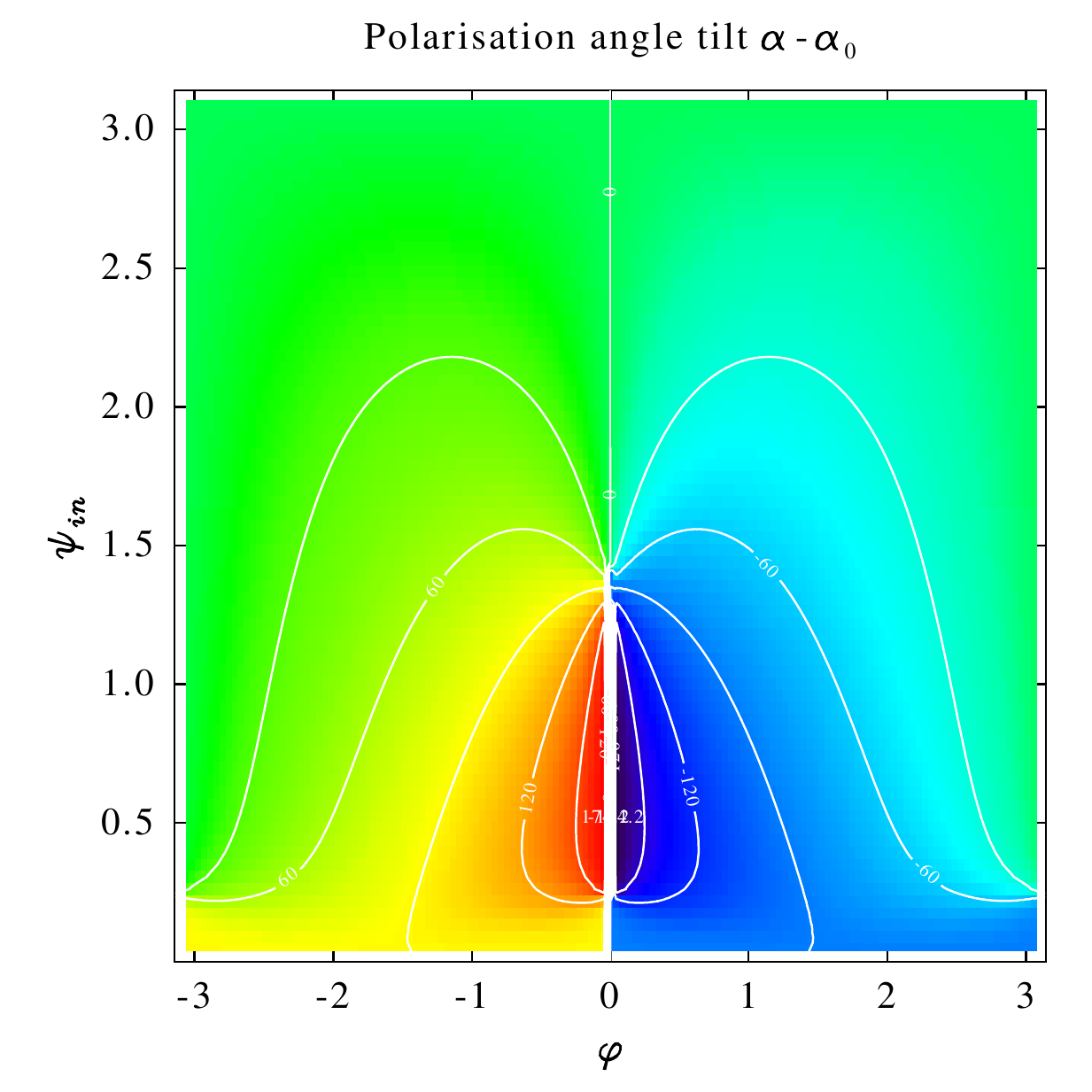}}\\
\parbox{1em}{\rotatebox{90}{$\vartheta$}}
\parbox{16.4cm}{
\includegraphics[viewport=32 20 345 335,clip,width=8.2cm]{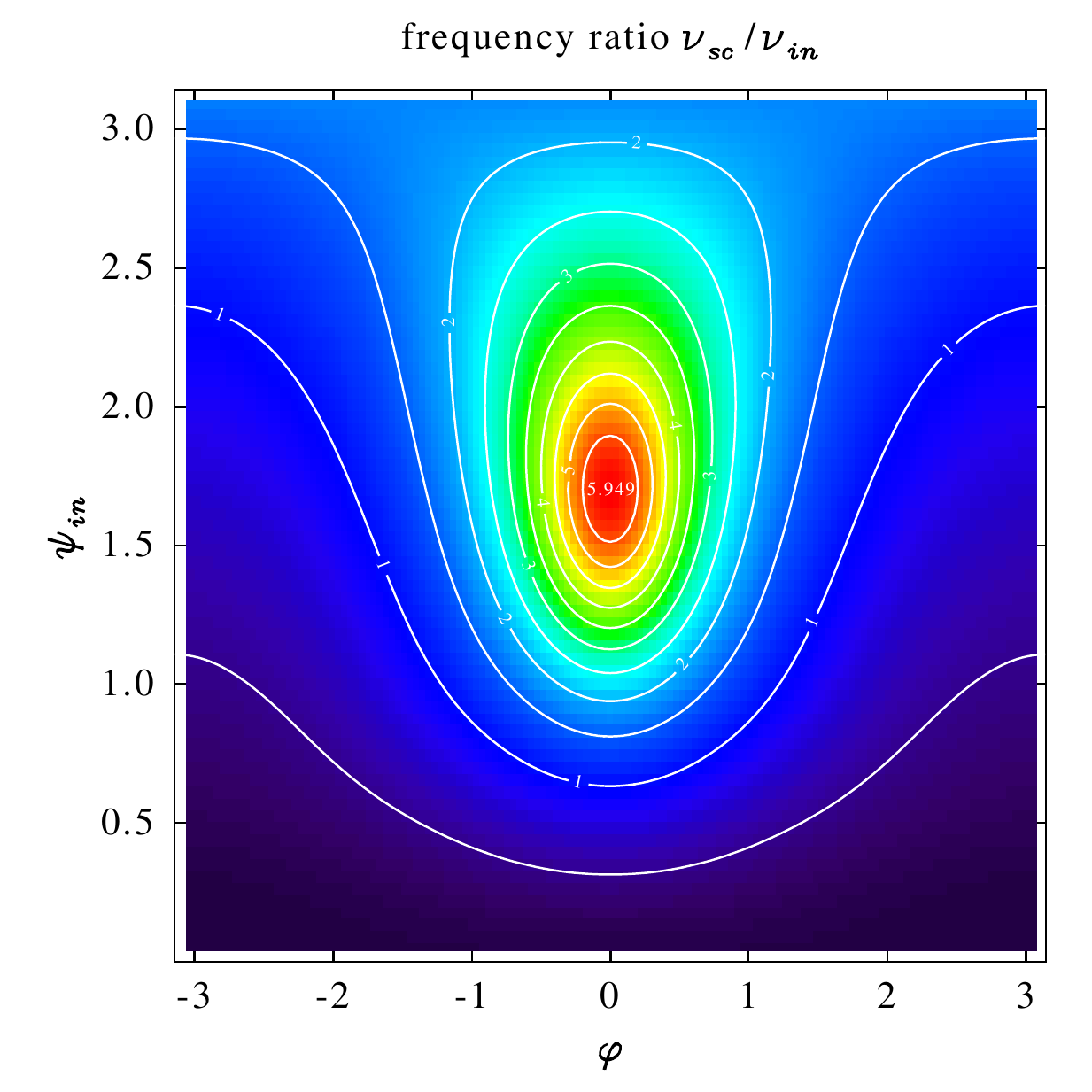}
\includegraphics[viewport=32 20 345 335,clip,width=8.2cm]{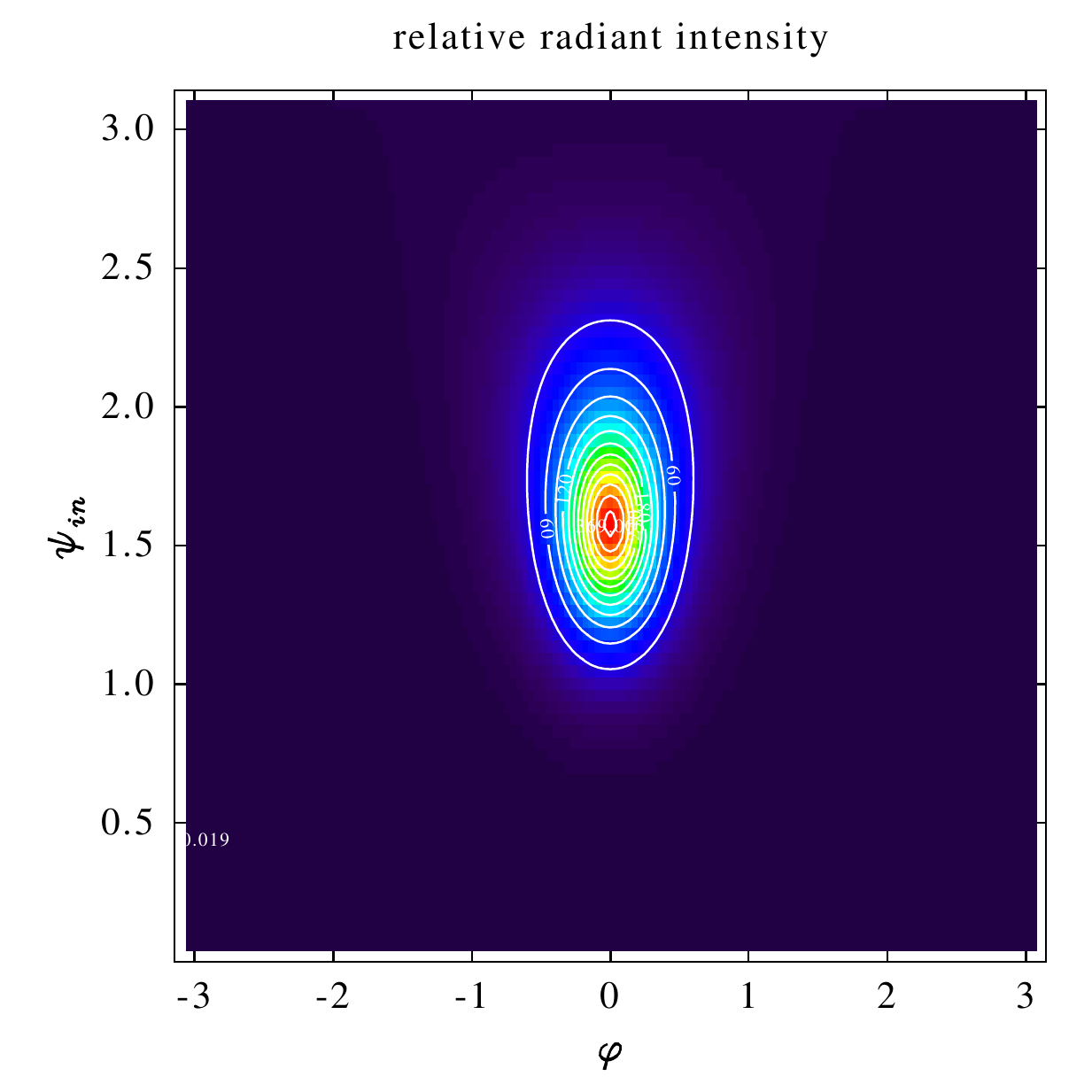}}\\
\hspace*{4.7cm}$\phi$\hspace*{8.1cm}$\phi$\\[-2ex]
\caption{Same as Fig.~\ref{Fig:FSun_03} but for $\beta=0.8$ From top left to
  bottom right: Degree of polarisation, deviation of the major polarisation
  direction from tangential in degrees, frequency blueshift factor and
  total radiant intensity amplification per electron.}
  \label{Fig:FSun_80}
\end{figure}

The above results for a single incident beam are applicable to the scattering
of Sun light at large distances $r$ when the solid view angle $\Omega(r)$
subtended by the solar disk at the scattering distance $r$ shrinks to a point.
For closer distances to the Sun, we have to integrate the directions
$\vect{\hat{k}}_\mathrm{in}$ in (\ref{Irr_Integ3}) over the solid angle cone
$\Omega(r)$ of the apparent Sun. The integrand is too involved to allow an
analytical integration, a numerical integration is however straight forward.
We use Gaussian-Legendre integration for the $\theta$ angle and Simpson in
azimuth.

In Figs.~\ref{Fig:FSun_03}, \ref{Fig:FSun_30} and \ref{Fig:FSun_80} we plot
the same parameters as above, now however for the scattering electron
at the finite distance $r=1.5 R_\odot$ and for a mean scattering angle of again
$\bar{\chi}=\pi/2$. The difference to the single beam case is the integration
of the incident radiation directions $\vect{\hat{k}}_\mathrm{in}$ over a
finite cone $\Omega$. At $r=1.5 R_\odot$ the visible solar disk subtends a
central cone angle of 41$^\circ$. For an electron at rest, we can derive from
(\ref{Irr_tan_Minnaert}) and (\ref{Irr_rad_Minnaert}) a polarisation degree
$P=0.62214$, a polarisation tilt $\alpha-\alpha_0=0$, a frequency shift of
unity and a total scattered radiant intensity per electron of $I_\mathrm{ref}=
I_\mathrm{tan}+I_\mathrm{rad} =5.4541\;10^{-30}~\mathrm{m}^2\;L_\odot$.

In Fig.~\ref{Fig:FSun_03}, we show the results for the same parameters as in
Fig.~\ref{Fig:Beam_03} again for $\beta=0.03$ but $r=1.5$ instead
$r\rightarrow\infty$.
The integration over $\Omega$ has very little effect on the
polarisation tilt, the frequency shift and the total intensity.
The maximum tilt of the polarisation
angle is somewhat intensified to $\pm 2.2^\circ$ for an electron velocity
direction normal to the mean scattering plane. The strongest modification is
found for the degree of polarisation (top left of Fig.~\ref{Fig:FSun_03}). It
varies between $P=0.603$ and 0.639. The deviation from the standard value for
an electron at rest is roughly proportional to $-\cos\vartheta$, i.e., to
the radial component of $-\uectg{\beta}$. This conforms
with Molodensky's qualitative argument \citep{Molodensky:1973} mentioned above:
the degree of polarisation decreases the faster the electron moves away
from the Sun because then the Sun appears bigger in the electron rest frame.
An inward moving electron sees a smaller apparent size and consequently a
enhanced polarisation.

\begin{figure}
\hspace*{\fill}
\includegraphics[viewport=190 50 650 480,clip,width=5cm]{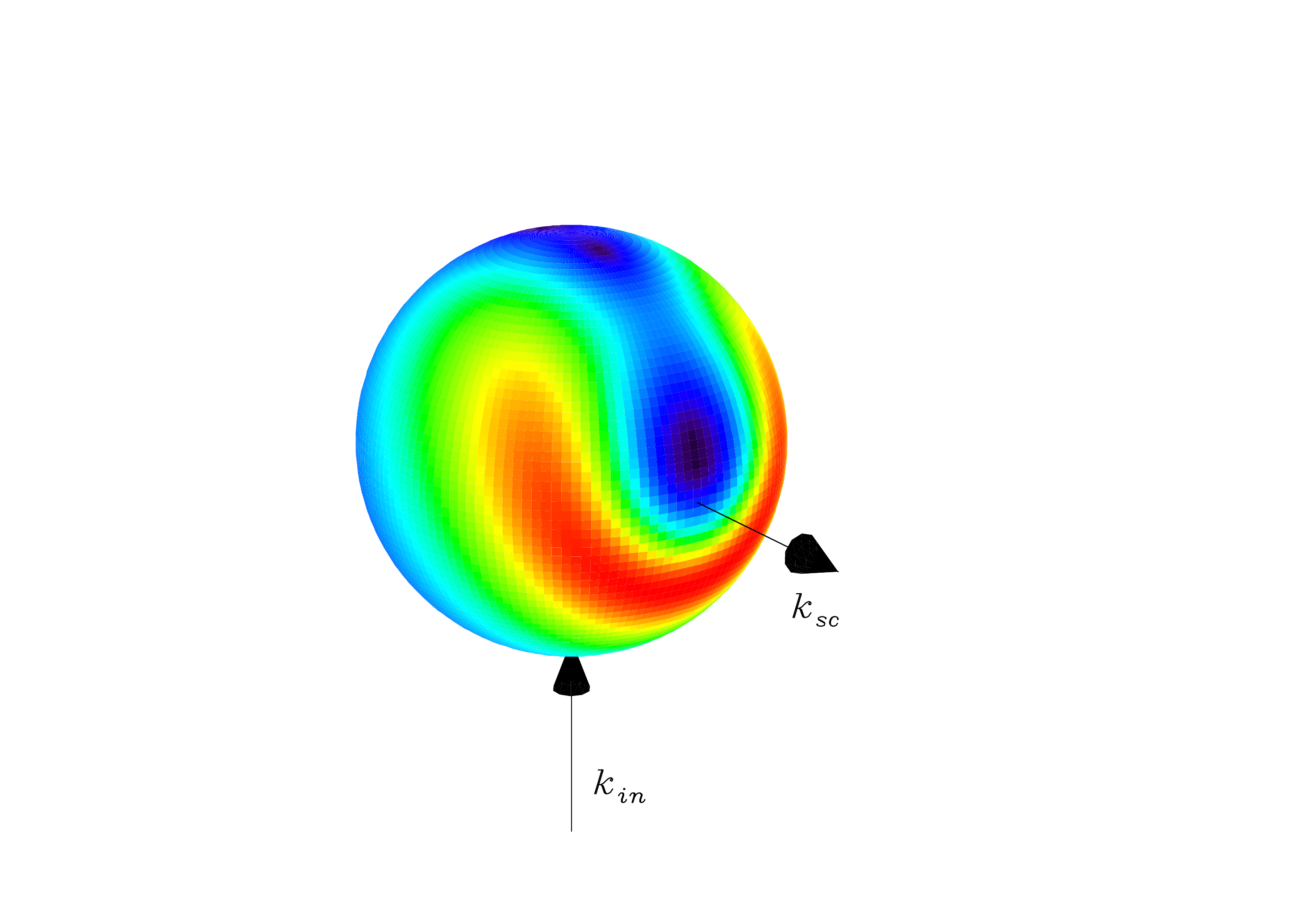}
\includegraphics[viewport=190 50 650 480,clip,width=5cm]{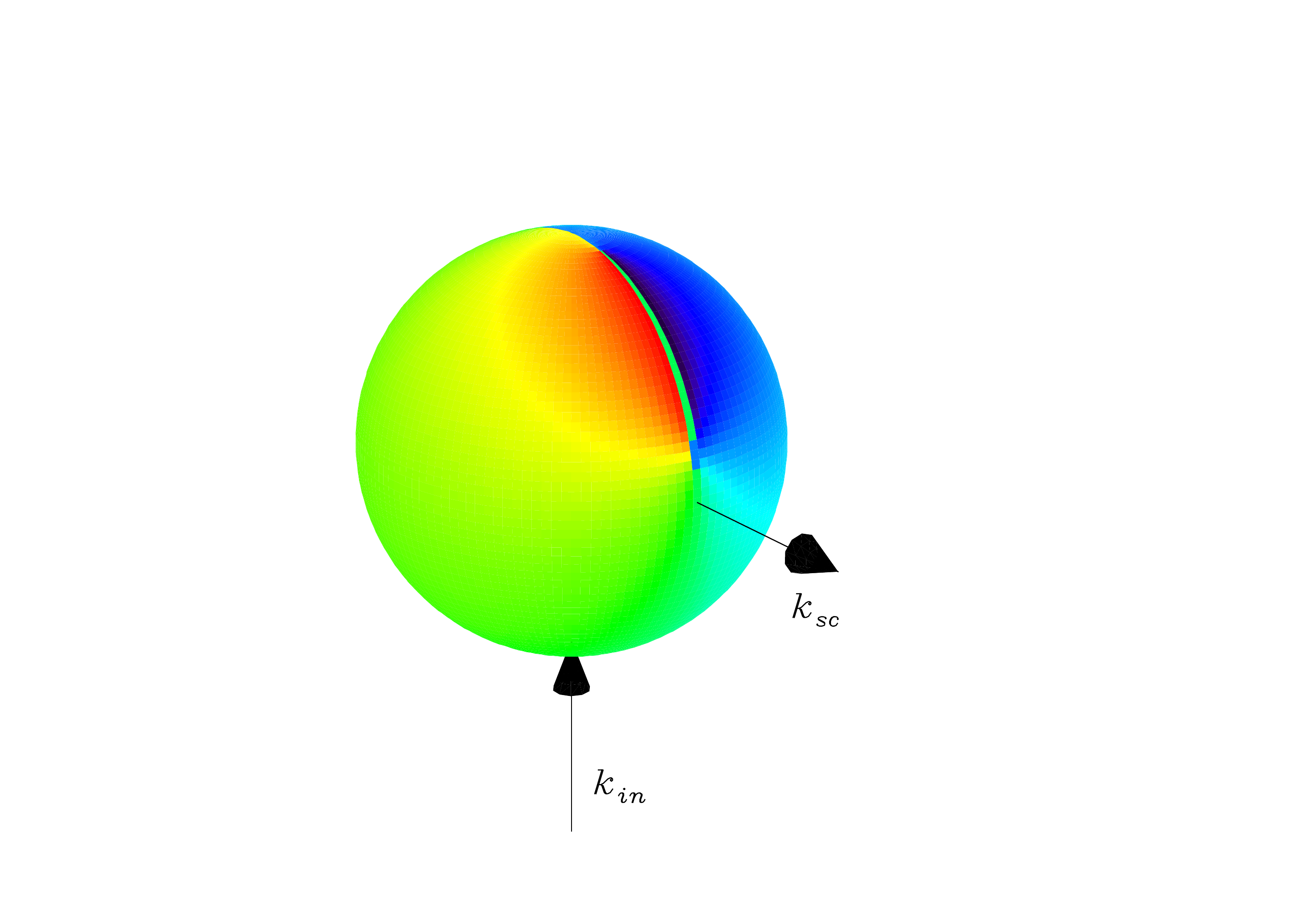}
\includegraphics[viewport=190 50 650 480,clip,width=5cm]{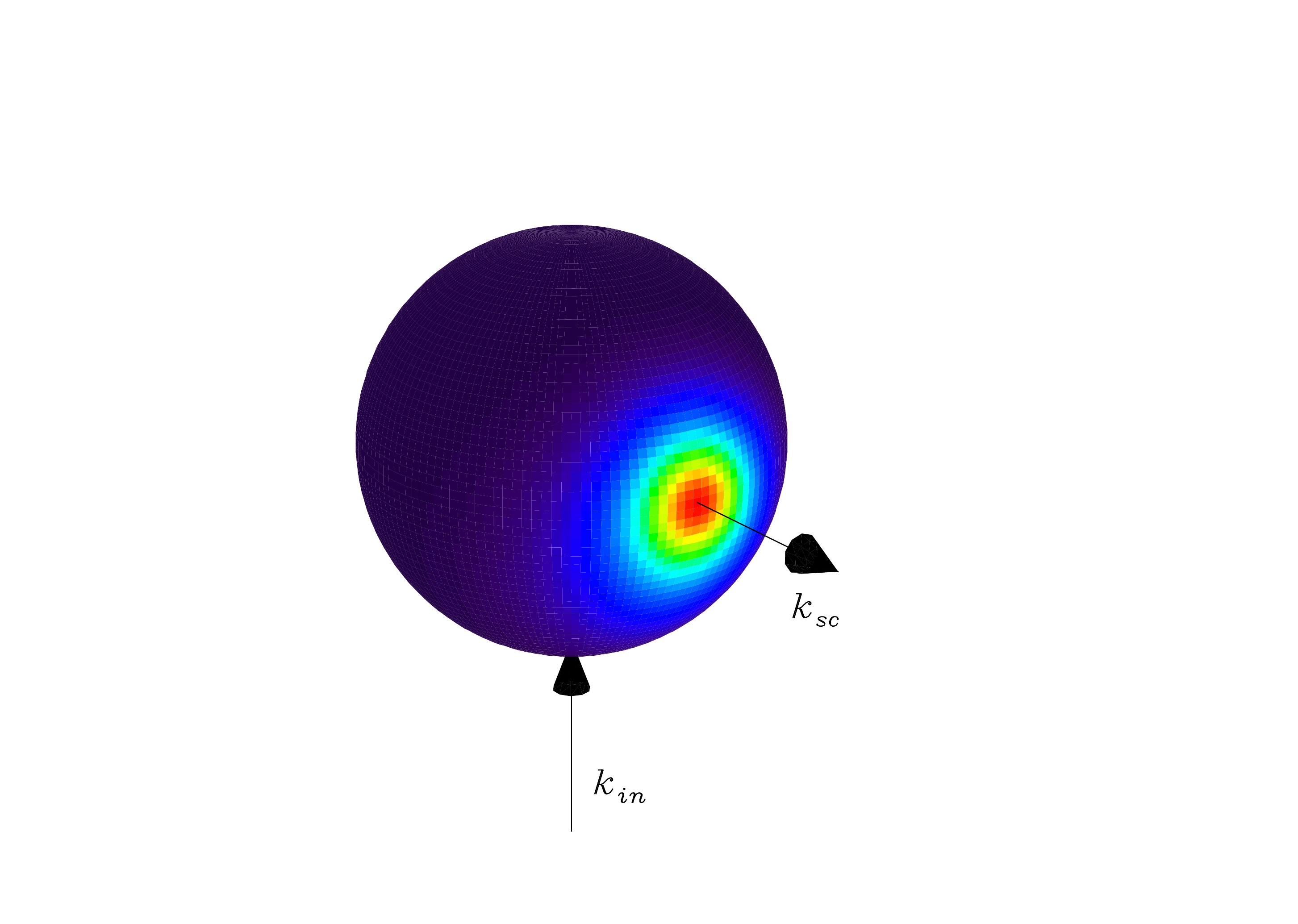}
\hspace*{\fill}
\caption{Results of Fig.~\ref{Fig:FSun_80} replotted on a 3D sphere to better
  illustrate the 3D variation with $\vect{\hat{\beta}}$ in space and their
  relation to the incident and scattered beam direction. From left to right:
  Degree of polarisation, deviation of the major polarisation direction from
  tangential and relative scattered total radiant intensity per electron. The
  colour code is the same as in Fig.~\ref{Fig:FSun_80}.}
  \label{Fig:FSun_80_3D}
\end{figure}


For higher values of $\beta=0.3$ and 0.8 the results are displayed in
Figs.~\ref{Fig:FSun_30} and \ref{Fig:FSun_80}. Again the forward beaming of
the scattering electron becomes increasingly important similarly to what we
found for the single incident beam. The distribution of the polarisation tilt
angle, the frequency shift and the total intensity amplification are similar
to Figs.~\ref{Fig:Beam_30} and \ref{Fig:Beam_80}.
The exception again is the degree of polarisation and it can again
qualitatively be explained by the Molodensky effect.
For outward directed $\uectg{\beta}$, i.e. $\vartheta>0$,
the apparent size of the Sun in the electron rest frame
is so much enhanced that $P$ becomes strongly reduced and
the scattered radiation is almost unpolarised.
In the reverse direction, the Sun's size appears
strongly reduced and the conditions resemble closely the single beam
case with a maximum polarisation of almost unity for $\uectg{\beta}$
in the scattering plane plane.
An electron at $r=1.5$ moving with $\beta=0.8$ away from the Sun sees the cone
angle of the solar disk increased from 41$^\circ$ to 98$^\circ$, in opposite
direction the apparent solar disk shrinks to 14$^\circ$. In
Fig.~\ref{Fig:FSun_80_3D} we have replotted the variation of the polarisation
degree, the tilt angle and the total intensity amplification with the electron
velocity direction on a 3D sphere.

The frequency shifts shown are always means over the visible solar
disk weighted by the respective radiance. However, each incoming beam
experiences its individual frequency shift which may well deviate from the
average. If the shift is so large that the scattered beam falls out of
the frequency band of the observing instrument, it will not be able to
contribute to the observation and the observed averages of the
angle and degree of polarisation, of frequency shift and
total intensity will change because of this omission.

The results described so far refer to only a single electron.
In order to derive the
full signal observed in a coronagraph pixel, we have to integrate the
radiant intensity per electron times the local number of electrons along the
line-of-sight as in (\ref{pow_LOS}). However, since the radiant intensity
matrix now also depends on $\vectg{\beta}$ we have also to sum over the
electron velocities weighted by the local distribution function
$f(\vectg{\beta})$ normalised to $\int f(\vectg{\beta})\,d^3\vectg{\beta}=1$.
For a polariser oriented along $\vect{\hat{p}}$, the radiant flux per pixel
observed at $\vect{r}_\mathrm{obs}$ will be
\begin{gather}
  \pow_\vect{\hat{p}}(\vect{r}_\mathrm{obs})
  =\frac{A_\mathrm{pixel}A_\mathrm{aperture}}{f^2}
  \int_\mathrm{LOS}\int_\mathrm{velocity~space}\mspace{-20mu}
  \vect{\hat{p}}\tp
  \vect{I}(\uect{k}\sca(\vect{r}),\vectg{\beta})
  \vect{\hat{p}}\;\;
  f(\vectg{\beta})\,d^3\vectg{\beta}\;
  N_e(\vect{r})d\ell
  \qquad\text{[W]}
\label{pow_LOS_beta}\end{gather}
where in the integral
$\vect{r}(\ell)=\vect{r}_\mathrm{obs}-\ell\vect{\hat{k}}_\mathrm{sc}$ denotes
the scattering site. The conventional result in (\ref{pow_LOS}) is recovered
if $f(\vectg{\beta})$ is confined to non-relativistic velocities. Whether we
retain a measurable relativistic effect therefore depends on the number of
energetic electrons along the line-of-sight compared to the cold core of the
distribution function.

But even if $f(\vectg{\beta})$ has a high energy tail, the velocity
integration in
(\ref{pow_LOS_beta}) has consequences. The antisymmetry of the polarisation
tilt angle with respect to the $\phi$-angle of $\vectg{\beta}$
cancels the collective tilt of the polarisation axis with respect to
$\vect{\hat{p}}_\mathrm{tan}$ in the case of an isotropic
electron velocity distribution function $f(\vectg{\beta})$.
Reversely, the observation of a deviation of the polarisation from the
tangential direction could be evidence for a relativistic electron beam.

\begin{figure}
\hspace*{\fill}
\includegraphics[viewport=15 5 350 335,clip,width=5.5cm]{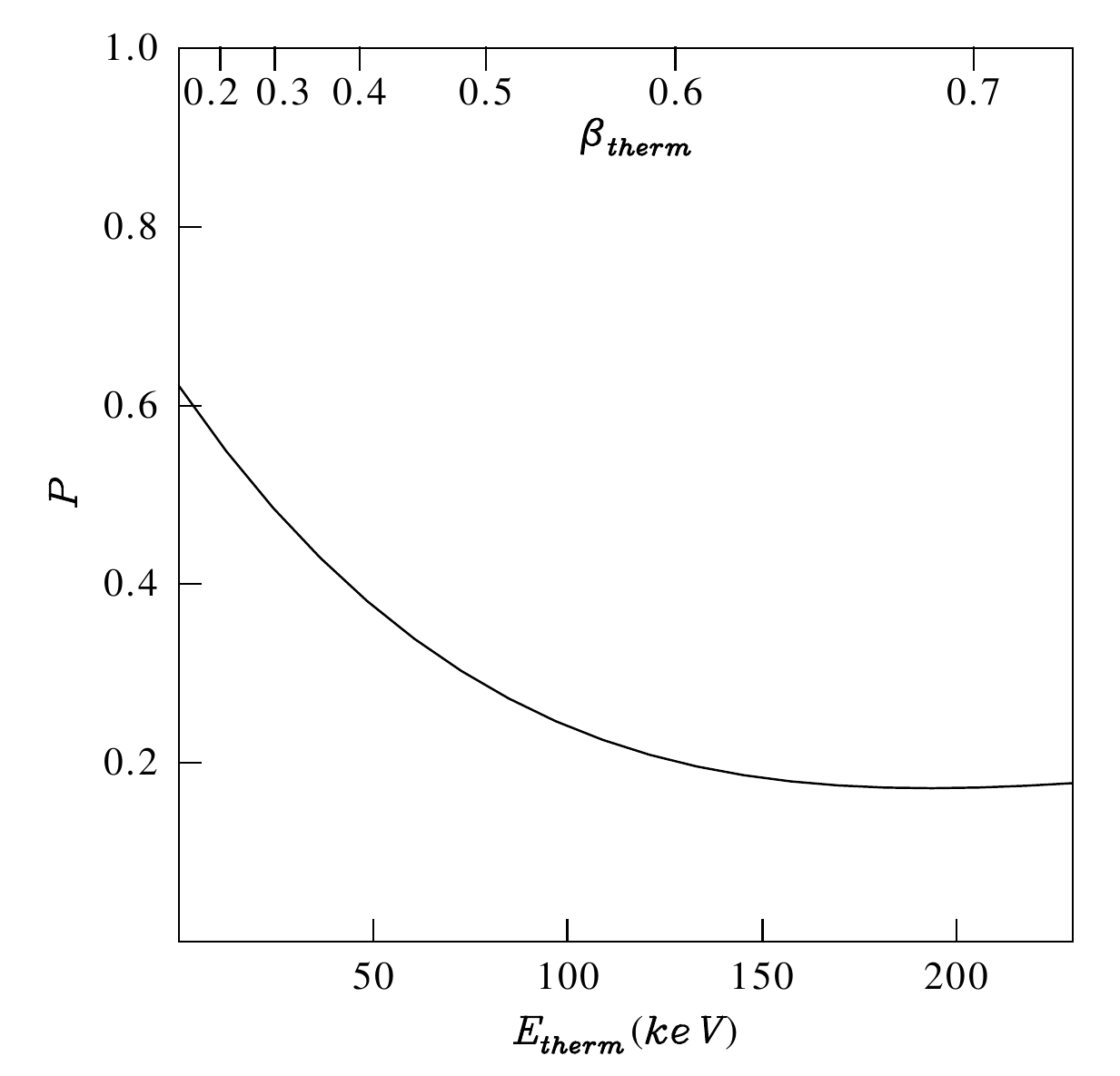}
\includegraphics[viewport=15 5 350 335,clip,width=5.5cm]{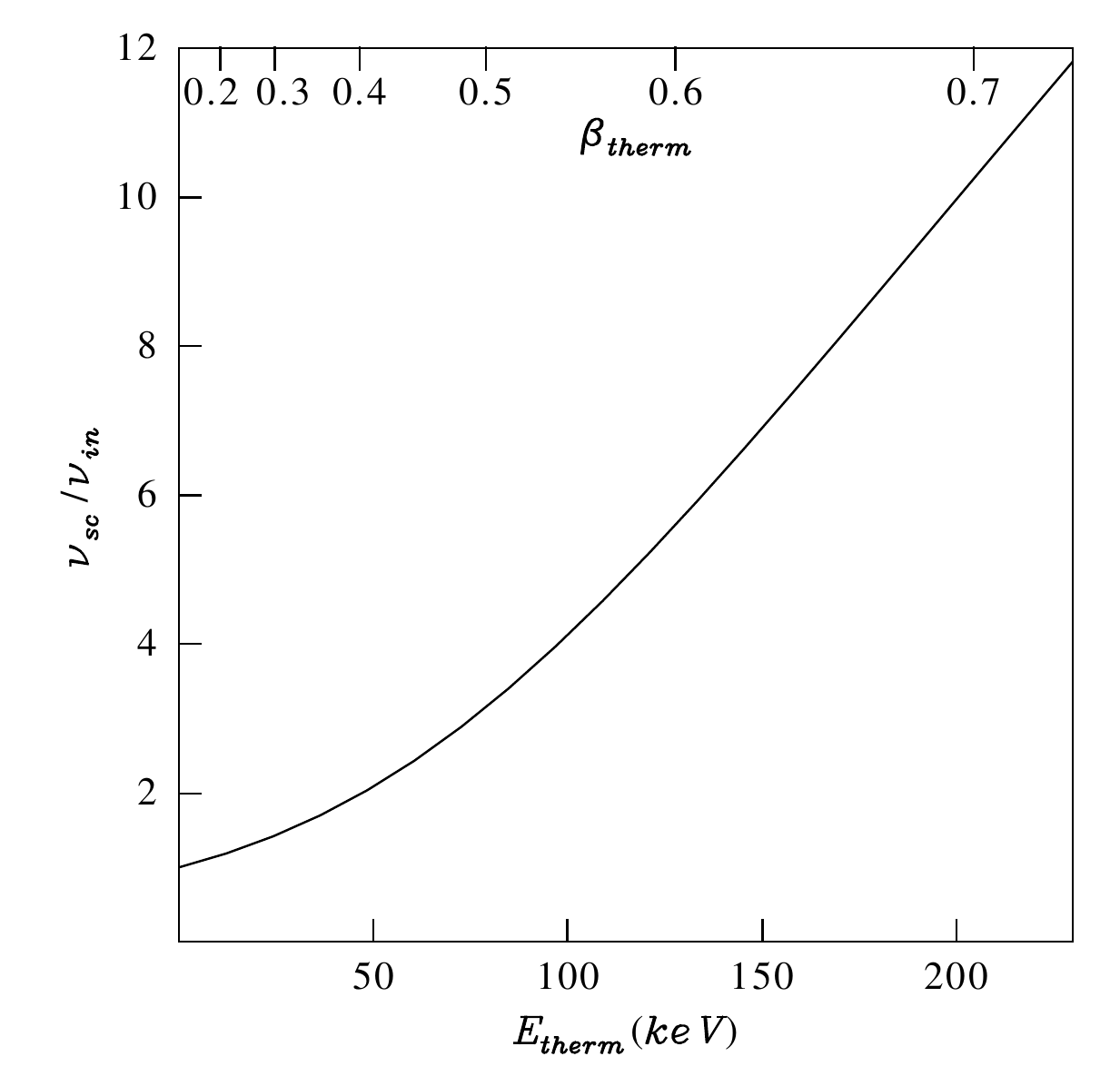}
\includegraphics[viewport=15 5 350 335,clip,width=5.5cm]{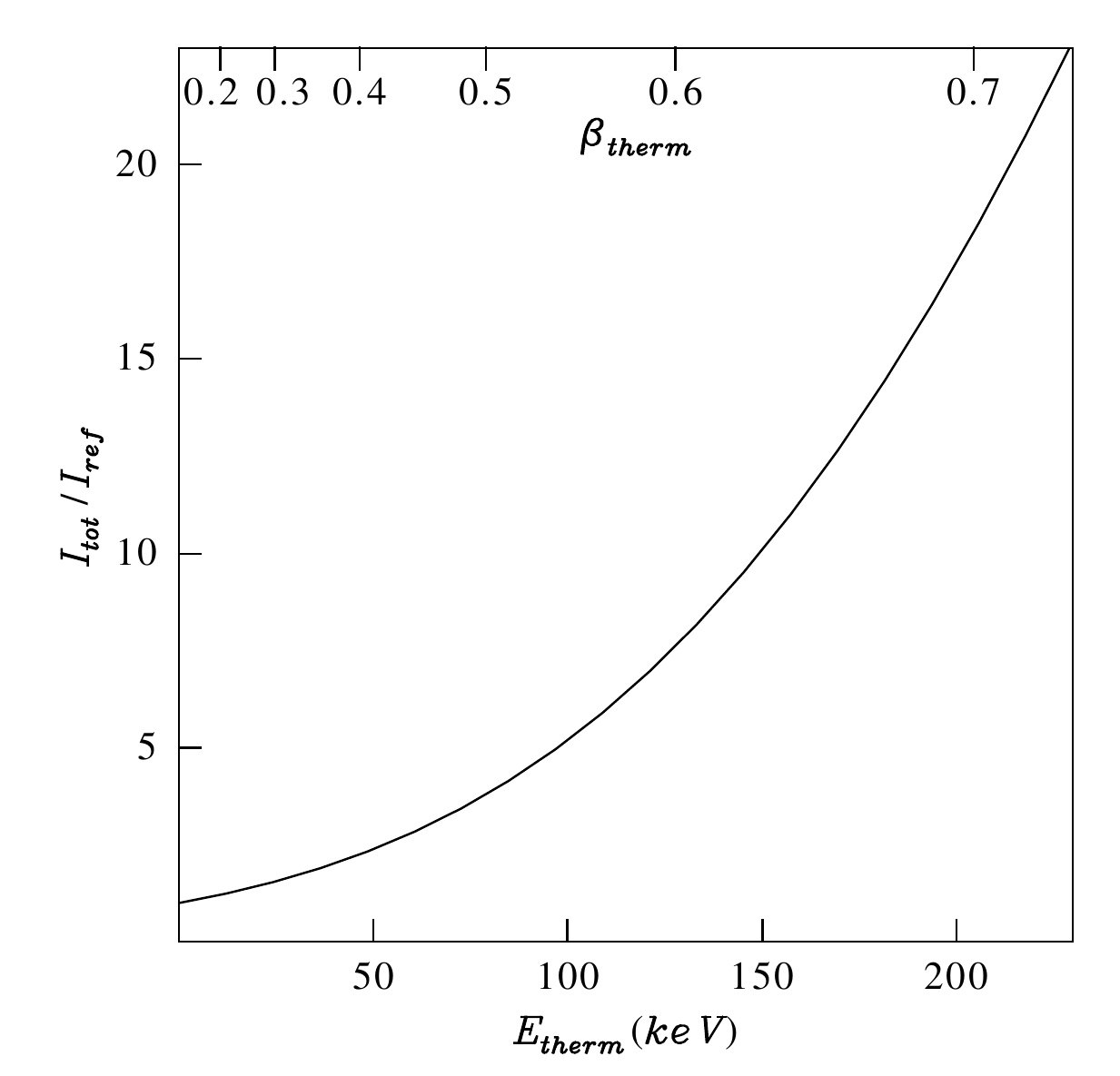}
\hspace*{\fill}
\caption{Degree of polarisation (left), effective frequency shift
  (centre) and radiant intensity amplification (right) 
  integrated over an isotropic J\"uttner distribution with thermal
  energy $E_\mathrm{therm}$. The thermal speed $\beta_\mathrm{therm}$
  corresponding to $E_\mathrm{therm}$ is marked at the top edge
  of the diagrams.
  The scattering site is assumed at $r=1.5
  R_\odot$ and the mean scattering angle is $\bar{\chi}=\pi/2$ as in
  Figs.~\ref{Fig:FSun_03}, \ref{Fig:FSun_30} and \ref{Fig:FSun_80}.
  \label{Fig:BetaIntegrared}}
\end{figure}

The relativistic effects on the degree of polarisation, on the frequency
shift and on the radiant intensity amplification
remain even for isotropic distribution functions. In
Fig.~\ref{Fig:BetaIntegrared} we show these three parameters if we
integrate over a relativistic J\"uttner distribution function of
various thermal energies. This distribution function is isotropic and
the relativistic extension of a Maxwell distribution function.
More details are given in the appendix in chapter~\ref{app:Juettner}.
The velocity-space integration results in a net depolarisation of
the scattered signal.
Recall that P=0.622 is the polarisation obtained for non-relativistic
electrons at $r=1.5$ and a mean scattering angle $\bar{\chi}=\pi/2$.
However the decrease saturates at $P\simeq 0.177$ which we again attribute
to the Molodensky effect:
for one half of the electron distribution the Sun's size becomes so huge
that these particles scatter practically unpolarised, while
for the other half the Sun reduces to a point source
contributing a polarisation like in the single beam case
in Fig.~\ref{Fig:Beam_03}.
Both the frequency shift and the intensity amplification
not only intensify in the forward direction 
of the scattered electron as we have seen above
but also their spherical average increases strongly with $\beta$.
The excess energy of the scattered photon is taken from the kinetic energy
of the electron. However, by using the Thomson cross section in the electron
rest frame we here neglect the corresponding energy loss for the electron
in the scattering process.

\section{Discussion and summary}

We have reformulated the classical Thomson scattering problem for the corona.
The final result is not new, but the new derivation may provide some new
understanding of the details of the process. The analytical integrals $A(r)$,
$B(r)$, $C(r)$ and $D(r)$ first derived by \cite{Schuster:1879} and
\cite{Minnaert:1930} are not related to Thomson scattering at all but they
already arise when the electric field correlation matrix
$\vect{\irr}_\mathrm{in}\propto\;\bra\vect{E}\vect{E}\tp\ket$ of the solar
incident irradiance at the scattering site is calculated. The subsequent
scattering process can be considered essentially as a simple projection of
$\vect{\irr}_\mathrm{in}$ into the direction toward the observer. Our general
result (\ref{Rin_sca2}) allows to easily calculate the Thomson scattered
radiant intensity per electron for incident radiation from the apparent solar
disk for all possible observed polarisation directions.
The details of our calculations are included in an extensive appendix.
It is therefore entirely transparent how they have to be extended if, e.g., the
limb darkening is required to include terms of higher than linear degree in
$\cos\zdis$, or the effects of a huge sun spot or a local polarised radiation
source on the solar surface are to be determined.

Each point along the observed line-of-sight contributes to the observed signal
$N_e$ times the local radiant intensity per electron. Due to the simple
nature of Thomson scattering, the tangential radiant intensity
polarised normal to the mean scattering plane does not depend on the
scattering angle. Only the radially polarised radiant intensity
$\rin_\mathrm{rad}$ depends on it.
It has been argued that the contribution from the Thomson sphere, defined 
as the surface where the mean scattering is $\bar{\chi}=\pi/2$, will dominate
the scattered signal from a given line-of-sight \citep{VourlidasHoward:2006}.
The relevance of the Thomson sphere for coronal Thomson scattering has been
extensively discussed by \cite{HowardDeforest:2012} and we can only reiterate
here their main arguments which are confirmed by our results.
The maximum of $\rin_\mathrm{tot}$ and $\rin_\mathrm{pol}$ on the Thomson
sphere is not due to peculiarities of the differential Thomson cross section 
but it is a consequence of the decrease of the entire line-of-sight integrand 
with distance $r$ from the solar centre.
This is due to the decrease of both, the incident solar irradiance and of the
scattering electron density with distance $r$.
For the total radiant intensity $\rin_\mathrm{tot}$ it is even the reverse,
since $\rin_\mathrm{rad}$ almost vanishes on the 
Thomson sphere it compensates in part the maximum of
$\rin_\mathrm{tan}$ along line-of-sight.
In this respect the term ``Thomson sphere'' is misleading.

For large elevation angles $\varepsilon$ from the Sun and an assumed
$N_e\propto r^-2$ dependence the line-of-sight integrand varies with
scattering angle $\bar{\chi}$ like $\propto 1-\cos^4\bar{\chi}$ for the total
and $\propto \sin^4\bar{\chi}$ for the tangentially polarised scatter signal
(see eqs.~\ref{Ktot}, \ref{Kpol} and also
\citep{HowardDeforest:2012,DeforestHoward:2013}). Therefore the decrease of
the line-of-sight integrand off the Thomson sphere is very moderate compared
to the gradient of the electron density at the boundaries of streamers,
current sheets and CMEs. These objects therefore may well be detectable in
coronagraph images even if they occur far off the Thomson sphere. Instructive
examples of forward calculations of line-of-sight integrals have, e.g., been
produced from MHD simulations of shocks by
\citet{XiongEtAl:2013a,XiongEtAl:2013b}.

The classical expression (\ref{pow_LOS}) of the observed scattered
power per pixel has been used in many papers to estimate the column 
mass of the plasma inside the cone of resolution subtended by a sensor
pixel.
Using suitable pre-event difference images and summing over all relevant 
pixels, the total mass of transients were estimated. The unknown variation 
of the electron density along the respective lines-of-sight in 
(\ref{pow_LOS}) was replaced in many of these studies by a 
$\delta$-distribution in the plane-of-sky.
\cite{VourlidasHoward:2006} and \cite{ColaninnoVourlidas:2009} pointed out 
that this leads to an underestimate of the CME mass when in reality it 
propagates at some angle with the
plane-of-sky. They suggested to centre the $\delta$-distribution rather at the
true propagation angle which can be estimated either from the location of the
source region on the solar surface, from a comparison of the observed
scattered power in different polarisations or from a triangulation from two
different directions as provided by the STEREO mission
\citep{HowardStereo:2008}.
In chapter~\ref{Sec:ObserveCME}, we show how these mass estimates are modified
further if a more realistic finite width of the CME mass distribution along the
line-of-sight is taken account of. If the true width is difficult to
estimate to sufficient precision, the variation of the line-of-sight integrals over
Gaussian density distributions with different widths $\delta\bar{\chi}$ as
presented in Figs.~\ref{Fig:MassErr} and ~\ref{Fig:MassErr_example} can still
be used to guess the error of the mass estimate. From the examples 
shown in these figures, it could well amount to 10\% or more, depending on 
the viewing geometry.

To demonstrate the versatility of our approach to calculate the Thomson
scattered radiant intensity, we have extended it to the case of relativistic
electrons. For coronal applications, similar calculations were initiated
\cite{Molodensky:1973}, but he only considers electron
velocity vectors $\vect{v}$ in the plane-of-the-sky and treats only a single
incident beam. \cite{NikoghossianKoutchmy:2001,NikoghossianKoutchmy:2002}
extended the electron velocity directions and also consider a finite size of
the apparent solar disk but they restrict their calculations to the scattered
total intensity $\rin_\mathrm{tot}$ and its net frequency shift. We think that
our approach is more intuitive and transparent than these previous
calculations and it yields the full set of Stokes parameters for the
scattered signal. The resulting expressions
(\ref{Radint_pp}) and (\ref{pkin_trans})
for the scattered radiant intensity per electron
is verified by an alternative but more
tedious derivation presented in the appendix.

The coronal scattering resembles in many aspects Thomson and inverse Compton
scattering processes in astrophysics which have received a growing interest in
recent years. An example are relativistic jets escaping from active galactic
nuclei which scatter their own synchrotron radiation as well as low energy
photons of the cosmic microwave background up to X-ray energies.
The polarisation of this X-ray radiation could help to distinguish how much 
each of the sources contribute \citep[e.g.,][]{McNamaraEtal:2009}. 

Thomson scattering is also of considerable interest in fusion plasma devices
as a diagnostic tool. In order to interpret the scattered signal the measured
response has been calculated by a number of authors
\citep[e.g.,][]{Hutchinson:2002,SegreZanza:2000,BeausangPrunty:2008}. As is
appropriate for laboratory plasmas, the calculations concentrate on
monochromatic, linearly polarised incident beams and assume that the electron
distribution is an isotropic Maxwellian or J\"uttnerian. Since in fusion plasmas the electrons
may reach several tens of keV
\footnote{For electrons the relation between velocity $v=c\beta$,
  Lorentz factor $\gamma$ and kinetic energy $E_\mathrm{kin}$ is
  $E_\mathrm{kin}/511\,\mathrm{keV}=\gamma-1$ and $\beta=\sqrt{1-1/\gamma^2}$},
relativistic effects are observed as a skewed
spectral spread with a net blue-shift of the scattered signal and
its depolarisation with increasing electron energy.

Whether similar deviations
from the non-relativistic standard can be observed in the corona depends
critically on the number of relativistic electrons compared to cold electrons
with $\beta<10^{-2}$ and on the precision of the observations.
For a pixel with resolution of $\Delta\theta=2''$ pointing close
to the limb so that the maximum electron density along the line-of-sight is
$N_e=10^{15} \mathrm{m}^{-3}$ we find approximately a total number of electrons
scattering into the pixel of
\[
N_e (200 R_\odot\Delta\theta)^2 R_\odot
= 0.4\;10^{-6} R_\odot^3 N_e
\simeq 10^{35}
\]
This is just two orders of magnitude larger than the $10^{33}$ electrons
at several keV required to produce an X-ray flare and an associated
type III radio burst \citep{AschwandenEtal:1995}.
Locally, the relativistic electrons may represent
10\% and more of the total electrons
\citep{KruckerEtal:2010,ChenPetrosian:2012}. The acceleration of these
electrons probably takes place well above the solar surface in above-the-loop-top
magnetic x-point configurations
\citep{KruckerEtal:2010,MannWarmuth:2011,CarleyEtal:2015}. The acceleration
sites therefore seem accessible to coronagraph observations and above some
large X-ray flares the number of relativistic electrons may be sufficient
to yield a detectable signal.
Note that the occurrence rate of flares has a power-law dependence
on energy \citep[e.g.,][]{LiGanFeng:2012}. 
More energetic flares occur less often but it seems that the most
energetic events observed so far were limited by statistical probability
rather than by a fundamental absolute upper energy boundary.

When evaluating the probability of detecting relativistic scattering effects
it also has to be kept in mind that a single relativistic electron amplifies
the scattered radiation considerably. The scattered radiant
intensity strongly increases with $\beta$ not only in their forward direction
due to beaming but also on average (see Fig.~\ref{Fig:BetaIntegrared}).
The intensification with $\beta$ eventually stops if Thomson scattering is no more
appropriate in the electron rest frame and has to be replaced by Compton
scattering.

Even for an electron energy of 0.03 keV which is just twice their thermal
energy in the corona we see small relativistic effects. Deviations from the
non-relativistic limit of the polarisation degree are of the order of a
few~\%, and the polarisation angle may be inclined by up to about $2^\circ$
away from the tangential direction. Deviations of this order of
magnitude may be produced all along the line-of-sight and
can be considered as the thermal noise of the corona in these
measurable parameters.
But observations which can uniquely be attributed to the Thomson scattering of
relativistic electrons are still scarce. A net frequency shift in coronal
threads above flaring active regions were predicted by
\cite{KoutchmyNikoghossian:2002} and later observed in a few cases as
deviations of a scaled colour index from unity which were derived from
colour-filtered coronagraph data \citep{KoutchmyNikoghossian:2005}.

Unexpectedly large degrees of polarisation have been observed by a number of authors
\citep{KoutchmySchatten:1971,Pepin:1970,SkomorovskyEtal:2012,QuEtal:2013}. In
these cases, the degree of polarisation exceeded the maximum non-relativistic
Thomson scatter value expected for the observed elongation
at a scattering angle $\bar{\chi}=\pi/2$.
As we have seen, an isotropic relativistic electron velocity distribution
should lead to a depolarisation of the observed
scattering signal. Enhancements of the polarisation degree beyond the
non-relativistic $\bar{\chi}=\pi/2$ maximum 
could be produced by relativistic electron beams with a velocity vector
near the scattering plane and a velocity component towards the
Sun. Alternatively, contaminations of the scattered white-light signal by coronal ion
emission lines could alter the degree of polarisation.
\cite{QuEtal:2013} favoured this explanation for large polarisation degrees 
they observed closely above the limb where non-relativistic Thomson scatter
should yield particularly low values for the polarisation degree.

Deviations of the polarisation orientation of the scattered signal from
tangential have also been reported only seldomly. This may be due to the fact
that a tangential polarisation is often firmly assumed and even used as a
premise to separate F and K corona \citep{Koutchmy:1994} or to eliminate a
polarisation bias due to seeing or instrumental effects
\citep{KulijanishviliKapanadze:2005}. Anomalous polarisation directions were
reported by \citep{Pepin:1970,ParkEtal:2001,SkomorovskyEtal:2012,QuEtal:2013}
while other studies explicitly found no deviation from tangential polarisation
within the measurement error
\citep{KoutchmyEtal:1993,KimEtal:1996,KulijanishviliKapanadze:2005}. All of
these studies were made during lunar eclipses and can at best represent a
snapshot of the state of the corona. On the other hand, years of space-borne
coronagraph data has been collected in recent years. A systematic scan of the
polarisation data from these collections by comparing the angle of
polarisation with the heliocentric azimuth angle of every pixel could yield
outliers which may be associated with nearby flares. For a few images such a
scan was produced from STEREO/SECCHI coronagraph data by
\citep{MoranEtal:2006}. However, they assumed the scattered signal to be
tangentially polarised and their incentive was to check the
polarimetric performance of their instrument rather than to search for true
observations of an anomalous polarisation.

\appendix
\renewcommand{\theequation}{\Alph{section}.\arabic{equation}}
\setcounter{equation}{0}

\section*{Appendix}

\section{Minnaert's coefficients}
\setcounter{equation}{0}
\label{App:MinnaertCoeff}


In this appendix we calculate the four integrals (\ref{IntegralsI_01}) and
(\ref{IntegralsJ_01}) relevant for the irradiance matrix elements at some
distance $r>R_\odot$ from the solar centre. We also derive the indefinite
integrals required for the case that the limb darkening of the surface
radiance $L(\cos\zdis)$ is expressed in a power series of $\cos\zdis$
involving powers higher than $n=1$. 

\subsection{Integration of the surface radiance}
\label{app:IntegrSurfRad}

The first step is a variable transformation from
the spherical integration angle $\theta$ to
$x=(r/R_\odot)\cos\theta$. Using $\sin\theta_\mathrm{max}=R_\odot/r$ from
(\ref{sinpsi}) and (\ref{cospsi}) we find for the transformed
lower integration boundary becomes
\begin{gather}
 x_\mathrm{min}=\frac{r}{R_\odot}\cos\theta_\mathrm{max}
               =\frac{r}{R_\odot}\sqrt{1-\sin^2\theta_\mathrm{max}}
               =\sqrt{(\frac{r}{R_\odot})^2-1}
               =\cot\theta_\mathrm{max}
\quad\text{and}\nonumber\\
  \frac{r}{R_\odot}
 =\sin^{-1}\theta_\mathrm{max}=\sqrt{1+x^2_\mathrm{min}}
\label{x_min_relations}\end{gather}
The cosine (\ref{cospsi}) of the local zenith angle of the beam from the solar
surface to the scattering site then is
\begin{equation*}
\cos\zdis=\sqrt{x^2-x^2_\mathrm{min}}
\end{equation*}
This variable transformation results in four type of integrals
which occur in (\ref{Irr_integral}), (\ref{IntegralsI_01}), (\ref{Irr3x3})
and (\ref{IntegralsJ_01}) 
\begin{gather}
 \begin{Bmatrix}I(r)\\J(r)\end{Bmatrix}
     =2\pi\int^1_{\cos\theta_\mathrm{max}} \mspace{-10mu}
       L(\sqrt{x^2-x_\mathrm{min}^2})  
      \begin{Bmatrix} 1 \\\cos^2\theta\end{Bmatrix}\;d\cos\theta
\nonumber\\        
     \;\stackrel{x=(r/R_\odot)\cos\theta}{=}\;
      2\pi\frac{R_\odot}{r}\int^{\sqrt{1+x^2_\mathrm{min}}}_{x_\mathrm{min}}
      L(\sqrt{x^2-x_\mathrm{min}^2})
      \begin{Bmatrix} 1 \\x^2\end{Bmatrix}\;dx
\label{I_transformed}\end{gather}
The solar limb darkening of the surface radiance can be represented by the
power series $L(\cos\zdis)=\sum_{k=0}^n b_k \cos^k\zdis$ where $\zdis$ is the
local zenith angle at the surface. We here follow the previous literature and
employ only terms up to $n=1$ and express the limb darkening by
$L(\cos\zdis)=(1-u)+u\cos\zdis$. Then (\ref{I_transformed}) results in the two
integrals $I_0$ and $J_0$ as coefficients of $1-u$ and two more integrals
$I_1$ and $J_1$ as coefficients of $u$. These integrals can be expressed by
elementary functions
\begin{gather}
 I_0(r)=2\pi\int^1_{\cos\theta_\mathrm{max}} \mspace{-10mu} d\cos\theta
       =2\pi (1-\cos\theta_\mathrm{max})
\label{I_0}\\
 J_0(r)=2\pi\int^1_{\cos\theta_\mathrm{max}} \mspace{-10mu}
        \cos^2\theta\; d\cos\theta
       =\frac{2\pi}{3}(1-\cos^3\theta_\mathrm{max})
\label{J_0}\\
 I_1(r)
   =2\pi\int^1_{\cos\theta_\mathrm{max}} \mspace{-10mu}
        \cos\zdis\; d\cos\theta
       =2\pi\frac{R_\odot}{r}\;
    \int^{\sqrt{1+x^2_\mathrm{min}}}_{x_\mathrm{min}}
          \sqrt{x^2-x_\mathrm{min}^2}\;dx
 \nonumber\\
 \stackrel{(\ref{I1_app})}{=}
     2\pi\frac{R_\odot}{r}\;\left[
      \frac{x}{2}\sqrt{x^2-x_\mathrm{min}^2}
     -\frac{x_\mathrm{min}^2}{2}
        \ln(x+\sqrt{x^2-x_\mathrm{min}^2})
         \right]^{\sqrt{1+x^2_\mathrm{min}}}_{x_\mathrm{min}}
 \nonumber\\
    =2\pi\frac{R_\odot}{r}\;\left[
      \frac{1}{2}\sqrt{1+x^2_\mathrm{min}}
     -\frac{x_\mathrm{min}^2}{2}
        \ln(\sqrt{1+x^2_\mathrm{min}}+1)
     +\frac{x_\mathrm{min}^2}{2}
        \ln(x_\mathrm{min})\right]
  \nonumber\\
    =\pi \;\left[1
     -\frac{\cos^2\theta_\mathrm{max}}{\sin\theta_\mathrm{max}}
        \ln(\frac{1+\sin\theta_\mathrm{max}}{\cos\theta_\mathrm{max}})\right]
  \label{I_1}\\   
 J_1(r)
   =\pi\int^1_{\cos\theta_\mathrm{max}} \mspace{-10mu}
        \cos\zdis\cos^2\theta\; d\cos\theta
       =2\pi(\frac{R_\odot}{r})^3\;
    \int^{\sqrt{1+x^2_\mathrm{min}}}_{x_\mathrm{min}}
       \sqrt{x^2-x_\mathrm{min}^2}\,x^2\;dx
\nonumber\\
 \stackrel{(\ref{J1_app})}{=}
    2\pi(\frac{R_\odot}{r})^3\;\left[
      \frac{x}{4}\sqrt{x^2-x_\mathrm{min}^2}^3
     +\frac{x x^2_\mathrm{min}}{8}\sqrt{x^2-x_\mathrm{min}^2}
     \right.\hspace*{6em}\nonumber\\\hspace*{13em}\left.
      -\frac{x_\mathrm{min}^4}{8}
         \ln(x+\sqrt{x^2-x_\mathrm{min}^2})
         \right]^{\sqrt{1+x^2_\mathrm{min}}}_{x_\mathrm{min}}
\nonumber\\
    =2\pi(\frac{R_\odot}{r})^3\;\left[
      \frac{1}{4}\sqrt{1+x^2_\mathrm{min}}
     +\frac{1}{8}x^2_\mathrm{min}\sqrt{1+x^2_\mathrm{min}}
     \right.\hspace*{6em}\nonumber\\\hspace*{6em}\left.
      -\frac{x_\mathrm{min}^4}{8}
        \big(\ln(\sqrt{1+x^2_\mathrm{min}}+1)
       -\ln x_\mathrm{min}\big)
         \right]
\nonumber\\
    =2\pi(\frac{R_\odot}{r})^2\;\left[
      \frac{1}{4}
     +\frac{1}{8}\cot^2\theta_\mathrm{max}
      -\frac{\cos^4\theta_\mathrm{max}}{8\sin^3\theta_\mathrm{max}}
        \ln\frac{1+\sin^{-1}\theta_\mathrm{max}}
                {\cot\theta_\mathrm{max}}
         \right]
\nonumber\\
    =2\pi\;\left[
      \frac{1}{4}\sin^2\theta_\mathrm{max}
     +\frac{1}{8}\cos^2\theta_\mathrm{max}
      -\frac{\cos^4\theta_\mathrm{max}}{8\sin\theta_\mathrm{max}}
        \ln\frac{1+\sin\theta_\mathrm{max}}
                {\cos\theta_\mathrm{max}}
         \right]
\nonumber\\
    =\frac{\pi}{2}\;\left[1
      -\frac{1}{2}\cos^2\theta_\mathrm{max}
      -\frac{\cos^4\theta_\mathrm{max}}{2\sin\theta_\mathrm{max}}
        \ln\frac{1+\sin\theta_\mathrm{max}}
                {\cos\theta_\mathrm{max}}
         \right]
\nonumber\\
    =\frac{\pi}{2}\;\left[1
      -\frac{\cos^2\theta_\mathrm{max}}{2}
      (1+\frac{\cos^2\theta_\mathrm{max}}{\sin\theta_\mathrm{max}}
        \ln\frac{1+\sin\theta_\mathrm{max}}
                {\cos\theta_\mathrm{max}})
         \right]
\label{J_1}\end{gather}
The indefinite integrals required above are derived in the subsequent
chapter~\ref{app:integrals} 
where we use the abbreviations $a=x_\mathrm{min}$, $u=\sqrt{x^2-a^2}$
and $w=\ln(x+u)$ \citep[see also][2.271.3 and 2.272.2]{GradshteynRyzhik:1980}.
The integral $I_0$ corresponds to the total irradiance of a Lambert
sphere with unit radiance. $I_1$ covers the corresponding linear term
in the series expansion of $L$ in $\cos\zdis$. The integrals $J_0$ and $J_1$
are needed for the expansion of the irradiance to a matrix.
In the following chapter~\ref{app:integrals} we also derive a recursion
for the indefinite integrals $I_n$ and $J_n$ required for degrees higher
than $n=1$.

The expressions commonly used are the following combinations
first introduced by \citep{Minnaert:1930} and
later popularised by \citep{Billings:1966}
\begin{gather}
C=\frac{1}{2\pi}(J_0+I_0) = \frac{1-\cos^3\theta_\mathrm{max}}{3}
                                 +1-\cos\theta_\mathrm{max}
\hspace*{4.0em}\nonumber\\\hspace*{2.5em}
          = \frac{4}{3}-\cos\theta_\mathrm{max}
           -\frac{1}{3}\cos^3\theta_\mathrm{max}
\label{Minnaert_C}\\
A = \frac{1}{2\pi}(3J_0-I_0)
  = 1 -\cos^3\theta_\mathrm{max} - 1 +\cos\theta_\mathrm{max}
\hspace*{8.0em}\nonumber\\\hspace*{6.5em}
          = \cos\theta_\mathrm{max}( 1 -\cos^2\theta_\mathrm{max})
          = \cos\theta_\mathrm{max}\sin^2\theta_\mathrm{max}
\label{Minnaert_A}\\[1ex]
 F = \frac{\cos^2\theta_\mathrm{max}}{\sin\theta_\mathrm{max}}
          \ln\frac{1+\sin\theta_\mathrm{max}}
               {\cos\theta_\mathrm{max}}\hspace*{4.5em}
\nonumber\\
D = \frac{1}{2\pi}(J_1+I_1) = \frac{1}{2}\left[\frac{1}{2}
      -\frac{1}{4}\cos^2\theta_\mathrm{max}
      -\frac{1}{4}\cos^2\theta_\mathrm{max} F
          + 1 - F \right]
\label{Minnaert_D}\\
  = \frac{1}{8}\;\left[6 -\cos^2\theta_\mathrm{max}
      -(4+\cos^2\theta_\mathrm{max}) F \right]
\nonumber\\
  = \frac{1}{8}\;\left[5 +\sin^2\theta_\mathrm{max}
      -(5-\sin^2\theta_\mathrm{max}) F \right]
\nonumber\\
B = \frac{1}{2\pi}(3J_1-I_1) = \frac{1}{2}\left[\frac{3}{2}
      -\frac{3}{4}\cos^2\theta_\mathrm{max}(1+F) - 1+F\right]
\nonumber\\
 = \frac{1}{8}\left[2
      - 3\cos^2\theta_\mathrm{max}
      +(4 - 3\cos^2\theta_\mathrm{max}) F \right]
\nonumber\\
 = -\frac{1}{8}\left[1
      - 3\sin^2\theta_\mathrm{max}
      -(1 + 3\sin^2\theta_\mathrm{max}) F \right]
\label{Minnaert_B}\end{gather}

\subsection{Indefinite integrals for $I_1$, $J_1$ and for further expansion
  terms of the surface radiance $L(\cos\zdis)$}
\label{app:integrals}

We here solve the indefinite integrals needed in the previous
chapter~\ref{app:IntegrSurfRad}.
In the calculations of \cite{Minnaert:1930} and in the main text, only
the first two terms in an expansion of the solar radiance $L$ in powers
$\cos\zdis$ up to linear term are taken account of. Expansion up to the
5$^\text{th}$ power have been determined by measurements
\citep[e.g.,][]{NeckelLabs:1994,Neckel:1996}. If these were to be included in
the calculation of the scattered radiant intensity per electron, indefinite
integrals of the type $I_m=\int u^m \,dx$ and $J_m=\int u^m x^2\,dx$ are
required. We will use the following abbreviations
\begin{gather*}
  u(x)=\sqrt{x^2-a^2},\qquad u'=\frac{x}{u}\\
  w(x)=\ln(x+u),      \qquad w'=\frac{1+u'}{x+u}=\frac{1}{x}
\end{gather*}
A transformation of the integration variable $x$ yields
\begin{gather}
  \int^X \frac{1}{u}\,dx \;\stackrel{x=a \cosh z}{=}\;
  \int^{\mathrm{arcosh}(X/a)} \frac{1}{\sinh z}\sinh z dz
\nonumber\\  
  = \int^{\mathrm{arcosh}(X/a)} dz = \mathrm{arcosh}\frac{X}{a}
  = \ln(X+u(X))-\ln(a)
\nonumber\\
\text{where we used}\qquad
dx=a\frac{d\cosh z}{dz}\;dz=a\sinh z\;dz
\nonumber\\
\text{and}\quad
u^2=x^2-a^2=a^2(\cosh^2 z -1)=a^2\sinh^2 z.
\nonumber\\
 \text{In short}\qquad \int \frac{1}{u}\,dx = w + \const
 \label{int1/u}
\end{gather}
Next we define more generally the indefinite integrals
(omitting $\const$ from now on)
\[
V_n=\int x^{2n}u\,dx, \qquad W_n=\int \frac{x^{2n}}{u}\,dx \qquad
\text{where from (\ref{int1/u})} \quad W_0 = w
\]
The integrals are related by the recurrence relations
\begin{gather}
W_n=\int \frac{x^{2n}}{u}\,dx
=\int (\frac{x^{2n-2}(x^2-a^2)}{u}+a^2\frac{x^{2n-2}}{u})\,dx
=\int (x^{2n-2}u + a^2 \int\frac{x^{2n-2}}{u})\,dx
\nonumber\\
=V_{n-1} + a^2 W_{n-1}
\label{W_n}\\
\text{and}\qquad (2n+1)V_n
= (2n+1)\int x^{2n}u\,dx
= \int (x^{2n+1})'u\,dx
= x^{2n+1}u - \int \frac{x^{2n+2}}{u}\,dx
\nonumber\\
= x^{2n+1}u - W_{n+1}
= x^{2n+1}u - V_{n}-a^2 W_{n}
\quad\text{or}
\nonumber\\
V_n = \frac{1}{2n+2}(x^{2n+1}u - a^2 W_n)
\label{V_n}\end{gather}
In case of a linear expansion of $L(\cos\zdis)$ we only need
the indefinite integrals $V_0$ for $I_1$ and $V_1$ for $J_1$.
By recurrence of (\ref{W_n}) and (\ref{V_n}) we obtain
successively
\begin{gather*}
  W_0=w, \qquad V_0=\frac{1}{2}(xu-a^2 w)
\\
  W_1=\frac{1}{2}(xu-a^2 w)+a^2 w=\frac{1}{2}(xu+a^2 w)
\\
  V_1=\frac{1}{4}(x^3u-\frac{a^2}{2}(xu+a^2w))
     =(\frac{x^2}{4}-\frac{a^2}{8})xu-\frac{a^4}{8}w
\\
  W_2=(\frac{x^2}{4}-\frac{a^2}{8})xu - \frac{a^2}{8}w
     +\frac{a^2}{2}(xu + a^2w)
     =(\frac{x^2}{4}+\frac{3a^2}{8})xu + \frac{3a^4}{8}w
\\
  V_2 = \frac{1}{6}(x^5u - a^2 (\frac{x^2}{4}+\frac{3a^2}{8})xu
      - \frac{3a^6}{8}w)
      = (\frac{x^4}{6} - \frac{a^2 x^2}{24} - \frac{3a^4}{48})xu
      - \frac{3a^6}{48}w
\\ \dots      
\end{gather*}
More can be derived in the same way.
From the web side \texttt{integrals.wolfram.com}
\begin{gather*}
  V_3 
     = \frac{x}{384}(48x^6-8a^2x^4-10a^4x^2-15a^6)u
     - \frac{15a^8}{384}w
\\
  V_4 
    = \frac{x}{3840}(384x^8-48a^2x^6-56a^4x^4-70a^6x^2-105a^8)u
    - \frac{105a^8}{3840}w
\\
  V_n 
    = \frac{x^{2n+1}u}{2n+1}\;
      \frac{_2F_1(-\frac{1}{2},n+\frac{1}{2},n+\frac{3}{2},\frac{x^2}{a^2})}
           {\sqrt{1-\frac{x^2}{a^2}}}
\end{gather*}
where
\[
_2F_1(a,b,c,x)=1+\frac{ab}{c}x + \frac{a(a+1)b(b+1)}{2c(c+1)}x^2
              +\frac{a(a+1)(a+2)b(b+1)(b+2)}{2c(c+1)(c+2)}x^3+\dots
\]              
is the hypergeometric function which reduces to a polynomial if $a$ or $b$ is
a negative integer. In our case it must be an infinite series because $u$ and $w$
cannot be expressed by a finite polynomial.

The final step is
\begin{gather*}
 I_m = \int u^m\,dx= \int (x^2-a^2)^{m/2}\,dx
 = \begin{cases}
    \sum\limits_{i=0}^{m/2} \begin{pmatrix} m/2 \\ i \end{pmatrix}
    (-a^2)^{i-m/2} \int x^{2i}\,dx\\
    \sum\limits_{i=0}^{(m-1)/2} \begin{pmatrix} (m-1)/2 \\ i \end{pmatrix}
    (-a^2)^{i-(m-1)/2} \int x^{2i}u\,dx 
  \end{cases}
\\ 
 = \begin{cases}
    \sum\limits_{i=0}^{m/2} \begin{pmatrix} m/2 \\ i \end{pmatrix}
    \displaystyle (-a^2)^{i-m/2} \frac{x^{2i+1}}{2i+1} & \text{if $m$ even}\\
    \sum\limits_{i=0}^{(m-1)/2} \begin{pmatrix} (m-1)/2 \\ i \end{pmatrix}
    (-a^2)^{i-(m-1)/2} V_i & \text{if $m$ odd}\\
  \end{cases}
 \\
 J_m = \int x^2 u^m\,dx= \int x^2 (x^2-a^2)^{m/2}\,dx
 = \begin{cases}
    \sum\limits_{i=0}^{m/2} \begin{pmatrix} m/2 \\ i \end{pmatrix}
    (-a^2)^{i-m/2} \int x^{2i+2}\,dx\\
    \sum\limits_{i=0}^{(m-1)/2} \begin{pmatrix} (m-1)/2 \\ i \end{pmatrix}
    (-a^2)^{i-(m-1)/2} \int x^{2i+2}u\,dx 
  \end{cases}
\\ 
 = \begin{cases}
    \sum\limits_{i=0}^{m/2} \begin{pmatrix} m/2 \\ i \end{pmatrix}
    \displaystyle (-a^2)^{i-m/2} \frac{x^{2i+3}}{2i+3} & \text{if $m$ even}\\
    \sum\limits_{i=0}^{(m-1)/2} \begin{pmatrix} (m-1)/2 \\ i \end{pmatrix}
    (-a^2)^{i-(m-1)/2} V_{i+1} & \text{if $m$ odd}\\
  \end{cases}
\end{gather*}
For the lowest powers we can write down the expressions explicitely
\begin{gather}
  I_1=V_0 =\int u\,dx
  = \frac{1}{2}xu-\frac{a^2}{2}w
\label{I1_app}\\ 
  I_2=\int u^2\,dx
  = \int(x^2-a^2)\,dx
  = \frac{x^3}{3}-a^2x
\nonumber\\  
 I_3=\int u^3\,dx
  = V_1-a^2 V_0
  = \frac{1}{4}(x^2-\frac{5a^2}{2})xu + \frac{3}{8}a^4 w
\nonumber\\  
  I_4=\int u^4\,dx
  = \int(x^4-2a^2x^2+a^4)\,dx
  = \frac{1}{5}x^5 - \frac{2}{3}a^2x^3 + a^4x
\nonumber\\  
 I_5= \int u^5\,dx
  = V_2-2a^2V_1+V_0
  = \frac{1}{48}(8x^4-26a^2x^2+33a^4)xu - \frac{15}{48}a^6 w
\nonumber\\
\dots
\nonumber\\  
  J_1=\int x^2u\,dx
  = V_1
  = \frac{1}{8}(2x^2-a^2)xu - \frac{1}{8}a^4 w
\label{J1_app}\\  
  J_2=\int x^2u^2\,dx
  = \int (x^4-a^2x^2)\,dx
  = x^3(\frac{1}{5}x^2-\frac{1}{3}a^2)
\nonumber\\
  J_3=\int x^2 u^3\,dx
  = V_2 - a^2V_1
  = \frac{1}{48}(8x^4-14a^2x^2+3a^4)xu + \frac{3}{48}a^6 w
\nonumber\\  
  J_4=\int x^2u^4\,dx
  = \int (x^6-2a^2x^4+a^4x^2)\,dx
  = x^3(\frac{1}{7}x^4-\frac{2}{5}a^2x^2+\frac{1}{3}a^4)
\nonumber\\
 J_5=\int x^2 u^5\,dx
  = V_3 - 2a^2V_2 + a^4V_1
  = \frac{1}{384}(48x^6-136a^2x^4+118a^4x^2-15a^6)xu - \frac{15}{384}a^8 w
\nonumber\\
\dots
\nonumber\end{gather}
From the web side \texttt{integrals.wolfram.com} the general expressions are
given as
\begin{gather*}
  I_m=\int u^m\,dx
    = xu^m \;\frac{_2F_1(\frac{1}{2},-m,\frac{3}{2},\frac{x^2}{a^2})}
           {(1-\frac{x^2}{a^2})^{m}}
\\  
 J_m=\int x^2 u^m\,dx
    = \frac{1}{3}x^3u^m \;\frac{_2F_1(\frac{3}{2},
     -\frac{m}{2},\frac{5}{2},\frac{x^2}{a^2})}
           {(1-\frac{x^2}{a^2})^{m/2}}
\end{gather*}
where $_2F_1$ is as above the hypergeometric function.

\subsection{Connections with van de Hulst's coefficients}
\label{app:VanDeHulstCoeff}

In the end there are just the two scattered radiant intensities per electron
$\rin_\mathrm{tan}$ and $\rin_\mathrm{rad}$ or
equivalently, their polarised and total combinations, which are relevant. Thus
the four coefficients $A$, $B$, $C$ and $D$ of Minnaert could for our purposes
also be reduced to only two. \cite{vandeHulst:1950} introduced two new
coefficients $\mathcal{A}$ and $\mathcal{B}$ which do the job, however, they
depend besides on $r$ also on $u$. They are used also in other papers
\citep[e.g.,][]{SaitoEtal:1970}.
Unfortunately, Van de Hulst's naming convention allows
them to be mixed up with Minnaert's $A$ and $B$. They are directly related to
the integrals (\ref{I_0}) to (\ref{J_1}) by
\citep[][eqs.~11 and 12.]{vandeHulst:1950}
\begin{gather}
 2\mathcal{A}+\mathcal{B}
 = \frac{(1-u)I_0 + u I_1}{\pi(1-u/3)}
 = \frac{(1-u)(3C-A) + u(3D-B)}{2(1-u/3)}
 \nonumber\\
 2\mathcal{A}-\mathcal{B}
 = \frac{(1-u)J_0 + u J_1}{\pi(1-u/3)}
 = \frac{(1-u)(A+C) - u(D+B)}{2(1-u/3)}
 \nonumber\\\text{or}\quad
 \mathcal{A}=\frac{(1-u)C+uD}{1-u/3},\quad
 \mathcal{B}=\frac{(1-u)(C-A)+u(D-B)}{1-u/3}
\label{VanHulstCoeff}\end{gather}
The numerator $1-u/3$ is
necessary because our coefficients are normalised to the radiance $\rad_\odot$
at solar centre while $\mathcal{A}$ and $\mathcal{B}$ are normalised to the
mean radiance
\[
\frac{I_\mathrm{in}(r)}{\pi}(\frac{r}{R_\odot})^2
\xrightarrow{r\rightarrow\infty}
 \bar{\rad}_\odot=(1-u/3)\,\rad_\odot
\]
of the solar disk, see (\ref{Irr_inf}). If we compare (\ref{VanHulstCoeff})
with (\ref{Irr_in_Minnaert}), we see that the irradiance matrix elements
$Q_\mathrm{in,xx}$ and $Q_\mathrm{in,zz}$ are proportional to $\mathcal{A}$
and $\mathcal{B}$, respectively.

\subsection{Limit $r \rightarrow\infty$}
\label{app:LimitInfty}

We here derive the asymptotic series expansions of Minnaert's coefficients
to be used for large $r$ because the exact expressions are numerically unstable
in this limit. This can be seen from the fact that some of the
terms have to be expanded to fifth order to retain the two
leading terms of the series expansion in $\theta_\mathrm{max}$.
These limiting formulae should preferably be used for
large $r$ in the numerical evaluation of the coefficients.
A similar expansion for the the Van de Hulst coefficients
$\mathcal{A}$ and $\mathcal{B}$ can be found in \citep{SaitoEtal:1970}.
The limit $r \rightarrow\infty$ implies
$\theta_\mathrm{max}\rightarrow\frac{R_\odot}{r}$
We first give the expansions of some subterms and then compose 
the final expressions from them.
\begin{gather*}
  \cos^2\theta_\mathrm{max}
      =1-\theta^2_\mathrm{max}
        +({\TS\frac{2}{24}}+\quart)\theta^4_\mathrm{max}
      =1-\theta^2_\mathrm{max}+\third\theta^4_\mathrm{max}\\[0.5ex]
  \sin^{-1}\theta_\mathrm{max}
      =\theta^{-1}_\mathrm{max}
        (1-\sixth\theta^2_\mathrm{max}
          +{\TS\frac{1}{120}}\theta^4_\mathrm{max})^{-1}
      =\theta^{-1}_\mathrm{max}
        (1+\sixth\theta^2_\mathrm{max}
          +({\TS\frac{1}{36}}-{\TS\frac{1}{120}})
              \theta^4_\mathrm{max})\\
      =\theta^{-1}_\mathrm{max}
        (1+\sixth\theta^2_\mathrm{max}
          +{\TS\frac{7}{360}}\theta^4_\mathrm{max})\\[1ex]
  \frac{1+\sin\theta_\mathrm{max}}{\cos\theta_\mathrm{max}}
 =\frac{1+\theta_\mathrm{max}
         -\sixth\theta^3_\mathrm{max}
         +{\TS\frac{1}{120}}\theta^5_\mathrm{max}}
        {1-\half\theta^2_\mathrm{max}
          +{\TS\frac{1}{24}}\theta^4_\mathrm{max}}
    \nonumber\\[0.5ex]
      =(1+\theta_\mathrm{max}
         -\sixth\theta^3_\mathrm{max}
         +{\TS\frac{1}{120}}\theta^5_\mathrm{max})
       (1+\half\theta^2_\mathrm{max}
         +(\quart-{\TS\frac{1}{24}})\theta^4_\mathrm{max})
    \nonumber\\[0.5ex]
      =(1+\theta_\mathrm{max}
         -\sixth\theta^3_\mathrm{max}
         +{\TS\frac{1}{120}}\theta^5_\mathrm{max})
       (1+\half\theta^2_\mathrm{max}
         +{\TS\frac{5}{24}}\theta^4_\mathrm{max})
    \nonumber\\[0.5ex]
      = 1+\theta_\mathrm{max}+\half\theta^2_\mathrm{max}
         +(\half-\sixth)\theta^3_\mathrm{max}
         +{\TS\frac{5}{24}}\theta^4_\mathrm{max}
         +({\TS\frac{1}{120}}-{\TS\frac{1}{12}}+{\TS\frac{5}{24}})
            \theta^5_\mathrm{max}
    \nonumber\\[0.5ex]
      =1+\theta_\mathrm{max}+\half\theta^2_\mathrm{max}
        +\third\theta^3_\mathrm{max}
        +{\TS\frac{5}{24}}\theta^4_\mathrm{max}
        +{\TS\frac{4}{30}}\theta^5_\mathrm{max}
    \nonumber\\[1ex]
  \ln(\frac{1+\sin\theta_\mathrm{max}}{\cos\theta_\mathrm{max}})
 =\ln(1+\theta_\mathrm{max}+\half\theta^2_\mathrm{max}
        +\third\theta^3_\mathrm{max}
        +{\TS\frac{5}{24}}\theta^4_\mathrm{max}
        +{\TS\frac{4}{30}}\theta^5_\mathrm{max})
    \nonumber\\
        =\theta_\mathrm{max}
        +(\half-\half)\theta^2_\mathrm{max}
        +(\third-\half+\third)\theta^3_\mathrm{max}
        +({\TS\frac{5}{24}}-\third-{\TS\frac{1}{8}}+\half-\quart)
              \theta^4_\mathrm{max}
    \nonumber\\[0.5ex]
        +({\TS\frac{4}{30}}-{\TS\frac{5}{24}}-\sixth+\third
          +\quart-\half+\fifth)\theta^5_\mathrm{max}
        =\theta_\mathrm{max}
        +\sixth\theta^3_\mathrm{max}
        +{\TS\frac{1}{24}}\theta^5_\mathrm{max}
    \nonumber\\[1ex]
 \frac{1}{\sin\theta_\mathrm{max}}
         \ln(\frac{1+\sin\theta_\mathrm{max}}{\cos\theta_\mathrm{max}})
      =(1+\sixth\theta^2_\mathrm{max}
         +{\TS\frac{7}{360}}\theta^4_\mathrm{max})
       \frac{1}{\theta_\mathrm{max}}
       (\theta_\mathrm{max}
       +\sixth\theta^3_\mathrm{max}
       +{\TS\frac{1}{24}}\theta^5_\mathrm{max})
    \nonumber\\[0.5ex]
      =(1+\sixth\theta^2_\mathrm{max}
         +{\TS\frac{7}{360}}\theta^4_\mathrm{max})
       (1+\sixth\theta^2_\mathrm{max}
       +{\TS\frac{1}{24}}\theta^4_\mathrm{max})
    \nonumber\\[1ex]
      = 1+{\TS\frac{2}{6}}\theta^2_\mathrm{max}
         +({\TS\frac{1}{24}}+{\TS\frac{1}{36}}+{\TS\frac{7}{360}})
             \theta^4_\mathrm{max}
      = 1+{\TS\frac{1}{3}}\theta^2_\mathrm{max}
         +{\TS\frac{8}{90}}\theta^4_\mathrm{max}
    \nonumber\\[1ex]
 \frac{\cos^2\theta_\mathrm{max}}{\sin\theta_\mathrm{max}}
         \ln(\frac{1+\sin\theta_\mathrm{max}}{\cos\theta_\mathrm{max}})
      =(1-\theta^2_\mathrm{max}+\third\theta^4_\mathrm{max})
       (1+{\TS\frac{1}{3}}\theta^2_\mathrm{max}
         +{\TS\frac{8}{90}}\theta^4_\mathrm{max})
    \nonumber\\[0.5ex]
      =(1-(1-\third)\theta^2_\mathrm{max}
         +(\third+{\TS\frac{8}{90}}-\third)\theta^4_\mathrm{max})
      =1-{\TS\frac{2}{3}}\theta^2_\mathrm{max}
        +{\TS\frac{4}{45}}\theta^4_\mathrm{max}
\end{gather*}
Using the above expansions for $r \rightarrow\infty$ and
$\theta_\mathrm{max}\rightarrow\frac{R_\odot}{r}$
we have the following limiting relations
\begin{gather}
 I_0(r)=2\pi (1-\cos\theta_\mathrm{max})
      \xrightarrow{r\rightarrow\infty}
 I_0^\infty=\pi(\theta_\mathrm{max}^2-{\TS\frac{1}{12}}\theta_\mathrm{max}^4)
\label{I_0_inf}\\[1ex]
 J_0(r)=\frac{2\pi}{3}(1-\cos^3\theta_\mathrm{max})
      \xrightarrow{r\rightarrow\infty}
 J_0^\infty =\frac{2\pi}{3}(1-(1-{\TS\frac{3}{2}}\theta^2_\mathrm{max}
                        +{\TS\frac{7}{8}}\theta^4_\mathrm{max}))
 \nonumber\\
    =\pi(\theta^2_\mathrm{max}-{\TS\frac{7}{12}}\theta^4_\mathrm{max})
\label{J_0_inf}\\[1ex]
I_1(r)=\pi\left[1
       -\frac{\cos^2\theta_\mathrm{max}}{\sin\theta_\mathrm{max}}
        \ln(\frac{1+\sin\theta_\mathrm{max}}{\cos\theta_\mathrm{max}})\right]
      \xrightarrow{r\rightarrow\infty}
 \nonumber\\
I_1^\infty
  =\pi\left[1-(1-{\TS\frac{2}{3}}\theta^2_\mathrm{max}
             +{\TS\frac{4}{45}}\theta^4_\mathrm{max})\right]
\nonumber\\
  =\pi({\TS\frac{2}{3}}\theta^2_\mathrm{max}
      -{\TS\frac{4}{45}}\theta^4_\mathrm{max})
 \label{I_1_inf}\\[1ex]
 J_1(r)=\frac{\pi}{2}\left[1
       -\frac{1}{2}\cos^2\theta_\mathrm{max}
       (1+\frac{\cos^2\theta_\mathrm{max}}{\sin\theta_\mathrm{max}}
         \ln\frac{1+\sin\theta_\mathrm{max}}
                 {\cos\theta_\mathrm{max}})\right]
       \xrightarrow{r\rightarrow\infty}
 \nonumber\\
 J_1^\infty=\frac{\pi}{2}\left[1
   -\half(1-\theta^2_\mathrm{max}+\third\theta^4_\mathrm{max})
         (2-{\TS\frac{2}{3}}\theta^2_\mathrm{max}
           +{\TS\frac{4}{45}}\theta^4_\mathrm{max})\right]
 \nonumber\\
   =\frac{\pi}{2}\left[1
   -(1-\theta^2_\mathrm{max}+\third\theta^4_\mathrm{max})
    (1-{\TS\frac{1}{3}}\theta^2_\mathrm{max}
      +{\TS\frac{2}{45}}\theta^4_\mathrm{max})\right]
 \nonumber\\
   =\frac{\pi}{2}\left[1
   -(1-(1+\third)\theta^2_\mathrm{max}
      +(\third+{\TS\frac{2}{45}}+\third)\theta^4_\mathrm{max})\right]
 \nonumber\\
   =\frac{\pi}{2}({\TS\frac{4}{3}}\theta^2_\mathrm{max}
                 -{\TS\frac{32}{45}}\theta^4_\mathrm{max})
   =\pi({\TS\frac{2}{3}}\theta^2_\mathrm{max}
       -{\TS\frac{16}{45}}\theta^4_\mathrm{max})
\label{J_1_inf}\end{gather}
We see that $I_0$ and $J_0$ and $I_1$ and $J_1$ have pairwise the same leading
term. For large $r$ we therefore find $J_0\rightarrow I_0$ and $J_1\rightarrow
I_1$. This could have been expected because then $\theta$ is bounded close to
zero and $\cos^2\theta$ in the integrand hardly deviates from unity.


The according limits for the Minnaert coefficients are
\begin{gather*}
C=\frac{1}{2\pi}(J_0+I_0) \xrightarrow{r\rightarrow\infty} C^\infty
 =\frac{1}{2\pi}(J^\infty_0+I^\infty_0)
 =\half(\theta^2_\mathrm{max}
        -{\TS\frac{7}{12}}\theta^4_\mathrm{max}
        +\theta^2_\mathrm{max}
        -{\TS\frac{1}{12}}\theta^4_\mathrm{max})
\nonumber\\
  = \theta^2_\mathrm{max}-{\TS\frac{1}{3}}\theta^4_\mathrm{max}
\nonumber\\
A = \frac{1}{2\pi}(3J_0-I_0) \xrightarrow{r\rightarrow\infty} A^\infty
  = \frac{1}{2\pi}(3J^\infty_0-I^\infty_0)
  = \half(3\theta^2_\mathrm{max}
        -3{\TS\frac{7}{12}}\theta^4_\mathrm{max}
         -\theta^2_\mathrm{max}
        +{\TS\frac{1}{12}}\theta^4_\mathrm{max})
\nonumber\\
  = \theta^2_\mathrm{max}-{\TS\frac{5}{6}}\theta^4_\mathrm{max}
\nonumber\\
F = 1-\frac{1}{\pi}I_1 \xrightarrow{r\rightarrow\infty} F^\infty
  = 1-\frac{1}{\pi}I^\infty_1
  = 1-{\TS\frac{2}{3}}\theta^2_\mathrm{max}
     +{\TS\frac{4}{45}}\theta^4_\mathrm{max}
\nonumber\\
D = \frac{1}{2\pi}(J_1+I_1)
\xrightarrow{r\rightarrow\infty} D^\infty=\frac{1}{2\pi}(J_1^\infty+I_1^\infty)
  =\half({\TS\frac{2}{3}}\theta^2_\mathrm{max}
        -{\TS\frac{16}{45}}\theta^4_\mathrm{max}
        +{\TS\frac{2}{3}}\theta^2_\mathrm{max}
        -{\TS\frac{4}{45}}\theta^4_\mathrm{max})
\nonumber\\
  ={\TS\frac{2}{3}}\theta^2_\mathrm{max}
  -{\TS\frac{2}{9}}\theta^4_\mathrm{max}
\nonumber\\
B = \frac{1}{2\pi}(3J_1-I_1)
\xrightarrow{r\rightarrow\infty} B^\infty=\frac{1}{2\pi}(3J_1^\infty-I_1^\infty)
  =\half(2\theta^2_\mathrm{max}
      -{\TS\frac{3\;16}{45}}\theta^4_\mathrm{max}
       -{\TS\frac{2}{3}}\theta^2_\mathrm{max}
       +{\TS\frac{4}{45}}\theta^4_\mathrm{max})
\nonumber\\
  ={\TS\frac{2}{3}}\theta^2_\mathrm{max}
  -{\TS\frac{22}{45}}\theta^4_\mathrm{max}
\nonumber\\
C-A=\frac{1}{\pi}(I_0-J_0)
\xrightarrow{r\rightarrow\infty}
(\TS\frac{5}{6}-\frac{1}{3})\theta^4_\mathrm{max}
=\theta^4_\mathrm{max}
\nonumber\\
D-B=\frac{1}{\pi}(I_1-J_1)
\xrightarrow{r\rightarrow\infty}
(\TS\frac{22}{45}-\frac{2}{9})\theta^4_\mathrm{max}
={\TS\frac{4}{15}}\theta^4_\mathrm{max}
\nonumber\end{gather*}

\subsection{Limit $r \rightarrow R_\odot$}
\label{app:LimitUnity}

The limit $r\rightarrow R_\odot$ implies $\theta_\mathrm{max}\rightarrow\pi/2$
and is less problematic to derive \citep[see also][]{Billings:1966}).
With the complementary angle $\epsilon=\pi/2-\theta_\mathrm{max}
\rightarrow 0$ we have
\begin{gather*}
C=\frac{1}{2\pi}(J_0+I_0)\xrightarrow{r\rightarrow R_\odot} C^{R_\odot} =\frac{4}{3}
\nonumber\\
A = \frac{1}{2\pi}(3J_0-I_0) \xrightarrow{r\rightarrow R_\odot} A^{R_\odot}
  = \sin\epsilon \simeq \epsilon
\nonumber\\
F= \frac{\cos^2\theta_\mathrm{max}}{\sin\theta_\mathrm{max}}
   \ln\frac{1+\sin\theta_\mathrm{max}}
           {\cos\theta_\mathrm{max}}
   \;\xrightarrow{\theta_\mathrm{max}\rightarrow\pi/2}\;
  -\epsilon^2 \ln\frac{\epsilon}{2}
   \;\xrightarrow{\epsilon\rightarrow 0}\; 0
\nonumber\\
D = \frac{1}{2\pi}(J_1+I_1) \xrightarrow{r\rightarrow R_\odot} D^{R_\odot}
  =\frac{1}{2}(\frac{1}{2}+1)=\frac{3}{4}
\nonumber\\
B = \frac{1}{2\pi}(3J_1-I_1) \xrightarrow{r\rightarrow R_\odot} B^{R_\odot}
  =-\frac{1}{8}(1-3)=\frac{1}{4}
\end{gather*}

\section{Abel transform}
\setcounter{equation}{0}
\label{App:AbelTrans}

The Abel transformation relates a scalar $P(r)$ which only depends on the
distance $r$ from the origin with a line integral $Q(\rho)$ on $P$. The
integration path is a straight line $\vect{c}(s,\rho)$ which passes the origin
at a distance $\rho$. The curve parameter $s$ along $\vect{c}$ is introduced
such that $|s|$ is the distance from the point closest to the origin on
$\vect{c}$ (see Fig.~\ref{Fig:ScaGeom_LOS}) so that $r=\sqrt{\rho^2+s^2}$.
Then
\begin{gather}
Q(\rho)
=\int_\vect{c} P(r)\;ds
=2\int_0^\infty P(\sqrt{\rho^2+s^2})\;ds
\nonumber\\
\stackrel{s=\sqrt{r^2-\rho^2}}{=}\;
 2\int_{|\rho|}^\infty P(r)\frac{r\,dr}{\sqrt{r^2-\rho^2}}
=\mathcal{A}(P)
\label{AbelTr_def}\end{gather}
Here, $P(r)$ needs to decrease faster than $r^{-1}$ at $r\rightarrow\infty$
for the integral to remain finite.
There are various equivalent forms for the inverse (proof by partial
integration; we abbreviate $Q'(\rho)=dQ/d\rho$)
\begin{gather}
P(r)=\mathcal{A}^{-1}(Q)
    =-\frac{1}{\pi}\int_{|r|}^\infty
        \frac{Q'(\rho)d\rho}{\sqrt{\rho^2-r^2}}
    =\frac{-1}{2\pi}\mathcal{A}(\frac{Q'(\rho)}{\rho})
\label{AbelInv_def}\\
    =-\frac{1}{r\pi} \frac{d}{dr}\int_r^\infty
    Q(\rho)\frac{d\rho}{\sqrt{\rho^2-r^2}} =-\frac{1}{\pi}
    \frac{d}{dr}\int_r^\infty
    \frac{Q(\rho)}{\rho}\frac{d\rho}{\sqrt{\rho^2-r^2}}
\nonumber\end{gather}
Often $Q(\rho)$ represents measurements like in our case it is proportional to
the observed image brightness profile with projected distance $\rho$ from the
Sun centre. Any noise in $Q(\rho)$ is strongly amplified due to the
differentiation of $Q$ and therefore the above direct inversion is prone to
noise. For practical inversions we therefore modify (\ref{AbelInv_def}) by
partial integration \citep{Yuan:2003}
\begin{gather}
P(r)=-\frac{1}{\pi}\int_{|r|}^\infty
      \frac{Q'(\rho)d\rho}{\sqrt{\rho^2-r^2}}
    =-\frac{1}{\pi}\big[
      \left.\frac{Q(\rho)}{\sqrt{\rho^2-r^2}}\right|_{\rho=|r|}^\infty
     +\int_{|r|}^\infty \frac{Q(\rho)\rho d\rho}{\sqrt{\rho^2-r^2}^3}
     \big]
\nonumber\\
    =-\frac{1}{\pi}\big[
     -Q(r)\lim_{\rho\rightarrow |r|}\frac{1}{\sqrt{\rho^2-r^2}}
     +\int_{|r|}^\infty \frac{Q(\rho)\rho d\rho}{\sqrt{\rho^2-r^2}^3}
     \big]
\nonumber\\
    =-\frac{1}{\pi}\big[
     +Q(r)\int_{|r|}^\infty\frac{d}{d\rho}\frac{1}{\sqrt{\rho^2-r^2}}\;d\rho
         +\int_{|r|}^\infty \frac{Q(\rho)\rho d\rho}{\sqrt{\rho^2-r^2}^3}
     \big]
\nonumber\\
   =-\frac{1}{\pi}\big[
     -Q(r)\int_{|r|}^\infty\frac{\rho d\rho}{\sqrt{\rho^2-r^2}^3}
         +\int_{|r|}^\infty \frac{Q(\rho)\rho d\rho}{\sqrt{\rho^2-r^2}^3}
     \big]
\nonumber\\
   =-\frac{1}{\pi}
    \int_{|r|}^\infty (Q(\rho)-Q(|r|))\frac{\rho d\rho}{\sqrt{\rho^2-r^2}^3}
\nonumber\\
   =-\frac{1}{2\pi}\mathcal{A}\big(
   \frac{Q(\rho)-Q(|r|)}{\rho^2-r^2}
   \big)
\label{AbelInv_dQ}\end{gather}
This way, we have got rid of the derivative at the price of a stronger
singularity at $\rho=|r|$.

In the next step we aim to avoid the infinite upper integral boundary
by transforming to the scattering angle $\bar{\chi}$ between the integration
path and the radial direction to $\vect{c}(s)$.
Starting from (\ref{AbelTr_def}) we have for the forward transform
\begin{gather}
Q(\rho)=\mathcal{A}(P)
= 2\int_{|\rho|}^\infty P(r)\frac{r\,dr}{\sqrt{r^2-\rho^2}}
\stackrel{r=x\,\rho}{=}
 2\rho\int_1^\infty P(x\,\rho)\frac{x\,dx}{\sqrt{x^2-1}}
\nonumber\\
\stackrel{x=1/t}{=}
 2\rho\int_0^1 P(\frac{\rho}{t})\frac{dt}{t^2\sqrt{1-t^2}}
\stackrel{t=\sin\bar{\chi}}{=}
 2\rho\int_0^{\pi/2} P(\frac{\rho}{\sin\bar{\chi}})\frac{d\bar{\chi}}{\sin^2\bar{\chi}}
\label{AbelTr_chi}\end{gather}
Effectively we have substituted $r=\rho/\sin\bar{\chi}$ by $\bar{\chi}$, the angle between
the radial vector to and the integration path .
$\vect{c}(s)$. Since only $\sin\bar{\chi}$ is involved, it does not matter if we
replace $\bar{\chi}$ by its complement $\pi-\bar{\chi}$ (one of the two equivalent
definitions could be viewed as the scattering angle). The inverse transformation
(\ref{AbelInv_def}) could equivalently be written as
\begin{gather}
P(r)=\mathcal{A}^{-1}(Q)
=-\frac{1}{\pi}\int_{|r|}^\infty
    Q'(\rho)\frac{d\rho}{\sqrt{\rho^2-r^2}}
\stackrel{\rho=xr}{=}
    -\frac{1}{\pi}\int_1^\infty
    Q'(xr)\frac{dx}{\sqrt{x^2-1}}
\nonumber\\
\stackrel{x=1/t}{=}
    -\frac{1}{\pi}\int_0^1
    Q'(\frac{r}{t})\frac{dt}{t\sqrt{1-t^2}}
\stackrel{t=\sin\bar{\chi}}{=}
    -\frac{1}{\pi}\int_0^{\pi/2}
    Q'(\frac{r}{\sin\bar{\chi}})\frac{d\bar{\chi}}{\sin\bar{\chi}}
\label{AbelInv_chi}\end{gather}
The inverse transformation as in (\ref{AbelInv_dQ}) gives after substitution of
$\rho$ by $\bar{\chi}$
\begin{gather*}
P(r)\stackrel{\rho=x|r|}{=}
    -\frac{1}{|r|\pi}\int_{1}^\infty
    [Q(x|r|)-Q(|r|)]\frac{x dx}{\sqrt{x^2-1}^3}
\\\stackrel{x=1/t}{=}
    -\frac{1}{|r|\pi}\int_{0}^1
    [Q(|r|/t)-Q(|r|)]\frac{dt}{\sqrt{1-t^2}^3}
\\\stackrel{t=\sin\bar{\chi}}{=}
    -\frac{1}{|r|\pi}\int_{0}^{\pi/2}
    [Q(\frac{|r|}{\sin\bar{\chi}})-Q(|r|)]\frac{d\bar{\chi}}{\cos^2\bar{\chi}}
\end{gather*}
The integrand is finite at both integral boundaries:
\begin{gather*}
\text{at}\;\bar{\chi}\rightarrow 0\qquad
\frac{Q(\frac{|r|}{\sin\bar{\chi}})-Q(|r|)}{\cos^2\bar{\chi}} \rightarrow
Q(\infty)-Q(|r|)=-Q(|r|)
\\
\text{at}\;\bar{\chi}\rightarrow \frac{\pi}{2}\qquad
\frac{Q(\frac{|r|}{\sin\bar{\chi}})-Q(|r|)}{\cos^2\bar{\chi}}
=\lim_{\epsilon=\pi/2-\bar{\chi}\rightarrow 0}
\frac{Q(\frac{|r|}{\cos\epsilon})-Q(|r|)}{\sin^2\epsilon}
\\
=\lim_{\epsilon\rightarrow 0}
\frac{1}{\sin^2\epsilon}
\sum_{i=1}^n \frac{|r|^i}{i!} Q^{(i)}(|r|)
(\frac{1}{\cos\epsilon}-1)^i
=\lim_{\epsilon\rightarrow 0}
\sum_{i=1}^n \frac{|r|^i}{i!} Q^{(i)}(|r|)
\frac{(1-\cos\epsilon)^i}{\cos^i\epsilon\sin^2\epsilon}
\\
=\lim_{\epsilon\rightarrow 0}
\sum_{i=1}^n \frac{|r|^i}{i!} Q^{(i)}(|r|)
\frac{(1-\cos\epsilon)^{i-1}}{\cos^i\epsilon(1+\cos\epsilon)}
=\sum_{i=1}^n \frac{|r|^i}{i!} Q^{(i)}(|r|)\frac{\delta_{i1}}{2}
=\frac{|r|}{2}Q'(|r|)
\end{gather*}

We can generalise the Abel transformation to scalar distributions which are not
azimuthally symmetric but also depend on some power of $|\sin\bar{\chi}|$. For
$\rho\ge 0$
\begin{gather}
Q_n(\rho)
=2\int_\rho^\infty P(r)|\sin\bar{\chi}|^n\frac{r\;dr}{\sqrt{r^2-\rho^2}}
=2\int_\rho^\infty P(r)(\frac{\rho}{r})^n\frac{r\;dr}{\sqrt{r^2-\rho^2}}
\nonumber\\
=2\rho^n\int_\rho^\infty \frac{P(r)}{r^n}\frac{r\;dr}{\sqrt{r^2-\rho^2}}
=\rho^n\mathcal{A}(\frac{P}{r^n})
\label{AbelTr_gen}\\
\frac{P(r)}{r^n}=-\frac{1}{\pi}\int_r^\infty
\frac{d}{d\rho}(\frac{Q_n}{\rho^n})\frac{d\rho}{\sqrt{\rho^2-r^2}}
=\mathcal{A}^{-1}(\frac{Q_n}{\rho^n})
\label{AbelInv_gen}\end{gather}

\subsection{Expansion in inverse powers}
\label{app:ExpInvPow}

We her consider inverse powers $Q(\rho)=\rho^{-\gamma}$, where $\gamma$ may be
any real positive number. Then $Q'(\rho)=-\gamma \rho^{-\gamma-1}$ and
\begin{gather}
P(r)=\frac{\gamma}{\pi}\int_0^{\pi/2}
      (\frac{r}{\sin\bar{\chi}})^{-\gamma-1}\frac{d\bar{\chi}}{\sin\bar{\chi}}
    =\frac{\gamma}{\pi} r^{-\gamma-1}
    \int_0^{\pi/2}\sin^\gamma\bar{\chi}\; d\bar{\chi}
\nonumber\\
    =\frac{\gamma}{2\pi} B(\frac{\gamma+1}{2},\frac{1}{2}) \;r^{-\gamma-1}
\label{AbelTr_power}\end{gather}
where we used the definition of the beta-function
[Gradshteyn\&Ryzhik 8.380+8.384]
\begin{gather}
  B(x,y)=2\int_0^1 t^{2x-1}(1-t^2)^{y-1}\;dt
        =\int_0^1 t^{x-1}(1-t)^{y-1}\;dt
 \\ = 2\int_0^{\pi/2} \sin^{2x-1}\bar{\chi} \cos^{2y-1}\bar{\chi}\;d\bar{\chi}
        = \frac{\Gamma(x)\Gamma(y)}{\Gamma(x+y)}
        = B(y,x)
\label{BetaFctn}\end{gather}
We especially need the beta function for the case where one of the two
arguments is 1/2. We have for $\gamma\in\mathbb{R}$ \citep[see also][]{vandeHulst:1950}
\begin{gather*}
  B(\frac{\gamma}{2},\frac{1}{2})
= 2\int_0^{\pi/2} \sin^{\gamma-1}\bar{\chi} \;d\bar{\chi}
        = \sqrt{\pi}\frac{\Gamma(\frac{\gamma}{2})}
                         {\Gamma(\frac{\gamma+1}{2})}
\stackrel{\text{for}\;\gamma\in\mathbb{N}_+}{=}
\begin{cases}
\frac{(\gamma-2)(\gamma-4)\dots 1}
     {(\gamma-1)(\gamma-3)\dots 2} \pi
     & \text{if}\;\gamma\;\text{odd}\\[1ex]
\frac{(\gamma-2)(\gamma-4)\dots 2}
     {(\gamma-1)(\gamma-3)\dots 3}
     & \text{if}\;\gamma\;\text{even}
 \end{cases}
\end{gather*}
The inversion of (\ref{AbelTr_power}) yields
\begin{gather*}
Q(\rho)=\frac{\gamma}{\pi} B(\frac{\gamma+1}{2},\frac{1}{2})\;\rho
\int_0^{\pi/2} (\frac{\rho}{\sin\bar{\chi}})^{-\gamma-1}\frac{d\bar{\chi}}{\sin^2\bar{\chi}}
\\
=\frac{\gamma}{\pi} B(\frac{\gamma+1}{2},\frac{1}{2})\;\rho^{-\gamma}
\int_0^{\pi/2} \sin^{\gamma-1}\bar{\chi}\,d\bar{\chi}
=\frac{\gamma}{2\pi} B(\frac{\gamma+1}{2},\frac{1}{2})
 B(\frac{\gamma}{2},\frac{1}{2})\;\rho^{-\gamma}
\end{gather*}
As a proof of consistency we have
\begin{gather*}
\frac{\gamma}{2\pi} B(\frac{\gamma+1}{2},\frac{1}{2})
                    B(\frac{\gamma}{2},\frac{1}{2})
= \frac{\gamma}{2\pi}
  \frac{\Gamma(\frac{\gamma+1}{2})\Gamma(\frac{1}{2})}
       {\Gamma(\frac{\gamma}{2}+1)}
  \frac{\Gamma(\frac{\gamma}{2})\Gamma(\frac{1}{2})}
       {\Gamma(\frac{\gamma+1}{2})}
= \frac{\gamma}{2}
  \frac{\Gamma(\frac{\gamma}{2})}
       {\Gamma(\frac{\gamma}{2}+1)}
=1
\end{gather*}
\begin{gather*}
\text{Hence we may fit $Q$ to an inverse power series}\quad
Q(\rho)=\sum_{\gamma>0} a_\gamma \rho^{-\gamma} \\
\text{and immediately have its transformation}\quad
P(r)= \sum_\gamma \frac{\gamma a_\gamma}{2\pi}
      B(\frac{\gamma+1}{2},\frac{\gamma}{2}) r^{-(\gamma+1)}
\end{gather*}
In the context of coronagraphy, $Q(\rho)$ is the brightness profile with
distance $\rho$ of the line-of-sight from the solar centre. It is observed
from an observer at a finite distance $r$ which contradicts the assumption
that the integration along the line-of-sight $\vect{c}(s)$ extends from
$-\infty$ to $\infty$. Coronagraph data are often given in terms of the
elongation $\varepsilon=\asin(\rho/R_\odot)$ instead of $\rho$ where $R_\odot$
is the solar radius. So call
$\tilde{Q}(\varepsilon)=Q(R_\odot\sin\varepsilon)$ the respective profile as
function of the elongation. This data for usually small $\varepsilon$ can be
extended to an infinite line-of-sight by adding the respective observation in
anti-Sun directions $\pi+\varepsilon$. Then
$Q(\rho)=\tilde{Q}(\varepsilon)+\tilde{Q}(\pi+\varepsilon)$ will give the data
suitable for the Abel transformation. For observations from 1 AU the part
$\tilde{Q}(\pi+\varepsilon)$ is usually negligible but for vantage points
closer to the Sun achieved by the future missions SOLAR ORBITER and SOLAR
PROBE it might matter.


\section{EM waves}
\setcounter{equation}{0}
\label{App:EMwaves}

Basically textbook wisdom rephrased here to set the convention for
some formulas which are used with different constants as in some of
the literature.
A good reference always is \citep{Jackson:1998}, for
polarisation and coherence \citep{Wolf:2007}.

\subsection{Complex wave representation}
\label{app:ComplexWave}

We denote by $\phi=\vect{\hat{k}}\tp\vect{r}-ckt$ the wave phase
and by $Z_i=Z'_i+iZ_i''$ the complex amplitude for one component $i$ of
the wave electric field. Then
\begin{gather*}
   E_i(\vect{r},t)
  =\Re[Z_i e^{i\phi}]
  =Z'_i\cos\phi-Z_i''\sin\phi
\\
  =\sqrt{{Z_i'}^2+{Z_i''}^2}(\frac{Z_i'}
  {\sqrt{{Z'_i}^2+{Z''_i}^2}}\cos\phi
  -\frac{Z_i''}{\sqrt{{Z_i'}^2+{Z_i''}^2}}\sin\phi)
\\ =\sqrt{Z_i'^2+Z_i''^2}(\cos\delta_i\cos\phi
                         -\sin\delta_i\sin\phi)
  =\sqrt{Z_i^*Z_i}\cos(\phi+\delta_i)
\end{gather*}
where $\delta_i=\atan(Z_i'/Z_i'')$. For the square of the electric field
component we then have
\begin{gather*}
   E_i^2(\vect{r},t)
  =Z_i'^2\cos^2\phi+Z_i''^2\sin^2\phi+2Z_i'Z_i''\cos\phi\sin\phi
\\
  <E_i^2(\vect{r},t)>
  =\frac{1}{2}(Z_i'^2+Z_i''^2)
  =\frac{1}{2}Z_i^*Z_i
\end{gather*}
for the average $\bra\dots\ket$ over the wave phase.
The total wave energy density is composed of the electric and
the magnetic field fluctuations in all directional components
\begin{gather*}
  w(\vect{r},t)
  = \frac{\epsilon_0}{2}\sum_i E_i^2(\vect{r},t)
  + \frac{1}{2\mu_0}\sum_i B_i^2(\vect{r},t)
  = \epsilon_0 \sum_i E_i^2(\vect{r},t),
\\
 \bra w\ket
  =\epsilon_0 \sum_i \bra E_i^2(\vect{r},t)\ket
  =\frac{\epsilon_0}{2}\sum_i Z_i^*Z_i
\end{gather*}
The energy density of the electric field fluctuations alone accounts for only half
this value.

\subsection{Polarisation}
\label{app:Polarisation}

We use the same notation as in the previous chapter and shorten the real
amplitude in component $i$ to $A_i=\sqrt{Z_i^*Z_i}$. By $\vect{\hat{e}}_1$ and
$\vect{\hat{e}}_1$ we designate an orthogonal polarisation base
normal to the wave propagation direction $\vect{\hat{k}}$.
The electric field vector is as in chapter~\ref{app:ComplexWave}
but in vector notation
\begin{gather*}
  \vect{E}(\vect{r},t)=\Re[\vect{Z}\;e^{i\phi}],\qquad
  \vect{Z}=A_1\,e^{i\delta_1}\vect{\hat{e}}_1
          +A_2\,e^{i\delta_2}\vect{\hat{e}}_2
\\
 E_i(\vect{r},t)
 =A_i\cos(\phi+\delta_i)
 =\Re(\vect{\hat{e}}\tp_i\vect{Z}\,e^{i\phi})
\end{gather*}
 where $A_1^2+A_2^2$ determines the wave energy,
       $A_1^2-A_2^2$ its ellipticity,
       $\delta_2-\delta_1$ the polarisation rotation angle and
       $\delta_2+\delta_1$ the wave phase.
Second order expressions of the electric field components are
\begin{gather*}
 E_1^2(\vect{r},t)+E_2^2(\vect{r},t)
= A_1^2 \cos^2(\phi+\delta_1)
 +A_2^2 \cos^2(\phi+\delta_2)
\\
  E_1(\vect{r},t) E_2(\vect{r},t)
= A_1 A_2
\cos(\phi+\delta_1)
\cos(\phi+\delta_2)
\\
= \half A_1 A_2
(\cos(2\phi+\delta_1+\delta_2)
+\cos(\delta_1-\delta_2))
\\
  E_1(\vect{r},t)E_2(\vect{r},t+\frac{\pi}{2ck})
=-A_1 A_2
\cos(\phi+\delta_1)
\sin(\phi+\delta_2)
\\
=-\,\half A_1 A_2
(\sin(2\phi+\delta_1+\delta_2)
-\sin(\delta_1-\delta_2))
\end{gather*}
Averaging over the wave phase gives the Stokes components
\begin{gather*}
S_I=\bra E_1^2(\vect{r},t)\ket + \bra E_2^2(\vect{r},t)\ket
 =\frac{A^2_1+A^2_2}{2}
\\   
 =\half(|\vect{\hat{e}}_1\tp\vect{Z}|^2
       +|\vect{\hat{e}}_2\tp\vect{Z}|^2)
\\ 
S_Q=\bra E_1^2(\vect{r},t)\ket-\bra E_2^2(\vect{r},t)\ket
  =\frac{A^2_1-A^2_2}{2}
\\
   =\half(|\vect{\hat{e}}_1\tp\vect{Z}|^2
         -|\vect{\hat{e}}_2\tp\vect{Z}|^2)
\\ 
\half S_U=\bra E_1(\vect{r},t) E_2(\vect{r},t)\ket
   =\frac{A_1 A_2}{2}\cos(\delta_1-\delta_2)
   =\half\Re[(\vect{\hat{e}}_1\tp\vect{Z})
             (\vect{\hat{e}}_2\tp\vect{Z}^*)]
\\ 
\half S_V=\bra E_1(\vect{r},t) E_2(\vect{r},t+\frac{\pi}{2ck})\ket
   =\frac{A_1 A_2}{2}\sin(\delta_1-\delta_2)
   =\half\Im[(\vect{\hat{e}}_1\tp\vect{Z})
             (\vect{\hat{e}}_2\tp\vect{Z}^*)]
\end{gather*}
All of these correlations are contained in the electric field correlation
matrix
\[
  \vect{R}(\Delta\vect{r},dt)
   =\bra\vect{E}(\vect{r},t)\vect{E}\tp(\vect{r}+\Delta\vect{r},t)\ket
  +i\bra\vect{E}(\vect{r},t)\vect{E}\tp(\vect{r}+\Delta\vect{r},t+dt)\ket
\]
To retrieve the Stokes parameters we need, according to the above relations,
just the correlation at $\Delta\vect{r}=0$ with $dt=0$ and $dt=\pi/2ck$. With
these arguments the above matrix yields the hermitian 3$\times$3
correlation $\vect{R}(0,0)+i\vect{R}(0,dt)$.
Note that $R_{12}(0,dt)=R_{21}(0,-dt)=-R_{21}(0,dt)$.
For waves propagating exclusively into direction $\vect{\hat{k}}$, the only
non-zero wave components are $E_1$ and $E_2$ and $\vect{\hat{e}}_1$ and
$\vect{\hat{e}}_2$ are orthogonal directions which span the polarisation
plane normal to $\vect{\hat{k}}$.
Then $\vect{R}\vect{\hat{k}}=0$ and the relevant 2$\times$2 submatrix in the
coordinates along $\vect{\hat{e}}_1$ and $\vect{\hat{e}}_2$ is the
coherency matrix \vect{J} \citep{Wolf:2007}. With the above relations 
\begin{gather}
  \vect{J}=
  \begin{pmatrix}
    R_{11}(0,0) &
    R_{12}(0,0)+iR_{12}(0,dt) \\ 
    R_{12}(0,0)-iR_{12}(0,dt) &
    R_{22}(0,0) \\
  \end{pmatrix}
\nonumber\\ 
= \frac{1}{2}\begin{pmatrix}
    A^2_1 &
    A_1A_2\cos(\delta_1-\delta_2) \\
    A_1A_2\cos(\delta_1-\delta_2) &
    A^2_2
  \end{pmatrix}
+\frac{i}{2}\begin{pmatrix}
    0 &
    A_1A_2\sin(\delta_1-\delta_2) \\
   -A_1A_2\sin(\delta_1-\delta_2) &
    0
  \end{pmatrix}
\nonumber\\ 
= \frac{1}{2}\begin{pmatrix}
    S_I+S_Q &
    S_U \\
    S_U &
    S_I-S_Q
  \end{pmatrix}
+ \frac{i}{2}\begin{pmatrix}
    0 &
    S_V \\
   -S_V  &
    0
  \end{pmatrix}
\label{StokesJ}\end{gather}

Each of the four polarisation parameters in (\ref{StokesJ}) can be associated
with one of the four Pauli spin matrices. The advantage of the matrix
formulation of the polarisation states is that we can easily transform them to
a new coordinate system. Consider, e.g., a rotated polarisation base
\[
 \rvecc{\vect{\hat{e}'}_1}{\vect{\hat{e}'}_2}
=\begin{pmatrix} c & -s\\ s & c \end{pmatrix}
 \rvecc{\vect{\hat{e}}_1}{\vect{\hat{e}}_2}
\]
where $c$ and $s$ are short-hand for $\cos\alpha$ and $\sin\alpha$,
respectively, for
the rotation angle $\alpha$ from $\vect{\hat{e}}_i$ to the
$\vect{\hat{e}'}_i$. A positive angle $\alpha$ rotates
$\vect{\hat{e}}_i$ to $\vect{\hat{e}'}_i$ in an anticlockwise sense.
Then the matrix elements (\ref{StokesJ}) in the rotated frame are
$\vect{\hat{e}'}_i{}\tp\vect{J}\vect{\hat{e}'}_j=J'_{ij}$ are
\begin{gather*}
\vect{J}'
=\begin{pmatrix} c & s\\ -s & c \end{pmatrix} \vect{J}
 \begin{pmatrix} c & -s\\ s & c \end{pmatrix}
\\
=\begin{pmatrix} c & s\\ -s & c \end{pmatrix} 
 \begin{pmatrix} J_{11}c+J_{12}s & -J_{11}s+J_{12}c \\
                 J_{21}c+J_{22}s & -J_{21}s+J_{22}c
  \end{pmatrix}
\\
=\begin{pmatrix}
 J_{11}c^2+J_{22}s^2+(J_{12}+J_{21})cs &
-(J_{11}-J_{22})cs+J_{12}c^2-J_{21}s^2 \\
-(J_{11}-J_{22})cs+J_{21}c^2-J_{12}s^2 &
J_{11}s^2+J_{22}c^2-(J_{12}+J_{21})cs &
\end{pmatrix}
\\
=\frac{1}{2}\begin{pmatrix}
S_I+S_Q(c^2-s^2) +2S_Ucs &
-2S_Qcs + S_U(c^2-s^2)+iS_V \\
-2S_Qcs + S_U(c^2-s^2)-iS_V &
S_I-S_Q(c^2-s^2)-2S_Ucs &
\end{pmatrix}
\\
= \frac{1}{2}\begin{pmatrix}
    S_I+S_Q' &
    S_U'+iS_V \\
    S_U'-iS_V &
    S_I-S_Q'
  \end{pmatrix}
\qquad\text{where we introduced}
\\
\rvecc{S_Q'}{S_U'}
=\rvecc{S_Q(c^2-s^2)+2S_Ucs}{S_U(c^2-s^2)-2S_Qcs}
=\rvecc{S_Q\cos 2\alpha +S_U\sin 2\alpha}
       {S_U\cos 2\alpha -S_Q\sin 2\alpha}
=\begin{pmatrix}
       \cos 2\alpha & \sin 2\alpha\\
      -\sin 2\alpha & \cos 2\alpha
  \end{pmatrix}
\rvecc{S_Q}{S_U}
\end{gather*}
Rotating the coordinate system about the $\vect{\hat{k}}$ axis by angle
$\alpha$ rotates the two linear polarisation parameters reversely by
$2\alpha$.
We may choose a special rotation angle $\alpha=\alpha^*$ such that the above
rotation eliminates the $S_U'$ component. This angle is obviously
\begin{equation}
  \tan 2\alpha^*=\frac{S_U}{S_Q}
\label{PolRotation}\end{equation}
The polarisation ellipse of the wave therefore has its major axis
in the direction of $\pm\vect{\hat{e}'}_1$ which is anticlockwise
rotated by angle $\alpha^*$ from $\pm\vect{\hat{e}}_1$.
Properties of the coherency matrix $\vect{J}$ which are
invariant with respect to a rotation by angle $\alpha$ express general
features of the wave polarisation. They are
\begin{align}
  \half\mathrm{trace}(\vect{J})
 &=S_I
  && \quad\text{beam irradiance}/c\epsilon_0
  \nonumber\\
  \mathrm{det}(\vect{J})
 &=S_I^2-S_Q^2-(S_U+iS_V)(S_U-iS_V)\nonumber\\
 &=S_I^2-(S_Q^2+S_U^2+S_V^2)
  &&\label{PolInvariants}\\
  P
 &=\sqrt{1-\frac{4\,\mathrm{det}(\vect{J})}{\mathrm{trace}(\vect{J})^2}}
  =\frac{\sqrt{S_U^2+S_Q^2+S_V^2}}{S_I}
  &&\quad\text{degree of polarisation}
\nonumber\end{align}
The power of the polarised and unpolarised components are
$S_IP$ and $S_I(1-P)$, respectively.
Since it is unnecessary for our purposes, we restrict ourselves to
equal-time correlations which excludes the circular polarisation $S_V$
and $\vect{J}$ will be real symmetric rather than hermitian.
 
Reversely, the Stokes parameters can be retrieved also from a set of fixed
polariser orientations. We assume measurements of
$\vect{\hat{p}}\tp\vect{J}\vect{\hat{p}}$ were made as in many coronagraphs
for three polariser positions $\vect{\hat{p}}_0=(1,0)$,
$\vect{\hat{p}}_{60}=(1/2,\sqrt{3}/2)$ and
$\vect{\hat{p}}_{120}=(-1/2,\sqrt{3}/2)$. Then
\begin{gather*}
  \vect{\hat{p}}_0\tp\vect{J}\vect{\hat{p}}_0 =S_I+S_Q
\\  
  \vect{\hat{p}}_{60}\tp\vect{J}\vect{\hat{p}}_{60}
  =\frac{1}{4}(S_I+S_Q)+\frac{\sqrt{3}}{2}S_U+\frac{3}{4}(S_I-S_Q)
\\
  \vect{\hat{p}}_{120}\tp\vect{J}\vect{\hat{p}}_{120}
  =\frac{1}{4}(S_I+S_Q)-\frac{\sqrt{3}}{2}S_U+\frac{3}{4}(S_I-S_Q)
\\  
 \begin{pmatrix}
   1 &   1  & 0          \\
   1 & -1/2 &  \sqrt{3}/2 \\
   1 & -1/2 & -\sqrt{3}/2 
 \end{pmatrix}
 \rveccc{S_I}{S_Q}{S_U}
=\rveccc{\vect{\hat{p}}_0\tp\vect{J}\vect{\hat{p}}_0}
        {\vect{\hat{p}}_{60}\tp\vect{J}\vect{\hat{p}}_{50}}
        {\vect{\hat{p}}_{120}\tp\vect{J}\vect{\hat{p}}_{120}}
\qquad\text{or}\\[1ex]
 3S_I =\vect{\hat{p}}_{0}\tp\vect{J}\vect{\hat{p}}_0
      +\vect{\hat{p}}_{60}\tp\vect{J} \vect{\hat{p}}_{60}
      +\vect{\hat{p}}_{120}\tp\vect{J}\vect{\hat{p}}_{120}\\
 3S_Q =2\vect{\hat{p}}_{0}\tp\vect{J}\vect{\hat{p}}_0
      -\vect{\hat{p}}_{60}\tp\vect{J} \vect{\hat{p}}_{60}
      -\vect{\hat{p}}_{120}\tp\vect{J}\vect{\hat{p}}_{120}\\
 \sqrt{3}S_U =\vect{\hat{p}}_{60}\tp \vect{J}\vect{\hat{p}}_{60}
             -\vect{\hat{p}}_{120}\tp\vect{J}\vect{\hat{p}}_{120}
\end{gather*}
The angle of the major polarisation axis relative to the orientation of
$\vect{\hat{p}}_{0}$ is then given by (\ref{PolRotation})
\citep[p. 97]{Billings:1966}
\begin{equation}
  \tan{2\alpha^*}=\frac{S_U}{S_Q}
  =\frac{\sqrt{3}( \vect{\hat{p}}_{60}\tp\vect{J}\vect{\hat{p}}_{60}
                 -\vect{\hat{p}}_{120}\tp\vect{J}\vect{\hat{p}}_{120})}
    {2\vect{\hat{p}}_{0}\tp\vect{J}\vect{\hat{p}}_0
      -\vect{\hat{p}}_{60}\tp\vect{J} \vect{\hat{p}}_{60}
      -\vect{\hat{p}}_{120}\tp\vect{J}\vect{\hat{p}}_{120}}
\label{PolAngle}\end{equation}


\subsection{Wiener-Kintchine}
\label{App:Wiener-Kintchine}

We follow \cite{Papoulis:1981} except that we consider the three-dimensional
spatially random field components instead of a random time series.
The Fourier transform of a random wave field taken over a limited
space volume $V(\vect{r})$ centred at $\vect{r}$ is
\begin{gather*}
 \vect{\tilde{E}}_{V(\vect{r})}(\vect{k})
 =e^{ickt} \int_{V(\vect{r})} \vect{E}(\vect{r}',t)
           \;e^{-i\vect{k}\tp\vect{r}'}\;d^3\vect{r'}
\\
\vect{E}(\vect{r}',t)
=\int e^{-ickt}\vect{\tilde{E}}_{V(\vect{r}')}(\vect{k})
\;e^{i\vect{k}\tp\vect{r}'} \frac{d^3\vect{k}}{(2\pi)^3}
\quad\text{for}\;\vect{r}'\in V(\vect{r})
\end{gather*}
The factor $e^{ickt}$ is meant to eliminate the fast oscillatory time
dependence in the spatial Fourier transform of $\vect{E}(\vect{r},t)$. 
The power spectrum of this truncated process is \citep{Papoulis:1981}
\[
 P_V(\vect{k})
=\frac{1}{V}\;\vect{\tilde{E}}_V\!\tp(\vect{k})\vect{\tilde{E}}_V^*(\vect{k})
\]
Due to the incoherence of the field at larger distances, the power of
$\vect{\tilde{E}}_V\!\tp(\vect{k})\vect{\tilde{E}}_V^*(\vect{k})$
does on average not increase with $V^2$ as the integration volumes involved but
only with $V$.
For the stochastic expectation value of this expression the exact size and
shape of $V(\vect{r})$ does not matter as long it is large enough. We assume
that $V$ is a cube of edge length $2L$.
Additionally, the correlation $\vect{E}\tp(\vect{r},t)\vect{E}(\vect{r'},t)$ is
statistically homogeneous and depends only on the distance vector
$\vect{r}-\vect{r'}$, i.e., its expectation value assumes the form
$\bra\vect{E}\tp(\vect{r},t)\vect{E}(\vect{r'},t)\ket=R(\vect{r}-\vect{r'})$.
We then have
\begin{gather*}
 \bra P_V\ket\!(\vect{k})
=\frac{1}{V}
\int_V'\int_V \vect{E}\tp(\vect{r},t) \;e^{-i\vect{k}\tp\vect{r}}
             \vect{E}(\vect{r'},t) \;e^{i\vect{k}\tp\vect{r'}}
             d^3\vect{r}d^3\vect{r'}
\\
=\frac{1}{(2L)^3}
 \int_V'\int_V \bra\vect{E}\tp(\vect{r},t)\vect{E}(\vect{r'},t)\ket
             \;e^{-i\vect{k}\tp(\vect{r}-\vect{r'})}
             d^3\vect{r}d^3\vect{r'}
\\
=\big[\prod_{i=1,3} \frac{1}{2L}
  \int_{-L}^L\int_{-L}^L dr_i\,dr_i'\;e^{-ik_i(r_i-r_i')}\,\big]
  R(\vect{r}-\vect{r'})
\\
=\big[\prod_{i=1,3} 
       \int_{-2L}^0 \frac{d(r_i+r'_i)}{4L}
       \int_{-2L-(r_i+r'_i)}^{2L+(r_i+r'_i)}d(r_i-r'_i)\;
       e^{-ik_i(r_i-r'_i)} \big] R(\vect{r}-\vect{r'})
 \\+\big[\prod_{i=1,3} \int_0^{2L} \frac{d(r_i+r'_i)}{4L}
    \int_{-2L+(r_i+r'_i)}^{2L-(r_i+r'_i)} d(r_i-r'_i)\;
    e^{-ik_i(r_i-r'_i)}\,\big] R(\vect{r}-\vect{r'})
\\
 \bra w\ket\!(\vect{k})
=\lim_{V\rightarrow\infty}\bra P_V\ket\!(\vect{k})
\\
=\big[\prod_{i=1,3} 
       \int_{-2L}^0 \frac{d(r_i+r'_i)}{4L}
       \int_{-\infty}^{\infty}d(r_i-r'_i)\;
       e^{-ik_i(r_i-r'_i)} \big] R(\vect{r}-\vect{r'})
 \\+\big[\prod_{i=1,3} \int_0^{2L} \frac{d(r_i+r'_i)}{4L}
       \int_{-\infty}^{\infty}d(r_i-r'_i)\;
    e^{-ik_i(r_i-r'_i)}\,\big] R(\vect{r}-\vect{r'})
\\
=\big[\prod_{i=1,3}\int_{-\infty}^{\infty}
     d(r_i-r'_i)\;\;e^{-ik_i(r_i-r'_i)}\,\big]
             R(\vect{r}-\vect{r'})
\\
   =\int_{-\infty}^{\infty}d(\vect{r}-\vect{r'})\;
             R(\vect{r}-\vect{r'})
             \;e^{-i\vect{k}\tp(\vect{r}-\vect{r'})}
\end{gather*}
A similar procedure applies if we replace the electric field correlation
$R(\vect{r}-\vect{r'})$ by the respective correlation matrix
\[
  \vect{R}(\vect{r}-\vect{r'})
   =\bra\vect{E}(\vect{r},t)\vect{E}\tp(\vect{r'},t)\ket
\]
which is related to the scalar correlation by
$R(\vect{r}-\vect{r'})=\mathrm{trace}(\vect{R}(\vect{r}-\vect{r'}))$.

\section{Some topics in special relativity}
\setcounter{equation}{0}
\label{App:SpecRel}

Just the essentials needed in section~\ref{Sec:ElecMotion}. The more complete
basics can be found in \citep{Jackson:1998,French:1968,Woodhouse:2003},
thorough discussions on special relativity in \citep{Born:2001}.

\subsection{Velocity addition and aberration}
\label{app:VelAberr}

\begin{figure}
\hspace*{\fill}
\parbox{7cm}{\includegraphics[width=7cm]{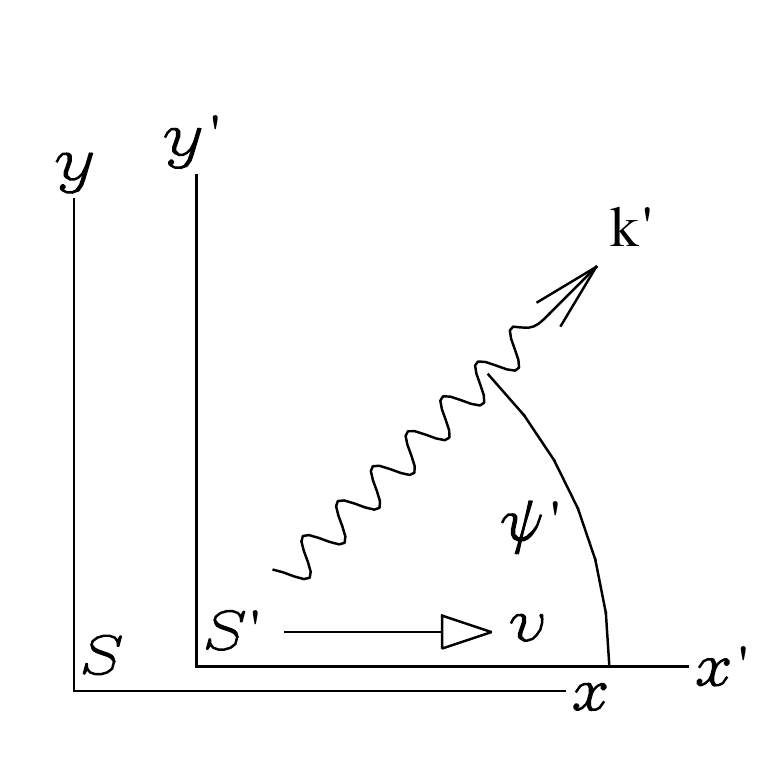}}
\hspace*{\fill}
\parbox{9cm}{\caption{Illustration of the coordinate systems used.
    The photon is observed in frame $S'$ propagating in direction
    $\vect{\hat{k}}'$ with the angle $\psi'$. The corresponding
    direction in reference frame $S$ is $\vect{\hat{k}}$ with the angle $\psi$
    with respect to the $\uect{x}$ axis. The relation between the angles 
    is given in (\ref{Aberr_c_}),(\ref{Aberr_s_}), (\ref{Aberr_c}) and
    (\ref{Aberr_s}).
    \label{Fig:SpecRelSys}}}
\hspace*{\fill}
\end{figure}
In the tradition of most textbooks we define two coordinate systems: orthogonal
coordinates $(x,y,z,t)$ in frame $S$ and coordinates $(x',y',z',t')$ in frame
$S'$. Similarly named axes are parallel but the origin of frame $S'$ moves 
with $\vect{v}=v\vect{\hat{x}}$ as seen in $S$.
Often $S$ is called the lab frame and $S'$ the (co)moving or rest
frame if it is attached to a particle. The selection seems arbitrary but
becomes unique by the definition of the relative velocity $\vect{v}$ between
the frames: $S'$ moves with $\vect{v}$ in $S$ but $S$ moves with $-\vect{v}$
in $S'$.
The Galilean transformations from the lab frame $S$ to the (co)moving
frame $S'$ and reverse read
\begin{gather}
  x'=x-vt,\quad y'=y,\quad z'=z,\quad t'=t
  \quad\text{or}\quad
  \vect{r}'=\vect{r}-\vect{v}t\;\qquad
\label{GalTrans_}\\
  x=x'+vt',\quad y=y',\quad z=z',\quad t=t'
  \quad\text{or}\quad
  \vect{r}=\vect{r}'+\vect{v}t
\label{GalTrans}
\nonumber  
\end{gather}
The Lorentz transformation between the two reference frames differs
from the Galilean transformation in the spatial
coordinates only by the Lorentz factor $\gamma=1/\sqrt{1-\beta^2}$ and a new
transformation for the time which causes simultaneity to depend on space.
\begin{gather}
  x'=\gamma\,(x-\beta ct),\quad y'=y,\quad z'=z,\quad ct'=\gamma\,(ct-\beta x)
\label{LorTrans_}
\\
  \text{or}\quad
  \vect{r}'=\gamma\vectg{\hat{\beta}}\uectg{\beta}{}\tp(\vect{r}
                         -\vectg{\beta}ct)
          +(\1-\vectg{\hat{\beta}}\uectg{\beta}{}\tp)\vect{r}
 =\vect{r}-\gamma\vectg{\beta}t
         - (1-\gamma)\vectg{\hat{\beta}}\uectg{\beta}{}\tp\vect{r}
\nonumber  
\\[0.5ex]
  x=\gamma\,(x'+\beta ct'),\quad y=y',\quad z=z',\quad ct=\gamma\,(ct'+\beta x')
\label{LorTrans}
\\
  \text{or}\quad
  \vect{r}=\gamma\vectg{\hat{\beta}}\uectg{\beta}{}\tp(\vect{r}'
                        +\vectg{\beta}ct')
            +(\1-\vectg{\hat{\beta}}\uectg{\beta}{}\tp)\vect{r}'
  =\vect{r}'+\gamma\vectg{\beta}ct'
            - (1-\gamma)\vectg{\hat{\beta}}\uectg{\beta}{}\tp\vect{r}'
\nonumber  
\end{gather}
Assume that in $S'$ a moving object is observed with velocity $\vect{u}'$ such
that
\[
  x'=u'_x t',\quad y'=u'_y t'
\]
In the Galilean framework, the object would move in $S$ with
$\vect{u}=\vect{u}'+v\vect{\hat{x}}$. Insertion of (\ref{LorTrans_}) instead
of (\ref{GalTrans_}) for the dashed coordinates yields the rules
for adding velocities relativistically.
\begin{gather}
  x'=u'_x t'
\quad\xrightarrow{S'\rightarrow S}\quad
  \gamma\,(x-\beta ct)=\frac{u'_x}{c} \gamma\,(ct-\beta x)
\nonumber\\\quad\text{reorder to}\quad 
  (1+ \beta \frac{u'_x}{c} ) x = (u'_x + \beta c) t 
\quad\text{or}\quad
  u_x=\frac{u'_x + v}{1+ \beta u'_x/c}
\label{Vadd_ux}\\  
  y'=u'_y t'
\quad\xrightarrow{S'\rightarrow S}\quad
  y=\frac{u'_y}{c} \gamma\,(ct-\beta x)
   =\frac{u'_y}{c} \gamma\,(ct-\beta u_x t)
   =\frac{u'_y}{c} \gamma\,(c-\frac{\beta (u'_x + v)}{1+ \beta u'_x/c}) t
\nonumber\\
   =\frac{u'_y}{c} \gamma\,\frac{c+ \beta u'_x
            - \beta (u'_x + v)}{1+ \beta u'_x/c} t
   = u'_y \gamma\,\frac{1-\beta^2}{1+ \beta u'_x/c} t
\nonumber\\
   = \frac{u'_y/\gamma}{1+ \beta u'_x/c} t
\quad\text{or}\quad
  u_y=\frac{u'_y/\gamma}{1+ \beta u'_x/c}
\label{Vadd_uy}\end{gather}
and similarly for $u_z$. The inverse transformation is
\begin{gather}
\text{from(\ref{Vadd_ux})}\quad
  u'_x+v
= u_x(1+ \beta u'_x/c)
= u_x+ \beta u_x u'_x/c
\quad\text{follows}\nonumber\\
 u_x-v
=u'_x- \beta u_x u'_x/c
=u'_x(1- \beta u_x/c)
\quad\text{or}\nonumber\\
u'_x=\frac{u_x-v}{1- \beta u_x/c}
\label{Vadd_ux_}\\
\text{from(\ref{Vadd_uy}) and (\ref{Vadd_ux_})}\quad
 u'_y/\gamma
= u_y(1+ \beta u'_x/c)
= u_y(1+ \beta\frac{(u_x-v)/c}{1- \beta u_x/c})
\nonumber\\
= u_y\frac{1- \beta u_x/c + \beta u_x/c-\beta v/c}{1- \beta u_x/c}
= u_y\frac{1-\beta^2}{1- \beta u_x/c}
\quad\text{or}\nonumber\\
u'_y=\frac{u_y/\gamma}{1- \beta u_x/c}
\label{Vadd_uy_}\end{gather}
The coordinate-free version (\ref{Vadd_ux_}) and (\ref{Vadd_uy_}) for a
transformation $S\rightarrow S'$ reads
\[
\vect{u}'=\frac{1}{1-\vectg{\beta}\tp\vect{u}/c}
  [(\1-\vectg{\hat{\beta}}\uectg{\beta}{}\tp) (\vect{u}-\vect{v})
  +\frac{1}{\gamma}\vectg{\hat{\beta}}\uectg{\beta}{}\tp\vect{u}]
\]

\begin{figure}
\hspace*{\fill}
\parbox{7cm}{\includegraphics[width=8cm]{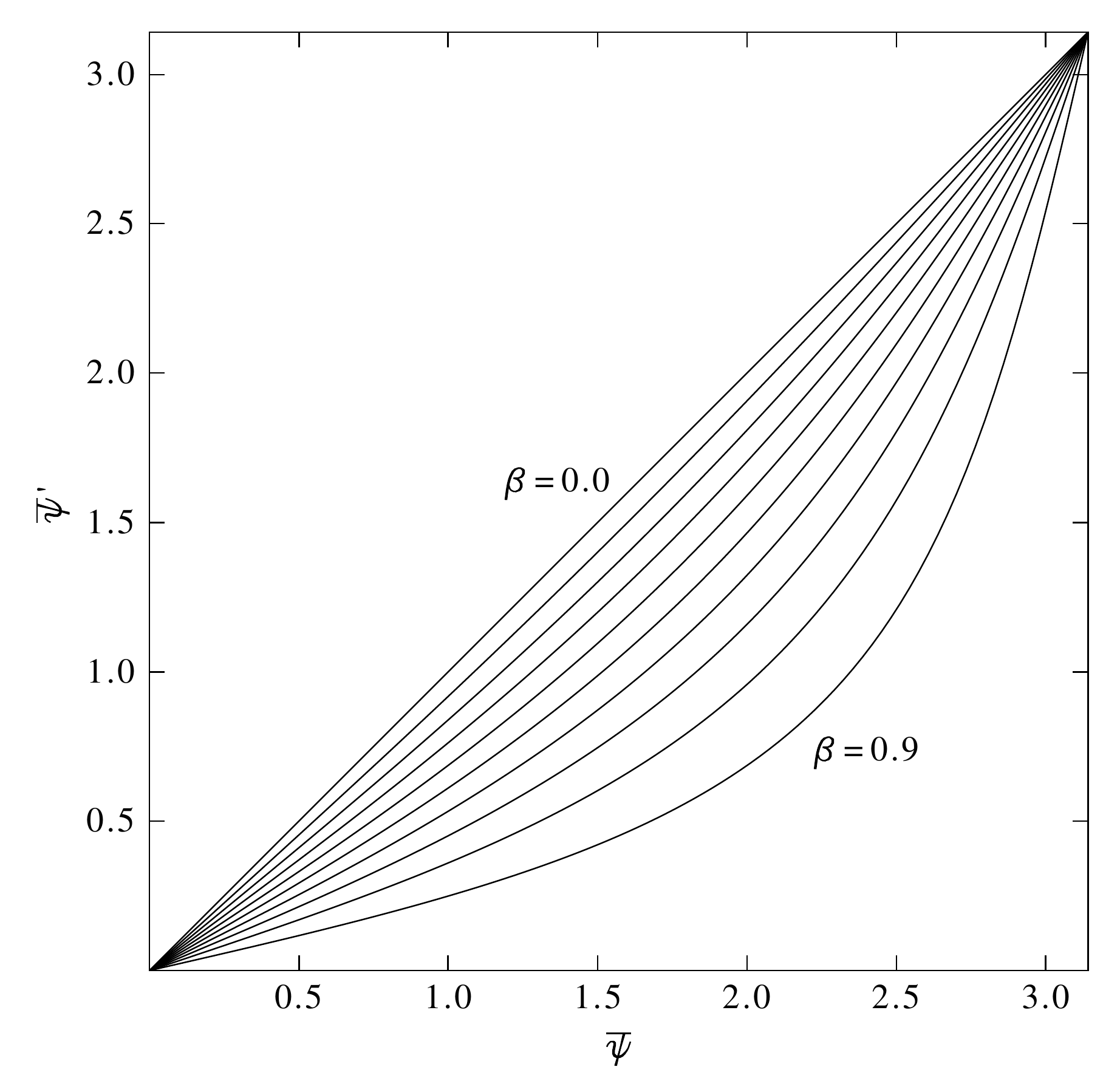}}
\hspace*{\fill}
\parbox{9cm}{\caption{Graphical representation of (\ref{Aberr_bc_}) to
    demonstrate the headlight effect: In the moving frame $S'$, the
    propagation angle $\bar{\psi}'=\pi-\psi'$ of an incoming photon is always
    smaller than the corresponding angle $\bar{\psi}=\pi-\psi$ in the frame
    $S$.
    \label{Fig:Aberration}}}
\hspace*{\fill}
\end{figure}

If the observed object is a photon we call the plane spanned by the velocity
$\vect{v}$ between $S$ and $S'$ and the propagation direction of the photon
the aberration plane. Without loss of generality, we can confine it
to the $\vect{\hat{x}}$, $\vect{\hat{y}}$ plane.
\footnote{The axes $\vect{\hat{x}}$ and $\vect{\hat{y}}$ differ from the
  Cartesian system used in the main text to integrate the solar irradiance as,
  e.g., in Fig.~\ref{Fig:ScaGeom3D}. We therefore will rename the axes
  $\vect{\hat{x}}$, $\vect{\hat{y}}$ and $\vect{\hat{z}}$ of this chapter
  defining the aberration plane to $\vectg{\hat{\beta}}$, $\vectg{\hat{\mu}}$
  and $\vectg{\hat{\nu}}$ in the main text.} With angle $\psi'$ in frame $S'$
between the photon velocity $\vect{u}'$ and $\vect{v}=v\vect{\hat{x}}$ in
system $S'$ we have $\vect{u}'=c(\cos\psi',\sin\psi')$ and (\ref{Vadd_ux},
\ref{Vadd_uy}) become
\[
  u_x=c\;\frac{\cos\psi' + \beta}{1+ \beta \cos\psi'},\quad
  u_y=c\;\frac{\sin\psi'/\gamma}{1+ \beta \cos\psi'}
\]
The velocity magnitude in system $S$ is, as expected
\begin{gather}
  \frac{u_x^2+u_y^2}{c^2}
=\frac{(\cos\psi' + \beta)^2+\sin^2\psi'/\gamma^2}{(1+ \beta \cos\psi')^2}
=\frac{\cos^2\psi' + 2\beta\cos\psi'+\beta^2 + \sin^2\psi'(1-\beta^2)}
      {(1+ \beta \cos\psi')^2}
\nonumber\\
=\frac{1 + 2\beta\cos\psi'+\beta^2(1-\sin^2\psi')}{(1+ \beta \cos\psi')^2}
=\frac{1 + 2\beta\cos\psi'+\beta^2\cos^2\psi'}{(1+ \beta \cos\psi')^2}
=1
\label{Aberr_norm}\end{gather}
and the angles at which the photon is seen in $S$ are
\begin{gather}
  \cos\psi=\frac{u_x}{c}=\frac{\cos\psi' + \beta}{1+ \beta \cos\psi'}
\label{Aberr_c_}\\
  \sin\psi=\frac{u_y}{c}=\frac{\sin\psi'/\gamma}{1+ \beta \cos\psi'}
\label{Aberr_s_}\end{gather}
This yields the aberration transformation of the angle of photon
propagation in the aberration plane. The inversion of
(\ref{Aberr_c_}) and (\ref{Aberr_s_}) is
\begin{gather}
   \cos\psi' + \beta
  =\cos\psi(1+ \beta \cos\psi')
\quad\text{reorder to}\quad
   \cos\psi' - \beta \cos\psi\cos\psi'
  =\cos\psi - \beta
\nonumber\\
  \text{or}\quad
   \cos\psi'=\frac{\cos\psi - \beta}{1 - \beta \cos\psi}
\label{Aberr_c}\\   
  \sin\psi'
 =\gamma\sin\psi(1+ \beta \cos\psi')
 =\gamma\sin\psi(1+ \beta \frac{\cos\psi - \beta}{1 - \beta \cos\psi})
\nonumber\\
 =\gamma\sin\psi \;\frac{1 - \beta \cos\psi + \beta(\cos\psi - \beta)}{1
   - \beta \cos\psi}
 =\gamma\sin\psi \;\frac{1 - \beta^2}{1 - \beta \cos\psi}
\nonumber\\
\text{or}\quad
  \sin\psi'
 =\frac{\sin\psi}{\gamma(1 - \beta \cos\psi)}
\label{Aberr_s}
\end{gather}
\begin{figure}
\hspace*{\fill}
  \includegraphics[width=5cm]{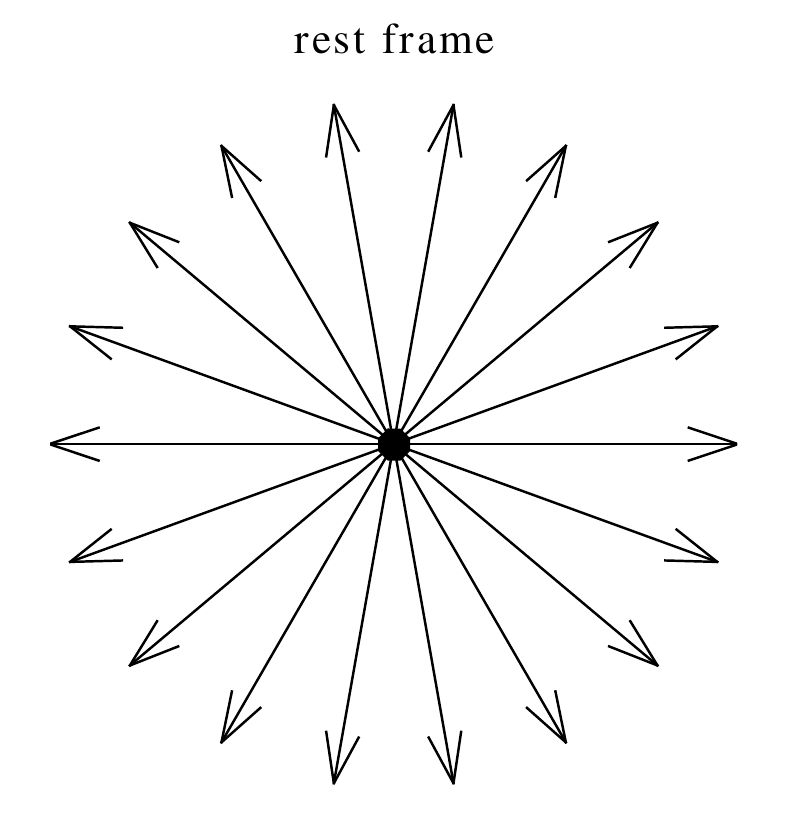}
\hspace*{3ex}
  \includegraphics[width=5cm]{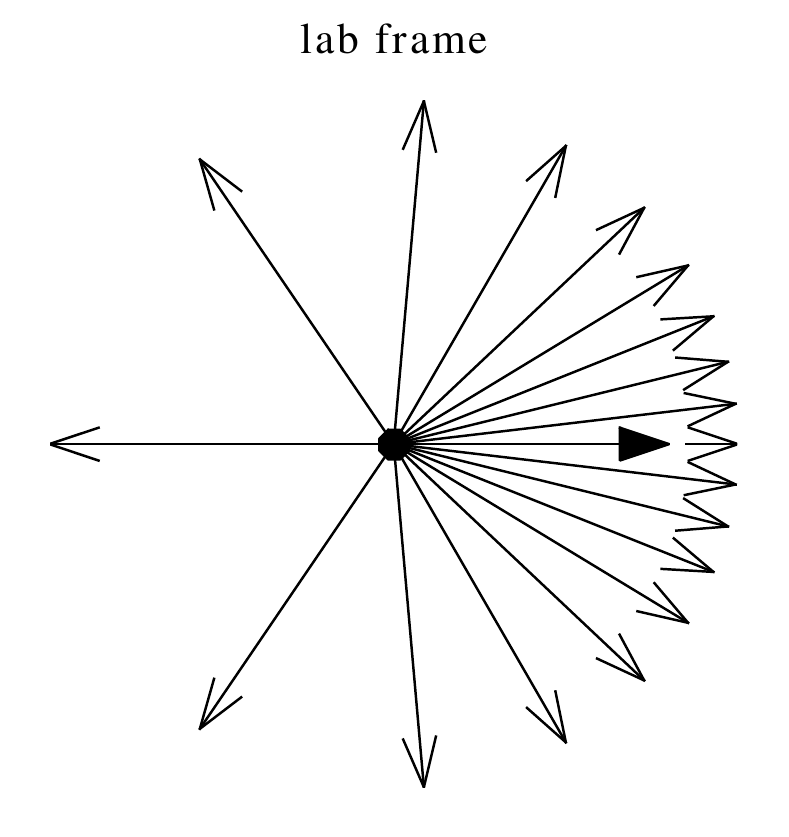}
\hspace*{\fill}\\
\hspace*{\fill}
  \includegraphics[width=5cm]{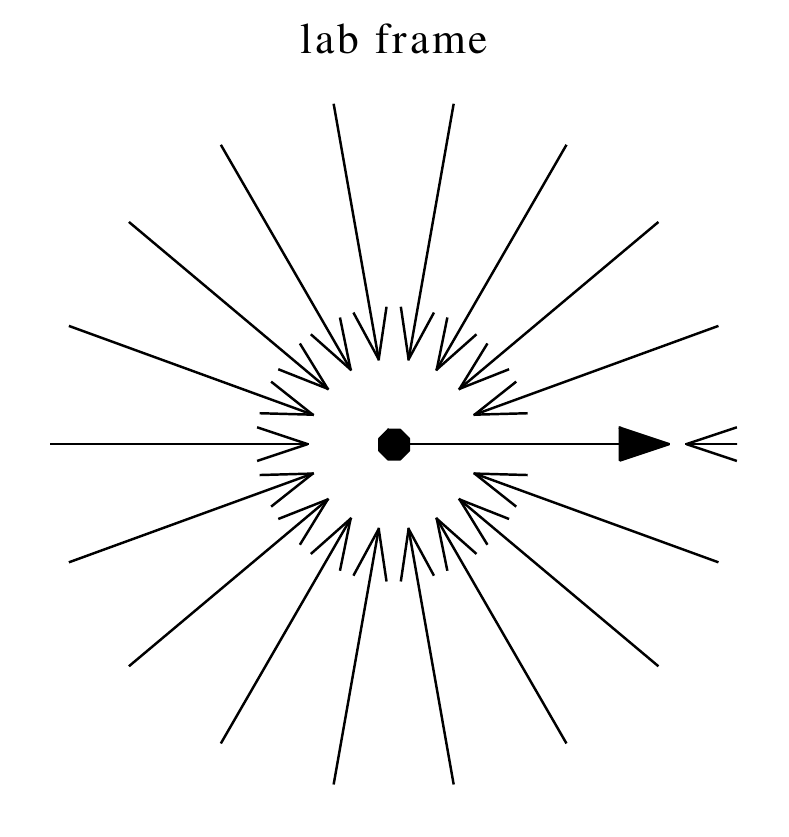}
\hspace*{3ex}
  \includegraphics[width=5cm]{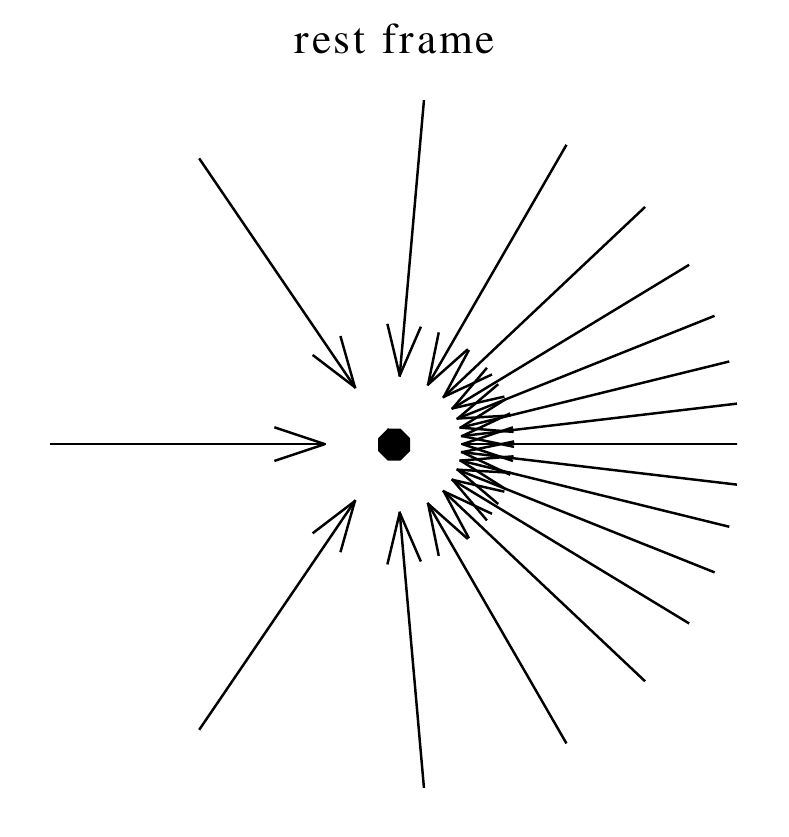}
\hspace*{\fill}
\caption{Illustration of the headlight effect. In the top row we assume a
  particle radiating isotropically in its own rest frame (top left). In the
  lab frame the particle is seen to move to the right and its radiation is
  beamed in forward direction (top right). A reversal of $\vect{k}$ and $\vect{v}$
  yields an equivalent forward concentration. In the bottom row we assume the
  particle is isotropically illuminated in the lab frame (bottom left).
  In the particle frame, the radiation is seen to come preferentially from
  the forward direction. \label{Fig:Headlight}}
\hspace*{\fill}
\end{figure}
We recall that $\psi$ and $\psi'$ are the respective angles between
$v\vect{\hat{x}}$ and the propagation direction of the photon (see
Fig.~\ref{Fig:SpecRelSys}). For observers it is more convenient to use the
angles $\bar{\psi}=\pi-\psi$ and $\bar{\psi}'=\pi-\psi'$ of the opposite
direction with $v\vect{\hat{x}}$, i.e., the direction in which incoming
photons are observed in $S$ and $S'$, respectively. The respective aberration
formulas are obtained from (\ref{Aberr_s_}), (\ref{Aberr_c}) and
(\ref{Aberr_s}) by replacing $\psi\rightarrow\bar{\psi}$,
$\psi'\rightarrow\bar{\psi}'$ and either $\cos,\sin \rightarrow -\cos,-\sin$
or equivalently $\beta\rightarrow -\beta$. Hence a reversal of the photon
propagation and of $\vect{v}$ does not change the angles.
Thus (\ref{Aberr_c_}) and (\ref{Aberr_c}) yield for incoming photons
a transformation between the cosines of the observation angles
\begin{equation}
  \cos\bar{\psi}=\frac{\cos\bar{\psi}' - \beta}{1 - \beta\cos\bar{\psi}'},
  \qquad
  \cos\bar{\psi}'=\frac{\cos\bar{\psi} + \beta}{1 + \beta\cos\bar{\psi}}
\label{Aberr_bc_}\end{equation}
This relation is shown in Fig.~\ref{Fig:Aberration}: we always have
$\bar{\psi}'\ge\bar{\psi}$. As a consequence, in the moving system $S'$, the
observed photons arrive from a more forward direction. This effect is
sometimes termed headlight or searchlight effect and is illustrated in
Fig.~\ref{Fig:Headlight}. When seen from $S'$, the moving system is $S$ has
velocity $-\vect{v}$ and aberration to a ``more forward'' direction in $S$
turns the photon direction back again.

In the literature you find aberration formulas with both sign conventions,
(\ref{Aberr_bc_}), (\ref{Aberr_c}) and (\ref{Aberr_c_}). On has to carefully
check how the relative system velocity $\vect{v}$ is defined and whether the
angle $\psi$ or $\bar{\psi}$ is implied. We consider (\ref{Aberr_c}) as the
``generic'' version because all variables on the right-hand-side are as
observed in the same system $S$ including $\beta$. The only left-hand-side
variable, angle $\psi'$, is observed in $S'$ In (\ref{Aberr_c_}) the
right-hand-side includes the angle $\psi'$ but also $\beta$,
the relative system velocity as observed in $S$. In order to cast it in a
consistent generic form, the sign of $\beta$ has to be reversed.

\subsection{Frequency shift}
\label{app:FreqShift}

The wave phase is a relativistic invariant scalar and we require the
phase difference between two space-time points in both frames $S$ and $S'$
to be the same
\[
 \vect{k}\tp\Delta\vect{r}-ck\Delta t
=\omega(\frac{\vect{\hat{k}}\tp\Delta\vect{r}}{c}-\Delta t)
=\omega'(\frac{\vect{\hat{k}'}{}\tp\Delta\vect{r}'}{c}-\Delta t')
=\vect{k}'{}\tp\Delta\vect{r}'-ck'\Delta t'
\]
Here $\omega$ and $\omega'$ are the frequencies and $\vect{\hat{k}}$ and
$\vect{\hat{k}}'$ the propagation directions observed in reference frame $S$
and $S'$, respectively. The end points of the space-time distance
$(\Delta{r'},\Delta t')$ are the
Lorentz transformations of the respective end points of $(\Delta{r},\Delta t)$.
Insertion of the transformations (\ref{LorTrans_}), (\ref{Aberr_c}) and
(\ref{Aberr_s}) for the dashed coordinate variables gives
\begin{figure}
\hspace*{\fill}
\parbox{7cm}{\includegraphics[width=8cm]{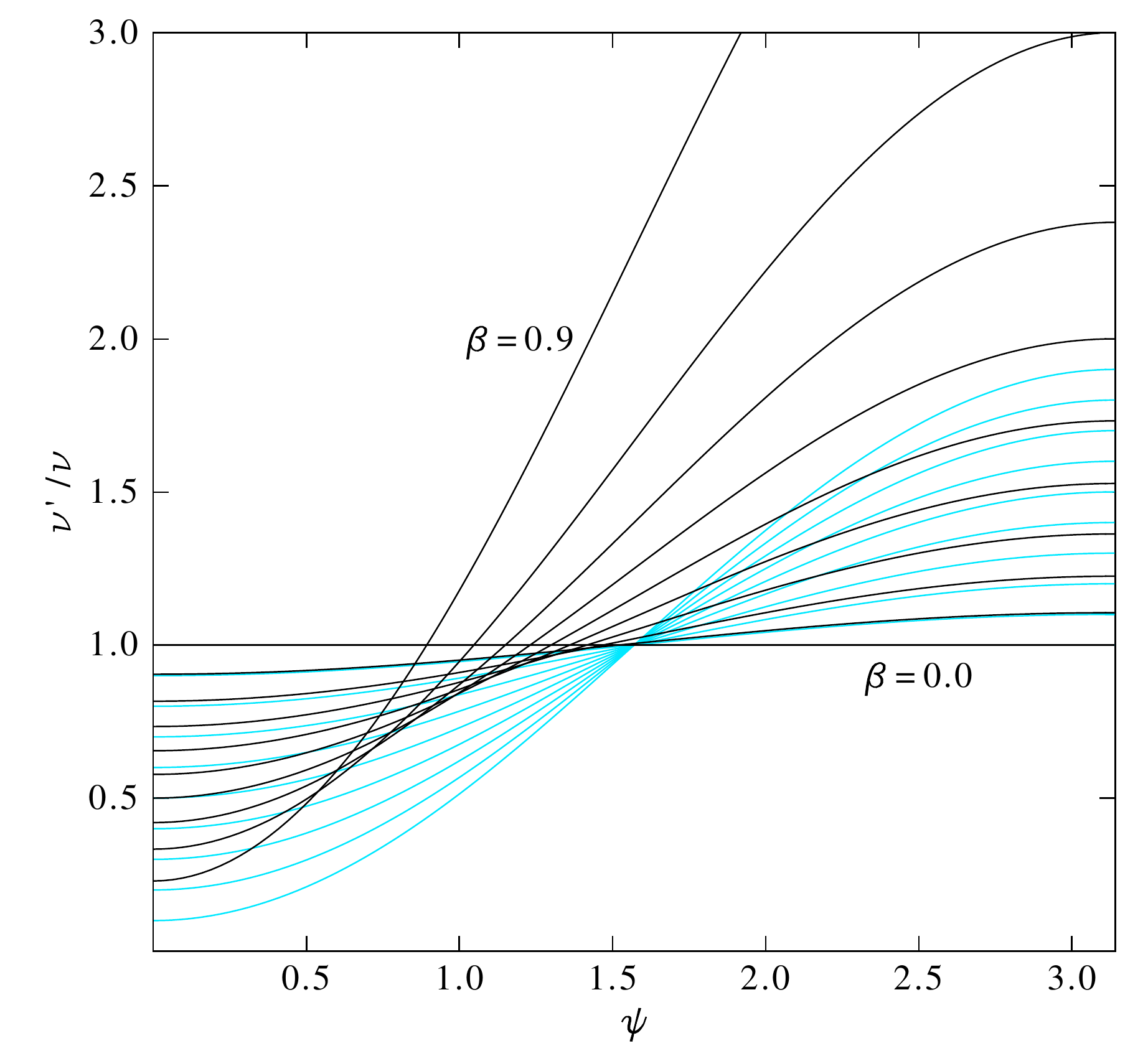}}
\hspace*{\fill}
\parbox{9cm}{\caption{Graphical representation of
    (\ref{FrqTran}) to demonstrate the ``Comptonisation'' effect:
    the relativistic Doppler-upshift for incident photons is larger
    in magnitude as the respective downshift for escaping photons.
    In contrast, the non-relativistic, classical Doppler shift is
    shown in light blue. It agrees with (\ref{FrqTran}) but with
    $\gamma$ set to unity.
    \label{Fig:FrqShift}}}
\hspace*{\fill}
\end{figure}

\begin{gather}
 \omega(\frac{\vect{\hat{k}}\tp\Delta\vect{r}}{c}-\Delta t)
=\omega'(\frac{\Delta x'\cos\psi'}{c}
        +\frac{\Delta y'\sin\psi'}{c}-\Delta t')
\nonumber\\ 
=\omega'\gamma\,[
       \frac{(\Delta x-\beta c\Delta t)\cos\psi'}{c}
     +\frac{\Delta y\sin\psi'}{c}
     -(\Delta t-\frac{\beta}{c}\Delta x)]
\nonumber\\         
=\omega'\frac{\gamma}{c}\,[
       (\Delta x-\beta c\Delta t)\,\frac{\cos\psi - \beta}
         {1 - \beta \cos\psi}
      +\Delta y\,\frac{\sin\psi}{\gamma(1 - \beta \cos\psi)}
      -(c\Delta t-\beta\Delta x)]
\nonumber\\
=\omega'\gamma\,
     [(1 +\beta \frac{\cos\psi - \beta}{1 - \beta \cos\psi})\Delta t
     -(\beta+\frac{\cos\psi - \beta}{1 - \beta \cos\psi})\Delta x
     -\frac{\sin\psi}{\gamma(1 - \beta \cos\psi)}\Delta y]
\nonumber\\
=\omega'\gamma\,[
     \frac{(1-\beta^2)\cos\psi}{1 - \beta \cos\psi}\Delta x
    +\frac{\sin\psi}{\gamma(1 - \beta \cos\psi)}\Delta y
    -\frac{1- \beta^2}{1 - \beta \cos\psi} \Delta t]
\nonumber\\
=\frac{\omega'}{\gamma(1 - \beta \cos\psi)}
     [\cos\psi\Delta x-\sin\psi\Delta y - \Delta t]
\nonumber\\\text{or}\quad
 \omega'=\omega\gamma(1 - \beta \cos\psi)
\label{FrqTran}\end{gather}
The reverse is immediately obtained if we use (\ref{Aberr_s_}) to
replace $\cos\psi$ by $\cos\psi'$
\begin{gather}
\omega=\frac{\omega'}{\gamma(1 - \beta \cos\psi)}
   =\frac{\omega'}{\gamma(1 - \beta \frac{\cos\psi' + \beta}
     {1+ \beta\cos\psi'})}
\nonumber\\
=\frac{\omega'}{\gamma} \;\frac{1+ \beta\cos\psi'}
      {1+ \beta\cos\psi' - \beta(\cos\psi' + \beta)}
=\frac{\omega'}{\gamma} \;\frac{1+ \beta\cos\psi'}{1- \beta^2}
\nonumber\\
=\omega'\gamma (1+ \beta\cos\psi')
\label{FrqTran_inv}\end{gather}
For convenience, we will introduce the Doppler shift factor $D$ as the
``generic'' version (\ref{FrqTran}) of the frequency transformation 
\begin{equation}
 D(\uect{k},\vectg{\beta})
 =\frac{1}{\gamma(1-\vectg{\beta}\tp\vect{\hat{k}})}
\label{FrqDoppler}\end{equation}
The convention is again that the arguments of $D$ including the
relative velocity of the frames are measured in the same frame.
If $\omega$ and $\vect{k}$ are the photon frequency and wave vector in
$S$ and $\vect{v}=c\vectg{beta}$ the velocity of $S'$ in measured in $S$
then the Doppler-shifted frequency in frame $S'$ is
\begin{equation}
\omega'=\frac{\omega}{D(\uect{k},\vectg{\beta})},\qquad
\text{or reversely}\quad
\omega=\frac{\omega'}{D(\uect{k}',-\vectg{\beta})}
\label{FrqTr}\end{equation}
In both cases the observables in frames $S$ and $S'$ are separated on
the different sides of the equations.
The minus sign in $D(\uect{k}',-\vectg{\beta})$ is needed because
in frame $S'$ the system $S$ of the left-hand-side moves with
$-\vect{v}$. We see immediately
\[
  D(\uect{k}',-\vectg{\beta})=\frac{1}{D(\uect{k},\vectg{\beta})},
  \qquad
  D(-\vect{\hat{k}},\vectg{\beta})=D(\uect{k},-\vectg{\beta}),
\]
provided the directions $\vect{\hat{k}}$ and $\vect{\hat{k}'}$ are
appropriately related by aberration.
The same frequency-shift conversion factor applies to the wave number
because $\omega=c k$ and $\omega'=c k'$ must hold in each frame
\begin{equation}
 k'=\frac{k}{D(\uect{k},\vectg{\beta})},\qquad
 k =\frac{k'}{D(\uect{k}',-\vectg{\beta})}
\label{WvnTran}
\end{equation}

\subsection{Representation of the wave phase by the wave 4-vector}
\label {App:four-vect}

Define the the 4-vectors position and wave vector for a photon by 
$\vectf{r}=\lvecc{ct}{\vect{r}}\tp$ and $\vectf{k}=(\omega/c,\vect{k})$. The
momentum is related to the respective 4-wave vector by
$\vectf{p}=\hbar\vectf{k}$.
All 4-vectors are transformed by the Lorentz
transformation (\ref{LorTrans}) which we now rewrite independently from a
coordinate system as a matrix to be multiplied to 4-vectors like $\vectf{r}$ or
$\vectf{k}$. Here the first column is to be multiplied with the time component
of a 4-vector, the second column has three components and is to be multiplied
with its space components. In this form, the Lorentz transformation is 
\begin{equation*}
\vectf{L}=\begin{pmatrix}
          \gamma & -\gamma\vectg{\beta}\tp \\
         -\gamma\vectg{\beta} &
 \boldsymbol{1}+(\gamma-1)\vectg{\hat{\beta}}\uectg{\beta}{}\tp
         \end{pmatrix}
\qquad
\vectf{L}^{-1}=\begin{pmatrix}
          \gamma & \gamma\vectg{\beta}\tp \\
          \gamma\vectg{\beta} &
 \boldsymbol{1}+(\gamma-1)\vectg{\hat{\beta}}\uectg{\beta}{}\tp
         \end{pmatrix}
\end{equation*}
The transformation of $\vectf{r}$ and $\vectf{k}$ yield
\begin{gather*}
\vectf{r}'=\vectf{L}\vectf{r}
=\rvecc{\gamma(ct-\vectg{\beta}\tp\vect{r})}
       {(\boldsymbol{1}-\vectg{\hat{\beta}}\uectg{\beta}{}\tp)\vect{r}
         +\gamma\vectg{\hat{\beta}}(\uectg{\beta}{}\tp\vect{r}-\beta ct)}
\\
\vectf{k}'=\vectf{L}\vectf{k}
=\rvecc{\gamma(\frac{\omega}{c}-\vectg{\beta}\tp\vect{k})}
       {(\boldsymbol{1}-\vectg{\hat{\beta}}\uectg{\beta}{}\tp)\vect{k}
         +\gamma\vectg{\hat{\beta}}(\uectg{\beta}{}\tp
                \vect{k}-\beta\frac{\omega}{c})}
\end{gather*}
The product of $\vectf{r}'$ and $\vectf{k}'$ yields the same wave phase as
$\vectf{k}\tp\vectf{r}$     
\begin{gather*}
\vectf{k}'{}\tp\vectf{r}'
=-\gamma^2(ct-\vectg{\beta}\tp\vect{r})
          (\frac{\omega}{c}-\vectg{\beta}\tp\vect{k})
\\
 +[(\boldsymbol{1}-\vectg{\hat{\beta}}\uectg{\beta}{}\tp)\vect{r}
 +\gamma\vectg{\hat{\beta}}(\uectg{\beta}{}\tp\vect{r}-\beta ct)
  [(\boldsymbol{1}-\vectg{\hat{\beta}}\uectg{\beta}{}\tp)\vect{k}
 +\gamma\vectg{\hat{\beta}}(\uectg{\beta}{}\tp
             \vect{r}-\beta\frac{\omega}{c})]
\\
=-\gamma^2(ct-\vectg{\beta}\tp\vect{r})
     (\frac{\omega}{c}-\vectg{\beta}\tp\vect{k}) 
 +\vect{k}\tp(\boldsymbol{1}-\vectg{\hat{\beta}}\uectg{\beta}{}\tp)
              \vect{r}
 +\gamma^2(\uectg{\beta}{}\tp\vect{r}-\beta ct)
          (\uectg{\beta}{}\tp\vect{r}-\beta\frac{\omega}{c})
\\
= -\gamma^2(\omega t +\vect{k}\tp\vectg{\beta}\vectg{\beta}\tp\vect{r})
  +\gamma^2(\frac{\omega}{c}\vectg{\beta}\tp\vect{r}
  + ct\vectg{\beta}\tp\vect{k})
\\ 
 +\vect{k}\tp(\boldsymbol{1}
 -\vectg{\hat{\beta}}\uectg{\beta}{}\tp)\vect{r}
 +\gamma^2(\vect{k}\vectg{\hat{\beta}}
           \uectg{\beta}{}\tp\vect{r}+\beta^2 \omega t)
          - \gamma^2(\frac{\omega}{c} \vectg{\beta}\tp\vect{r}
          + ct \vectg{\beta}\tp\vect{k})
\\
=-\gamma^2(1-\beta^2)\omega t
 +\vect{k}\tp(\boldsymbol{1}-\vectg{\hat{\beta}}
               \uectg{\beta}{}\tp)\vect{r}
 +\gamma^2(1-\beta^2)\vect{k}\vectg{\hat{\beta}}\uectg{\beta}{}\tp\vect{r}
\\
=-\omega t+\vect{k}\tp\vect{r}
=\vectf{k}\tp\vectf{r}
\end{gather*}
Hence the Lorentz transformation does not effect 4-vector products. 
Moreover, we find again the laws for frequency shift and wave vector aberration
in the expression of the transformed $\vectf{k}'$.
The first row in $\vectf{k}'$ yields the frequency shift (\ref{FrqTran})
\[
\omega'=\gamma(\omega-c\beta k\cos\psi)
       =\gamma(1-\beta \cos\psi) \omega
       =\frac{\omega}{D(\uect{k},\vectg{\beta})} 
\]
The second row yields the wave vector transformation
\begin{equation}
\vect{k}'
=\vect{k}+(\gamma-1)\vectg{\hat{\beta}}\uectg{\beta}{}\tp\vect{k}
         -\gamma\frac{\omega}{c}\vectg{\beta}
=\vect{k}+(\gamma-1)\vectg{\hat{\beta}}\uectg{\beta}{}\tp\vect{k}
         -\gamma k\vectg{\beta}
\label{KvecTran1}\end{equation}
For the Jacobian of this wave vector transformation we find
\begin{gather}
\frac{d^3\vect{k}'}{d^3\vect{k}}
 =  \boldsymbol{1}+(\gamma-1)\frac{\vectg{\beta}\vectg{\beta}\tp}{\beta^2}
  -\gamma\vectg{\beta}\vect{\hat{k}}\tp
\quad\text{with determinant}  
\nonumber\\
\mathrm{det}(\frac{d^3\vect{k}'}{d^3\vect{k}})=\gamma(1-\beta\cos\psi)
=\frac{1}{D(\uect{k},\vectg{\beta})}
\label{WvnJacobian}\end{gather}
To derive the determinant, we can
without restriction let $\vectg{\beta}$ point along $\vect{\hat{x}}$ 
and place $\vect{k}$ into the ($x,y$) plane.

Further manipulating (\ref{KvecTran1}) we have
\begin{equation}
\vect{k}'
=(\boldsymbol{1}-\vectg{\hat{\beta}}\uectg{\beta}{}\tp)\vect{k}
 +\gamma\vectg{\hat{\beta}} (\uectg{\beta}{}\tp\vect{k} - \beta k)
=(\boldsymbol{1}-\vectg{\hat{\beta}}\uectg{\beta}{}\tp)\vect{k}
 +\gamma k \vectg{\hat{\beta}} (\cos\psi - \beta)
\label{KvecTran2}\end{equation}
We see for once that the projection of the wave vector perpendicular to
$\vectg{\hat{\beta}}$ remains unaffected by the transformation
\begin{equation}
 (\boldsymbol{1}-\vectg{\hat{\beta}}\uectg{\beta}{}\tp)\,\vect{k}'=
 (\boldsymbol{1}-\vectg{\hat{\beta}}\uectg{\beta}{}\tp)\,\vect{k}
\label{Aberr_kperp}\end{equation}
and for the component along $\vectg{\hat{\beta}}$ we find
\begin{gather*}
\vect{k}'{}\tp\vectg{\hat{\beta}}
=k'\cos\psi'
=\gamma k (\cos\psi - \beta)
 \\\text{and finally}\quad
\cos\psi'
=\gamma \frac{k}{k'} (\cos\psi - \beta)
=\gamma D(\uect{k},\vectg{\beta}) (\cos\psi - \beta)
=\frac{\cos\psi - \beta}{1-\beta\cos\psi}
\end{gather*}
where we used (\ref{WvnTran}) in the last step to reobtain
the aberration law (\ref{Aberr_c}).

The aberration formulas (\ref{Aberr_s}) and (\ref{Aberr_c}) can be
rewritten independent from a coordinate system
\begin{align}
\vect{\hat{k}'}{}\tp\vectg{\hat{\beta}}
  =\gamma D(\uect{k},\vectg{\beta})
    (\vect{\hat{k}}\tp\vectg{\hat{\beta}}-\beta),&\quad
\vect{\hat{k}}\tp\vectg{\hat{\beta}}
  =\gamma D(\uect{k}',-\vectg{\beta})
    (\vect{\hat{k}'}{}\tp\vectg{\hat{\beta}}+\beta)
\nonumber\\\text{or}\quad
\vect{k'}{}\tp\vectg{\hat{\beta}}
  =\gamma(\vect{k}\tp\vectg{\hat{\beta}}-k\beta),&\quad
\vect{k}\tp\vectg{\hat{\beta}}
  =\gamma(\vect{k'}{}\tp\vectg{\hat{\beta}}+k'\beta)
\label{Aberr_kpara}\end{align}

\subsection{Transformation of the wave field}
\label{app:EFieldTransform}

While aberration and frequency shift are derived in many standard
textbooks on special relativity, the Lorentz transformation of the photon
polarisation is seldomly considered. In fact, \citep{CockeHolm:1972}
seems to be the first paper devoted to this problem. More recent
treatments with applications to Gamma-ray bursts can be found in
\citep{LyutikovEtal:2003, Nalewajko:2009}.

Consider a linearly polarised wave, either polarised in the aberration plane
or normal to it. Recall that the aberration plane is spanned by the relative
speed $\vect{v}$ of reference frame $S'$ seen in $S$ and the propagation
directions $\vect{\hat{k}}'$ and $\vect{\hat{k}}$ of the photon in $S'$
and $S$, respectively.
The aberration plane is identical in both frames and a polarisation
in the plane or normal to it should be equally well identified in
both systems. The mapping factors of the two polarisation amplitudes is,
however, not immediately obvious. 

We define a local right-handed, orthogonal coordinate system
$\vect{\hat{e}}_1$, $\vect{\hat{e}}_2$ and $\vect{\hat{e}}_3=\vect{\hat{k}}$
attached to the photon in reference frame $S$. The photon electric field
amplitude can be specified by
\begin{gather*}
\vect{E}_\vect{k}(\vect{r},t)
=E_{\vect{k},1}(\vect{r},t)\vect{\hat{e}}_1
+E_{\vect{k},2}(\vect{r},t)\vect{\hat{e}}_2
\end{gather*}
The respective magnetic wave field is
$\vect{B}_\vect{k}=\vect{\hat{k}}/c\times\vect{E}_\vect{k}$
or 
\begin{gather*}
c\vect{B}_\vect{k}(\vect{r},t)
=-E_{\vect{k},2}(\vect{r},t)\vect{\hat{e}}_1
 +E_{\vect{k},1}(\vect{r},t)\vect{\hat{e}}_2
\end{gather*}
We want to transform the photon to the frame $S'$ which moves with
$\vect{v}=v\vect{\hat{x}}=c\beta\vect{\hat{x}}$. We use the freedom to rotate
the polarisation directions $\vect{\hat{e}}_1$ and $\vect{\hat{e}}_2$ such
that the aberration plane is spanned by $\vect{\hat{e}}_1$ and
$\vect{\hat{e}}_3=\vect{\hat{k}}$ and $\vect{\hat{e}}_2$ is normal to the
plane. With angle $\psi$ defined as in Fig.~\ref{Fig:SpecRelSys}, we can
decompose
\begin{align}
\vect{\hat{k}}=\cos\psi\vect{\hat{x}}+\sin\psi\vect{\hat{y}},
\quad
&\vect{\hat{e}}_1=\sin\psi\vect{\hat{x}}-\cos\psi\vect{\hat{y}},
\quad
\vect{\hat{e}}_2=-\vect{\hat{z}}
\label{wavesys}\\
\vect{\hat{x}}=\cos\psi\vect{\hat{k}}+\sin\psi\vect{\hat{e}}_1,
\quad
&\vect{\hat{y}}=\sin\psi\vect{\hat{k}}-\cos\psi\vect{\hat{e}}_1
\nonumber\end{align}
In this rotated frame the wave field is
\begin{gather*}
\vect{E}_\vect{k}
=E_{\vect{k},1}(\sin\psi\vect{\hat{x}}
               -\cos\psi\vect{\hat{y}})
-E_{\vect{k},2}\vect{\hat{z}}
\\  
c\vect{B}_\vect{\hat{k}}
=-E_{\vect{k},2}(\sin\psi\vect{\hat{x}}
                -\cos\psi\vect{\hat{y}})
 -E_{\vect{k},1}\vect{\hat{z}}
\end{gather*}
The electric field is transformed into $S'$ by
\citep[][p.~558]{Jackson:1998}
\begin{align*}
  E'_{\vect{k}',x}=E_{\vect{k},x}
   =\sin\psi E_{\vect{k},1}
  &=\gamma(1-\beta\cos\psi)\sin\psi'\;E_{\vect{k},1}
   =\frac{\sin\psi'}{D(\uect{k},\vectg{\beta})}E_{\vect{k},1}
\nonumber\\  
  E'_{\vect{k}',y}
  =\gamma(E_{\vect{k},y}-c\beta B_{\vect{k},z})
  &=-\gamma(\cos\psi -\beta) E_{\vect{k},1}
   =-\frac{\cos\psi'}{D(\uect{k},\vectg{\beta})}E_{\vect{k},1}
\nonumber\\  
  E'_{\vect{k}',z}
  =\gamma(E_{\vect{k},z}+c\beta B_{\vect{k},y})
  &=-\gamma(1-\beta\cos\psi)E_{\vect{k},2}
  =-\frac{E_{\vect{k},2}}{D(\uect{k},\vectg{\beta})}
\end{align*}
where for the last steps we used (\ref{Aberr_s}) and (\ref{Aberr_c})
for $E'_{\vect{k}',x}$ and $E'_{\vect{k}',y}$, respectively. In the moving
frame $S'$, we have the wave field component $E'_{\vect{k}',z}$ normal to the
aberration plane and the vector
$E'_{\vect{k}',x}\vect{\hat{x}}+E'_{\vect{k}',y}\vect{\hat{y}}$ in the
aberration plane. The transformation $S\longrightarrow S'$ can thus be
summarised by
\begin{align}
  \text{normal to aberration plane:}\quad
  \vect{\hat{e}}_2\tp \vect{E}_{\vect{k}} = E_{\vect{k},2}
  \quad&\xrightarrow{S\rightarrow S'}\quad
  \vect{\hat{e}}_2\tp \vect{E}'_{\vect{k}'}
   =\frac{E_{\vect{k},2}}{D(\uect{k},\vectg{\beta})}
\label{aberrE_z}\\  
  \text{in the aberration plane:}
  \hspace*{3.5em}
  (\boldsymbol{1}-\vect{\hat{e}}_2\vect{\hat{e}}_2\tp)\vect{E}_\vect{k}
 &= E_{\vect{k},1} \vect{\hat{e}}_1
  =( \sin\psi\vect{\hat{x}}-\cos\psi\vect{\hat{y}})
   E_{\vect{k},1}
  \hspace*{10em}
\nonumber\\
  \quad\xrightarrow{S\rightarrow S'}\quad
  (\boldsymbol{1}-\vect{\hat{e}}_2\vect{\hat{e}}_2\tp)\vect{E}'_{\vect{k}'}
 &= (\sin\psi\vect{\hat{x}}-\gamma(\cos\psi -\beta)\vect{\hat{y}})
 E_{\vect{k},1}
 \nonumber\\
 &=( \sin\psi'\vect{\hat{x}}-\cos\psi'\vect{\hat{y}})
  \frac{E_{\vect{k},1}}{D(\uect{k},\vectg{\beta})}
\label{aberrE_xy}
\end{align}
In the last step we used (\ref{Aberr_s_}) and (\ref{Aberr_c_}).
Hence the field components in the aberration plane and normal to it are
both modified in strength by the same factor $D^{-1}(\uect{k},\vectg{\beta})$
and the aberrated field vector is tilted exactly so that it becomes
normal to the aberrated propagation direction $\vect{\hat{k}'}$.

\subsection{Transformations of irradiance and radiance}
\label{app:SpecRel_radiance}

The energy density, the Poynting flux of a monochromatic wave
and the irradiance transform according to
(\ref{aberrE_z}) and (\ref{aberrE_xy}) from
the previous section as
\begin{gather*}
 W'_\vect{k'}
 = \frac{W_\vect{k}}{D^2(\vect{\hat{k}},\vectg{\beta})},\quad
 \vect{S}'_{\vect{k}'}
 = \frac{\vect{S}_\vect{k}}{D^2(\vect{\hat{k}},\vectg{\beta})}
\\  
  \irr'(\vect{r}',t')
  = c\epsilon_0 \bra\vect{E}'(\vect{r}',t')\tp\vect{E}'(\vect{r}',t')\ket = 
  \frac{\irr(\vect{r},t)}{D^2(\vect{\hat{k}},\vectg{\beta})}
\end{gather*}
where $\vect{k}$ and $\vect{k}'$ are related by aberration.
For the spectral energy density and the spectral Poynting flux of a spectral
distribution of waves we have to take into account that wave vectors transform
differently depending on their propagation direction.
Using (\ref{WvnTran}) we have for an element in wave vector space
\[
 d^3\vect{k}'
 = {k'}^2 dk' d\Omega'
 = \frac{k^2 d k}{D^3(\vect{\hat{k}},\vectg{\beta})} d\Omega'
\]
For the transformation of the solid angle we find since the azimuthal angle
$\phi=\phi'$ remains invariant so that
\begin{gather}
 d\Omega'=\sin\psi' d\psi' d\phi'
=\frac{d\cos'\psi}{d\cos\psi} \sin\psi d\psi d\phi
=\frac{d}{d\cos\psi}(\frac{\cos\psi-\beta}{1-\beta\cos\psi})
\;d\Omega
\nonumber\\
=(\frac{1}{1-\beta\cos\psi}
 +\frac{\beta(\cos\psi-\beta)} {(1-\beta\cos\psi)^2})\; d\Omega
=\frac{(1-\beta\cos\psi) + \beta(\cos\psi-\beta)}
  {(1-\beta\cos\psi)^2} \;d\Omega
\nonumber\\
=\frac{1-\beta^2}{(1-\beta\cos\psi)^2}\; d\Omega
=D^2(\vect{\hat{k}},\vectg{\beta}) \;d\Omega
\label{dOmegaTran}\end{gather}
In total, the element in wave vector space therefore transforms as
\begin{equation}
 d^3\vect{k}'
 = \frac{k^2 d k d\Omega}{D(\uect{k},\vectg{\beta})} 
 = \frac{d^3\vect{k}}{D(\uect{k},\vectg{\beta})} 
\label{d3kTran}\end{equation}
which agrees with (\ref{WvnJacobian}) derived by different means above.
Since $\erg(\vect{k}) d^3\vect{k}$ has to transform as $W_\vect{k}$,
we conclude
\[
\erg'(\vect{k}') d^3\vect{k}'
= \frac{\erg(\vect{k}) d^3\vect{k}}{D^2(\vect{\hat{k}},\vectg{\beta})}
= \frac{\erg(\vect{k})}{D(\uect{k},\vectg{\beta})}\;
  \frac{d^3\vect{k}}{D(\uect{k},\vectg{\beta})}
\quad\text{or}\quad
\erg'(\vect{k}')=\frac{\erg(\vect{k})}{D(\uect{k},\vectg{\beta})}
\]
Similar reasoning leads to the transformed radiance
\citep[see][for a different derivation]{WeiskopfEtal:1999}
\begin{equation}
 \rad'(\vect{\hat{k}'})
 =\int \erg(k'\vect{\hat{k}'})\; {k'}^2\,dk'
 =\int \frac{\erg(k\vect{\hat{k}})}{D(\uect{k},\vectg{\beta})}
  \;\frac{k^2\,dk}{D^3(\vect{\hat{k}},\vectg{\beta})}
 =\frac{\rad(\vect{\hat{k}})}{D^4(\vect{\hat{k}},\vectg{\beta})}
\label{RadTran}\end{equation}
while the irradiance
\begin{equation}
 \irr'=\int \rad'(\vect{\hat{k}'}) d\Omega'
  =\int \frac{\rad(\vect{\hat{k}})}{D^4(\vect{\hat{k}},\vectg{\beta})}
        \;D^2(\vect{\hat{k}},\vectg{\beta}) d\Omega
  =\frac{\irr}{D^2(\vect{\hat{k}},\vectg{\beta})}      
\label{IrrTran}\end{equation}
transforms like the field energy density as we have seen already above.
Let $(ct_\mathrm{em},\vect{r}_\mathrm{em})$ be the event of photon emission
and $(ct_\mathrm{in},\vect{r}_\mathrm{in})$ the event of its detection.
Then we have for the world line between these two events
\[
c^2(t_\mathrm{in}-t_\mathrm{em})^2
-|\vect{r}_\mathrm{in}-\vect{r}_\mathrm{em}|^2=
(c\Delta t)^2- d^2 = 0
\]
in all frames. Hence the travel time $\Delta t$ of a photon transforms
in the same way as the distance $d$ between the (retarded) position of
the source and the detector. From (\ref{LorTrans_}) we have between
two events in the moving system
\[
 c\Delta t'=\gamma(c\Delta t - \vectg{\beta}\tp\Delta\vect{r})
\]
For the events photon emission and detection we have a spatial distance of
$\Delta\vect{r}=\vect{\hat{k}}\,d=\vect{\hat{k}}c\Delta t$. This yields the
transformation of the travel time $\Delta t$ and the same transformation
for the distance $d$
\begin{equation}
  c\Delta t'=\frac{c\Delta t}{D(\uect{k},\vectg{\beta})},\qquad
  d'=\frac{d}{D(\uect{k},\vectg{\beta})}
\label{TravelTran}\end{equation}
Accordingly, the radiant intensity of a point source transforms like
\begin{equation}
  I'(\vect{\hat{k}'})
 =Q'(\vect{r}'_\mathrm{em}+d'\vect{\hat{k}'}) {d'}^2
 =\frac{Q(\vect{r}_\mathrm{em}+\vect{\hat{k}}\,d)}
           {D^2(\vect{\hat{k}},\vectg{\beta})}
 \frac{d^2}{D^2(\vect{\hat{k}},\vectg{\beta})}
 =\frac{I(\vect{\hat{k}})}{D^4(\vect{\hat{k}},\vectg{\beta})}
\label{RinTran}\end{equation}
The above relations have been derived previously and discussed in detail
\citep{McKinley:1979,McKinley:1980,EriksenGron:1992,Kraus:2000}.
Their derivation is, however, different from our approach by associating.
There, the radiance is associated by a particle stream of photons,
the world lines of which are transformed between frames
and the resulting count rates on a detector surface 
in the different frames are compared.

\subsection{Relativistic equilibrium velocity distribution}
\label{app:Juettner}

Both relativistic invariance and thermodynamic equilibrium are difficult to
reconcile. Since J\"uttner's first publication on this topic a century ago \citep{Juettner:1911},
it has been discussed in a series of papers and textbook contributions
without that a final consensus seems to have emerged. A full account of the
topic is therefore beyond the scope of this manuscript. A through discussion
on the state of art can be found in \citep{Debbasch:2008}.
Here, we follow essentially the approach of \cite{Lehmann:2006}.

Define $\vectf{r}=\lvecc{ct}{\vect{r}}$ and $\vectf{p}=\lvecc{E}{c\vect{p}}$ as
4-vectors for position and momentum of an electron. The space component of the momentum is
related to velocity by $\vect{p}=\gamma m_e c\vectg{\beta}$. In order to obtain
a relativistic invariant distribution function, all dependencies on $\vectf{r}$
and $\vectf{p}$ should be wrapped into invariant 4-vector products. We neglect
interactions and we can therefore just consider a single particle
distribution. For free particles an invariant extension of Maxwell's
distribution is
\begin{gather*}
 f(\vect{p}) \; d^3\vect{r} d^3 c\vect{p}
=  \frac{d\Gamma}{Z(T,\vectf{u})}
   \exp(-\frac{\vectf{u}\tp\vectf{p}}{k_B T})
\label{RelDist_0}\\
d\Gamma=
\int d\compf{r}^0 \int_0^\infty d\compf{p}_0\; 
\prod_{\alpha=1}^3 d\compf{r}^\alpha d\compf{p}_\alpha
\delta(g_1(\vectf{r},\vectf{p}))
\delta(g_2(\vectf{r},\vectf{p}))
\nonumber\end{gather*}
where $\vectf{u}=\lvecc{1}{\vect{u}/c}$ is the velocity 4-vector of the system
so that $\vectf{u}\tp\vectf{p}$ is invariant, 
$T$ is the system temperature and $Z(T,\vectf{u})$ the partition function.
The distribution function should be homogeneous and stationary, i.e.,
it should not depend on $\vectf{r}$ at all.
In the phase space element $d\Gamma$, all 8 components of
$\vectf{r}$ and $\vectf{p}$ are independent variables.
The $\delta$-functions in the phase space element $d\Gamma$
aim at reducing the 8-dimensional phase space
to the usual 6 dimensions by constraining time and energy.
The choice for $g_1$ is obvious, it must enforce
the electron's energy-momentum relation
\begin{gather*}
  g_1=\vectf{p}\tp\vectf{p}-m_e^2c^4
  =E^2-c^2\vect{p}^2=E^2-m_e^2c^4(\gamma^2\vectg{\beta}^2+1)
  =E^2-\gamma^2m_e^2c^4
\end{gather*}
For $g_1$ several variants could be considered:
\begin{align*}
 \text{Lehmann's invariant}\quad
 && g_2^\mathrm{L} &= \vectf{r}\tp\vectf{p}-m_e^2c^3\tau
 =cEt-c\vect{r}\tp\vect{p}-m_e c^2\tau
\\  
 & && =m_e c^2\gamma (ct-\vect{r}\tp\vectg{\beta})-m_e c^3\tau
\\
 \text{Inhomogeneous invariant}\quad
 && g_2^\mathrm{I} &= \vectf{r}\tp\vectf{r}-(c\tau)^2
      =(ct)^2-\vect{r}^2-(c\tau)^2
\\
 \text{J\"uttner non-invariant}\quad
 && g_2^\mathrm{J} &= \frac{R_0}{\gamma}-c\tau
      =\frac{ct}{\gamma}-c\tau
\\
 \text{Plain non-invariant}\quad
 && g_2^\mathrm{P} &= R_0-c\tau=c(t-\tau)
\end{align*}
For the first three cases, the condition $g_2=0$ restricts the $\vectf{r}$
integration which is finally required to determine the partition function
to a time-like hyperbola in Minkowski space which intersects the time axis at
$ct=c\tau$. Hence parameter $\tau$ is the eigentime of a particle regardless
of its velocity.
In a frame where the particle has velocity $\vectg{\beta}$ and starts from the
origin it is located at $(ct,\vect{r})$ after its own time $\tau$. Even though
all three $g_2$ have the same roots in Minkowski space and enforce the same
constraint on the four components of $\vectf{r}$, they intersect the root
with different first derivative which inflicts differences for the
integration. The fourth case is clearly not invariant as it ignores
the differences between time and eigentime.

When integrating over the energy and time constraints, we have to obey the
following rules. If $x_i$ are the roots of $g(x)$
\[
 \int_0^\infty dx\;\delta(g(x)) h(x)=\sum_i\frac{1}{g'(x_i)} h(x_i)
\]
$g_1$ has two roots $\compf{p}_0=E$ but the negative root is discarded by
the restriction to positive free energies. Abbreviating
$\exp(-\vectf{u}\tp\vectf{p}/k_B T)=h(\vectf{p})$ we find
\begin{gather*}
 \int_0^\infty d\compf{p}_0\;\delta(g_1) h(\vectf{p})
=\int_0^\infty dE\;\delta(E^2-\gamma^2m_e^2c^4)\, h(E,c\vect{p})
\\
=\int_0^\infty dE\; \frac{\delta(E-\gamma m_ec^2)}{2E} h(E,c\vect{p})
=\frac{h(\gamma m_ec^2,c\vect{p})}{2\gamma m_ec^2}
\end{gather*}
The integration over $\compf{r}_0$ yields for the different $g_1$ (we omit $h$
because it is independent on $\vectf{r}$)
\begin{gather*}
 \int_0^\infty d\compf{r}_0\;\delta(g_2) 
=\int_0^\infty d(ct)\;\delta(g_2) 
=\begin{cases}
 \DS \frac{1}{E}=\frac{1}{\gamma m_e c^2}   &
    \text{for}\;g_2^\mathrm{L}\\[1.5ex]
 \DS \frac{1}{2\sqrt{\vect{r}^2+(c\tau)^2}} &
     \text{for}\;g_2^\mathrm{I}\\[1ex]
 \hspace*{3em}\gamma & \text{for}\;g_2^\mathrm{J}\\
 \hspace*{3em}   1   & \text{for}\;g_2^\mathrm{P}
\end{cases}
\end{gather*}
We obviously have to discard $g_L^\mathrm{I}$ because it introduces a space
dependence into the distribution function. To show dependencies more clearly
we express $\gamma$ in terms of $\vect{p}$ by
\begin{equation}
\gamma^2(\vect{p})=1+\gamma^2\beta^2=1+\frac{\vect{p}^2}{m_e^2c^2}
\label{gammaP}
\end{equation}
Insertion into (\ref{RelDist_0}) gives for the three remaining cases
\begin{gather*}
 f(\vect{p}) \; d^3\vect{r} d^3 \vect{p}
=  \frac{d^3\vect{p}\;d^3\vect{r}}{Z(T,\vectf{u})}
   \exp(-\frac{\vectf{u}\tp(\gamma(\vect{p}) m_ec^2,c\vect{p})}{k_B T})
\begin{cases}
 \DS \frac{1}{2(\gamma(\vect{p}) m_e c^2)^2} &
     \text{for}\;g_2^\mathrm{L}\\[2ex]
 \DS \hspace*{2em} \frac{1}{2 m_e c^2}       &
     \text{for}\;g_2^\mathrm{J}\\[2ex]
 \DS \hspace*{1em}\frac{1}{2\gamma(\vect{p}) m_e c^2 }    &
     \text{for}\;g_2^\mathrm{P}
\end{cases}
\\
=  \frac{d^3\vect{p}\;d^3\vect{r}}{Z(T,\vectf{u})}
   \exp(-\frac{c\sqrt{\rule{0em}{2ex}(m_ec)^2+\vect{p}^2}-\vect{u}\tp\vect{p}}
              {k_B T})
 \begin{cases}
\DS \frac{1}{2c^2((m_ec)^2+\vect{p}^2)}
        & \text{for}\;g_2^\mathrm{L}\\[2ex]
\DS \hspace*{2.5em}\frac{1}{2 m_e c^2}
        & \text{for}\;g_2^\mathrm{J}\\[2ex]
\DS \frac{1}{2c\sqrt{\rule{0em}{1.5ex}(m_ec)^2+\vect{p}^2}}
        & \text{for}\;g_2^\mathrm{P}
\end{cases}
\end{gather*}
We see that depending on the choice of $g_2$ we obtain slightly different
distribution functions.
For $g_2^L$ we have relativistic invariance while J\"uttner's
original distribution for $g_2^J$ is manifestly stationary because
it commutes with the single particle Hamiltonian.
In numerical simulations, a relativistic gas of virtual particles seems
to approach the original J\"uttner distribution.
\citep{MontakhabEtal:2009}. We will therefore use the
J\"uttner distribution for our calculations of Fig.~\ref{Fig:BetaIntegrared}.

Note that when we want to use the electron velocity $\vectg{\beta}$ instead of
the momentum $\vect{p}$ as variable in the distribution function we have to 
insert $\vect{p}=m_e c\gamma \vectg{\beta}$ and also convert the momentum
space element to the velocity space element
\begin{gather*}
  f(\vect{p}) \;d^3\vect{r}d^3\vect{p}
= f(\frac{m_e c\vectg{\beta}}{\sqrt{1-\beta^2}})\;
 |\det(\frac{d^3\vect{p}}{d^3\vectg{\beta}})|
 \;d^3\vect{r}\,d^3\vectg{\beta}
\\
\frac{d^3\vect{p}}{d^3\vectg{\beta}}
=m_e c \gamma (\1+\gamma^2\vectg{\beta}\vectg{\beta}\tp)
\\
\det(\frac{d^3\vect{p}}{d^3\vectg{\beta}})
=(m_e c \gamma)^3(1+\gamma^2\beta^2)
=(m_e c)^3 \gamma^5
\end{gather*}

It remains to calculate the respective partition function which guarantees
the normalisation and it is also an important means to derive thermodynamic
equilibrium quantities. We restrict its
evaluation to the rest frame of the
system, i.e., for $\vect{u}=0$. Introducing
$\rho^2=\vect{p}^2/(m_ec)^2$ and $\Theta=m_ec^2/k_B T$
we obtain for the three different cases
\begin{gather*}
  Z(T,\vectf{u})
=  \int d^3\vect{p}\;d^3\vect{r}
   \exp(-\frac{c\sqrt{\rule{0em}{1.5ex}(m_ec)^2+\vect{p}^2}}{k_B T})
 \begin{cases}
\DS \frac{1}{2c^2((m_ec)^2+\vect{p}^2)}
     & \text{for}\;g_2^\mathrm{L}\\[2ex]
\DS \hspace*{2.5em} \frac{1}{2 m_e c^2}
     & \text{for}\;g_2^\mathrm{J}\\[2ex]
\DS \frac{1}{2c\sqrt{\rule{0em}{1.5ex}(m_ec)^2+\vect{p}^2}}
     & \text{for}\;g_2^\mathrm{P}
\end{cases}
\\
= \frac{4\pi V}{2 m_e c^2} \int_0^\infty d\rho\,
 \begin{Bmatrix}
 \DS  \frac{\rho^2}{m_ec^2\,(1+\rho^2)}\\[2.3ex]
         \rho^2\\[0.8ex]
 \DS  \frac{\rho^2}{\sqrt{\rule{0em}{1.5ex}1+\rho^2}}
 \end{Bmatrix}
   \exp(-\Theta\sqrt{1+\rho^2})
\\
= \frac{4\pi V}{2 m_e c^2} 
\begin{cases}
\DS  \frac{1}{m_ec^2}G(\Theta)\quad\text{with}\;G'(\Theta)
  =\frac{1}{\Theta} K_1(\Theta)    & \text{for}\;g_2^\mathrm{L}\\[1.5ex]
\DS  \frac{1}{\Theta} K_2(\Theta)
  =\quart(K_3(\Theta)-K_1(\Theta)) & \text{for}\;g_2^\mathrm{J}\\[1.5ex]
\DS  \frac{1}{\Theta} K_1(\Theta)=\half(K_2(\Theta)-K_0(\Theta)) &
  \text{for}\;g_2^\mathrm{P}
\end{cases}
\end{gather*}
To evaluate the integrals we used the integral representation
of the modified Bessel functions of the second kind
\citep[8.432.3]{GradshteynRyzhik:1980} 
\begin{gather}
  K_\nu(z)=K_{-\nu}(z)
 =\frac{\sqrt{\pi}}{\Gamma(\nu+\half)}(\frac{z}{2})^\nu
  \int_1^\infty e^{-zy}(y^2-1)^{\nu-\half}\;dy
\label{Kn0}\\\text{with derivative}\quad
  K'_\nu(z)
 = \frac{\nu}{z} K_\nu(z)
 -\frac{\sqrt{\pi}}{\Gamma(\nu+\half)}(\frac{z}{2})^\nu
   \int_1^\infty e^{-zy}y(y^2-1)^{\nu-\half}\;dy
\label{dKn0}\end{gather}
With the recurrence relations
\citep[8.486.10+11]{GradshteynRyzhik:1980}
\begin{gather}
  K_{\nu+1}(z)-K_{\nu-1}(z) =  2\,\frac{\nu}{z} K_\nu(z)
  \nonumber\\
  K_{\nu+1}(z)+K_{\nu-1}(z) = -2\,K'_\nu(z)
  \nonumber\\
  K'_\nu \mp \frac{\nu}{z} K_\nu = \mp K_{\nu\pm 1}
  \nonumber\\
\text{we find from (\ref{dKn0})}\qquad
  -(K'_\nu(z)-\frac{\nu}{z} K_\nu(z))=K_{\nu+1}(z)
\nonumber\\  
 =\frac{\sqrt{\pi}}{\Gamma(\nu+\half)}(\frac{z}{2})^\nu
  \int_1^\infty e^{-zy}y(y^2-1)^{\nu-\half}\;dy
\label{Kn1}\end{gather}
From (\ref{Kn0}), (\ref{Kn1}) and $\Gamma{3/2}=\sqrt{\pi}/2$
for $\nu=1$
\begin{gather*}
  K_1(z)=z\int_1^\infty e^{-zy}\sqrt{y^2-1}\;dy
\\
  K_2(z)=z\int_1^\infty e^{-zy}y\sqrt{y^2-1}\;dy
\end{gather*}
So that
\begin{gather*}
 \int_0^\infty
 \frac{\rho^2}{\rule{0em}{2.6ex}\sqrt{\rule{0em}{2ex}1+\rho^2}^{\,\alpha}}
 \,e^{-z\sqrt{1+\rho^2}} \,d\rho
 \;\stackrel{\rho=\sinh\,t}{=}\;
 \int_0^\infty \frac{\sinh^2 t}{\cosh^\alpha t}\,e^{-z\cosh\,t} \,d(\sinh\,t)\\
=\int_0^\infty \sinh^2 t \cosh^{1-\alpha} t    \,e^{-z\cosh\,t} \,dt
 \;\stackrel{\cosh\,t=y}{=}\;
 \int_1^\infty  \sqrt{y^2-1}\; y^{1-\alpha}\;e^{-zy}\,dy\\
=\begin{cases}
  \DS G(z)\;\text{with}\;G'(z)=\frac{1}{z} K_1(z) & \text{for}\;\alpha=2\\
  \DS \hspace*{3em}\frac{1}{z} K_2(z) & \text{for}\;\alpha=0\\[1.5ex]
  \DS \hspace*{3em}\frac{1}{z} K_1(z) & \text{for}\;\alpha=1\\
\end{cases}
\end{gather*}
For $G(z)$ we did not find a simple expression. Taking account of the fact
that $G(z)\rightarrow 0$ for $z\rightarrow\infty$
we can write $G(z)$ as \citep[p.20]{Rosenheinrich:2015}
\begin{gather*}
  G(z)=\int_z^\infty \frac{K_1(z')}{z'} dz'
\\  
  =zK_0(z)+K_1(z)+\frac{\pi z}{2}(K_0(z)L_1(z)+K_1(z)L_0(z))
\end{gather*}
where 
\[
L_{n}(z) =(\frac{z}{2})^{n+1}
  \sum_{k=0}^{\infty} \frac{1}{\Gamma(k+3/2)\Gamma(k+n+3/2)}(\frac{z}{2})^{2k}
\]
is the modified Struve function.

\subsection{Lorentz transformation of space-cones and spheres}
\label{app:SpecRel_cones}

Due to the mixture of time and space coordinates, a Lorentz transformation
(\ref{LorTrans_}) of extended objects from one frame to another is often not
intuitive. Even more so is the instantaneous view which observers have on an
object from different frames because different travel times of photons from
different parts of the object have to be taken into into account. As a
consequence, objects may appear considerably deformed from their rest frame
shape. In this section we want to elucidate how the Sun appears to a
relativistic coronal electron.

An exception from the above mentioned relativistic deformation are space-cones
peaked at the observer. Such a 3D space-cone (not to be confused with a 4D
light-cone) with its apex at $\vect{r}_\mathrm{obs}$ is given in its rest frame
$S$ by points $\vect{r}$ which satisfy
\[
 (\vect{r}-\vect{r}_\mathrm{obs})\tp\vect{\hat{a}}_\theta
=|\vect{r}-\vect{r}_\mathrm{obs}|\cos\theta
\]
where $\vect{\hat{a}}_\theta$ is the unit direction of the cone axis of a cone
of opening angle $\theta$. Photons received by the observer at $t=t\inc$ were
emitted from $\vect{r}$ at the retarded time
$t=t\inc-|\vect{r}-\vect{r}_\mathrm{obs}|/c$. Therefore the space-time events
$(ct,\vect{r})$ of emitting photons from the cone surface which all arrive at
$(ct\inc,\vect{r}_\mathrm{obs})=0$ are connected by the linear relation
\begin{equation}
 ct\cos\theta+\vect{r}\tp\vect{\hat{a}}_\theta=0
\label{RelCone}\end{equation}
If the observer collects the photons in a camera, he would observe the image
of a circle with radius $\theta$ and the centre in direction
$\vect{\hat{a}}_\theta$. Aside from their geometrical meaning, we may consider
$\cos\theta$ and $|\vect{\hat{a}}_\theta|$ as general coefficients in
(\ref{RelCone}) which represents a space-like cone as long as
$|\vect{\hat{a}}_\theta|>|\cos\theta|$.

\begin{figure}
\hspace*{\fill}
\includegraphics[width=11cm]{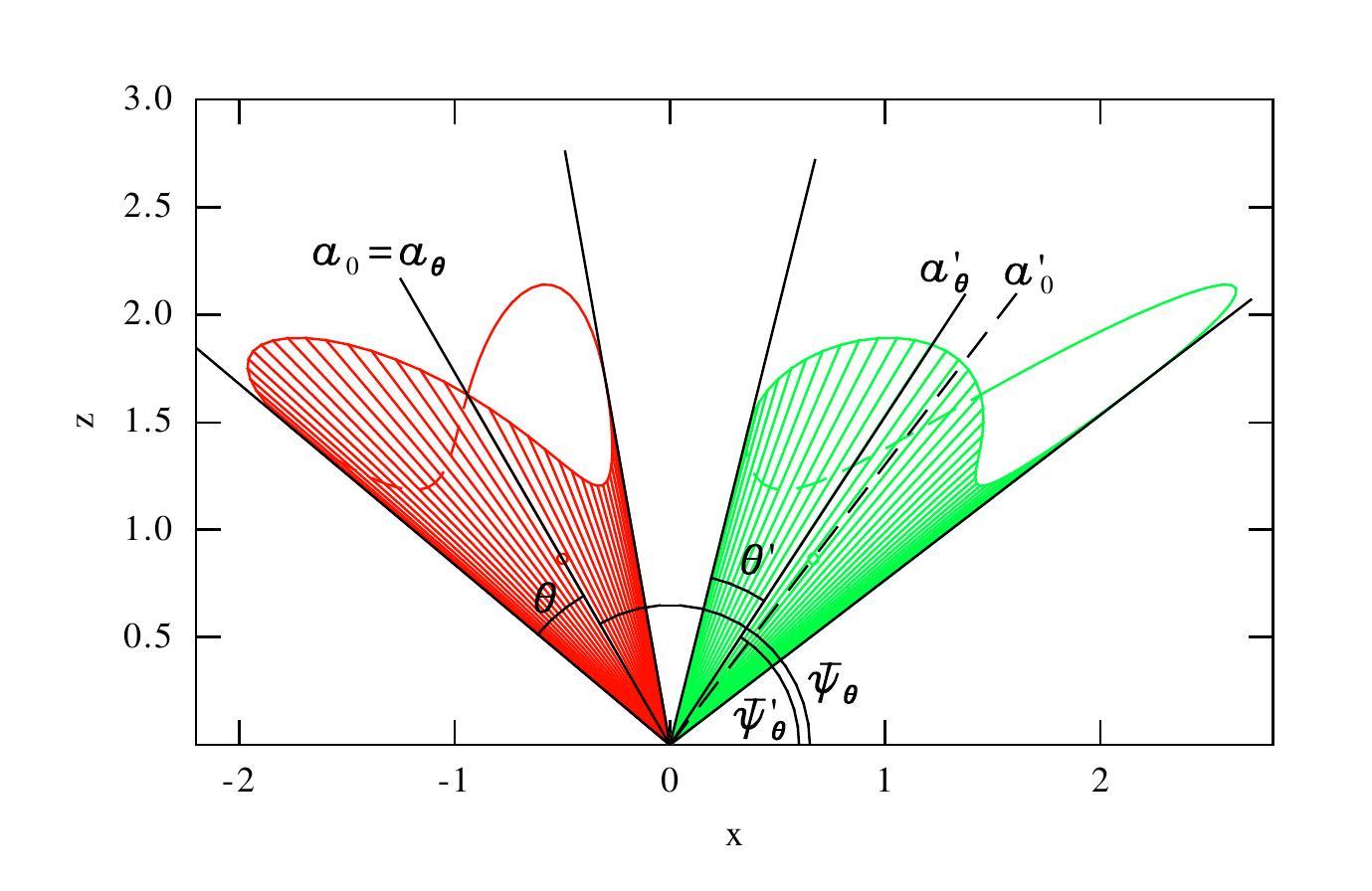}
\hspace*{\fill}
\caption{Lorentz transformation of a space cone from its rest fame (red) to a
  moving frame with $\beta=0.85$ to the right (green) for the observers in both
  frames at $\vect{r}=\vect{r}'=0$ at the time of observation.
  Here, $\vect{\hat{a}}_\theta$ and $\vect{\hat{a}}_0$ are the directions of
  the cone centres for a finite opening angle $\theta$ and of a degenerated cone
  of zero opening, both aligned in the rest frame.
  For the observers, the finite width cones
  appear as circles centred in directions $\vect{\hat{a}}_\theta$ (rest frame)
  and $\vect{\hat{a}}'_\theta$ (moving frame).
  In the moving frame, the cone
  centre $\vect{\hat{a}}'_\theta$ is not aligned any more with the
  aberrated direction $\vect{\hat{a}}_0'$ of the cone centre
  (indicated by a dashed line).
   \label{Fig:SpceCone}}
\hspace*{\fill}
\end{figure}

Consider another observer in a moving frame $S'$ at $\vect{r}'_\mathrm{obs}=0$
such that the origins of $S$ and $S'$ are the same at time $ct'=ct=0$. Then
the observer in the moving frame receives the same photons at $ct'=0$ as the
observer in $S$ at $ct=0$. However, in $S'$ they have different travel times
and were emitted form different coordinate positions.
Applying the Lorentz transformation to (\ref{RelCone}) must give
without explicit calculation
\begin{equation}
 ct'\cos\theta'+\vect{r}'{}\tp\vect{\hat{a}'}_\theta=0
\label{RelCone_}\end{equation}
because the linear structure of (\ref{RelCone}) is maintained due to the
linearity of the Lorentz transformation (\ref{LorTrans_}). So we obtain yet
another cone in the new frame and the moving observer will receive the
image of another circle provided the yet unknown coefficients in
(\ref{RelCone_}) obey $\cos\theta'<|\vect{\hat{a}'}_\theta|$.

To obtain the explicit expressions for $\vect{\hat{a}'}_\theta$ and
$\cos\theta'$, we follow \citep{Boas:1961}, apply the Lorentz transformation
to (\ref{RelCone}) and reorder the terms to obtain a form as in
(\ref{RelCone_}). For convenience, we assume that in the rest frame
the vector $\vect{\hat{a}}_\theta$ lies in the $x,z$ plane and forms
an angle $\bar{\psi}_\theta$ with the $\uect{x}$ axis. Then
$\vect{\hat{a}}_\theta=(\cos\bar{\psi}_\theta,0,\sin\bar{\psi}_\theta)$
and from (\ref{RelCone_}).
\begin{gather}
  ct\cos\theta + x\cos\bar{\psi}_\theta + z\sin\bar{\psi}_\theta=0
 \quad\xrightarrow{S\rightarrow S'}
\nonumber\\ 
 0=\gamma(ct'+\beta x')\cos\theta
  +\gamma(x'+\beta ct') \cos\bar{\psi}_\theta + z'\sin\bar{\psi}_\theta
\nonumber\\
  =\gamma(\cos\theta+\beta\cos\bar{\psi}_\theta) ct'
  +\gamma(\cos\bar{\psi}_\theta+\beta\cos\theta)x' + z'\sin\bar{\psi}_\theta
\nonumber
\intertext{Comparison with (\ref{RelCone_}) gives except for a
  common constant $b$}
b \cos\theta'=\gamma(\cos\theta+\beta\cos\bar{\psi}_\theta),\quad
b \cos\bar{\psi}'_\theta=\gamma(\cos\bar{\psi}_\theta+\beta\cos\theta),\quad
b \sin\bar{\psi}'_\theta=\sin\bar{\psi}_\theta,\quad
\label{RelConeAngl}\\
1=\cos^2\bar{\psi}'_\theta+\sin^2\bar{\psi}'_\theta
\quad\text{yields}\quad
b^2=\gamma^2(\cos\bar{\psi}_\theta+\beta\cos\theta)^2+\sin^2\bar{\psi}_\theta
\label{b2_def1}\end{gather}
With this normalisation we readily find that $|\cos\theta'|\le 1$.
We show that $b^2>b^2\cos^2\theta'$ or
\begin{gather*}
  b^2-\gamma^2(\cos\theta+\beta\cos\bar{\psi}_\theta)^2
 =\gamma^2(\cos\bar{\psi}_\theta+\beta\cos\theta)^2+\sin^2\bar{\psi}_\theta
 -\gamma^2(\cos\theta+\beta\cos\bar{\psi}_\theta)^2
 \\
 = \gamma^2\cos^2\bar{\psi}_\theta+\gamma^2\beta^2\cos^2\theta
 +2\gamma^2\beta\cos\bar{\psi}_\theta\cos\theta + \sin^2\bar{\psi}_\theta
\\-\gamma^2\cos^2\theta-\gamma^2\beta^2\cos^2\bar{\psi}_\theta
    -2\gamma^2\beta\cos\theta\cos\bar{\psi}_\theta
\\    
 =\gamma^2(1-\beta^2)\cos^2\bar{\psi}_\theta+\gamma^2(\beta^2-1)\cos^2\theta
 +\sin^2\bar{\psi}_\theta
\\    
 =1-\cos^2\theta = \sin^2\theta \ge 0
\end{gather*}
and we can also write the normalisation (\ref{b2_def1}) alternatively as
\begin{equation}
b^2=\gamma^2(\cos\theta+\beta\cos\bar{\psi}_\theta)^2+\sin^2\theta
\label{b2_def2}\end{equation}

Obviously, the images which both observers take with this setup both show a
circle, centred in direction $\vect{\hat{a}}_\theta$ or
$\vect{\hat{a}}'_\theta$ and with angular diameter $\theta$ or $\theta'$,
respectively. From (\ref{RelConeAngl}) and (\ref{b2_def2}) we have for the
transformed cone
\begin{gather}
\cos\theta'=\frac{(\cos\theta+\beta\cos\bar{\psi}_\theta)}
    {\sqrt{(\cos\theta+\beta\cos\bar{\psi}_\theta)^2+\sin^2\theta/\gamma^2}}
\nonumber\\
\cos\bar{\psi}'_\theta=\frac{\cos\bar{\psi}_\theta+\beta\cos\theta}
    {\sqrt{(\cos\theta+\beta\cos\bar{\psi}_\theta)^2+\sin^2\theta/\gamma^2}}
\label{cospsi_cone}\end{gather}
The situation is illustrated in Fig.~\ref{Fig:SpceCone}.
Note that the direction
$\vect{\hat{a}}'_\theta=(\cos\bar{\psi}'_\theta,0,\sin\bar{\psi}'_\theta)$
of the cone centre in the moving frame is slightly shifted with respect to
the aberrated direction of the cone axis, which we obtain if we compare
the above results with a cone of opening $\theta=0$ but the same
direction $\vect{\hat{a}}_0=\vect{\hat{a}}_\theta$ in the rest frame.
The aberrated direction $\vect{\hat{a}}'_0$ of this degenerated cone
axis is then
\[
\cos\bar{\psi}'_0=\frac{\cos\bar{\psi}_0+\beta}
            {1+\beta\cos\bar{\psi}_0}
\]
which is, as expected, the aberration (\ref{Aberr_bc_}) for incoming photons.
We will refer to this phenomenon as aberration shift of cone centres. It
causes any space curve which projects as a circle to an observer in the rest
frame to be seen by a moving observer from the same perspective as an
aberrated circle with its apparent centre shifted with respect to the
aberrated centre of the circle in the rest frame. This shift increases with
the opening angle $\theta$ of the cone, or, equivalently with the radius of
the observed circle and vanishes naturally if $\theta$ approaches zero
(see Fig.~\ref{Fig:AbbLatCent} below).

\begin{figure}
\hspace*{\fill}
\includegraphics[width=11cm]{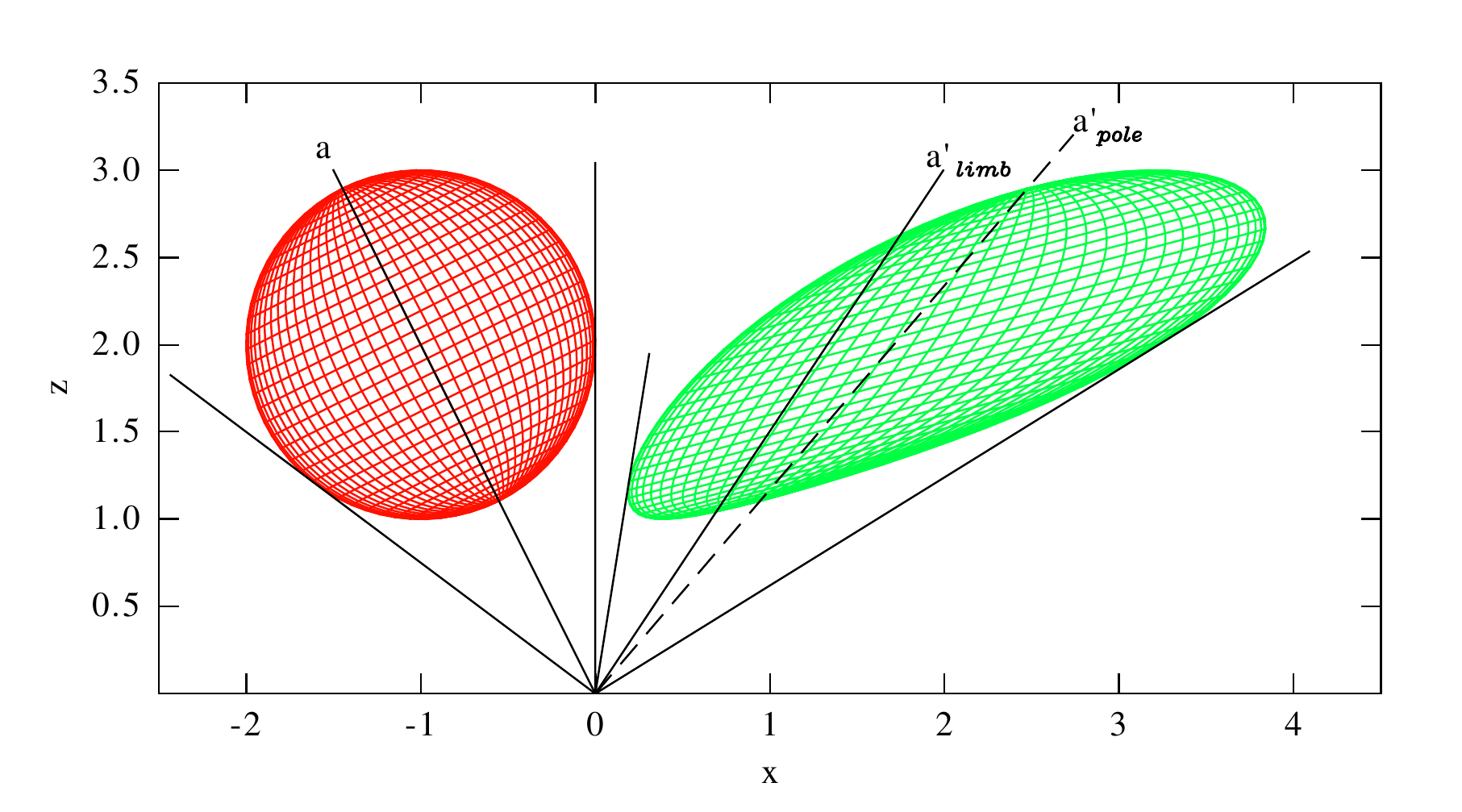}
\hspace*{\fill}\\
\hspace*{\fill}
\includegraphics[width=11cm]{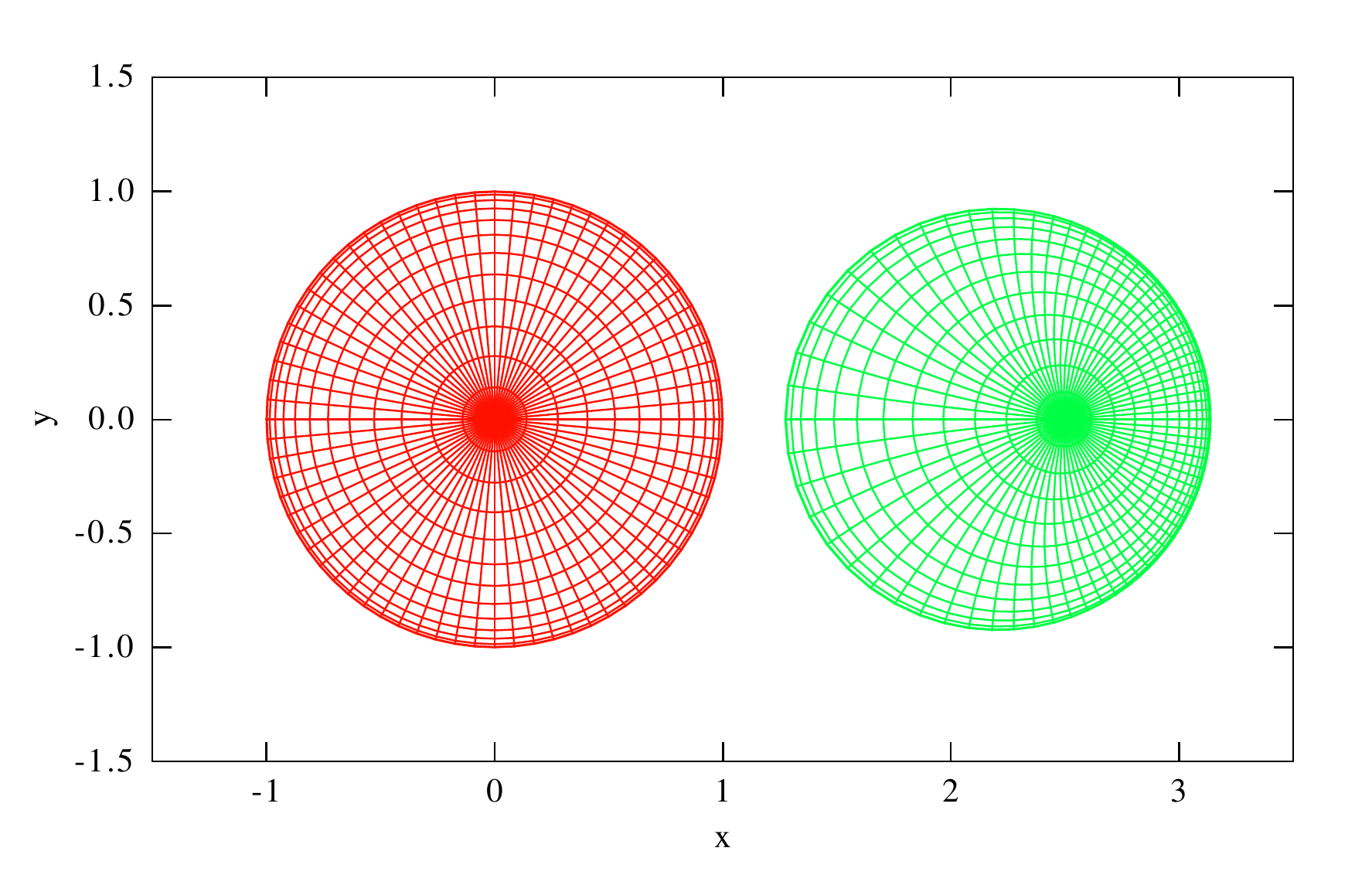}
\hspace*{\fill}
\caption{Lorentz transformation of a sphere from its rest frame (red) to a
  moving frame with $\beta=0.85$ along the $\uect{x}$-axis (green).
  The upper figure
  shows the transformed geometries from a side view with the observers at
  $(ct,\vect{r})=(ct',\vect{r}')=0$. The bottom panel shows the images the
  two observers would obtain in projective geometry with a camera oriented
  in direction to the centre of the respective limb circle.
  \label{Fig:SphereShape}}
\hspace*{\fill}
\end{figure}

We will now apply these results to the surface of a sphere with radius
$R_\odot$. A natural coordinate system on the sphere's surface in its rest
frame has a $\uect{z}$-axis along the direction from the observer to the
centre of
the sphere which we will again call $\vect{\hat{a}}$. Latitude circles on the
sphere are then obtained as intersections of the sphere surface with cones
defined above with the cone opening angle varying from $\theta=0$ (observer
pole) to $\theta=\asin(R_\odot/r)$ (observer limb) where $r$ is the distance
of the observer from the centre of the sphere in the rest frame. Note that
distance $r$ and angle $\theta$ correspond to the notation in
Fig.~\ref{Fig:ScaGeom3D} if we assume that the observer is located at the
scattering site. In Fig.~\ref{Fig:SphereShape} we give an example of how this
sphere transforms into the frame of a moving observer located at the origin
at the time of the observation. The upper diagram shows the sphere in its
rest frame (red) and in the moving frame (green).
The transformed sphere has been obtained by a Lorentz
transformation of the space-time events $(ct=-|\vect{r}|,\vect{r})$ at
which the received photons were emitted from the surface of the sphere.
For the top perspective in Fig.~\ref{Fig:SphereShape} the sphere was
assumed transparent so that a transform also for the backside of the sphere
is obtained.
Here $\vect{\hat{a}}$ is the direction to the sphere centre and to the
observer pole in the rest frame. It transforms to
$\vect{\hat{a}}'_\mathrm{pole}=\vect{\hat{a}}'_0$ by aberration. The largest
latitude (limb) circle corresponds to a cone the axis of which transforms from
$\vect{\hat{a}}$ in the rest frame to a slightly different direction
$\vect{\hat{a}}'_\mathrm{limb}=\vect{\hat{a}}'_{\theta=\asin(r/R_\odot)}$.

\begin{figure}
\hspace*{\fill}
\parbox{7cm}{\includegraphics[width=8cm]{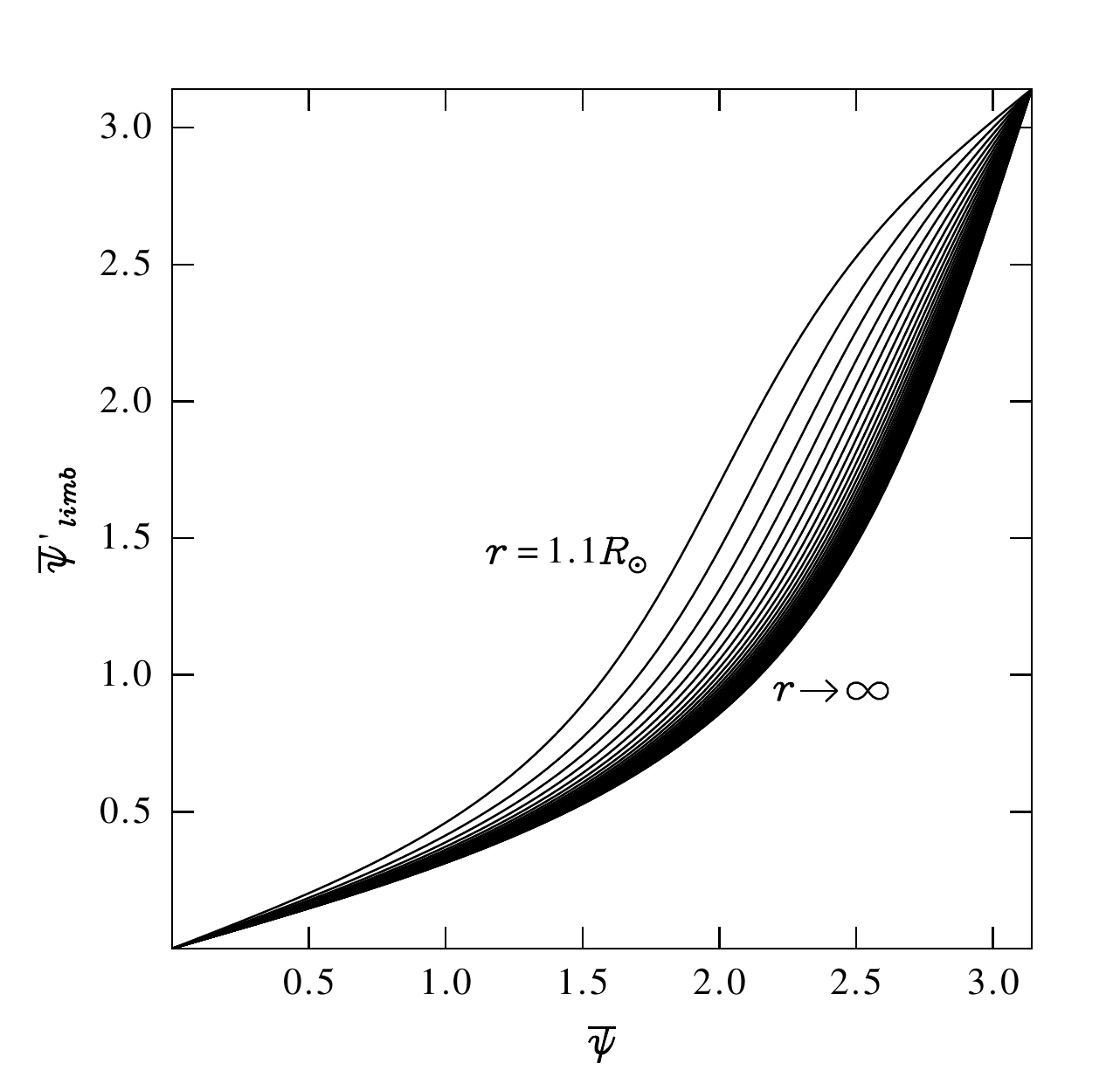}}
\hspace*{\fill}
\parbox{9cm}{\caption{Apparent direction of the centre of the limb circle
    versus its direction in the rest frame seen by an observer moving with
    $\beta=0.85$. $\bar{\psi}$ and $\bar{\psi}'$ are the angles of the centre
    direction with respect to the motion of the moving observer. This modified
    aberration for the centre of finite sized spheres is given for various
    distances from the sphere. The leftmost curve is for $r=1.1 R_\odot$,
    further curves to the right have $r$ increased in steps of $0.1 R_\odot$.
    In the limit of large $r$ (rightmost curves), the standard aberration is
    approached.\label{Fig:AbbLatCent}}}
\hspace*{\fill}
\end{figure}

Ignoring the difference between $\vect{\hat{a}}'_\mathrm{pole}$ and
$\vect{\hat{a}}'_\mathrm{limb}$ for a while, we see that the visible shape of
a sphere (limb) transforms into another sphere with different apparent radius
and seen in an other direction which, especially for observer with a large
distance and small $\cos\theta_\mathrm{max}$ is close to the aberrated
direction of its sphere's centre. Both observers see the same surface section
of the sphere which is not surprising because they collect exactly the same
photons in their respective image. However, since both observers see the
sphere in a different direction, the observations may be interpreted as if the
sphere is rotated in the moving frame. This so called Terrell rotation has
first been described by \citep{Terrell:1959,Penrose:1959} and treated in more
detail by \citep{BurkeStrode:1991} for an infinitely distant observer. For a
distant observer, $\cos\theta$ is very close to unity for all latitude
circles and the rotation appears to be solid.

For a close observer the deviation between $\vect{\hat{a}}'_\mathrm{pole}$ and
$\vect{\hat{a}}'_\mathrm{limb}$ matters. It is a direct consequence of the
aberration shift of cone centres discussed above and it is also directly
visible in the loss of concentricity of the latitude circles in the image of
the moving observer (bottom of Fig.~\ref{Fig:SphereShape})\footnote{Strictly
  speaking, each latitude circle projects into a circle only if the projection
  axis is along the direction $\vect{\hat{a}}'_\theta$ towards its apparent
  centre.
  Since these directions differ slightly for each latitude, i.e. with
  $\theta$, all except one
  latitude circle may be slightly deformed, depending on the camera model used
  by the observer.}.
In Fig.~\ref{Fig:AbbLatCent} we show the modified aberration law
(\ref{cospsi_cone}) of the limb centre $\vect{\hat{a}}'_\theta$
for $\theta=\asin(r/R_\odot)$ at various distances $r$ and for $\beta=0.85$.

\begin{figure}[ht!]
\hspace*{\fill}
\includegraphics[viewport=10 40 345 330,width=7.000cm]
  {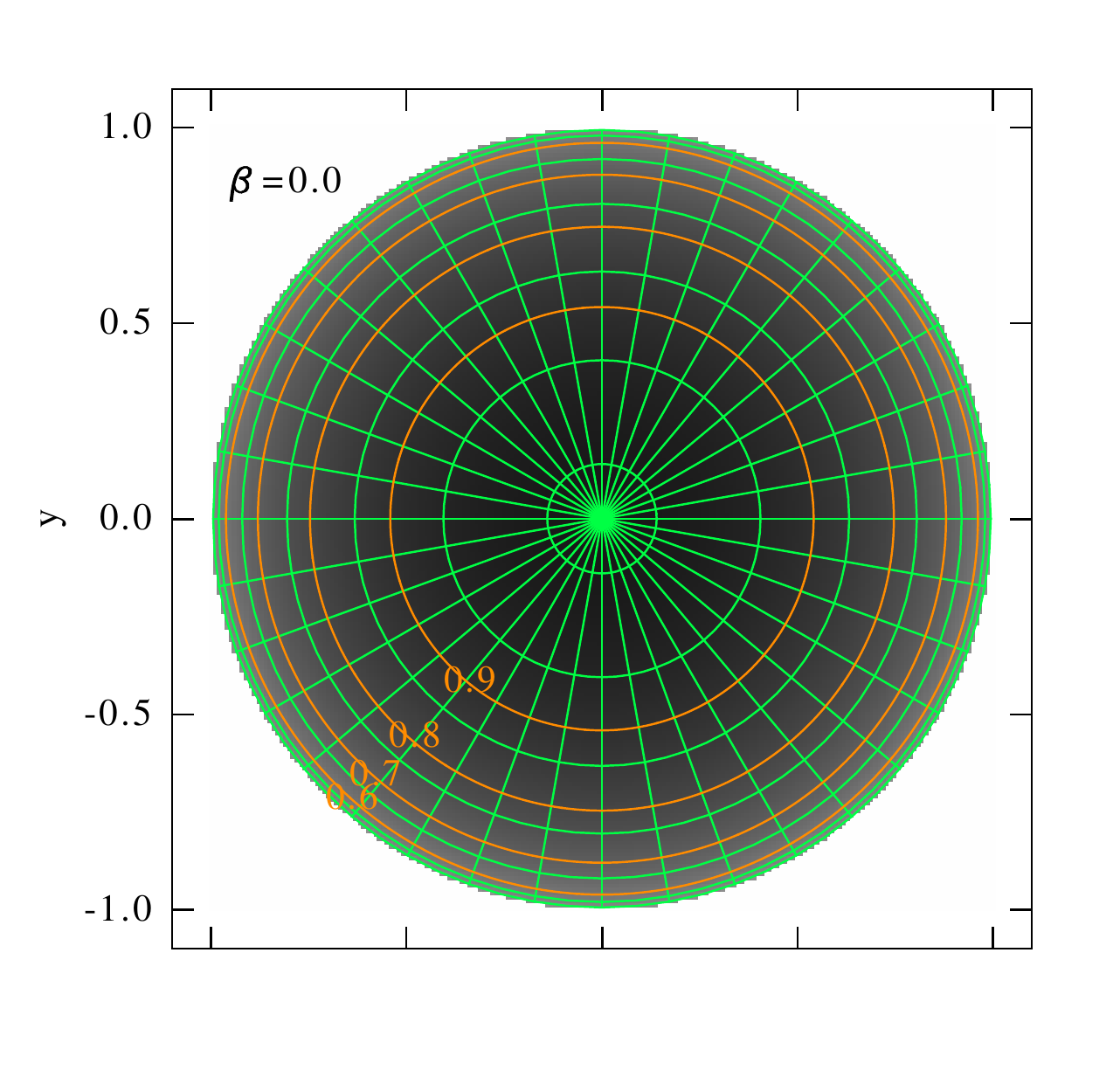}
\hspace*{0.5em}
\includegraphics[viewport=50 40 345 330,width=6.159cm]
  {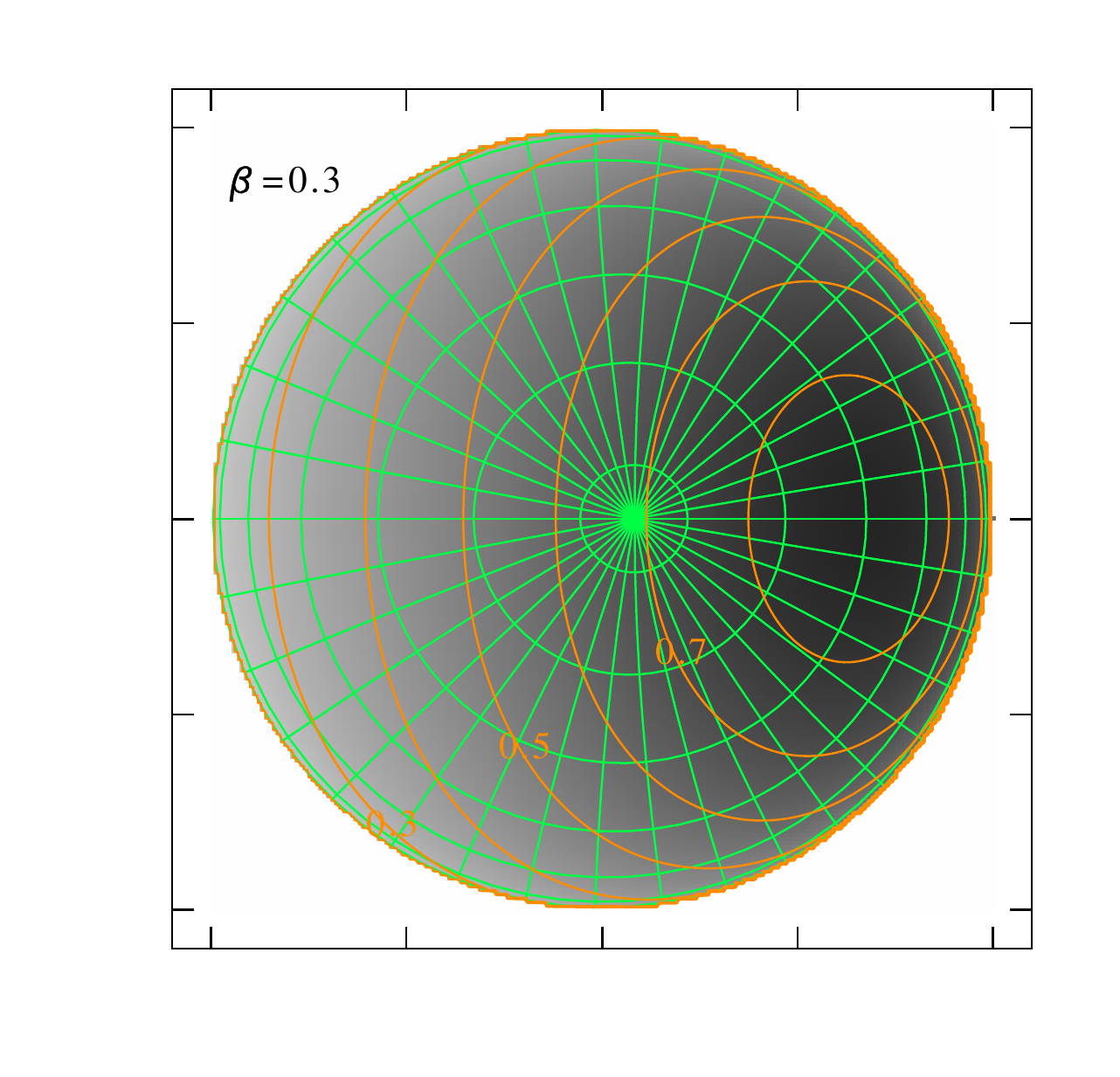}
\hspace*{\fill}
\\[0.5ex]
\hspace*{\fill}
\includegraphics[viewport=10 0 345 330,width=7.000cm]
  {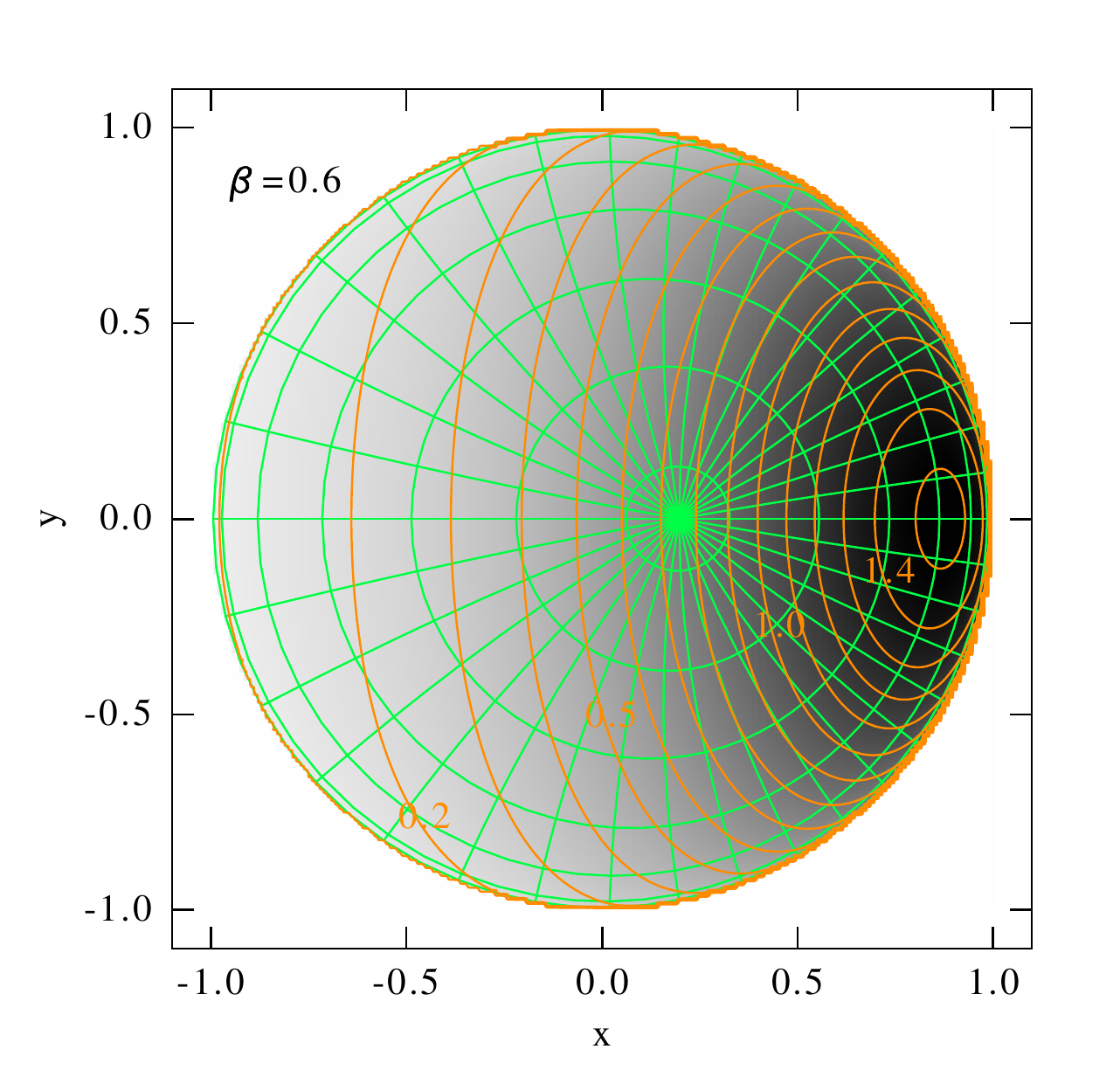}
\hspace*{0.5em}
\includegraphics[viewport=50 00 345 330,width=6.159cm]
  {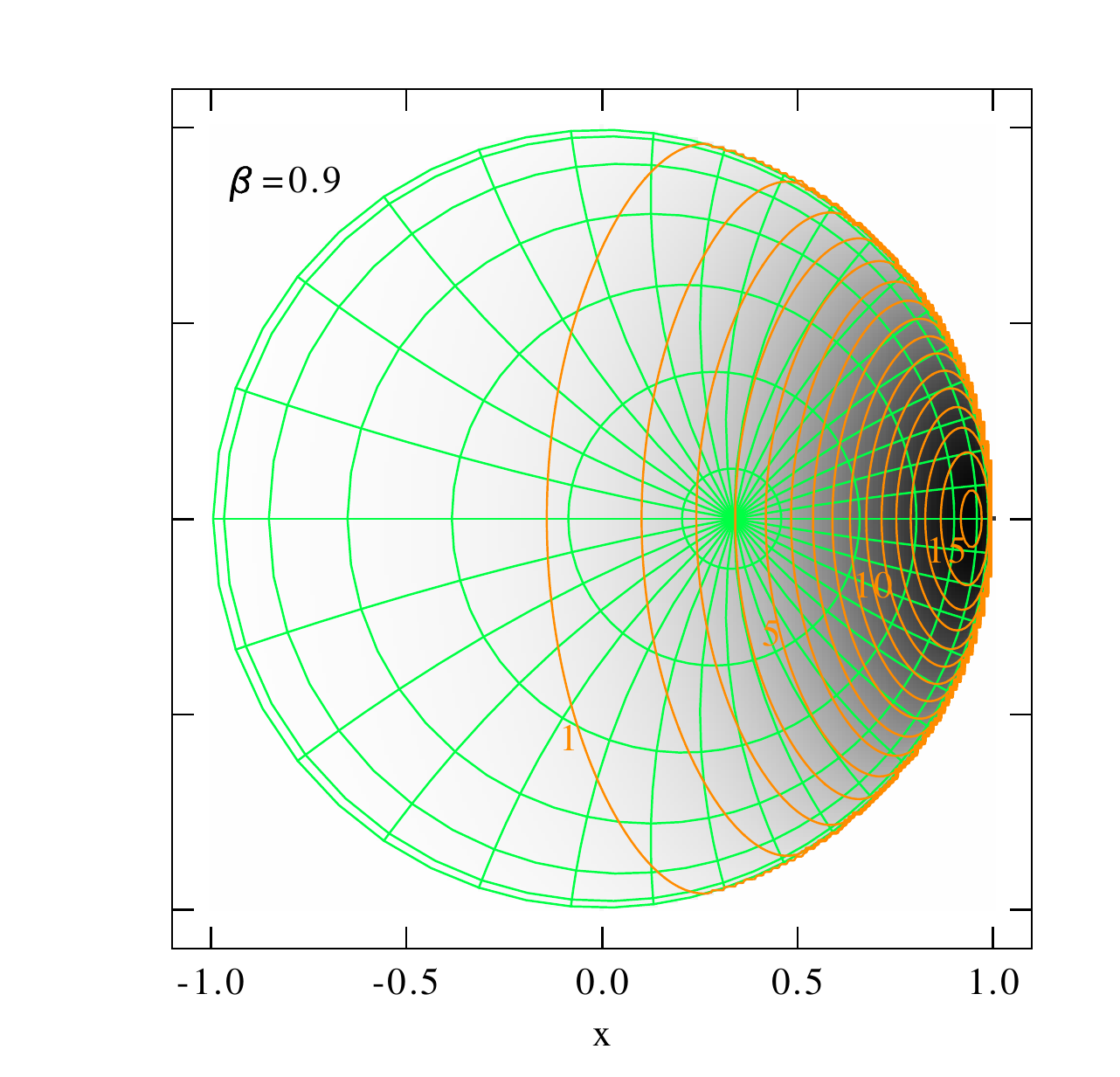}
\hspace*{\fill}
  \caption{Radiance distribution of the Sun as seen from a moving observer.
    The radiance in the Sun's rest frame includes limb darkening as in
    (\ref{Rad_Sun}). The position and velocity direction of the observer is
    the same as in Fig.~\ref{Fig:SphereShape} except that $\beta$ is varied as
    indicated. The radiance is shown in reverse grey scale with contours
    superposed in orange. The maximum radiance values are 1.0, 0.8581, 1.5271,
    16.2358 for the values of $\beta=$ 0, 0.3, 0.6 and 0.9, respectively. We
    have also superposed the spherical grid (green) with the pole pointing to
    the observer in rest frame of the sphere in as in
    Fig.~\ref{Fig:SphereShape}. The grid appears less distorted than the
    radiance distribution. \label{Fig:SphereRad}}
\end{figure}

The distortion of the visible surface of the sphere has consequences for
distribution of the surface radiance. Compared to the rest frame, the radiance
is enhanced inversely proportional to the area ratio of the surface grid
elements between rest frame and moving frame as, e.g., shown in in the lower
diagram of Fig.~\ref{Fig:SphereShape} \citep[see also][]{Kraus:2000}. This
apparent distortion of the sphere's surface is expressed in the transformation
of a space angular element $d\Omega'/d\Omega=D^2(\vect{\hat{k}},\vectg{\beta})$
(\ref{dOmegaTran}). In addition, the photon electric field is also transformed
so that apparent surface radiance effectively transforms to the moving frame
as in (\ref{RadTran}). As a result, the apparent
radiance distribution in the moving frame is even more shifted to the
forward hemisphere compared to the
observer pole and its maximum value becomes for our geometry drastically
enhanced for $\beta>0.4$. Recall that distortion of the visible surface not
only increases with increasing $\beta$ but also with decreasing distance $r$
to the sphere.

\subsection{Derivation of (\ref{pkin_trans})}
\label{app:Derv_pkin_trans}

In this chapter we prove (\ref{pkin_trans}) using the definitions
(\ref{Aberr_nin}) to (\ref{Aberr_numu}) for the two orthogonal
polarisation $\uect{p}_1$ and $\uect{p}_2$.

We insert the definitions (\ref{Aberr_kin}), (\ref{PolBase})
and (\ref{PolBase_}) to obtain for $i=1$
\[
 \uect{p}_1\tp\uect{k}\inc
=\sin\psi\inc \uectg{\nu}\sca\tp\uectg{\mu}\inc,\qquad
 \uect{p}'{\tp_1}\uect{k}'\inc
=\sin\psi'\inc \uectg{\nu}\sca\tp\uectg{\mu}\inc
\]
since $\uectg{\nu}\sca\perp\uectg{\beta}$. The ratio between
$\sin\psi'\inc$ and $\sin\psi\inc$ is given by the aberration
relation (\ref{Aberr_s}). Using definition (\ref{FrqDoppler}),
we have
\begin{gather*}
 \uect{p}'{\tp_1}\uect{k}'\inc
=\sin\psi'\inc \uectg{\nu}\sca\tp\uectg{\mu}\inc
=\frac{\sin\psi\inc}{\gamma(1-\beta\cos\psi\inc)}
 \uectg{\nu}\sca\tp\uectg{\mu}\inc
= D(\uect{k}\inc,\vectg{\beta})\;\uect{p}\tp_1\uect{k}\inc
\end{gather*}
This relation is in agreement with (\ref{pkin_trans}) because
$\uect{p}_1=\uect{p}'_1=\uectg{\nu}\sca$ is perpendicular to $\uectg{\beta}$.

For $i=2$ the derivation is somewhat more involved because
$\uect{p}_2$ is not perpendicular $\uectg{\beta}$. We again insert
(\ref{Aberr_kin}), (\ref{PolBase}) and (\ref{PolBase_}) and
replace the aberrated angles by (\ref{Aberr_c}) and (\ref{Aberr_s}).
\begin{gather}
\uect{p}_2\tp\uect{k}\inc
=\sin\psi\inc\cos\psi\sca\uectg{\mu}\inc\tp\uectg{\mu}\sca
-\sin\psi\sca\cos\psi\inc
\label{p2kin_0}\\[1ex]
\uect{p}'{_2\tp}\uect{k}'\inc
=(\frac{\cos\psi\inc-\beta}{1-\beta\cos\psi\inc}\uectg{\beta}
 +\frac{\gamma^{-1}\sin\psi\inc}{1-\beta\cos\psi\inc}\uectg{\mu}\inc)\tp
 (\frac{\cos\psi\sca-\beta}{1-\beta\cos\psi\sca}\uectg{\mu}\sca
 -\frac{\gamma^{-1}\sin\psi\sca}{1-\beta\cos\psi\sca}\uectg{\beta})
\nonumber\\
\stackrel{\uectg{\mu}\tp\uectg{\beta}=0}{=}\;\;
 \frac{\sin\psi\inc(\cos\psi\sca-\beta)
       \uectg{\mu}\inc\tp\uectg{\mu}\sca
       -\sin\psi\sca(\cos\psi\inc-\beta)}
  {\gamma(1-\beta\cos\psi\sca)(1-\beta\cos\psi\inc)}
\nonumber\\
=\frac{\sin\psi\inc\cos\psi\sca \uectg{\mu}\inc\tp\uectg{\mu}\sca
       -\sin\psi\sca\cos\psi\inc
       -\beta\sin\psi\inc \uectg{\mu}\inc\tp\uectg{\mu}\sca
       +\sin\psi\sca\beta}
  {\gamma(1-\beta\cos\psi\sca)(1-\beta\cos\psi\inc)}
\nonumber\\
\stackrel{\text(\ref{p2kin_0})}{=}
\frac{\uect{p}_2\tp\uect{k}\inc
       -\beta\sin\psi\inc \uectg{\mu}\inc\tp\uectg{\mu}\sca
       +\sin\psi\sca\beta}
  {\gamma(1-\beta\cos\psi\sca)(1-\beta\cos\psi\inc)}
\nonumber\\
=\frac{\uect{p}_2\tp\uect{k}\inc}
      {\gamma(1-\beta\cos\psi\inc)}
+\frac{\uect{p}_2\tp\uect{k}\inc
      -\uect{p}_2\tp\uect{k}\inc(1-\beta\cos\psi\sca)
       -\beta\sin\psi\inc \uectg{\mu}\inc\tp\uectg{\mu}\sca
       +\sin\psi\sca\beta}
  {\gamma(1-\beta\cos\psi\sca)(1-\beta\cos\psi\inc)}
\nonumber\\
=\frac{\uect{p}_2\tp\uect{k}\inc}
      {\gamma(1-\beta\cos\psi\inc)}
+\frac{\uect{p}_2\tp\uect{k}\inc\beta\cos\psi\sca
       -\beta\sin\psi\inc \uectg{\mu}\inc\tp\uectg{\mu}\sca
       +\sin\psi\sca\beta}
  {\gamma(1-\beta\cos\psi\sca)(1-\beta\cos\psi\inc)}
\nonumber\\\stackrel{\text{(\ref{FrqDoppler})}}{=}
D(\uect{k}\inc,\vectg{\beta})\big(\uect{p}_2\tp\uect{k}\inc
+\frac{\uect{p}_2\tp\uect{k}\inc\beta\cos\psi\sca
       -\beta\sin\psi\inc \uectg{\mu}\inc\tp\uectg{\mu}\sca
       +\sin\psi\sca\beta}
    {1-\beta\cos\psi\sca}\big)
\label{p2kin_1}\end{gather}
In the last steps we have split off the first term in (\ref{pkin_trans}). The
second term on the right-hand-side has already the correct denominator
compared with (\ref{pkin_trans}) so that it only remains to analyse the
numerator of the second term in (\ref{p2kin_1}). Inserting the expressions
for $\uect{p}_2\tp\uect{k}\inc$ and $\uectg{\mu}\inc\tp\uectg{\mu}\sca$ we
find
\begin{gather*}
  \uect{p}_2\tp\uect{k}\inc\beta\cos\psi\sca
       -\beta\sin\psi\inc \uectg{\mu}\inc\tp\uectg{\mu}\sca
       +\sin\psi\sca\beta
\\ \stackrel{\text{(\ref{p2kin_0})}}{=}       
  (\sin\psi\inc\cos\psi\sca\uectg{\mu}\inc\tp\uectg{\mu}\sca
-\sin\psi\sca\cos\psi\inc)
        \beta\cos\psi\sca
       -\beta\sin\psi\inc \uectg{\mu}\inc\tp\uectg{\mu}\sca
       +\sin\psi\sca\beta
\\
 =\sin\psi\inc\cos\psi\sca\uectg{\mu}\inc\tp\uectg{\mu}\sca\beta\cos\psi\sca
-\sin\psi\sca\cos\psi\inc\beta\cos\psi\sca
       -\beta\sin\psi\inc \uectg{\mu}\inc\tp\uectg{\mu}\sca
       +\sin\psi\sca\beta
\\
 =\beta\sin\psi\inc(\cos^2\psi\sca-1)
 \uectg{\mu}\inc\tp\uectg{\mu}
 +\beta\sin\psi\sca(1-\cos\psi\inc\cos\psi\sca)
\\ \stackrel{\text{(\ref{Aberr_mumu})}}{=}       
 -\beta\frac{\sin\psi\inc\sin^2\psi\sca}{\sin\psi\inc\sin\psi\sca}
  \uect{k}\inc\tp(\1-\uectg{\beta}\uectg{\beta}{}\tp)\uect{k}\sca
 +\beta\sin\psi\sca(1-\cos\psi\inc\cos\psi\sca)
\\ 
=-\beta\sin\psi\sca(\cos\chi-\cos\psi\inc\cos\psi\sca)
 +\beta\sin\psi\sca(1-\cos\psi\inc\cos\psi\sca)
\\ 
=\beta\sin\psi\sca(1-\cos\chi)
\stackrel{\text{(\ref{PolBase})}}{=}       
-(1-\cos\chi)\uect{p}_2\tp\uectg{\beta}
\end{gather*}
Inserting the numerator in (\ref{p2kin_1}) we obtain (\ref{pkin_trans})
\[
 \uect{p}'{_2\tp}\uect{k}'\inc
=D(\uect{k}\inc,\vectg{\beta})\big(\uect{p}_2\tp\uect{k}\inc
 -\frac{1-\cos\chi}{1-\beta\cos\psi\sca}
      \uect{p}_2\tp\uectg{\beta}\big)
\]

\section{Alternative derivation of the scattered radiant intensity}
\label{App:Derv_altern}

\newcommand{\vdist}{\ensuremath{\boldsymbol{\ell}}}
\newcommand{\udist}{\ensuremath{\boldsymbol{\hat{\ell}}}}

This derivation follows the approach used in the plasma physics and plasma
diagnostics literature \citep[e.g.,][]{SegreZanza:2000,Hutchinson:2002}.
It does not use any transformations but rests entirely
in the observer frame. The setup used in lab experiments differs largely from
the situation of Thomson scattering in the corona. In particular, an
unpolarised incident beam is not used in the lab and is not treated in related
scattering calculations I am aware of.

\subsection{Lienard-Wiechert potential}
\setcounter{equation}{0}
\label{app:LienardWichert}

We start with the electromagnetic potential of an accelerated electron with
orbit $\vect{r}(t)$ which passes the site of the scattering which we
called $\vect{r}$
in the main text at the retarded time $t\ret$. The coordinates $(t,\vect{x})$
denote time and location at which the scattered field is observed.
All times and coordinates are in the observer frame, the dash in this chapter
just marks integration variables.
The potential is \citep[e.g.,][]{Jackson:1998}
\begin{gather*}
  \phi(t,\vect{x}) = \frac{e}{4\pi\epsilon_0}
  \int_{-\infty}^t dt' \frac{1}{|\vect{x}-\vect{r}(t')|}
  \delta(t-t'-\frac{|\vect{x}-\vect{r}(t')|}{c})
\\  
  \vect{A}(t,\vect{x}) = \frac{e\mu_0}{4\pi}
  \int_{-\infty}^t dt' \frac{\vect{v}(t')}{|\vect{x}-\vect{r}(t')|}
  \delta(t-t'-\frac{|\vect{x}-\vect{r}(t')|}{c})
\end{gather*}
By the $\delta$ function integration, we can treat $t'$ as an ordinary
variable and move the evaluation of the retarded time to the final evaluation
of the integral. The retarded time is only given implicitly by the
intersection of the particle world line $(ct',\vect{r}(t'))$with
the backwards light cone $c(t'-t)-|\vect{x}-\vect{r}(t')|=0$
from the observation event at $(ct,\vect{x})$.
Formally, it is the solution for $t'$ of $t'=t-|\vect{x}-\vect{r}(t')|/c$ .
By $\vectg{\beta}=\dot{\vect{r}}/c$ we again denote the particle velocity.
To keep the formulas short, we introduce some abbreviations
\begin{align*}
  \vdist(t',\vect{x})&=\vect{x}-\vect{r}(t') &&
    \text{distance vector from emitting particle to observer}\\
  \beta_\dist(t') &= \vectg{\beta}(t')\tp\udist(t') &&
  \text{projection of $\vectg{\beta}$ along the direction of $\vdist$}\\
  T(t',\vect{x})&=|\vect{x}-\vect{r}(t')|/c &&
    \text{travel time}\\
  \vartheta(t',t,\vect{x})&=t-t'-T(t',\vect{x})    &&
     \text{$\delta$-function argument}\\
  t\ret \quad &\;\;|\quad \vartheta(t\ret,t,\vect{x})=0 &&
  \text{retarded time}
\end{align*}
With $\mu_0=1/c^2\epsilon_0$ and the above abbreviations we can write
the potential concisely as
\begin{gather*}
  \frac{\phi}{c}(t,\vect{x}) = \frac{e}{4\pi\epsilon_0 c^2}
  \int_{-\infty}^t dt' \frac{1}{T(t',\vect{x})}
  \delta(\vartheta(t',t,\vect{x}))
\\  
  \vect{A}(t,\vect{x}) = \frac{e}{4\pi\epsilon_0 c^2}
  \int_{-\infty}^t dt' \frac{\vectg{\beta}(t')}{T(t',\vect{x})}
  \delta(\vartheta(t',t,\vect{x}))
\end{gather*}
To derive the wave field of the moving charge we have to differentiate with
respect to $\vect{x}$ and $t$. Later we will also have to differentiate with
respect to $t'$. The dependence on these arguments is made clear above by
listing all arguments explicitly. Here a list of partial derivatives we
will need
\begin{align}
  \dpar{}{t'}T(t',\vect{x})
 =&\dpar{}{t'}\frac{|\vect{x}-\vect{r}(t')|}{c}
 =-\frac{(\vect{x}-\vect{r}(t'))\tp}{|\vect{x}-\vect{r}(t')|}
     \frac{\vect{\dot{r}}(t')}{c}
\nonumber\\
 =&-\udist{}\tp(t')\vectg{\beta}(t')=-\beta_\ell(t')
\label{dtd_T}\\ 
  c\grd_\vect{x} T(t',\vect{x})
 =&\grd_\vect{x} |\vect{x}-\vect{r}(t')|
 =\frac{\vect{x}-\vect{r}(t')}{|\vect{x}-\vect{r}(t')|}
 =\udist(t',\vect{x})
\label{grd_T}\\ 
  c\grd_\vect{x} \frac{1}{T(t',\vect{x})}
 =&c^2\grd_\vect{x} \frac{1}{|\vect{x}-\vect{r}(t')|}
 =-c^2\frac{\vect{x}-\vect{r}(t')}{|\vect{x}-\vect{r}(t')|^3}
 =-\frac{\udist(t',\vect{x})}{T^2(t',\vect{x})}
\label{grd_invT}\\[0.5ex] 
  \dpar{}{t}\vartheta(t',t,\vect{x})
 =&\dpar{}{t}\!(t-t'-T(t',\vect{x})) = 1
\label{dtd_vartheta}\\
  \dpar{}{t'}\vartheta(t',t,\vect{x})
 =&\dpar{}{t'}\!(t-t'-T(t',\vect{x}))
 =-1-\dpar{}{t'}T(t',\vect{x})\\
 =&-1+\udist{}\tp(t')\vectg{\beta}(t')
 =-(1-\beta_\dist(t'))
 =-\kappa(t')
\label{dt_vartheta}\\
  c\grd_\vect{x}\vartheta(t',t,\vect{x})
  =& c\grd_\vect{x}(t-t'-T(t',\vect{x}))
  =-c\grd_\vect{x} T(t',\vect{x})
  =-\udist(t',\vect{x})
\label{grd_vartheta}\end{align}
With these rules the electric field becomes
\begin{gather*}
  \vect{E}(t,\vect{x})=
  -c\grd_\vect{x}\frac{\phi}{c}-\dpar{}{t}\!\vect{A}
\\
\stackrel{\text{(\ref{grd_invT})}}{=}
\frac{e}{4\pi\epsilon_0 c^2}
  \int_{-\infty}^t dt' \big[
           \frac{\udist(t',\vect{x})}{T^2(t',\vect{x})}\;
           \delta(\vartheta(t,\vect{x},t'))
           -\frac{1}{T(t',\vect{x})}\,
           c\grd_\vect{x}\delta(\vartheta(t,\vect{x},t'))\\
           -\frac{\vectg{\beta}(t')}{T(t',\vect{x})}
           \dpar{}{t}\delta(\vartheta(t,\vect{x},t'))
           \big]
\\  
= \frac{e}{4\pi\epsilon_0 c^2}
  \int_{-\infty}^t dt' \big[
           \frac{\udist(t',\vect{x})}{T^2(t',\vect{x})}\;
           \delta(\vartheta(t,\vect{x},t'))
        -\frac{\delta'(\vartheta(t,\vect{x},t'))}{T(t',\vect{x})}
           (c\grd_\vect{x}+\vectg{\beta}(t')\dpar{}{t})\vartheta(t,\vect{x},t')
           \big]
\\  
\stackrel{\text{(\ref{dt_vartheta},\ref{grd_vartheta})}}{=}
 \frac{e}{4\pi\epsilon_0 c^2}
  \int_{-\infty}^t dt' \big[
           \frac{\udist(t',\vect{x})}{T^2(t',\vect{x})}\;
           \delta(\vartheta(t,\vect{x},t'))
        -\frac{\delta'(\vartheta(t,\vect{x},t'))}{T(t',\vect{x})}
           (-\udist(t',\vect{x})+\vectg{\beta}(t'))
           \big]
\\  
=\frac{e}{4\pi\epsilon_0 c^2}
  \int_{-\infty}^t dt' \big[
           \frac{\udist(t',\vect{x})}{T^2(t',\vect{x})}\;
           \delta(\vartheta(t,\vect{x},t'))
        +\frac{\udist(t',\vect{x})-\vectg{\beta}(t')}{T(t',\vect{x})}
           \delta'(\vartheta(t,\vect{x},t'))
           \big]
\end{gather*}
Next we replace the integration variable $t'$ by $\vartheta(t,\vect{x},t')$.
Then
the particle time $t'$ becomes a function of $t,\vect{x}$ and $\vartheta$. The
Jacobian of this transformation has the inverse magnitude of
$|\dpar{}{t'}\vartheta(t,\vect{x},t')|=|\kappa(t',\vect{x})|$, see
(\ref{dtd_vartheta}). Note that since $\beta_\ell\le\beta<1$,
$\kappa=1-\beta_\ell$ is always positive. We can therefore substitute
$dt'=|\kappa|^{-1}\;d\vartheta=\kappa^{-1}\;d\vartheta$. The old integration
boundaries $t'=-\infty\dots t$ map to $\vartheta=-T(t,\vect{x})\dots\infty$.
Since $T>0$ the $\delta$-function gives exactly one contribution at
$\vartheta=0$ and the lower boundary does not really matter. We 
can replace it by $-\infty$.
\begin{gather*}
  \vect{E}(t,\vect{x})
= \frac{e}{4\pi\epsilon_0 c^2}
    \int_{-\infty}^\infty  \frac{d\vartheta}{\kappa(t'(\vartheta),\vect{x})}
    \big[\frac{\udist(t'(\vartheta),\vect{x})}
              {T^2(t'(\vartheta),\vect{x})} \delta(\vartheta)
          +\frac{\udist(t'(\vartheta),\vect{x})-\vectg{\beta}(t'(\vartheta))}
              {T(t'(\vartheta),\vect{x})} \delta'(\vartheta)
     \big]
\end{gather*}
Next we partially integrate the 2nd term to get rid of the derivative
of the $\delta$ function and change the derivative with respect to $\vartheta$
back to a derivative with respect to $t'$:
\[
  \int d\vartheta \;[\frac{\udist-\vectg{\beta}}{\kappa T}]
  \;\frac{d\delta}{d\vartheta}
=-\int d\vartheta \;\delta\;\frac{d}{d\vartheta}
  [\frac{\udist-\vectg{\beta}}{\kappa T}]
=-\int d\vartheta \;\delta\; \frac{1}{(-\kappa)}\frac{d}{dt'}
  [\frac{\udist-\vectg{\beta}}{\kappa T}]
\]
Gives
\begin{equation}
  \vect{E}(t,\vect{x})
= \frac{e}{4\pi\epsilon_0 c^2}
    \int_{-\infty}^\infty d\vartheta
   \frac{\delta(\vartheta)}{\kappa(t'(\vartheta),\vect{x})} 
     \big[\frac{\udist(t'(\vartheta),\vect{x})}
              {T^2(t'(\vartheta),\vect{x})} 
           +\frac{d}{dt'}\big(
          \frac{\udist(t',\vect{x})-\vectg{\beta}(t')}
               {\kappa(t',\vect{x})T(t',\vect{x})}\big)_{t'=t'(\vartheta)}
           \big]
\label{Efield1}\end{equation}
For the derivative with respect to $t'$ in (\ref{Efield1}) we use
(\ref{dtd_T}) and (\ref{dtd_vartheta}) and
(omitting the arguments)
\begin{gather*}
     \frac{d}{dt'}\vectg{\beta}=\vectg{\dot\beta}
\qquad     
     \frac{d}{dt'}\vdist=\frac{d}{dt'}(\vect{x}-\vect{r})
     =-c\vectg{\beta}
\\
     \frac{d}{dt'}\frac{1}{T}
     =-\frac{1}{T^2}\frac{d}{dt'}T
     \stackrel{\text{(\ref{dtd_T})}}{=}
     \frac{1}{T^2}\beta_\dist
\\
     \frac{d}{dt'}\udist
    =\frac{d}{dt'}\frac{\vdist}{\dist}
    =-\frac{1}{T}\vectg{\beta}
     +\frac{\vdist}{c}\;\frac{d}{dt'}\frac{1}{T}
    =-\frac{1}{T}\vectg{\beta}
    + \frac{\vdist}{c}\;\frac{\beta_\dist}{T^2}
    =-\frac{1}{T}\vectg{\beta}
     +\udist\;\frac{1}{T}\beta_\dist\\
    = \frac{1}{T}(\beta_\dist\udist-\vectg{\beta})
    = -\frac{1}{T}(\1-\udist\udist{}\tp)\vectg{\beta}
\\
      \frac{d}{dt'}\kappa
    =\frac{d}{dt'}(1-\beta_\dist)
    =-\frac{d}{dt'}\udist{}\tp\vectg{\beta}
    =-\udist{}\tp\vectg{\dot\beta}
    +\frac{1}{T}\vectg{\beta}\tp(1-\udist
                \udist{}\tp)\vectg{\beta}
    =-\dot\beta_\dist+\frac{1}{T}(\beta^2-\beta_\dist^2)
\\
    \frac{d}{dt'}\frac{1}{\kappa}
   =-\frac{1}{\kappa^2}\frac{d}{dt'}\kappa
   =\frac{1}{\kappa^2}(\dot\beta_\dist+\frac{1}{T}(\beta_\dist^2-\beta^2))
\end{gather*}
The square bracket in the integrand of (\ref{Efield1}) then gives
\begin{gather*}
  \frac{\udist}{T^2} +
   \frac{d}{dt'}\frac{\udist-\vectg{\beta}}{\kappa\,T}
\\
 = \underbrace{\frac{\kappa^2\udist}{\kappa^2 T^2}}%
_{\text{1}}
 + \frac{1}{\kappa T}
    [\underbrace{\frac{\beta_\dist\udist-\vectg{\beta}}{T}}%
_{\text{2}}
    -\underbrace{\vectg{\dot\beta}}%
_{\text{3}}]
 +\underbrace{\frac{\udist-\vectg{\beta}}{\kappa}\,\frac{\beta_\dist}{T^2}}%
_{\text{4}}
 + \frac{\udist-\vectg{\beta}}{T}\,
   \frac{1}{\kappa^2}(
   \underbrace{\dot\beta_\dist}%
_{\text{5}}
  +\underbrace{\frac{1}{T}(\beta_\dist^2-\beta^2)}%
_{\text{6}})
\end{gather*}
We first
order the terms according to powers of $T^{-1}$ and secondly according to the
vector coefficient\\[-4ex]
\begin{gather*}
  \frac{\udist}{T^2} +
   \frac{d}{dt'}\frac{\udist-\vectg{\beta}}{\kappa\,T}
 = \frac{1}{\kappa^2 T}[
   -\overbrace{\kappa\vectg{\dot\beta}}^{3}
   +\overbrace{(\udist-\vectg{\beta})\dot\beta_\dist}^{5}
   ]
\\ 
 + \frac{1}{\kappa^2 T^2}[
    \overbrace{\kappa^2\udist}^{1}
   +\overbrace{\kappa(\beta_\dist\udist-\vectg{\beta})}^{2}
   +\overbrace{\kappa(\udist-\vectg{\beta})\beta_\dist}^{4}
   +\overbrace{(\udist-\vectg{\beta})(\beta_\dist^2-\beta^2)}^{6}]
 \\ \text{term}\propto \frac{1}{\kappa^2 T}: \qquad
  -\kappa\vectg{\dot\beta}+(\udist-\vectg{\beta})\dot\beta_\dist
 =(\udist-\vectg{\beta})\dot\beta_\dist - (1-\beta_\dist)\vectg{\dot\beta}
 \\
 =(\udist-\vectg{\beta}) \udist{}\tp\vectg{\dot\beta}
 - \udist{}\tp(\udist-\vect{\beta})\vectg{\dot\beta}
 =\udist\times(\udist-\vectg{\beta})\times\vectg{\dot\beta}
 \\ \text{term}\propto \frac{1}{\kappa^2 T^2}: \qquad
    \kappa^2\udist
   +\kappa(\beta_\dist\udist-\vectg{\beta})
   +\kappa(\udist-\vectg{\beta})\beta_\dist
   +(\udist-\vectg{\beta})(\beta_\dist^2-\beta^2)
\\
  =(\kappa^2+2\kappa\beta_\dist+\beta_\dist^2-\beta^2)\udist
  -(\kappa+\kappa\beta_\dist +\beta^2_\dist-\beta^2)\vectg{\beta}
\\
  =((\kappa+\beta_\dist)^2-\beta^2)\udist
  -(\kappa(1+\beta_\dist) +\beta^2_\dist-\beta^2)\vectg{\beta}
\\
  =(1-\beta^2)\udist
  -(1-\beta_\dist^2 +\beta^2_\dist-\beta^2)\vectg{\beta}
\\
   = (1-\beta^2)\udist
    -(1-\beta^2)\vectg{\beta}
   = \frac{1}{\gamma^2}(\udist-\vectg{\beta})
\end{gather*}
Inserting the two final terms for the square bracket of (\ref{Efield1}), we find
\begin{gather*}
  \vect{E}(t,\vect{x})
= \frac{e}{4\pi\epsilon_0 c^2}
    \int_{-\infty}^{\infty}  \frac{d\vartheta}{\kappa^3}
    \big[
      \frac{\udist\times(\udist-\vectg{\beta})
        \times\vectg{\dot\beta}}{T} 
   +\frac{\udist-\vectg{\beta}}{\gamma^2T^2}
    \big]\delta(\vartheta)
\end{gather*}
where $\delta(\vartheta)$ fixes $t'$ to the retarded time $t\ret$.
Since the retarded time can be explicitly calculated only in few cases
we can execute the integration only symbolically by writing
\begin{equation}
  \vect{E}(t,\vect{x})
= \frac{e}{4\pi\epsilon_0 c^2}
    \big[
     \frac{\udist\times(\udist-\vectg{\beta})
        \times\vectg{\dot\beta}}{\kappa^3 T} 
    +\frac{\udist-\vectg{\beta}}{\gamma^2\kappa^3 T^2}
    \big]_{t\ret}
\label{AccChargeField}\end{equation}
The first term depends on the acceleration of the particle and is the
radiative part of the field. It decreases with $T^{-1}=c/\dist$ from the
retarded position of the particle. In the far field only this term is
important. Obviously, the radiating part of the field is perpendicular
to the retarded direction $\udist(t\ret)$ from the source.
This part will be discussed further below as the source of the scattered
wave field.

\subsection{Electrostatic part}

The second term in (\ref{AccChargeField}) is the electrostatic field component
decreasing more rapidly with $T^{-2}=(c/\dist)^2$. The static field is
largely longitudinal. For a particle with a constant, non-accelerated velocity
this is the only term in (\ref{AccChargeField}). We assume the
charged particle moves with constant $\vect{v}=c\vectg{\beta}$ along the
trajectory $\vect{r}(t)=\vect{r}_0+\vect{v}t$.
Then the field observed at $(t,\vect{x})$ is directed along
\begin{gather*}
 \udist(t\ret)-\vectg{\beta}
=\frac{\vect{x}-\vect{r}_0-\vect{v}t\ret}{\dist(t\ret)}-\frac{\vect{v}}{c}
=\frac{\vect{x}-\vect{r}_0-\vect{v}t\ret-\vect{v}d(t\ret)/c}{\dist(t\ret)}
\\
=\frac{\vect{x}-\vect{r}_0-\vect{v}t\ret-\vect{v}T(t\ret)}{\dist(t\ret)} 
=\frac{\vect{x}-\vect{r}_0-\vect{v}t}{\dist(t\ret)} 
=\frac{\vdist(t)}{\dist(t\ret)} 
\end{gather*}
This has not the retarded but the actual direction from the source.
The field  therefore becomes
\begin{equation}
  \vect{E}_\mathrm{stat}(t,\vect{x})
= \frac{e}{4\pi\epsilon_0 c^2}
    \big[
    \frac{\udist-\vectg{\beta}}{\gamma^2\kappa^3 T^2}
    \big]_{t\ret}
= \frac{e}{4\pi\epsilon_0}
  \frac{\vdist(t)}{\gamma^2\kappa^3 \dist^3(t\ret)}
\label{statEfield}\end{equation}
The denominator can be evaluated as follows:
Be $d$ the minimum distance of the particle path from the observer at
$\vect{x}$ and the point $\vect{r}_0$ of closest approach passed
by the particle at $t=0$. Then
\citep{RybickiLightman:1979}
\begin{gather}
  \vdist(t)=\vect{x}-\vect{r(t)}=(-vt,d)
  \quad\text{in a suitably rotated coordinate system.}
  \nonumber\\
  \text{Then}\quad \dist^2(t)=d^2+v^2 t^2 \quad\text{and}
  \nonumber\\
  \dist^2(t\ret)
  =d^2+v^2 t\ret^2
  =d^2+v^2(t-\frac{\dist(t\ret)}{c})^2
  =\dist^2(t) - 2 \beta v t\;\dist(t\ret) +\beta^2 \dist^2(t\ret)
  \nonumber\\
\text{give}\quad
  \dist^2(t\ret) +2\gamma^2\beta v t\;\dist(t\ret)
  -\gamma^2\dist^2(t)=0
  \nonumber\\\text{or}\quad
  \dist(t\ret)
  =-\gamma^2\beta v t
  +\sqrt{\rule{0mm}{2ex}\gamma^4\beta v^2 t^2+\gamma^2\dist^2(t)}
  =-\gamma^2\beta v t
  +\gamma\sqrt{\rule{0mm}{2ex}(\gamma\beta v t)^2 + \dist^2(t)}
\label{retdist}\\
\text{Next evaluate}\quad
 \gamma^2\kappa\dist(t\ret)
=\gamma^2(\dist(t\ret) - \vectg{\beta}\tp\vdist(r\ret))
=\gamma^2(\dist(t\ret) + \beta vt\ret)
\nonumber\\
=\gamma^2(\dist(t\ret) + \beta v(t-\frac{\dist(t\ret)}{c}))
=\gamma^2(\dist(t\ret) + \beta vt -\beta^2\dist(t\ret))
=\dist(t\ret) + \gamma^2\beta vt
\label{kapparetdist}\\
\text{compare (\ref{retdist}) and (\ref{kapparetdist})}\quad
 \gamma^2\kappa\dist(t\ret)
 =\gamma\sqrt{\rule{0mm}{2ex}(\gamma\beta v t)^2+\dist^2(t)}
\nonumber\end{gather}
Insertion in (\ref{statEfield}) yields the electrostatic field
in a form where the retarded time does not occur any more
\begin{gather}
  \vect{E}_\mathrm{stat}(t,\vect{x})
= \frac{e}{4\pi\epsilon_0}
  \frac{\gamma\vdist(t)}{\rule{0mm}{2ex}((\gamma\beta v t)^2+\dist^2(t))^{3/2}}
= \frac{e}{4\pi\epsilon_0}
  \frac{\gamma\vdist(t)}{\rule{0mm}{2ex}((\gamma^2\beta^2+1) (v t)^2+d^2)^{3/2}}
\nonumber\\  
= \frac{e}{4\pi\epsilon_0}
  \frac{\gamma\vdist(t)}{\rule{0mm}{2ex}((\gamma v t)^2+d^2)^{3/2}}
\label{statEfield2}\end{gather}

\subsection{Acceleration of a point charge in a wave field}
\label{app:AccChargeEF}

For the radiating part of the field in (\ref{AccChargeField}) we specify the
particle acceleration as being the result of an incident electromagnetic wave
which propagates in direction $\uect{k}\inc$ and has the wave field
$\vect{E}\inc(t,\vect{x})$ and $c\vect{B}\inc(t,\vect{x})=
\uect{k}\inc\times\vect{E}\inc(t,\vect{x})$.
Like above, we will also abbreviate the projection of
$\vectg{\beta}$ along $\uect{k}\inc$ by
$\beta\inc=\vectg{\beta}\tp\uect{k}\inc=\beta\cos\psi\inc$.
The relativistic equation of motion of an electron with rest mass $m_e$ is
\citep[e.g.,][]{Jackson:1998}
\begin{gather}
 \Dpar{}{t}\vect{p} = \Dpar{}{t}\gamma m_e\vect{v}
 =-e(\vect{E}\inc+\vect{v}\times\vect{B}\inc)
 =-e(\vect{E}\inc+\vectg{\beta}\times c\vect{B}\inc)
 =-e[(1-\beta\inc)\1+\uect{k}\inc\vectg{\beta}\tp]\vect{E}\inc
\nonumber\\
 \text{or}\quad\Dpar{}{t}\gamma\vectg{\beta}
 =-\frac{e}{m_e c}
  [(1-\beta\inc)\1+\uect{k}\inc\vectg{\beta}\tp]\vect{E}\inc
\nonumber\\
 \text{Using}\quad\Dpar{}{t}\gamma=
 \Dpar{}{t}\frac{1}{\sqrt{1-\beta^2}}
=\frac{\vectg{\beta}\tp\vectg{\dot{\beta}}}{\sqrt{1-\beta^2}^3}
=\gamma^3\vectg{\beta}\tp\vectg{\dot{\beta}}
\quad\text{we find}\nonumber\\ 
 \Dpar{}{t}\gamma\vectg{\beta}
=\dot\gamma\vectg{\beta}+\gamma\vectg{\dot\beta}
=\gamma^3\vectg{\beta}\;\vectg{\beta}\tp\vectg{\dot\beta}
+\gamma\vectg{\dot\beta}
=\gamma(\gamma^2\vectg{\beta}\;\vectg{\beta}\tp\vectg{\dot\beta}
       +\vectg{\dot\beta})
\nonumber\\       
=-\frac{e}{m_e c}
  [(1-\beta\inc)\1+\uect{k}\inc\vectg{\beta}\tp]\vect{E}\inc
\label{RelMom1}
\end{gather}
We first solve for the projection along $\vectg{\beta}$.
A scalar multiplication of (\ref{RelMom1}) with $\vectg{\beta}$ yields
\[
\gamma^2\beta^2\;\vectg{\beta}\tp\vectg{\dot\beta}
  +\vectg{\beta}\tp\vectg{\dot\beta}
=(\gamma^2\beta^2+1)   \vectg{\beta}\tp\vectg{\dot\beta}
=\gamma^2 \vectg{\beta}\tp\vectg{\dot\beta}
=-\frac{e}{\gamma m_e c}\vectg{\beta}\tp\vect{E}\inc
\]
Insertion of $\gamma^2\vectg{\beta}\tp\vectg{\dot\beta}$ in
(\ref{RelMom1}) gives
\begin{gather}
  \vectg{\dot\beta}
=-\frac{e}{\gamma m_e c}
  [(1-\beta\inc)\1+\uect{k}\inc\vectg{\beta}\tp]\vect{E}\inc
 -\gamma^2\vectg{\beta}\;\vectg{\beta}\tp\vectg{\dot\beta}
\nonumber\\ 
=-\frac{e}{\gamma m_e c}
  [(1-\beta\inc)\1+\uect{k}\inc\vectg{\beta}\tp]\vect{E}\inc
 +\frac{e}{\gamma m_e c}\vectg{\beta}\;\vectg{\beta}\tp\vect{E}\inc
\nonumber\\
=-\frac{e}{\gamma m_e c} ((1-\beta\inc)\1
  +(\uect{k}\inc-\vectg{\beta})\vectg{\beta}\tp] \vect{E}\inc
\label{ElecAcc}\end{gather}
Hence the acceleration in the wave field is not just parallel
to the electric field as in the non-relativistic case. There is also
a component in the direction of wave propagation from the wave Lorentz force
which transfers some momentum from the photon to the electron. There is also a
breaking component along $-\vectg{\beta}$ which prevents the velocity to
exceed $\beta=1$.

\subsection{Scattered wave field}

For the scattering problem, only the radiative term in (\ref{AccChargeField})
counts where the acceleration $\vectg{\dot{\beta}}$ is replaced by
(\ref{ElecAcc}). The electric field (\ref{AccChargeField}) is then
called $\vect{E}\sca$ and the direction $\udist$ corresponds to the
propagation direction $\uect{k}\sca$ of the scattered wave.
Accordingly, $\beta_\dist$ will be renamed in
$\beta\sca=\vectg{\beta}\tp\uect{k}\sca=\beta\cos\psi\sca$.
Moreover, we abbreviate
\[
\vect{h}\inc=\uect{k}\inc-\vectg{\beta},\qquad
\vect{h}\sca=\uect{k}\sca-\vectg{\beta}
\]
Note the vectors $\vect{h}\inc$ and $\vect{h}\sca$ are not unit vectors
any more. In particular
\begin{gather*}
 \vect{h}\sca\tp\uect{k}\sca=1-\beta\sca,\qquad
 \vect{h}\sca\tp\uect{k}\inc=\cos\chi-\beta\inc,\qquad
 \vect{h}\sca\tp\uectg{\beta}=\beta\sca-\beta^2\\
 \vect{h}\sca\tp\vect{h}\inc=\cos\chi-\beta\inc-\beta\sca+\beta^2
\end{gather*}
and similarly for ``sc'' replaced by ''in''.
The angle $\chi=\acos\;\uect{k}\sca\tp\uect{k}\inc$ is the scattering angle.
Insertion of (\ref{ElecAcc}) into the radiation part of 
(\ref{AccChargeField}) we obtain
\begin{gather}
  \vect{E}_\mathrm{sc}(t,\vect{x})
= \frac{e}{4\pi\epsilon_0 c^2}
    \big[
      \frac{\vect{\hat{k}}_\mathrm{sc}
      \times\vect{h}\sca
      \times\vectg{\dot\beta}}{\kappa^3 T} 
    \big]_{t\ret}
\nonumber\\    
= \frac{-e^2}{4\pi\epsilon_0 m_ec^2}
    \big[
      \frac{\vect{\hat{k}}_\mathrm{sc}
     \times\vect{h}\sca
     \times[(1-\beta_\mathrm{in})\vect{E}_\mathrm{in}
            +\vectg{\beta}\tp\vect{E}_\mathrm{in}\;
            \vect{h}\inc]
           }{\gamma \kappa^3 cT} 
    \big]_{t\ret}
\nonumber\\    
= -r_e\big[\frac{[(1-\beta_\mathrm{sc})\vectg{1}
     -\vect{h}\sca\vect{\hat{k}}_\mathrm{sc}\tp]\;
     [(1-\beta_\mathrm{in})\vectg{1}
      +\vect{h}\inc\vectg{\beta}\tp]\;\vect{E}_\mathrm{in}}
      {\gamma \kappa^3 \dist} 
    \big]_{t\ret}
= -r_e\big[\frac{\vect{S}\vect{A}\vect{E}_\mathrm{in}}
                {\gamma \kappa^3 \dist}
    \big]_{t\ret}
\label{Esca}\end{gather}
where we used the classical electron radius $r_e=e^2/4\pi\epsilon_0 m_ec^2$
and the distance $\dist=cT$ between the scattering site $\vect{r}(t\ret)$
and $\vect{x}$. In the last step we introduced the operators
\[
  \vect{A}=[(1-\beta_\mathrm{in})\vectg{1}
              +\vect{h}\inc\vectg{\beta}\tp]
 \quad\text{and}\quad
  \vect{S}=[(1-\beta_\mathrm{sc})\vectg{1}
              -\vect{h}\sca\vect{\hat{k}}_\mathrm{sc}\tp],
\]
which describe the particle acceleration as a result of the wave electric
field and the scattered field due to the accelerated particle motion,
respectively. With their help the correlation matrix of the
scattered electric field can be obtained from
\[
  \big(\frac{\gamma\kappa^3 \dist}{r_e}\big)^2
  \vect{E}_\mathrm{sc}\vect{E}_\mathrm{sc}\tp
 =\vect{S}\vect{A} \;\vect{E}_\mathrm{in} (\vect{S}\vect{A} \;
  \vect{E}_\mathrm{in})\tp
 =\vect{S}\vect{A} \;\vect{E}_\mathrm{in} \vect{E}_\mathrm{in}\tp
  \vect{A}\tp\vect{S}\tp
\]
where for an unpolarised incident field
\[
 \vect{E}_\mathrm{in} \vect{E}_\mathrm{in}\tp
=[\vectg{1}-\vect{\hat{k}}\inc\vect{\hat{k}}\inc\tp]\,E\inc^2
\]
The actions of $\vect{S}\dots\vect{S}\tp$
and $\vect{A}\dots\vect{A}\tp$ will be treated separately.

\subsection{$\vect{A}\vect{E}_\mathrm{in} \vect{E}_\mathrm{in}\tp\vect{A}\tp$}

In this section we derive the the acceleration correlation matrix,
essentially the 3D correlation matrix of $\vectg{\dot\beta}$ due to
the incident wave field $\vect{E}_\mathrm{in}$ of an unpolarised wave.
\begin{gather*}
  \vect{E}_\mathrm{in} \vect{E}_\mathrm{in}\tp \vect{A}\tp
=[\vectg{1}-\vect{\hat{k}}_\mathrm{in}\vect{\hat{k}}_\mathrm{in}\tp]\;
 [(1-\beta_\mathrm{in})\vectg{1} + \vectg{\beta}\vect{h}\inc\tp]\,E\inc^2
\\
= [(1-\beta_\mathrm{in})(\vectg{1} - \vect{\hat{k}}\inc\vect{\hat{k}}\inc\tp)
 + \vectg{\beta}\vect{h}\inc\tp - \beta\inc\vect{\hat{k}}\inc\vect{h}\inc\tp]
  \,E\inc^2
\end{gather*}
We know that
$\vect{A} \vect{E}_\mathrm{in} \vect{E}_\mathrm{in}\tp \vect{A}\tp$ must
be symmetric and can be decomposed in the four relevant tensorial
components:
$\vectg{1}$,
$\vectg{\beta}\vectg{\beta}\tp$,
$\uect{h}\inc\uect{h}\inc\tp$
$\vectg{\beta}\uect{h}\inc+\uect{h}\vectg{\beta}\inc\tp$
Operating $\vect{A}$ on $\vect{E}_\mathrm{in} \vect{E}_\mathrm{in}\tp
\vect{A}\tp$ from above gives
\begin{gather*}
 \vect{A} \vect{E}_\mathrm{in} \vect{E}_\mathrm{in}\tp \vect{A}\tp
= [(1-\beta_\mathrm{in})\vectg{1}
  +\vect{h}\inc\vectg{\beta}\tp]
  [(1-\beta_\mathrm{in})(\vectg{1} - \vect{\hat{k}}\inc\vect{\hat{k}}\inc\tp)
    + \vectg{\beta}\vect{h}\inc\tp
    - \beta\inc\vect{\hat{k}}\inc\vect{h}\inc\tp]
\\
=(1-\beta_\mathrm{in})^2(\vectg{1} - \vect{\hat{k}}\inc\vect{\hat{k}}\inc\tp)
+(1-\beta_\mathrm{in})\vectg{\beta}\vect{h}\inc\tp
-(1-\beta_\mathrm{in})\beta\inc\uect{k}\inc\vect{h}\inc\tp\\
+(1-\beta_\mathrm{in})(\vect{h}\inc\vectg{\beta}\tp
             -\beta\inc\vect{h}\inc\uect{k}\inc\tp)
+\beta^2    \vect{h}\inc\vect{h}\inc\tp
-\beta^2\inc\vect{h}\inc\vect{h}\inc\tp
\\
 =(1-\beta_\mathrm{in})^2
  (\vectg{1} - \vect{\hat{k}}\inc\vect{\hat{k}}\inc\tp)
 +(1-\beta_\mathrm{in})
  (\vect{h}\inc\vectg{\beta}\tp+\vectg{\beta}\vect{h}\inc\tp)\\
 -(1-\beta_\mathrm{in})\beta\inc
  (\vect{h}\inc\vect{\hat{k}}\inc\tp+\vect{\hat{k}}\inc\vect{h}\inc\tp)
 +(\beta^2-\beta^2\inc)\vect{h}\inc\vect{h}\inc\tp
\end{gather*}
Insert $\vect{\hat{k}}\inc=\vect{h}\inc+\vectg{\beta}$ and order
according to tensor element
\begin{align*}
\text{coeff of}\; &\1
                  & \text{is}\; &
  (1-\beta_\mathrm{in})^2 \\
\text{coeff of}\; &\vect{h}\inc\vect{h}\inc\tp
                  &\text{is}\; &
 -(1-\beta_\mathrm{in})^2 - 2(1-\beta\inc)\beta\inc+\beta^2-\beta^2\inc\\
&&
=&-(1-\beta_\mathrm{in})(1-\beta_\mathrm{in} + 2\beta\inc)+\beta^2-\beta^2\inc\\
&&
=&-(1-\beta_\mathrm{in})(1+\beta\inc)+\beta^2-\beta^2\inc
 =-1+\beta^2\\
\text{coeff of}\; &(\vectg{\beta}\vect{h}\inc\tp+\vect{h}\inc\vectg{\beta}\tp)
                  &\text{is}\; &
  -(1-\beta_\mathrm{in})^2 +(1-\beta\inc) -(1-\beta\inc)\beta\inc=0\\
\text{coeff of}\; &\vectg{\beta}\vectg{\beta}\tp
                  & \text{is}\; &
  -(1-\beta_\mathrm{in})^2
\end{align*}
This gives finally
\begin{equation}
 \vect{A} \vect{E}_\mathrm{in} \vect{E}_\mathrm{in}\tp \vect{A}\tp
=(1-\beta_\mathrm{in})^2(\1-\vectg{\beta}\vectg{\beta}\tp)
-(1-\beta^2)\vect{h}\inc\vect{h}\inc\tp
\label{AEEA}\end{equation}

\subsection{$\vect{S}\dots\vect{S}\tp$}

Next we will deal with
\[
  \vect{S}=(1-\beta\sca)\1 -\vect{h}\sca\uect{k}\sca\tp
\]
Be $\vect{\hat{p}}$ a polarisation direction $\perp$ to $\uect{k}\sca$. Then
$\vect{\hat{p}}\tp\vect{E}\sca\vect{E}\sca\tp\vect{\hat{p}}$ is the
power of the field polarised in this polarisation direction. From the above
it can be written as
\begin{gather*}
    \vect{\hat{p}}\tp\vect{E}\sca\vect{E}\sca\tp\vect{\hat{p}}
   =\big(\frac{r_e}{\gamma\kappa^3 d}\big)^2
    \vect{\hat{p}}\tp\vect{S}\vect{A}\vect{E}\inc
    \vect{E}\inc\tp\vect{A}\tp\vect{S}\tp\vect{\hat{p}}
\end{gather*}
For $\vect{A}\vect{E}\sca\vect{E}\sca\tp\vect{A}\tp$ we have a concise
form (\ref{AEEA}), so consider its modification by multiplication from
left and right with
\begin{gather*}
  \vect{\hat{p}}\tp\vect{S}
 =\vect{\hat{p}}\tp[(1-\beta\sca)\1 -\vect{h}\sca\uect{k}\sca\tp]
\\
 =(1-\beta\sca)\vect{\hat{p}}\tp
 -\vect{\hat{p}}\tp\vect{h}\sca\uect{k}\sca\tp
 =(1-\beta\sca)\vect{\hat{p}}\tp
 +\vect{\hat{p}}\tp\uectg{\beta}\uect{k}\sca\tp
 =(1-\beta\sca)\vect{\hat{p}}\tp
 +\beta\pol\uect{k}\sca\tp
\end{gather*}
We made here use of the fact that $\vect{\hat{p}}\tp\uect{k}\sca=0$
and introduced $\beta\pol=\uect{p}\tp\vectg{\beta}$.
Written explicitly
\begin{gather*}
   \frac{1}{E\inc^2}\;
   \big(\frac{\gamma\kappa^3 \dist}{r_e}\big)^2
    \vect{\hat{p}}\tp\vect{E}\sca\vect{E}\sca\tp\vect{\hat{p}}
\\    
= [(1-\beta\sca)\vect{\hat{p}}\tp+\beta\pol\uect{k}\sca\tp]\;
  [(1-\beta\inc)^2[\1-\uectg{\beta}\uectg{\beta}{}\tp]-(1-\beta^2)\vect{h}\inc\vect{h}\inc\tp]\;
  [(1-\beta\sca)\vect{\hat{p}}+\beta\pol\uect{k}\sca]
\\[0.5ex]
\begin{aligned}
= (1-\beta\sca)\vect{\hat{p}}\tp\;
  &[(1-\beta\inc)^2[\1-\uectg{\beta}\uectg{\beta}{}\tp]  -  (1-\beta^2)\vect{h}\inc\vect{h}\inc\tp]\;
  (1-\beta\sca)\vect{\hat{p}}\\
+ (1-\beta\sca)\vect{\hat{p}}\tp\;
  &[(1-\beta\inc)^2[\1-\uectg{\beta}\uectg{\beta}{}\tp]  -  (1-\beta^2)\vect{h}\inc\vect{h}\inc\tp]\;
  \beta\pol\uect{k}\sca\\
+ \beta\pol\uect{k}\sca\tp\;
  &[(1-\beta\inc)^2[\1-\uectg{\beta}\uectg{\beta}{}\tp]  -  (1-\beta^2)\vect{h}\inc\vect{h}\inc\tp]\;
  (1-\beta\sca)\vect{\hat{p}}\\
+ \beta\pol\uect{k}\sca\tp\;
  &[(1-\beta\inc)^2[\1-\uectg{\beta}\uectg{\beta}{}\tp]  -  (1-\beta^2)\vect{h}\inc\vect{h}\inc\tp]\;
  \beta\pol\uect{k}\sca
\end{aligned}
\\[0.5ex]
\begin{aligned}
= &(1-\beta\inc)^2(1-\beta\sca)^2
  &&\vect{\hat{p}}\tp[\1-\uectg{\beta}\uectg{\beta}{}\tp]
  \vect{\hat{p}}\\
+ &(1-\beta\inc)^2(1-\beta\sca)\beta\pol
  &&\vect{\hat{p}}\tp[\1-\uectg{\beta}\uectg{\beta}{}\tp]
  \uect{k}\sca\\
+ &(1-\beta\inc)^2(1-\beta\sca)\beta\pol
  &&\uect{k}\sca\tp[\1-\uectg{\beta}\uectg{\beta}{}\tp]
  \vect{\hat{p}}\\
+ &(1-\beta\inc)^2\beta^2\pol
  &&\uect{k}\sca\tp[\1-\uectg{\beta}\uectg{\beta}{}\tp]
  \uect{k}\sca
\end{aligned}
\hspace{2em}
\begin{aligned}
- &(1-\beta^2)(1-\beta\sca)^2
  &&\vect{\hat{p}}\tp[\vect{h}\inc\vect{h}\inc\tp]\vect{\hat{p}}\\
- &(1-\beta^2)(1-\beta\sca)\beta\pol
  &&\vect{\hat{p}}\tp[\vect{h}\inc\vect{h}\inc\tp]\uect{k}\sca\\
- &(1-\beta^2)(1-\beta\sca)\beta\pol
  &&\uect{k}\sca\tp[\vect{h}\inc\vect{h}\inc\tp]\vect{\hat{p}}\\
- &(1-\beta^2)\beta^2\pol
  &&\uect{k}\sca\tp[\vect{h}\inc\vect{h}\inc\tp]\uect{k}\sca
\end{aligned}
\\[0.5ex]
\begin{aligned}
= &(1-\beta\inc)^2(1-\beta\sca)^2
  &&(1-\beta^2\pol)\\
+ &(1-\beta\inc)^2(1-\beta\sca)\beta\pol
  &&(-2\beta\pol\beta\sca)\\
+ &(1-\beta\inc)^2\beta^2\pol
  &&(1-\beta^2\sca)
\end{aligned}
\hspace{2em}
\begin{aligned}
- &(1-\beta^2)(1-\beta\sca)^2
  &&(\vect{\hat{p}}\tp\uect{k}\inc-\beta\pol)^2\\
- &(1-\beta^2)(1-\beta\sca)\beta\pol
  &&2(\vect{\hat{p}}\tp\uect{k}\inc-\beta\pol)(\cos\chi-\beta\sca)\\
- &(1-\beta^2)\beta^2\pol
  &&(\cos\chi-\beta\sca)^2
\end{aligned}
\end{gather*}
The coefficient of $(1-\beta\inc)^2$ gives
\begin{gather*}
   (1-\beta\sca)^2(1-\beta\pol^2)
 -2(1-\beta\sca)\beta^2\pol\beta\sca
  +(1-\beta^2\sca)\beta^2\pol
\\
   (1-\beta\sca)^2
  -[\underbrace{(1-\beta\sca)^2
 +2(1-\beta\sca)\beta\sca
  -(1-\beta^2\sca)}_{%
 +\cancelto{\scriptstyle a}{1}
 -\cancelto{\scriptstyle b}{2\beta\sca}
 +\cancelto{\scriptstyle c}{\beta^2\sca}
 +\cancelto{\scriptstyle b}{2\beta\sca}
 -\cancelto{\scriptstyle c}{2\beta^2\sca}
 -\cancelto{\scriptstyle a}{1}
 +\cancelto{\scriptstyle c}{\beta^2\sca}
 =0}]\beta^2\pol
=(1-\beta\sca)^2
\end{gather*}
The coefficient of $-(1-\beta^2)$ gives
\begin{gather*}
   (1-\beta\sca)^2(\vect{\hat{p}}\tp\uect{k}\inc-\beta\pol)^2
 +2(1-\beta\sca)\beta\pol(\vect{\hat{p}}\tp\uect{k}\inc-\beta\pol)
                (\cos\chi-\beta\sca)
  +\beta^2\pol(\cos\chi - \beta\sca)^2
\\
=[(1-\beta\sca)(\vect{\hat{p}}\tp\uect{k}\inc-\beta\pol)
              +\beta\pol(\cos\chi - \beta\sca)]^2
\\
=[(1-\beta\sca)\vect{\hat{p}}\tp\uect{k}\inc
  -\cancelto{\scriptstyle a}{(1-\beta\sca)\beta\pol}
  +\beta\pol(\cos\chi-1)
  -\cancelto{\scriptstyle a}{\beta\pol(\beta\sca-1)}]^2
\\
=[ (1-\beta\sca)\vect{\hat{p}}\tp\uect{k}\inc
  -(1-\cos\chi)\beta\pol ]^2
\end{gather*}
Recall that with the abbreviations introduced above
\begin{gather*}
\kappa=1-\beta\sca,\quad
D(\uect{k}\sca,\vectg{\beta})=\frac{1}{\gamma(1-\beta\sca)},\quad
D(\uect{k}\inc,\vectg{\beta})=\frac{1}{\gamma(1-\beta\inc)}
\end{gather*}
We finally obtain
\begin{gather*}
    [\vect{\hat{p}}\tp\vect{E}\sca\vect{E}\sca\tp\vect{\hat{p}}]
    (\vect{x},t)
=\frac{r^2_e}{\gamma^2\kappa^6 \dist^2}
 [(1-\beta\inc)^2(1-\beta\sca)^2\\\left.
 -(1-\beta^2) \big( (1-\beta\sca)\vect{\hat{p}}\tp\uect{k}\inc
              -(1-\cos\chi)\beta\pol \big)^2 ]
  \;E\inc^2\;\right|_{t\ret}
\\ 
=\left.\frac{r^2_e}{\dist^2}
\frac{(1-\beta\inc)^2}{\gamma^2(1-\beta\sca)^4}
[1 -\frac{1}{\gamma^2(1-\beta\inc)^2}\big( \vect{\hat{p}}\tp\uect{k}\inc
      -\frac{1-\cos\chi}{1-\beta\sca}\beta\pol \big)^2 ]
\;E\inc^2\;\right|_{t\ret}
\\ 
=\left.\frac{r^2_e}{\dist^2}
\frac{D^4(\uect{k}\sca,\vectg{\beta})}{D^2(\uect{k}\inc,\vectg{\beta})}
[1-D^2(\uect{k}\inc,\vectg{\beta})\big( \vect{\hat{p}}\tp\uect{k}\inc
      -\frac{1-\cos\chi}{1-\beta\sca}\beta\pol \big)^2 ]
\;E\inc^2\;\right|_{t\ret}
\\
=\left.\frac{r^2_e}{\dist^2}
\frac{D^4(\uect{k}\sca,\vectg{\beta})}{D^2(\uect{k}\inc,\vectg{\beta})}
[1-\big( D(\uect{k}\inc,\vectg{\beta})\vect{\hat{p}}\tp(\uect{k}\inc
      -\frac{1-\cos\chi}{1-\beta\sca}\vectg{\beta})\big)^2 ]
\;E\inc^2\;\right|_{t\ret}
\end{gather*}
All times on the right-hand-sides are retarded times. The electron position
is at $\vect{r}(t\ret)$ and all quantities refer to the observer frame.

For the comparison with (\ref{Radint_pp}) we have to recall that the incident
field fluctuations are related to the incident irradiance $c\epsilon_0
E\inc^2=\irr\inc=\rad d\Omega(\uect{k}\inc)$ and the scattered radiant
intensity is $c\epsilon_0\;
\dist^2\;\vect{\hat{p}}\tp\vect{E}\sca\vect{E}\sca\tp\vect{\hat{p}}
=\dist^2\;\vect{\hat{p}}\tp\vect{Q}\sca\vect{\hat{p}}$.

\section*{Acknowledgments}

I owe many thanks to Serge Koutchmy for making me aware of the relativistic
effects in Thomson scattering and for many helpful comments.
I am also grateful to Marilena Mierla and Li Feng for inspiring discussions and
careful corrections of an earlier version of this manuscript.

\bibliography{/home/binhest/tex/inp/journalabbr_long,ThomsonScatt}

\ifx\undefined\allcaps\def\allcaps#1{#1}\fi
\begin{thebibliography}{91}
\providecommand{\natexlab}[1]{#1}
\providecommand{\url}[1]{\texttt{#1}}
\expandafter\ifx\csname urlstyle\endcsname\relax
  \providecommand{\doi}[1]{doi: #1}\else
  \providecommand{\doi}{doi: \begingroup \urlstyle{rm}\Url}\fi

\bibitem[{Antonucci} and {Miller}(1985)]{AntonnucciMiller:1985}
R.~R.~J. {Antonucci} and J.~S. {Miller}.
\newblock {Spectropolarimetry and the nature of NGC 1068}.
\newblock \emph{Astrophysical Journal}, 297:\penalty0 621--632, October 1985.
\newblock \doi{10.1086/163559}.

\bibitem[{Aschwanden} et~al.(1995){Aschwanden}, {Benz}, {Dennis}, and
  {Schwartz}]{AschwandenEtal:1995}
M.~J. {Aschwanden}, A.~O. {Benz}, B.~R. {Dennis}, and R.~A. {Schwartz}.
\newblock {Solar Electron Beams Detected in Hard X-Rays and Radio Waves}.
\newblock \emph{Astrophysical Journal}, 455:\penalty0 347--365, December 1995.
\newblock \doi{10.1086/176582}.

\bibitem[{Badalyan} et~al.(1993){Badalyan}, {Livshits}, and
  {Sykora}]{BadalianEtal:1993}
O.~G. {Badalyan}, M.~A. {Livshits}, and J.~{Sykora}.
\newblock {Polarization of the white-light corona and its large-scale structure
  in the period of solar cycle maximum}.
\newblock \emph{Solar Physics}, 145:\penalty0 279--290, June 1993.
\newblock \doi{10.1007/BF00690656}.

\bibitem[{Beausang} and {Prunty}(2008)]{BeausangPrunty:2008}
K.~V. {Beausang} and S.~L. {Prunty}.
\newblock {An analytic formula for the relativistic Thomson scattering spectrum
  for a Maxwellian velocity distribution}.
\newblock \emph{Plasma Physics and Controlled Nuclear Fusion}, 50\penalty0
  (9):\penalty0 095001 (10pp), September 2008.
\newblock \doi{10.1088/0741-3335/50/9/095001}.

\bibitem[{Billings}(1966)]{Billings:1966}
D.~E. {Billings}.
\newblock \emph{{A guide to the solar corona}}.
\newblock Academic Press, New York, 1966.

\bibitem[{Boas}(1961)]{Boas:1961}
M.~L. {Boas}.
\newblock {Apparent Shape of Large Objects at Relativistic Speeds}.
\newblock \emph{American Journal of Physics}, 29:\penalty0 283--286, May 1961.
\newblock \doi{10.1119/1.1937751}.

\bibitem[Born(2001)]{Born:2001}
M.~Born.
\newblock \emph{Die Relativit\"atstheorie Einsteins. Kommentiert und erweitert
  von J.~Ehlers und M.~P\"ossel}.
\newblock Springer, 2001.

\bibitem[{Bowles}(1958)]{Bowles:1958}
K.~L. {Bowles}.
\newblock {Observations of vertical incidence scatter from the ionosphere at 41
  Mc/s}.
\newblock \emph{Physical Review Letters}, 1:\penalty0 454--455, 1958.

\bibitem[{Burke} and {Strode}(1991)]{BurkeStrode:1991}
J.~R. {Burke} and F.~J. {Strode}.
\newblock {Classroom exercises with the Terrell effect}.
\newblock \emph{American Journal of Physics}, 59:\penalty0 912--915, October
  1991.
\newblock \doi{10.1119/1.16670}.

\bibitem[{Carley} et~al.(2015){Carley}, {Reid}, {Vilmer}, and
  {Gallagher}]{CarleyEtal:2015}
E.~P. {Carley}, H.~{Reid}, N.~{Vilmer}, and P.~T. {Gallagher}.
\newblock {Low frequency radio observations of bi-directional electron beams in
  the solar corona}.
\newblock \emph{Astronomy and Astrophysics}, 581:\penalty0 A100, September
  2015.
\newblock \doi{10.1051/0004-6361/201526251}.

\bibitem[{Chen} and {Petrosian}(2012)]{ChenPetrosian:2012}
Q.~{Chen} and V.~{Petrosian}.
\newblock {Impulsive Phase Coronal Hard X-Ray Sources in an X3.9 Class Solar
  Flare}.
\newblock \emph{Astrophysical Journal}, 748:\penalty0 33, March 2012.
\newblock \doi{10.1088/0004-637X/748/1/33}.

\bibitem[{Cocke} and {Holm}(1972)]{CockeHolm:1972}
W.~J. {Cocke} and D.~A. {Holm}.
\newblock {Lorentz Transformation Properties of the Stokes Parameters}.
\newblock \emph{Nature Physical Science}, 240:\penalty0 161--162, December
  1972.
\newblock \doi{10.1038/physci240161b0}.

\bibitem[{Colaninno} and {Vourlidas}(2009)]{ColaninnoVourlidas:2009}
R.~C. {Colaninno} and A.~{Vourlidas}.
\newblock {First Determination of the True Mass of Coronal Mass Ejections: A
  Novel Approach to Using the Two STEREO Viewpoints}.
\newblock \emph{Astrophysical Journal}, 698:\penalty0 852--858, June 2009.
\newblock \doi{10.1088/0004-637X/698/1/852}.

\bibitem[{Debbasch}(2008)]{Debbasch:2008}
F.~{Debbasch}.
\newblock {Equilibrium distribution function of a relativistic dilute perfect
  gas}.
\newblock \emph{Physica A Statistical Mechanics and its Applications},
  387:\penalty0 2443--2454, April 2008.
\newblock \doi{10.1016/j.physa.2007.10.076}.

\bibitem[{DeForest} et~al.(2013){DeForest}, {Howard}, and
  {Tappin}]{DeforestHoward:2013}
C.~E. {DeForest}, T.~A. {Howard}, and S.~J. {Tappin}.
\newblock {The Thomson Surface. II. Polarization}.
\newblock \emph{Astrophysical Journal}, 765:\penalty0 44, March 2013.
\newblock \doi{10.1088/0004-637X/765/1/44}.

\bibitem[{Ellis} et~al.(2005){Ellis}, {Dogariu}, {Ponomarenko}, and
  E.]{EllisEtal:2005}
J.~{Ellis}, A.~{Dogariu}, S.~{Ponomarenko}, and {Wolf} E.
\newblock {Degree of polarization of statistically stationary electromagnetic
  fields}.
\newblock \emph{Optics Communications}, 248:\penalty0 333--337, 2005.
\newblock \doi{10.1016/j.optcom.2004.12.050}.

\bibitem[{Eriksen} and {Gr{\o}n}(1992)]{EriksenGron:1992}
E.~{Eriksen} and O.~{Gr{\o}n}.
\newblock {The observed intensity of a radiating moving object}.
\newblock \emph{European Journal of Physics}, 13:\penalty0 210--214, September
  1992.
\newblock \doi{10.1088/0143-0807/13/5/002}.

\bibitem[{Fiocco} and {Thompson}(1963)]{FioccoThompson:1963}
G.~{Fiocco} and E.~{Thompson}.
\newblock {Thomson Scattering of Optical Radiation from an Electron Beam}.
\newblock \emph{Physical Review Letters}, 10:\penalty0 89--91, 1963.

\bibitem[French(1968)]{French:1968}
A.~P. French.
\newblock \emph{Special Relativity}.
\newblock The M.I.T. Introductory Physics Series. W. W. Norton, 1968.

\bibitem[{Gradshteyn} and {Ryzhik}(1980)]{GradshteynRyzhik:1980}
I.~S. {Gradshteyn} and I.~M. {Ryzhik}.
\newblock \emph{{Table of integrals, series and products}}.
\newblock Academic press, 1980.

\bibitem[{Harvey}(2015)]{Harvey:2014}
J.~W. {Harvey}.
\newblock {Two centuries of Solar Polarimetry}.
\newblock In K.~N. {Negendra}, S.~{Bagnulo}, B.~{Centeno}, and M.~J.
  {Mart\'ines Gonz\'ales}, editors, \emph{Polarimetry: From Sun to Stars and
  stellar environments}, number 305 in Proceedings of the IAU Symposium, pages
  2--11, 2015.
\newblock \doi{10.1017/S1743921315004457}.

\bibitem[{Hayes} et~al.(2001){Hayes}, {Vourlidas}, and
  {Howard}]{HayesEtal:2001}
A.~P. {Hayes}, A.~{Vourlidas}, and R.~A. {Howard}.
\newblock {Deriving the electron density of the solar corona from the inversion
  of total brightness measurements}.
\newblock \emph{Astrophysical Journal}, 548:\penalty0 1081--1086, February
  2001.
\newblock \doi{10.1086/319029}.

\bibitem[{Hoffman}(2015)]{Hoffman:2014}
J.~L. {Hoffman}.
\newblock {Polarimetry as a wondow into supernova explosions and progenitors}.
\newblock In K.~N. {Negendra}, S.~{Bagnulo}, B.~{Centeno}, and M.~J.
  {Mart\'ines Gonz\'ales}, editors, \emph{Polarimetry: From Sun to Stars and
  stellar environments}, number 305 in Proceedings of the IAU Symposium, pages
  269--274, 2015.
\newblock \doi{10.1017/S1743921315004883}.

\bibitem[{Howard} et~al.(2008){Howard}, {Moses}, {Vourlidas}, {Newmark},
  {Socker}, {Plunkett}, {Korendyke}, {Cook}, {Hurley}, {Davila}, {Thompson},
  {St Cyr}, {Mentzell}, {Mehalick}, {Lemen}, {Wuelser}, {Duncan}, {Tarbell},
  {Wolfson}, {Moore}, {Harrison}, {Waltham}, {Lang}, {Davis}, {Eyles},
  {Mapson-Menard}, {Simnett}, {Halain}, {Defise}, {Mazy}, {Rochus}, {Mercier},
  {Ravet}, {Delmotte}, {Auchere}, {Delaboudiniere}, {Bothmer}, {Deutsch},
  {Wang}, {Rich}, {Cooper}, {Stephens}, {Maahs}, {Baugh}, {McMullin}, and
  {Carter}]{HowardStereo:2008}
R.~A. {Howard}, J.~D. {Moses}, A.~{Vourlidas}, J.~S. {Newmark}, D.~G. {Socker},
  S.~P. {Plunkett}, C.~M. {Korendyke}, J.~W. {Cook}, A.~{Hurley}, J.~M.
  {Davila}, W.~T. {Thompson}, O.~C. {St Cyr}, E.~{Mentzell}, K.~{Mehalick},
  J.~R. {Lemen}, J.~P. {Wuelser}, D.~W. {Duncan}, T.~D. {Tarbell}, C.~J.
  {Wolfson}, A.~{Moore}, R.~A. {Harrison}, N.~R. {Waltham}, J.~{Lang}, C.~J.
  {Davis}, C.~J. {Eyles}, H.~{Mapson-Menard}, G.~M. {Simnett}, J.~P. {Halain},
  J.~M. {Defise}, E.~{Mazy}, P.~{Rochus}, R.~{Mercier}, M.~F. {Ravet},
  F.~{Delmotte}, F.~{Auchere}, J.~P. {Delaboudiniere}, V.~{Bothmer},
  W.~{Deutsch}, D.~{Wang}, N.~{Rich}, S.~{Cooper}, V.~{Stephens}, G.~{Maahs},
  R.~{Baugh}, D.~{McMullin}, and T.~{Carter}.
\newblock {Sun Earth Connection Coronal and Heliospheric Investigation
  (SECCHI)}.
\newblock \emph{Space Science Reviews}, 136:\penalty0 67--115, April 2008.
\newblock \doi{10.1007/s11214-008-9341-4}.

\bibitem[{Howard} and {DeForest}(2012)]{HowardDeforest:2012}
T.~A. {Howard} and C.~E. {DeForest}.
\newblock {The Thomson Surface. I. Reality and Myth}.
\newblock \emph{Astrophysical Journal}, 752:\penalty0 130, June 2012.
\newblock \doi{10.1088/0004-637X/752/2/130}.

\bibitem[Hutchinson(2002)]{Hutchinson:2002}
I.~H. Hutchinson.
\newblock \emph{Principles of Plasma Diagnostics}.
\newblock Cambridge University Press, 2nd edition, 2002.

\bibitem[Jackson(1998)]{Jackson:1998}
J.~D. Jackson.
\newblock \emph{Classical electrodynamics}.
\newblock Wiley, 3rd edition, 1998.

\bibitem[{J{\"u}ttner}(1911)]{Juettner:1911}
F.~{J{\"u}ttner}.
\newblock {Das Maxwellsche Gesetz der Geschwindigkeitsverteilung in der
  Relativtheorie}.
\newblock \emph{Annalen der Physik}, 339:\penalty0 856--882, 1911.
\newblock \doi{10.1002/andp.19113390503}.

\bibitem[{Kemp} et~al.(1987){Kemp}, {Henson}, {Steiner}, and
  {Powell}]{KempEtal:1987}
J.~C. {Kemp}, G.~D. {Henson}, C.~T. {Steiner}, and E.~R. {Powell}.
\newblock {The optical polarization of the sun measured at a sensitivity of
  parts in ten million}.
\newblock \emph{Nature}, 326:\penalty0 270--273, March 1987.
\newblock \doi{10.1038/326270a0}.

\bibitem[{Kim} et~al.(1996){Kim}, {Bougaenko}, {Belenko}, {Koutchmi},
  {Matsuura}, and {Picazzio}]{KimEtal:1996}
I.~S. {Kim}, O.~I. {Bougaenko}, I.~A. {Belenko}, S.~{Koutchmi}, O.~T.
  {Matsuura}, and E.~{Picazzio}.
\newblock The coronograph-polarimeter: An algorithm for creation of solar
  corona polarization images.
\newblock \emph{Radiophysics and Quantum Electronics}, 39:\penalty0 865--868,
  October 1996.

\bibitem[{Koutchmy}(1994)]{Koutchmy:1994}
S.~{Koutchmy}.
\newblock {Coronal physics from eclipse observations}.
\newblock \emph{Advances in Space Research}, 14:\penalty0 29--39, April 1994.
\newblock \doi{10.1016/0273-1177(94)90156-2}.

\bibitem[{Koutchmy} and {Lamy}(1985)]{KoutchmyLamy:1985}
S.~{Koutchmy} and P.~L. {Lamy}.
\newblock {The F-corona and the circum-solar dust evidences and properties}.
\newblock In R.~H. {Giese} and P.~{Lamy}, editors, \emph{IAU Colloq. 85:
  Properties and Interactions of Interplanetary Dust}, volume 119 of
  \emph{Astrophysics and Space Science Library}, pages 63--74, 1985.
\newblock \doi{10.1007/978-94-009-5464-9\_14}.

\bibitem[{Koutchmy} and {Nikoghossian}(2002)]{KoutchmyNikoghossian:2002}
S.~{Koutchmy} and A.~G. {Nikoghossian}.
\newblock {Coronal linear threads: W-L radiation of supra-thermal streams}.
\newblock \emph{Astronomy and Astrophysics}, 395:\penalty0 983--989, December
  2002.
\newblock \doi{10.1051/0004-6361:20021269}.

\bibitem[{Koutchmy} and {Nikoghossian}(2005)]{KoutchmyNikoghossian:2005}
S.~{Koutchmy} and A.~G. {Nikoghossian}.
\newblock {Analysis of the radiation of coronal suprathermal streams}.
\newblock \emph{Astrophysics}, 48:\penalty0 62--67, January 2005.
\newblock \doi{10.1007/s10511-005-0007-6}.

\bibitem[{Koutchmy} and {Schatten}(1971)]{KoutchmySchatten:1971}
S.~{Koutchmy} and K.~H. {Schatten}.
\newblock {Observations and Discussions Concerning `High Polarization Features
  in the Solar Corona}.
\newblock \emph{Solar Physics}, 17:\penalty0 117--128, March 1971.
\newblock \doi{10.1007/BF00152866}.

\bibitem[{Koutchmy} et~al.(1993){Koutchmy}, {Molodenskii}, {Nikol'Skii}, and
  {Filippov}]{KoutchmyEtal:1993}
S.~{Koutchmy}, M.~M. {Molodenskii}, G.~M. {Nikol'Skii}, and B.~P. {Filippov}.
\newblock {Measuring the polarization of the solar corona}.
\newblock \emph{Astronomy Reports}, 37:\penalty0 286--290, May 1993.

\bibitem[{Kraus}(2000)]{Kraus:2000}
U.~{Kraus}.
\newblock {Brightness and color of rapidly moving objects: The visual
  appearance of a large sphere revisited}.
\newblock \emph{American Journal of Physics}, 59:\penalty0 56--60, 2000.

\bibitem[{Krucker} et~al.(2010){Krucker}, {Hudson}, {Glesener}, {White},
  {Masuda}, {Wuelser}, and {Lin}]{KruckerEtal:2010}
S.~{Krucker}, H.~S. {Hudson}, L.~{Glesener}, S.~M. {White}, S.~{Masuda}, J.-P.
  {Wuelser}, and R.~P. {Lin}.
\newblock {Measurements of the Coronal Acceleration Region of a Solar Flare}.
\newblock \emph{Astrophysical Journal}, 714:\penalty0 1108--1119, May 2010.
\newblock \doi{10.1088/0004-637X/714/2/1108}.

\bibitem[{Kulijanishvili} and {Kapanadze}(2005)]{KulijanishviliKapanadze:2005}
V.~I. {Kulijanishvili} and N.~G. {Kapanadze}.
\newblock {Polarization and Physical Properties of the August 11, 1999
  White-Light Corona}.
\newblock \emph{Solar Physics}, 229:\penalty0 45--62, June 2005.
\newblock \doi{10.1007/s11207-005-3521-0}.

\bibitem[{Lehmann}(2006)]{Lehmann:2006}
E.~{Lehmann}.
\newblock {Covariant equilibrium statistical mechanics}.
\newblock \emph{Journal of Mathematical Physics}, 47\penalty0 (2):\penalty0
  023303 (18pp), February 2006.
\newblock \doi{10.1063/1.2165771}.

\bibitem[{Levasseur-Regourd} et~al.(2001){Levasseur-Regourd}, {Mann}, {Dumont},
  and {Hanner}]{LevasseurEtal:2001}
A.-C. {Levasseur-Regourd}, I.~{Mann}, R.~{Dumont}, and M.~S. {Hanner}.
\newblock {Optical and Thermal Properties of Interplanetary Dust}.
\newblock In E.~{Gr{\"u}n}, B.~A.~S. {Gustafson}, S.~{Dermott}, and
  H.~{Fechtig}, editors, \emph{{Interplanetary Dust}}, Astronomy and
  Astrophysics Library, pages 69--94. Springer, Berlin, 2001.

\bibitem[{Li} et~al.(2012){Li}, {Gan}, and {Feng}]{LiGanFeng:2012}
Y.~P. {Li}, W.~Q. {Gan}, and L.~{Feng}.
\newblock {Statistical Analyses on Thermal Aspects of Solar Flares}.
\newblock \emph{Astrophysical Journal}, 747:\penalty0 133, March 2012.
\newblock \doi{10.1088/0004-637X/747/2/133}.

\bibitem[{Llebaria} et~al.(2010){Llebaria}, {Loirat}, and
  {Lamy}]{LlebariaEtal:2010}
A.~{Llebaria}, J.~{Loirat}, and P.~{Lamy}.
\newblock {Restitution of multiple overlaid components on extremely long series
  of solar corona images}.
\newblock In \emph{Computational Imaging VIII}, volume 7533 of
  \emph{Proceedings of the SPIE}, page 75330Y (8pp), January 2010.
\newblock \doi{10.1117/12.838738}.

\bibitem[{Lyutikov} et~al.(2003){Lyutikov}, {Pariev}, and
  {Blandford}]{LyutikovEtal:2003}
M.~{Lyutikov}, V.~I. {Pariev}, and R.~D. {Blandford}.
\newblock {Polarization of Prompt Gamma-Ray Burst Emission: Evidence for
  Electromagnetically Dominated Outflow}.
\newblock \emph{Astrophysical Journal}, 597:\penalty0 998--1009, November 2003.
\newblock \doi{10.1086/378497}.

\bibitem[{Mann} and {Warmuth}(2011)]{MannWarmuth:2011}
G.~{Mann} and A.~{Warmuth}.
\newblock {Budget of energetic electrons during solar flares in the framework
  of magnetic reconnection}.
\newblock \emph{Astronomy and Astrophysics}, 528:\penalty0 A104, April 2011.
\newblock \doi{10.1051/0004-6361/201014389}.

\bibitem[{McKinley}(1979)]{McKinley:1979}
J.~M. {McKinley}.
\newblock {Relativistic transformations of light power}.
\newblock \emph{American Journal of Physics}, 47:\penalty0 602--605, 1979.
\newblock \doi{10.1119/1.11762}.

\bibitem[{McKinley}(1980)]{McKinley:1980}
J.~M. {McKinley}.
\newblock {Relativistic transformation of solid angle}.
\newblock \emph{American Journal of Physics}, 48:\penalty0 612--614, August
  1980.
\newblock \doi{10.1119/1.12329}.

\bibitem[{McNamara} et~al.(2009){McNamara}, {Kuncic}, and
  {Wu}]{McNamaraEtal:2009}
A.~L. {McNamara}, Z.~{Kuncic}, and K.~{Wu}.
\newblock {X-ray polarization in relativistic jets}.
\newblock \emph{Monthly Notices of the Royal Astronomical Society},
  395:\penalty0 1507--1514, May 2009.
\newblock \doi{10.1111/j.1365-2966.2009.14608.x}.

\bibitem[{Mierla} et~al.(2011){Mierla}, {Inhester}, {Rodriguez}, {Gissot},
  {Zhukov}, and {Srivastava}]{MierlaEtal:2011}
M.~{Mierla}, B.~{Inhester}, L.~{Rodriguez}, S.~{Gissot}, A.~{Zhukov}, and
  N.~{Srivastava}.
\newblock {On 3D reconstruction of coronal mass ejections: II. Longitudinal and
  latitudinal width analysis of 31 August 2007 event}.
\newblock \emph{Journal of Atmospheric and Solar-Terrestrial Physics},
  73:\penalty0 1166--1172, June 2011.
\newblock \doi{10.1016/j.jastp.2010.11.028}.

\bibitem[{Minnaert}(1930)]{Minnaert:1930}
M.~{Minnaert}.
\newblock {On the continuous spectrum of the corona and its polarisation}.
\newblock \emph{Zeitschrift f{\"u}r Astrophysik}, 1:\penalty0 209--236, 1930.

\bibitem[{Molodensky}(1973)]{Molodensky:1973}
M.~M. {Molodensky}.
\newblock {On an Anomalous Polarization of the Corona}.
\newblock \emph{Solar Physics}, 28:\penalty0 465--475, February 1973.
\newblock \doi{10.1007/BF00152317}.

\bibitem[{Montakhab} et~al.(2009){Montakhab}, {Ghodrat}, and
  {Barati}]{MontakhabEtal:2009}
A.~{Montakhab}, M.~{Ghodrat}, and M.~{Barati}.
\newblock {Statistical thermodynamics of a two-dimensional relativistic gas}.
\newblock \emph{Physical Review E}, 79\penalty0 (3):\penalty0 031124 (5pp),
  March 2009.
\newblock \doi{10.1103/PhysRevE.79.031124}.

\bibitem[{Moran} and {Davila}(2004)]{MoranDavila:2004}
T.~G. {Moran} and J.~M. {Davila}.
\newblock {Three-Dimensional Polarimetric Imaging of Coronal Mass Ejections}.
\newblock \emph{Science}, 305:\penalty0 66--71, July 2004.
\newblock \doi{10.1126/science.1098937}.

\bibitem[{Moran} et~al.(2006){Moran}, {Davila}, {Morrill}, {Wang}, and
  {Howard}]{MoranEtal:2006}
T.~G. {Moran}, J.~M. {Davila}, J.~S. {Morrill}, D.~{Wang}, and R.~{Howard}.
\newblock {Solar and Heliospheric Observatory/Large Angle Spectrometric
  Coronagraph Polarimetric Calibration}.
\newblock \emph{Solar Physics}, 237:\penalty0 211--222, August 2006.
\newblock \doi{10.1007/s11207-006-0147-9}.

\bibitem[{Nalewajko}(2009)]{Nalewajko:2009}
K.~{Nalewajko}.
\newblock {Polarization of synchrotron emission from relativistic reconfinement
  shocks}.
\newblock \emph{Monthly Notices of the Royal Astronomical Society},
  395:\penalty0 524--530, May 2009.
\newblock \doi{10.1111/j.1365-2966.2009.14559.x}.

\bibitem[{Neckel}(1996)]{Neckel:1996}
H.~{Neckel}.
\newblock {On the wavelength dependency of solar limb darkening
  ($\lambda\lambda$ 303 TO 1099 nm)}.
\newblock \emph{Solar Physics}, 167:\penalty0 9--23, 1996.

\bibitem[{Neckel} and {Labs}(1994)]{NeckelLabs:1994}
H.~{Neckel} and D.~{Labs}.
\newblock {Solar limb darkening 1986-1990 ($\lambda\lambda$ 303 to 1099 nm)}.
\newblock \emph{Solar Physics}, 153:\penalty0 91--114, 1994.

\bibitem[{Nikoghossian} and {Koutchmy}(2001)]{NikoghossianKoutchmy:2001}
A.~G. {Nikoghossian} and S.~{Koutchmy}.
\newblock {Interpretation of the Radiation of Coronal Suprathermal Streams. I.}
\newblock \emph{Astrophysics}, 44:\penalty0 528--535, October 2001.
\newblock \doi{10.1023/A:1014261224500}.

\bibitem[{Nikoghossian} and {Koutchmy}(2002)]{NikoghossianKoutchmy:2002}
A.~G. {Nikoghossian} and S.~{Koutchmy}.
\newblock {On Interpretation of the Radiation of Coronal Suprathermal Streams.
  II}.
\newblock \emph{Astrophysics}, 45:\penalty0 489--496, October 2002.
\newblock \doi{10.1023/A:1021863314630}.

\bibitem[{Oudmaijer}(2007)]{Oudmaijer2007}
R.~D. {Oudmaijer}.
\newblock {Spectropolarimetry and the Study of Circumstellar Disks}.
\newblock \emph{Astrophysics and Space Science Proceedings}, 1:\penalty0
  83--104, 2007.
\newblock \doi{10.1007/978-1-4020-5425-9\_5}.

\bibitem[Papoulis(1981)]{Papoulis:1981}
A.~Papoulis.
\newblock \emph{Probability, Random Variables and Stochastic Processes}.
\newblock McGraw-Hill Series in System Siences. McGraw-Hill, 1981.

\bibitem[{Park} et~al.(2001){Park}, {Kim}, {Bugaenko}, {Divlekeev}, {Popov},
  and {Dermenjiev}]{ParkEtal:2001}
Y.-D. {Park}, I.~S. {Kim}, O.~I. {Bugaenko}, M.~I. {Divlekeev}, V.~V. {Popov},
  and V.~N. {Dermenjiev}.
\newblock {The plane of polarization of the solar coronal emission on August
  11, 1999}.
\newblock \emph{Astronomy Reports}, 45:\penalty0 729--737, September 2001.
\newblock \doi{10.1134/1.1398922}.

\bibitem[{Penrose}(1959)]{Penrose:1959}
R.~{Penrose}.
\newblock {The apparent shape of a relativisically moving sphere}.
\newblock \emph{Proceedings of the Cambridge Philosophical Scociety},
  55:\penalty0 137--139, 1959.

\bibitem[{Pepin}(1970)]{Pepin:1970}
T.~J. {Pepin}.
\newblock {Observations of the Brightness and Polarization of the Outer Corona
  during the 1966 November 12 Total Eclipse of the Sun}.
\newblock \emph{Astrophysical Journal}, 159:\penalty0 1067, March 1970.
\newblock \doi{10.1086/150384}.

\bibitem[{Prunty}(2014)]{Prunty:2014}
S.~L. {Prunty}.
\newblock {A primer on the theory of Thomson scattering for high-temperture
  fusion plasmas}.
\newblock \emph{Physica Scripta}, 89:\penalty0 128001 (44pp), November 2014.
\newblock \doi{doi:10.1088/0031-8949/89/12/128001}.

\bibitem[{Qu} et~al.(2013){Qu}, {Deng}, {Dun}, {Chang}, {Zhang}, {Cheng},
  {Allington-Smith}, {Murray}, {Qu}, {Xue}, and {Ma}]{QuEtal:2013}
Z.~Q. {Qu}, L.~H. {Deng}, G.~T. {Dun}, L.~{Chang}, X.~Y. {Zhang}, X.~M.
  {Cheng}, J.~{Allington-Smith}, G.~{Murray}, Z.~N. {Qu}, Z.~K. {Xue}, and
  L.~{Ma}.
\newblock {On the Combination of Imaging-polarimetry with Spectropolarimetry of
  Upper Solar Atmospheres during Solar Eclipses}.
\newblock \emph{Astrophysical Journal}, 774:\penalty0 71, September 2013.
\newblock \doi{10.1088/0004-637X/774/1/71}.

\bibitem[{Qu{\'e}merais} and {Lamy}(2002)]{QuemeraisLamy:2002}
E.~{Qu{\'e}merais} and P.~{Lamy}.
\newblock {Two-dimensional electron density in the solar corona from inversion
  of white light images - Application to SOHO/LASCO-C2 observations}.
\newblock \emph{Astronomy and Astrophysics}, 393:\penalty0 295--304, October
  2002.
\newblock \doi{10.1051/0004-6361:20021019}.

\bibitem[{Rayleigh}(1871)]{Rayleigh:1871}
W.~J.~Lord {Rayleigh}.
\newblock {On the scattering of light by small particles}.
\newblock \emph{Philosophical Magazine}, 41:\penalty0 447--454, 1871.

\bibitem[{Rosenheinrich}(2015)]{Rosenheinrich:2015}
W.~{Rosenheinrich}.
\newblock {Tables of some indefinite integrals of Bessel functions}, September
  2015.
\newblock URL \url{http://www.eah-jena.de/\~{}rsh/Forschung/Stoer/besint.pdf}.
\newblock accessed 2016 Mar 19 19:54:13 CET.

\bibitem[{Rybicki} and {Lightman}(1979)]{RybickiLightman:1979}
G.~B. {Rybicki} and A.~P. {Lightman}.
\newblock \emph{{Radiative processes in astrophysics}}.
\newblock Wiley-Interscience, New York, 1979.

\bibitem[{Saito} et~al.(1970){Saito}, {Makita}, {Nishi}, and
  {Hata}]{SaitoEtal:1970}
K.~{Saito}, M.~{Makita}, K.~{Nishi}, and S.~{Hata}.
\newblock {A non-spherical axisymmetric model of the solar K corona of the
  minimum type}.
\newblock \emph{Annals of the Tokyo Astronomical Observatory}, 12:\penalty0
  53--120, 1970.

\bibitem[{Schuster}(1879)]{Schuster:1879}
A.~{Schuster}.
\newblock {On the polarisation of the Solar Corona}.
\newblock \emph{Monthly Notices of the Royal Astronomical Society},
  40:\penalty0 35--56, December 1879.

\bibitem[{Segre} and {Zanza}(2000)]{SegreZanza:2000}
S.~E. {Segre} and V.~{Zanza}.
\newblock {Polarization of radiation in incoherent Thomson scattering by high
  temperature plasma}.
\newblock \emph{Physics of Plasma}, 7:\penalty0 2677--2684, 2000.
\newblock \doi{10.1063/1.874110}.

\bibitem[{Skomorovsky} et~al.(2012){Skomorovsky}, {Trifonov}, {Mashnich},
  {Zagaynova}, {Fainshtein}, {Kushtal}, and {Chuprakov}]{SkomorovskyEtal:2012}
V.~I. {Skomorovsky}, V.~D. {Trifonov}, G.~P. {Mashnich}, Y.~S. {Zagaynova},
  V.~G. {Fainshtein}, G.~I. {Kushtal}, and S.~A. {Chuprakov}.
\newblock {White-Light Observations and Polarimetric Analysis of the Solar
  Corona During the Eclipse of 1 August 2008}.
\newblock \emph{Solar Physics}, 277:\penalty0 267--281, April 2012.
\newblock \doi{10.1007/s11207-011-9910-7}.

\bibitem[{Sunyaev} and {Titarchuk}(1985)]{SunyaevTitarchuk:1985}
R.~A. {Sunyaev} and L.~G. {Titarchuk}.
\newblock {Comptonization of low-frequency radiation in accretion disks Angular
  distribution and polarization of hard radiation}.
\newblock \emph{Astronomy and Astrophysics}, 143:\penalty0 374--388, February
  1985.

\bibitem[{Terrell}(1959)]{Terrell:1959}
J.~{Terrell}.
\newblock {Invisibility of the Lorentz contraction}.
\newblock \emph{Physical Reviews}, 116:\penalty0 1041--145, 1959.

\bibitem[{Thomson}(1907)]{Thomson:1907}
J.~J. {Thomson}.
\newblock \emph{{The corpuscular theory of matter}}.
\newblock Constable, London, 1907.

\bibitem[Unknown(1879)]{Ranyard:1879}
Unknown.
\newblock {Chapter XLII: Polariscopic Observations}.
\newblock \emph{Memoirs of the Royal Astronomical Society}, 41:\penalty0
  255--335, 1879.
\newblock author may be A.C. Ranyard.

\bibitem[{Van de Hulst}(1950)]{vandeHulst:1950}
H.~C. {Van de Hulst}.
\newblock {The electron density of the solar corona}.
\newblock \emph{Bulletin Astronomical Institute of the Netherlands},
  11:\penalty0 135, February 1950.

\bibitem[{Vink} et~al.(2002){Vink}, {Drew}, {Harries}, and
  {Oudmaijer}]{VinkEtal:2002}
J.~S. {Vink}, J.~E. {Drew}, T.~J. {Harries}, and R.~D. {Oudmaijer}.
\newblock {Probing the circumstellar structure of Herbig Ae/Be stars}.
\newblock \emph{Monthly Notices of the Royal Astronomical Society},
  337:\penalty0 356--368, November 2002.
\newblock \doi{10.1046/j.1365-8711.2002.05920.x}.

\bibitem[{Vourlidas} and {Howard}(2006)]{VourlidasHoward:2006}
A.~{Vourlidas} and R.~A. {Howard}.
\newblock {The Proper Treatment of Coronal Mass Ejection Brightness: A New
  Methodology and Implications for Observations}.
\newblock \emph{Astrophysical Journal}, 642:\penalty0 1216--1221, May 2006.
\newblock \doi{10.1086/501122}.

\bibitem[{Wang} and {Wheeler}(2008)]{WangWheeler:2008}
L.~{Wang} and J.~C. {Wheeler}.
\newblock {Spectropolarimetry of Supernovae}.
\newblock \emph{Annual Review of Astron and Astrophys}, 46:\penalty0 433--474,
  September 2008.
\newblock \doi{10.1146/annurev.astro.46.060407.145139}.

\bibitem[{Weiskopf} et~al.(1999){Weiskopf}, {Kraus}, and
  {Ruder}]{WeiskopfEtal:1999}
D.~{Weiskopf}, U.~{Kraus}, and H.~{Ruder}.
\newblock {Searchlight and Doppler effects in the visualization of special
  relativity: a corrected derivation of the transformation of radiance}.
\newblock \emph{ACM Transactions on Graphics}, 18:\penalty0 278--292, July
  1999.

\bibitem[{Wolf}(2007)]{Wolf:2007}
E.~{Wolf}.
\newblock \emph{{Introduction to the Theory of Coherence and Polarization of
  Light}}.
\newblock Cambridge University press, 2007.

\bibitem[{Wolf} and {Henning}(1999)]{WolfHenning:1999}
S.~{Wolf} and T.~{Henning}.
\newblock {AGN polarization models}.
\newblock \emph{Astronomy and Astrophysics}, 341:\penalty0 675--682, January
  1999.

\bibitem[{Wood} and {Brown}(1994)]{BrownWood:1994}
K.~{Wood} and J.~C. {Brown}.
\newblock {Effect of electron thermal motions on Thomson scattered line
  profiles from hot circumstellar envelopes}.
\newblock \emph{Astronomy and Astrophysics}, 291:\penalty0 202--208, November
  1994.

\bibitem[{Wood} et~al.(1993){Wood}, {Brown}, and {Fox}]{WoodEtal:1993}
K.~{Wood}, J.~C. {Brown}, and G.~K. {Fox}.
\newblock {Polarimetric line profiles from optically thin Thomson scattering
  circumstellar envelopes}.
\newblock \emph{Astronomy and Astrophysics}, 271:\penalty0 492--500, April
  1993.

\bibitem[Woodhouse(2003)]{Woodhouse:2003}
N.~M.~J. Woodhouse.
\newblock \emph{Special Relativity}.
\newblock Springer, 2003.

\bibitem[{Xiong} et~al.(2013{\natexlab{a}}){Xiong}, {Davies}, {Bisi}, {Owens},
  {Fallows}, and {Dorrian}]{XiongEtAl:2013a}
M.~{Xiong}, J.~A. {Davies}, M.~M. {Bisi}, M.~J. {Owens}, R.~A. {Fallows}, and
  G.~D. {Dorrian}.
\newblock {Effects of Thomson-Scattering Geometry on White-Light Imaging of an
  Interplanetary Shock: Synthetic Observations from Forward Magnetohydrodynamic
  Modelling}.
\newblock \emph{Solar Physics}, 285:\penalty0 369--389, July
  2013{\natexlab{a}}.
\newblock \doi{10.1007/s11207-012-0047-0}.

\bibitem[{Xiong} et~al.(2013{\natexlab{b}}){Xiong}, {Davies}, {Feng}, {Owens},
  {Harrison}, {Davis}, and {Liu}]{XiongEtAl:2013b}
M.~{Xiong}, J.~A. {Davies}, X.~{Feng}, M.~J. {Owens}, R.~A. {Harrison}, C.~J.
  {Davis}, and Y.~D. {Liu}.
\newblock {Using Coordinated Observations in Polarized White Light and Faraday
  Rotation to Probe the Spatial Position and Magnetic Field of an
  Interplanetary Sheath}.
\newblock \emph{Astrophysical Journal}, 777:\penalty0 32, November
  2013{\natexlab{b}}.
\newblock \doi{10.1088/0004-637X/777/1/32}.

\bibitem[{Yuan}(2003)]{Yuan:2003}
Z.-G {Yuan}.
\newblock {Filtered Abel transform and its application in combustion
  diagnostics}.
\newblock \emph{NASA/CR-2003-212121}, page (11pp), March 2003.

\end{thebibliography}

\end{document}